\documentclass[prb,twocolumn,superscriptaddress,showpacs,preprintnumbers,floatfix,eqsecnum]{revtex4-2}

\usepackage{amsmath,amssymb,bm}
\usepackage{graphicx}

 \pdfoutput=1

\begin{document}

\title{Thermoelectric transport and current noise through a multilevel Anderson impurity: 
  Three-body Fermi-liquid corrections in quantum dots and magnetic alloys}

 \author{Yoshimichi Teratani}
 \affiliation{
 Department of Physics, Osaka City University, Sumiyoshi-ku, 
 Osaka 558-8585, Japan }

\affiliation{Nambu Yoichiro Institute of Theoretical and Experimental Physics, 
Osaka Metropolitan University, Osaka 558-8585, Japan}

 \author{Kazuhiko Tsutsumi}
 \affiliation{
 Department of Physics, Osaka City University, Sumiyoshi-ku, 
 Osaka 558-8585, Japan }

\affiliation{Nambu Yoichiro Institute of Theoretical and Experimental Physics, 
Osaka Metropolitan University, Osaka 558-8585, Japan}

 \author{Kaiji Motoyama}
 \affiliation{
 Department of Physics, Osaka City University, Sumiyoshi-ku, 
 Osaka 558-8585, Japan }
 \affiliation{
 Department of Physics, Osaka Metropolitan University, Sumiyoshi-ku, 
 Osaka 558-8585, Japan }

\author{Rui Sakano}
\affiliation{Department of Physics, Keio University, 
3-14-1 Hiyoshi, Kohoku-ku, Yokohama, Kanagawa 223-8522, Japan}

 \author{Akira Oguri}
 \affiliation{
 Department of Physics, Osaka City University, Sumiyoshi-ku, 
 Osaka 558-8585, Japan }
\affiliation{Nambu Yoichiro Institute of Theoretical and Experimental Physics, 
Osaka Metropolitan University, Osaka 558-8585, Japan}

\date{\today}

\begin{abstract}
We present a comprehensive Fermi liquid description 
for thermoelectric transport and current noise,
applicable to multilevel quantum dots (QD) 
and magnetic alloys (MA) without electron-hole or time-reversal symmetry.
Our formulation for the low-energy transport 
is based on an Anderson model with $N$ discrete impurity levels, 
 and is asymptotically exact at low energies, up to the next-leading order terms  
 in power expansions with respect to temperature $T$ and bias voltage $eV$. 
The expansion coefficients can be expressed in terms of the Fermi liquid parameters,  
which include the three-body correlation functions 
defined with respect to the equilibrium ground state 
in addition to the linear susceptibilities and 
the occupation number $N_d^{}$ of impurity electrons.  
We apply this formulation to the SU($N$) symmetric QD and MA, 
and calculate the correlation functions for $N=4$ and $6$,  
using  numerical renormalization group approach. 
The three-body correlations are shown to be determined by a single parameter 
over a wide range of electron fillings $1 \lesssim N_d^{} \lesssim N-1$  
for strong Coulomb interactions $U$,  
and they also exhibit the plateau structures 
due to the SU($N$) Kondo effects  at integer values of $N_d^{}$.
We find that the Lorenz number $L=\kappa/(T \sigma)$ for QD and MA, 
defined as the ratio of the thermal conductivity $\kappa$  
to the electrical conductivity $\sigma$,
deviates from the universal Wiedemann-Franz value $\pi^2/(3e^2)$
as the temperature increases from $T=0$, 
showing the $T^2$ dependence, the coefficient for which 
depends on the three-body correlations away from half filling.
Furthermore,  we find that the current noise for the SU(4) quantum dots and that for SU(6)  
show a pronounced difference at the quarter $N_d^{}/N=1/4$ and $3/4$ fillings.  
In particular, the linear noise for $N=4$ exhibits a flat peak while the peak for $N=6$ 
shows a round shape, reflecting the fact that, at these filling points,  
the SU($N$) Kondo effects occur for $N \equiv 0$  (mod $4$),
whereas the intermediate-valence fluctuations occur for $N \equiv 2$  (mod $4$).   
We also demonstrate the role of three-body correlations on 
the nonlinear current noise and the other transport coefficients.
\end{abstract}


\maketitle

\section{Introduction}
\label{sec:introduction}

The Kondo effect is a fascinating 
many-body effect\cite{Kondo-Resistance-Minimum, HewsonBook},
taking place in dilute magnetic alloys (MA), 
quantum dots (QD), and other novel quantum systems such as 
ultracold atomic gases \cite{TakahashiColdGasKondo} 
and quark matter \cite{QCD_Kondo}.
It was shown in the 1970s 
that the low-energy behavior of the Kondo systems can be 
described by a quantum impurity version of the  Fermi liquid (FL) theory 
\cite{NozieresFermiLiquid,YamadaYosida2,YamadaYosida4,ShibaKorringa,Yoshimori}. 
In particular,  using the numerical renormalization group (NRG) 
approach \cite{WilsonRMP,KWW1,KWW2}, 
Wilson {\it et al.\/} demonstrated  
that the low-lying excited states of the Kondo and the Anderson models 
exhibit a one-to-one correspondence with the excitations 
of the renormalized quasiparticles. 

The quasiparticles are asymptotically free in the equilibrium ground state,  
where the perturbations from the environment or reservoirs,  
which may depend on external parameters 
such as  frequency $\omega$, temperature $T$, bias voltages $eV$, etc., are absent.  
As these perturbations are adiabatically switched on, 
the quasiparticles capture 
the damping rate of order $\omega^2$, $T^2$, and $(eV)^2$ 
through the residual interactions 
and this significantly affects the transport properties 
\cite{NozieresFermiLiquid,YamadaYosida2,YamadaYosida4,ShibaKorringa,Yoshimori,Hershfield1,ao2001PRB}.  
When the electron-hole or time-reversal symmetry is broken 
by a potential or external fields,  
the quasiparticles also capture the energy shift of order  
 $\omega^2$, $T^2$, and $(eV)^2$, i.e., 
the corrections in  the same order as the ones due to the damping rate.   
The contributions of these higher-order energy shifts, however, had not been fully 
understood until very recently.

It has recently been clarified that these higher-order energy shifts of the quasiparticles   
can be described exactly in terms of the three-body correlations 
between impurity electrons. 
The proof was given in two different ways, which complement each other.
One is  given by  Mora {\it et al.\/} and Filippone {\it et al.\/} 
\cite{MoraSUnKondoI,MoraSUnKondoII,MoraMocaVonDelftZarand,FilipponeMocaWeichselbaumVonDelftMora},  extending Nozi\`{e}res' description \cite{NozieresFermiLiquid}   
 that is based on an invariance against the 
``{\it floating of Kondo resonance on the Fermi sea}". 
The other is based on the higher-order Fermi liquid relations 
\cite{ao2017_1,ao2017_2_PRB,ao2017_3_PRB},  
which can be derived from the Ward identities 
for the second derivatives of the self-energy,  
 extending Yamada-Yosida's field-theoretical approach 
\cite{YamadaYosida2,YamadaYosida4,ShibaKorringa,Yoshimori}.   
These proofs enabled one to express  
the next-leading order terms of the transport coefficients 
in terms of three-body correlation functions,     
and these formulations have been applied to 
the nonlinear currents, current noise, and thermocurrent 
through quantum impurities without electron-hole or time-reversal symmetry
\cite{KarkiKiselev,KarkiMoraVonDelfKiselex,KarkiKiselevII,MocaMoraWeymannZarand,TerataniSakanoOguri2020,ao2021,TsutsumiTerataniSakanoAO2021prb,TsutsumiSUNpaper}.

The purpose of this paper is twofold. 
The first one is to extend the latest version of the FL theory for treating 
 the thermoelectric transport coefficients 
of multilevel quantum dots and magnetic alloys  
described by the Anderson model with $N$ arbitrary impurity levels. 
The second one  is to demonstrate 
how the next-leading order terms of various transport coefficients 
vary with the impurity level $\epsilon_d^{}$ and  Coulomb interaction $U$.
Specifically, for the second purpose, 
 we consider quantum dots 
and magnetic alloys having an SU($N$) symmetry, 
and calculate the three-body correlation functions for $N=4$ and $6$ 
using the NRG approach.
There have been a number of intensive works which 
 theoretically studied low-energy transport:   
 the nonlinear current 
\cite{GlazmanRaikh,NgLee,Hershfield1,WingreenMeir,MeirWingreen,Izumida1998,izumida2001kondo,ao2001PRB,Aligia2012,Aligia2014,FilipponeMocaWeichselbaumVonDelftMora},
nonlinear current noise \cite{GogolinKomnikPRL,Sela2006,Golub,SelaMalecki,MoraSUnKondoII,SakanoFujiiOguri,KarkiMoraVonDelfKiselex}, 
and thermoelectric transport\cite{CostiThermo,MoraSUnKondoI,Costimatthermopower,ao2017_3_PRB}.
However, the parameter space for quantum impurities  
is so huge that many parts are still left  unexplored. 
In this paper,   we explore the whole region of the electron fillings, 
$0 \leq N_d^{} \leq N$,  
in which the occupation number $N_d^{}$ of electrons in the impurity levels varies 
continuously across the various SU($N$) Kondo 
and intermediate valence regimes.

One of the advantages of the Kondo systems realized in quantum dots
is that the information about the many-body quantum states    
can be probed in such a highly tunable way 
\cite{Goldhaber-Gordon1998nature,Goldhaber-Goldon1998PRL,Cronenwett-Oosterkamp-Kouwenhoven1998,vanderWiel2000}.  
 For instance, recent experiments have  succeeded in directly probing 
 the Kondo screening cloud \cite{KondoCloud2020}.
Furthermore, low-energy Fermi liquid behaviors 
have experimentally been confirmed for nonequilibrium currents 
\cite{GrobisGoldhaber-Gordon,ScottNatelson}, 
current noise \cite{Heiblum,Delattre2009,KobayashiKondoShot,Ferrier2016,Hata2021}, 
and themocurrent \cite{Costi2022,Svilansexperimentthermopower}.  
Internal degrees of freedoms of quantum impurities also 
bring an interesting variety to the Kondo effect. 
The systems having the SU($4$) symmetry 
have been realized, for instance, in the multiorbital semiconductor quantum dots and  
carbon nanotube (CNT) quantum dots \cite{RMP-Kouwenhoven}, 
and have intensively been investigated theoretically \cite{Izumida1998,Borda2003,Choi2005,Eto2005,Sakano2006,Anders2008,SU4_Kondo_ferro_Weymann,MoraSUnKondoI,MoraSUnKondoII,FilipponeZarandMoraTKsu4,MantelliMocaZarandGrifoni,TerataniSakanoOguri2020,TerataniPRB2020,KarkiKiselev} and experimentally \cite{Sasaki2000,Jarillo-Herrero2005,Babic2004,Makarovski2007,Cleuziou2013,Ferrier2016}.  
Realization of  the SU($N$) Kondo effects 
for various $N>2$ has also been 
proposed by several authors, 
using triple quantum dots \cite{Kuzumenko2006,SU3KondoLopez} 
 and CNT \cite{SU12Kuzmenko}. 
In this paper, we provide a comprehensive FL view of  
the low-energy transport for SU(4) and SU(6) symmetric QD and MA.

Our results reveal the fact that 
the three-body correlations exhibit plateau structures,  
caused by the SU($N$) Kondo effects occurring 
at integer fillings $N_d^{}=1$, $2$, $\ldots, N-1$.  
We also calculate the Lorenz number $L=\kappa/(T \sigma)$ for QD and MA, 
defined as the ratio of the thermal conductivity $\kappa$  
to the electrical conductivity $\sigma$,  
and show how it deviates from the universal Wiedemann-Franz value $\pi^2/(3e^2)$
as the temperature increases from $T=0$.
Furthermore, we demonstrate 
how  the three-body correlations affect 
 the order $T^2$ electrical conductivity, the  order $T^3$ thermal conductivity, 
 and the order $(eV)^3$  nonlinear current and current noise. 
At quarter  $N_d^{}/N=1/4$  and 
three-quarters $N_d^{}/N=3/4$ fillings,  
the SU($N$) Kondo effects occur for $N \equiv 0$  (mod $4$)   
while the intermediate valence fluctuations occur for $N \equiv 2$  (mod $4$).
We show that this dependence on (mod $4$) causes a pronounced difference appearing   
in the peak structures of linear noise for $N=4$ and $6$.

This paper is organized as follows. 
In Sec.\ \ref{sec:formulation}, 
we describe the multilevel Anderson impurity model for quantum dots 
and magnetic alloys. 
Sections \ref{sec:FL_transport_next_leading_order} 
and \ref{sec:thermoelectric_transport_QD} are 
 devoted to the FL descriptions for the next-leading order terms of 
electrical and thermoelectric transport coefficients for quantum dots, 
applicable to arbitrary $N$ impurity-level structures. 
In Sec.\ \ref{sec:FL_parameters_For_SUN_Anderson}, 
the low-energy transport formulas  
for the SU($N$) quantum dots are described 
 in terms of the five FL parameters.
We present the NRG results for the three-body correlation functions 
in Sec.\ \ref{sec:NRG_results_three_body}.  
The results  for  nonlinear current,  current noise, and 
 thermoelectric transport through quantum dots are discussed 
in Secs.\ \ref{sec:NRG_CvCs} and \ref{sec:NRG_CtCkappa}.   
Section \ref{sec:transport_dilute_alloys_SUN} is devoted to the 
 FL description of the three-body correlations in 
thermoelectric transport of dilute magnetic alloys.  
In Sec.\ \ref{sec:NRG_MA}, we discuss the 
results for the electrical and thermal resistivities of the SU(4) and SU(6) 
magnetic alloys. 
Summary is given in Sec.\ \ref{sec:summary}. 
In Appendix, we provide details of the derivations for the transport formulas  
and additional NRG results 
for the FL parameters in the SU($N$) cases for $N=2$, $4$, and $6$, 
for comparison.

\section{Model}
\label{sec:formulation}

We consider  a multi-orbital Anderson impurity 
coupled to two noninteracting leads on the left ($L$) and right ($R$):  
$\mathcal{H} =
\mathcal{H}_d + \mathcal{H}_c  + \mathcal{H}_\mathrm{T}$,  
\begin{align}
\mathcal{H}_d =& \  
\sum_{\sigma=1}^N
\epsilon_{d\sigma}^{}\, n_{d\sigma}^{}  
+
\frac{1}{2} \sum_{\sigma \neq \sigma'} U_{\sigma\sigma'}^{} \, 
n_{d\sigma}^{} n_{d\sigma'}^{} \;, 
\label{eq:H_d_N}
\\ 
\mathcal{H}_c =& \   
\sum_{j=L,R} \sum_{\sigma=1}^N 
\int_{-D}^D  \!\! d\epsilon\,  \epsilon\, 
c^{\dagger}_{\epsilon j \sigma} c_{\epsilon j \sigma}^{},
\\
\mathcal{H}_\mathrm{T} =& \   
- \sum_{j=L,R} \sum_{\sigma=1}^N   v_{j}^{}
\left( \psi_{j,\sigma}^\dag d_{\sigma}^{} + 
d_{\sigma}^{\dag} \psi_{j,\sigma}^{} \right) .
\label{Hami_seri_part}
\end{align}
Here, the level index runs over $\sigma=1, 2, \ldots, N$.  
The inter-electron interaction $U_{\sigma\sigma'}^{}$ 
generally depends on $\sigma$ and $\sigma'$,    
with the requirements $U_{\sigma'\sigma}^{}=U_{\sigma\sigma'}^{}$ 
for $\sigma'\neq \sigma$. 
For  $N=2$,   it describes  the usual single-orbital 
Anderson model for spin $1/2$ fermions. 
The operator 
$d^{\dag}_{\sigma}$ creates an impurity electron 
with spin $\sigma$ 
in the impurity level of energy $\epsilon _{d\sigma}^{}$, and  
$n_{d\sigma}^{} \equiv d^{\dag}_{\sigma} d^{}_{\sigma}$. 
Conduction electrons in the two lead at  $j=L$ and $R$ 
obey  the  anti-commutation relation 
$
\{ c^{\phantom{\dagger}}_{\epsilon j \sigma}, 
c^{\dagger}_{\epsilon'j'\sigma'}
\} = \delta_{jj'} \,\delta_{\sigma\sigma'}   
\delta(\epsilon-\epsilon')$. 
The linear combination of the conduction electrons, 
$\psi^{}_{j, \sigma} \equiv  \int_{-D}^D d\epsilon \sqrt{\rho_c^{}} 
\, c^{\phantom{\dagger}}_{\epsilon j \sigma}$  with $\rho_c^{}=1/(2D)$, 
couples to the impurity level. 
The bare level width due to the tunnel couplings 
is given by $\Delta \equiv \Gamma_L + \Gamma_R$ with 
$\Gamma_{j} = \pi \rho_c^{} v_{j}^2$.  
We consider the parameter region where 
the half band-width $D$  
is much grater than the other energy scales, 
$D \gg \max( U_{\sigma\sigma'}, \Delta, |\epsilon_{d\sigma}^{}|, T, |eV|)$. 
We use a unit $k_B^{}=1$ throughout this paper.

Low-energy properties of the Anderson model 
 can be described in terms of a set of Fermi liquid parameters  
defined with respect to the equilibrium ground state, i.e.,   
the phase shift $\delta_\sigma^{}$, 
the linear susceptibilities $\chi_{\sigma\sigma'}^{}$ 
and the three-body correlation functions 
$\chi_{\sigma\sigma'\sigma''}^{[3]}$ between impurity 
electrons; see Appendix \ref{sec:FL_parameters} for details. 
The phase shift is a primary parameter that reflects 
the charge distribution of impurity levels,  
through the Friedel sum rule:  
 $\langle n_{d\sigma}^{}\rangle 
 \xrightarrow{\,T\to 0\,}\delta_\sigma^{}/\pi$.     
In this paper, we explore the low-energy transport 
of quantum dots and magnetic alloys 
in the whole region of the electron fillings,  
\begin{align}
N_{d}^{}\,\equiv\,  \sum_{\sigma=1}^{N} 
\langle n_{d\sigma}^{} \rangle \,.
\end{align}
The current conservation, 
 which follows from the Heisenberg equation of motion for 
the occupation number  $n_{d\sigma}^{}$,   
\begin{align}
\frac{\partial  n_{d\sigma}^{}}{\partial t} 
+ \widehat{J}_{R,\sigma} - \widehat{J}_{L,\sigma}\, =\, 0 \, 
\label{eq:current_conservation}
\end{align}
also plays an essential role in the Fermi liquid description, 
 through the Ward identities.
Here,  $\widehat{J}_{L,\sigma}$ represents 
the current flowing from the left lead to the dot, 
and  $\widehat{J}_{R,\sigma}$  the current from the dot to the right lead:  
\begin{align}
\widehat{J}_{L,\sigma} =& \     -i\,  v_L 
\left(
\psi^{\dagger}_{L\sigma} d^{}_{\sigma} 
-d^{\dagger}_{\sigma} \psi^{}_{L\sigma} \right) , 
\\
\widehat{J}_{R,\sigma} =& \  + i\, v_R 
\left(
\psi^{\dagger}_{R\sigma} d^{}_{\sigma} 
-d^{\dagger}_{\sigma} \psi^{}_{R\sigma}\right) .
\end{align}

In the next section, we will give a brief overview of the low-energy 
expansion formulas for nonlinear current 
and current noise, obtained previously 
in Refs.\  \onlinecite{ao2021,TsutsumiSUNpaper},    
to show how the formulas can be expressed in terms of 
the FL parameters, including the three-body correlations. 
This is for comparison with the formulas for thermal electric transport coefficients, 
which we extend to multilevel quantum dots and magnetic alloys in this paper. 
Specifically, we will describe the derivation 
of the formulas for multilevel quantum dots and magnetic alloys  
in Sec.\ \ref{sec:thermoelectric_transport_QD}  
and Appendix \ref{sec:thermoelectric_transport_general_MA}, 
respectively.

\section{Fermi liquid description for nonlinear current noise of QD}
\label{sec:FL_transport_next_leading_order}

We consider the nonequilibrium steady state 
under a finite bias voltage  $eV \equiv \mu_L-\mu_R$, 
applied between the two leads by setting the chemical potentials 
of the left and right leads to be $\mu_L$ and  $\mu_R$, respectively.
The retarded Green's function plays 
a central role in the microscopic description 
of the Fermi liquid transport: 
\begin{align}
G_\sigma^r(\omega)=& \
-i\int_{0}^{\infty}dt\,e^{i(\omega+i0^+) t}\,\Bigl\langle\,\Bigl\{d_\sigma^{}(t),\,d_\sigma^\dagger\Bigr\} \Bigr\rangle_{V} \,, 
\label{eG}
\\
A_{\sigma}^{}(\omega) \, \equiv & \  
-\frac{1}{\pi} \,\mathrm{Im}\,G_{\sigma}^{r}(\omega)\;.
\label{eq:spectral_function}
\end{align}
Here, $\langle \cdots\rangle_{V}$ 
represents a nonequilibrium steady-state average 
taken with the statistical density matrix, 
which is constructed at finite bias voltages $eV$ and temperatures $T$, 
using the Keldysh formalism \cite{MeirWingreen,Hershfield1,Caroli}.

\subsection{Differential conductance $dJ/dV$}

The nonequilibrium current $J$ through quantum dots   can be expressed 
in terms of the spectral function $A_{\sigma}^{}(\omega)$ 
\cite{MeirWingreen,Hershfield1}:
\begin{align}
J \,= & \   \frac{e}{h}
\sum_{\sigma}  
\int_{-\infty}^{\infty} \! d\omega\,  
\bigl[\,f_L(\omega)-f_R(\omega) \,\bigr]\, 
\mathcal{T}_{\sigma}(\omega)  , 
\label{eq:current_formula}
\\
\mathcal{T}_{\sigma}(\omega) \, = & \  
\frac{4\Gamma_L \Gamma_R}{\Gamma_L+\Gamma_R} \,\pi  
A_{\sigma} (\omega) \,.
\label{eq:transmissionPB}
\end{align}
Here,  $f_{j}(\omega) \equiv  f(\omega-\mu_{j}^{})$  for $j=L$, $R$,  
with  $f(\omega)=[e^{\omega/T}+1]^{-1}$  the Fermi function. 
Specifically,  in this paper,  we consider the case 
where the chemical potentials are applied in a symmetric way:
\begin{align}
\mu_L \, = \, -\mu_R  \ \equiv \  \frac{1}{2}\,eV \,, 
\label{eq:chemical_potentials}
\end{align}
choosing the Fermi level at equilibrium to be the origin 
of one-particle energies $E_F=0$. 
Note that the role of bias and tunneling asymmetries, $(\mu_L+\mu_R)/2 \neq E_F$ 
and $\Gamma_L \neq \Gamma_R^{}$, has been precisely   
discussed in Refs.\ \onlinecite{TsutsumiTerataniSakanoAO2021prb,TsutsumiSUNpaper}.

The nonlinear current $J$ can be expanded 
up to the next-leading order terms,  
using the low-energy asymptotic form of    
the spectral function  $A_{\sigma} (\omega)$,  
which has been obtained 
up to terms of order $\omega^2$, $(eV)^2$, and $T^2$ \cite{ao2021},   
 as shown  
in Appendix \ref{sec:Low_energy_asymptotic_form_A}. 
In particular,  for symmetric junctions with  $\Gamma_L=\Gamma_R$ 
 and $\mu_L=-\mu_R=eV/2$, 
the low-energy expansion of the differential conductance takes the form, 
\begin{align}
\frac{dJ}{dV}  =
\frac{e^2}{h}  \sum_{\sigma=1}^N 
\left[\,
\sin^2 \delta_{\sigma}  
-  c_{T,\sigma}^{} \left(\pi T\right)^2
-  c_{V,\sigma}^{} \left(eV \right)^2 \,+\, \cdots 
\right].  
\label{eq:conductance_ct_cv_general}
\end{align}
Here, the coefficients $c_{T,\sigma}^{}$ and $c_{V,\sigma}^{}$  
of the next-leading order terms can be expressed in terms of 
the phase shift $\delta_\sigma^{}$, 
the linear susceptibilities $\chi_{\sigma\sigma'}^{}$, 
and the static three-body correlation functions 
$\chi_{\sigma\sigma'\sigma''}^{[3]}$ 
defined in Appendix \ref{sec:FL_parameters}: 
\begin{align}
&
\!\!\!\!
c_{T,\sigma}^{} =   
\frac{\pi^2}{3} 
\Biggl[\,
-
\biggl(
\chi_{\sigma\sigma}^2
+ 
2\sum_{\sigma'(\neq \sigma)}\chi_{\sigma\sigma'}^2
\biggr) 
\cos 2 \delta_{\sigma}
\nonumber \\
& \qquad \qquad 
\, + 
\biggl( 
\chi_{\sigma\sigma\sigma}^{[3]} +
\sum_{\sigma'(\neq \sigma)} \chi_{\sigma\sigma'\sigma'}^{[3]}
\biggr)
\frac{\sin 2\delta_{\sigma}}{2\pi}\,
\Biggr],
\label{eq:c_T_multi}
\\
&
\!\!\!\!
c_{V,\sigma}^{} = 
\frac{\pi^2}{4}
\Biggl[\,
-
\biggl( \chi_{\sigma\sigma}^2 +  
5\sum_{\sigma'(\neq \sigma)}\chi_{\sigma\sigma'}^2
\biggr)
\cos 2 \delta_{\sigma} 
\nonumber \\
& \qquad \qquad 
+
\biggl(
\chi_{\sigma\sigma\sigma}^{[3]} +
3\sum_{\sigma'(\neq \sigma)} \chi_{\sigma\sigma'\sigma'}^{[3]}  
\biggr)
\frac{\sin 2\delta_{\sigma}}{2\pi} 
\,\Biggr].
\label{eq:c_V_multi}
\end{align}

\subsection{Nonlinear current noise $S_\mathrm{noise}^{\mathrm{QD}}$}
\label{subsec:NRG_noise}

We also 
study the current noise  
\cite{Hershfield2,TerataniSakanoOguri2020,ao2021},   
defined by  
\begin{align}
S_\mathrm{noise}^\mathrm{QD}  = \,
e^2 \sum_{\sigma\sigma'}\int_{-\infty}^{\infty} \!\!   dt  \,
\left\langle 
\delta \widehat{J}_{\sigma}(t) \, \delta \widehat{J}_{\sigma'}(0) 
+\delta \widehat{J}_{\sigma'}(0) \, \delta \widehat{J}_{\sigma}(t)
\right\rangle_{V}^{} .
\label{eq:S_noise}
\end{align}
Here,
$\delta \widehat{J}_{\sigma}(t)  \equiv 
\widehat{J}_\sigma (t) 
- \bigl\langle  \widehat{J}_\sigma (0)  \bigr\rangle_{V}^{}
$ represents fluctuations of the symmetrized current: 
\begin{align}
\widehat{J}_{\sigma}
\equiv  & \   
\frac{\Gamma_L \widehat{J}_{R,\sigma} 
	+\Gamma_R \widehat{J}_{L,\sigma}}{\Gamma_L+\Gamma_R} \;.
\label{eq:symmetrized_current_def}
\end{align}
Behavior of the current noise in the low-energy Fermi liquid regime 
has been studied by several authors, taking into account 
the three-body correlations \cite{MoraMocaVonDelftZarand,MoraSUnKondoII}.
In a previous work,  
we have derived a general formula for 
the current noise through the multilevel Anderson impurity model 
up to terms of order $|eV|^3$  
for symmetric junctions  $\Gamma_L=\Gamma_R$  
and $\mu_L=-\mu_R= eV/2$ \cite{TerataniSakanoOguri2020,ao2021}: 
\begin{align}
S_\mathrm{noise}^{\mathrm{QD}} =   
\frac{2e^2}{h} 
|eV| 
\sum_{\sigma=1}^{N}
\left[\frac{\sin^2 2\delta_\sigma^{}}{4} \,  
+ c_{S,\sigma}^{} \left(eV\right)^2 +  \cdots 
\right] .
\end{align}
The coefficient $c_{S,\sigma}^{}$ for the next-leading order term 
has been calculated 
by taking into account all components of the Keldysh vertex function 
together with 
the low-energy asymptotic form 
of $A_{\sigma}^{}(\omega)$. 
It has been shown to be expressed in the following form,    
\begin{align}
\!\! 
c_{S,\sigma}^{} = & \    
\frac{\pi^2}{12} 
\Biggl[
\,  
\cos 4 \delta_{\sigma}\,
\chi_{\sigma\sigma}^2 
+ \bigl( 2+3\cos 4 \delta_{\sigma} \bigr)
\sum_{\sigma'(\neq \sigma)}\chi_{\sigma\sigma'}^2
\nonumber 
\\
& 
\quad \ \  +\,  
4\sum_{\sigma' (\neq \sigma)} \cos 2\delta_{\sigma}^{} \cos 2\delta_{\sigma'}^{}
\,  \chi_{\sigma\sigma'}^2
\nonumber \\
& 
\quad \ \   +\,3 \sum_{\sigma'(\neq \sigma)}
\sum_{\sigma''(\neq \sigma,\sigma')}
\sin 2\delta_{\sigma}^{} \,\sin 2\delta_{\sigma'}^{}
\chi_{\sigma\sigma''}^{}  \chi_{\sigma'\sigma''}^{}
\nonumber \\
& \quad \ \   
-\,
\biggl(
\chi_{\sigma\sigma\sigma}^{[3]}
+ 3 \sum_{\sigma'(\neq \sigma)} \chi_{\sigma\sigma'\sigma'}^{[3]}
\biggr)\,\frac{\sin 4\delta_{\sigma}}{4\pi}\,
\Biggr] . 
\label{eq:c_S_multi}
\end{align}

\section{Fermi liquid description for thermoelectric transport of QD}
\label{sec:thermoelectric_transport_QD}

In this section, we discuss thermoelectric transport through 
multilevel quantum dots in the linear-response regime,  
the low-energy behaviors of which can be deduced from 
the asymptotic form of the spectral function $A_{\sigma}(\omega)$ given  
in Appendix \ref{sec:Low_energy_asymptotic_form_A}. 

The linear conductance $g$, 
thermopower $\mathcal{S}_\mathrm{QD}^{}$
and thermal conductance $\kappa_\mathrm{QD}^{}$ of a quantum dot 
 can be expressed in the form 
\cite{GuttmanBergman,CostiZlatic2010,MocaThermo,KarkiKiselev,Aligia2018thermo}:
\begin{align}
g\, \equiv   & \ 
\left. \frac{dJ}{dV} \right|_{eV=0}^{} 
 \ = \ 
\frac{e^2}{h}
\sum_\sigma \mathcal{L}_{0,\sigma}^{\mathrm{QD}} \,, 
\label{eq:linear_conductance_QD} \\
\mathcal{S}_\mathrm{QD}^{}\, =&\ \frac{-1}{|e|T} 
\frac{\sum_{\sigma} \mathcal{L}_{1,\sigma}^{\mathrm{QD}}}
{\sum_{\sigma} \mathcal{L}_{0,\sigma}^{\mathrm{QD}}} \,, 
\label{eq:thermopower_QD}\\ 
\kappa_\mathrm{QD}^{} \,= & \  
\frac{1}{h\, T}
\left[\, 
\sum_{\sigma}
\mathcal{L}_{2,\sigma}^{\mathrm{QD}} 
- 
\frac{ \left(
\sum_{\sigma}
\mathcal{L}_{1,\sigma}^{\mathrm{QD}}\right)^2}{\sum_{\sigma}
\mathcal{L}_{0,\sigma}^{\mathrm{QD}}} \,\right] \,. 
\label{eq:thermal_coefficients_QD}
\end{align}
Here,  $\mathcal{L}_{n,\sigma}^\mathrm{QD}$ 
for $n=0,1$, and $2$ is defined at $eV=0$ 
 with respect to the thermal equilibrium, as   
\begin{align}
\mathcal{L}_{n,\sigma}^\mathrm{QD} = 
\int_{-\infty}^{\infty}  
d\omega\, 
\omega^n\, 
\mathcal{T}_{\sigma}(\omega) \,
\left( -
\frac{\partial f(\omega)}{\partial \omega}
\right) ,
\label{eq:Ln_QD}
\end{align}
with  $\mathcal{T}_{\sigma}(\omega) $,  the transmission probability 
defined in Eq.\ \eqref{eq:transmissionPB}.
Note that  thermal conductance  $\kappa_\mathrm{QD}^{}$ is the linear-response 
coefficient of the heat current $J_Q^{}=   \kappa_\mathrm{QD}^{}\, \delta T$,        
flowing from the high-temperature side toward the low-temperature side, 
with $\delta T$ the temperature difference between the two sides.

The thermoelectric coefficients 
$\mathcal{L}_{n,\sigma}^{\mathrm{QD}}$ 
can be calculated,  
by substituting the low-energy asymptotic form of  $A_{\sigma}(\omega)$, 
given in  Eq.\ \eqref{eq:A_including_T_eV_N_orbital}  
into  $\mathcal{T}_{\sigma}(\omega)$.
At low temperatures, the component $\mathcal{L}_{0,\sigma}^\mathrm{QD}$ 
for $n=0$, which  determines $g$, is given by 
\begin{align}
\mathcal{L}_{0,\sigma}^{\mathrm{QD}}
\,=  & \   
\sin^2 \delta_{\sigma}  
\,-\,   c_{T,\sigma}^{}\, \left(\pi T\right)^2 
\, + \, O(T^4) \,. 
\label{eq:L_0_sigma_QD}
\end{align}
Here, $c_{T,\sigma}^{}$ is the coefficient  
that we have already described in Eq.\ \eqref{eq:c_T_multi}.
The next component, $\mathcal{L}_{1,\sigma}^{\mathrm{QD}}$ for $n=1$, 
takes the form  
\begin{align}
\mathcal{L}_{1,\sigma}^{\mathrm{QD}}
\,=& \   
\frac{\pi \Delta}{3} \rho_{d\sigma}' \left(\pi T\right)^2
\, + \, O(T^4) 
\,.
\label{eq:L_1_sigma_QD}
\end{align}
Here,  $\rho_{d\sigma}'$  is 
the derivative of the density of states with respect to the frequency $\omega$,   
which can also be written in terms of the phase shift $\delta_{\sigma}^{}$ 
and diagonal linear susceptibility $\chi_{\sigma\sigma}^{}$, 
as shown in Eq.\ \eqref{eq:rho_d_omega_2}.
Thus, the leading-order term of  thermopower for quantum dots is given by
\begin{align}
\mathcal{S}_\mathrm{QD}^{}\, =& \  
-\frac{\pi^2}{3} \, 
\frac{\sum_{\sigma} \rho_{d\sigma}'}{\sum_{\sigma}\rho_{d\sigma}^{}}
\,\frac{T}{|e|} 
\, + \, O(T^3) 
\,.
\end{align}
The thermal conductance $\kappa_\mathrm{QD}^{}$ depends on the other component  
  $\mathcal{L}_{2,\sigma}^\mathrm{QD}$ for $n=2$, the low-energy asymptotic form 
of which is given by  
\begin{align}
\!\! 
\mathcal{L}_{2,\sigma}^{\mathrm{QD}}
\,= & \    
\frac{\left(\pi T\right)^2 }{3}
\Bigl[ \,\sin^2 \delta_{\sigma} \, 
\,+\,a_{2,\sigma}^{\mathrm{QD}}   
\left(\pi T\right)^2   \,+\, O(T^4) \,\Bigr],
\label{eq:L_2_sigma_QD}
\nonumber \\
\!\! 
a_{2,\sigma}^{\mathrm{QD}} \,\equiv & \   
\frac{7\pi^2}{5}
\Biggl[\,
\cos 2\delta_{\sigma}
\left(
\chi_{\sigma\sigma}^2
+ 
\frac{6}{7} 
\sum_{\sigma'(\neq \sigma)}\chi_{\sigma\sigma'}^2
\right)
\nonumber \\
&   
- \frac{\sin 2\delta_{\sigma}}{2\pi}
\, 
\left(
\chi_{\sigma\sigma\sigma}^{[3]}
+ 
\frac{5}{21}
\sum_{\sigma'(\neq \sigma)}
\chi_{\sigma\sigma'\sigma'}^{[3]}
\right)
\,\Biggr] .
\end{align}
Therefore, the thermal conductance can be calculated 
 up to terms of order $T^{3}$,  
by substituting these asymptotic forms into Eq.\ \eqref{eq:thermal_coefficients_QD}: 
\begin{align}
& 
\!\!\!    
\kappa_{\mathrm{QD}}^{}\,= \, 
\frac{\pi^{2}\,T}{3\,h}\,\sum_{\sigma=1}^{N}
\left[\,\sin^{2}\delta_{\sigma}
\,-\,c_{\kappa,\sigma}^{\mathrm{QD}}\,\left(\pi T\right)^{2}\,+\,\cdots \right],
\label{eq:thermal_conductance_T_cube}
\\ 
& \!\!\! 
 c_{\kappa,\sigma}^{\mathrm{QD}}\,= \,  
-
a_{2,\sigma}^{\mathrm{QD}} 
\,+\,\frac{\pi^2}{3}\,
\frac{\left( \displaystyle \mathstrut 
\frac{1}{N}\sum_{\sigma''}\,\chi_{\sigma'' \sigma''}^{}\,
\sin\,2\delta_{\sigma''}\right)^{2}}
{\bigl( \overline{\sin^2 \delta} \bigr)_\mathrm{AM}^{}}  
\,.
\label{eq:c_kappa_QD_general}
\end{align}
Here, the arithmetic mean (AM) of the phase shifts is defined by 
\begin{align}
\bigl( \overline{\sin^2 \delta} \bigr)_\mathrm{AM}^{}
\,\equiv \, 
\frac{1}{N}\,\sum_{\sigma}\,\sin^{2}\delta_{\sigma}^{} \,. 
\label{eq:AM_sin2delta}
\end{align}

Furthermore,  the Lorenz number
  $L_{\mathrm{QD}} \equiv \kappa_{\mathrm{QD}}^{}/( g\, T)$ 
can be calculated up to terms of order $T^2$: 
\begin{align}
L_{\mathrm{QD}} 
\, =  & \  \frac{\pi^{2}}{3\,e^{2}}\,\left[\,1\,-\,
\frac{c_{L}^{\mathrm{QD}}}{\bigl( \overline{\sin^2 \delta} \bigr)_\mathrm{AM}^{}} 
\,
\left(\pi T\right)^{2} \ + \  O(T^4) \,\right] , 
\label{eq:L_QD_general_formula}
\\
c_{L}^{\mathrm{QD}}\,= & \ 
\frac{1}{N}\, \sum_{\sigma}\left(c_{\kappa,\sigma}^{\mathrm{QD}}
\,-\,c_{T,\sigma}^{}\right) \,. 
\label{eq:C_L_QD_general_def}
\end{align}
The Wiedemann-Franz law holds between the leading-order terms  
 of the linear conductance $g$ 
and the thermal conductance $\kappa_{\mathrm{QD}}^{}$, 
such that  the Lorenz number approaches the universal value  
$L_{\mathrm{QD}} \xrightarrow{T\to 0}\pi^{2}/(3e^{2})$ 
in the low-temperature limit. 
The Lorenz number deviates from this universal value  
 as temperature increases, exhibiting the $T^2$ dependence.

\section{Three-body correlations in the SU($N$) symmetric Fermi liquid}
\label{sec:FL_parameters_For_SUN_Anderson}

This Hamiltonian $\mathcal{H}$, 
defined in Sec.\ \ref{sec:formulation}, 
has an SU($N$) symmetry   
in the case at which the impurity states become degenerate 
$\epsilon_{d\sigma}^{} \equiv \epsilon_{d}^{}$ for all  $\sigma$ 
and the Coulomb interaction is  isotropic  
$U_{\sigma\sigma'} \equiv U$  for all  $\sigma$ and $\sigma'$.

In the atomic limit  $v_{j}^{} \to  0$ of the SU($N$) symmetric case, 
the total number of impurity electrons $N_{d}$ takes an integer value 
$M$ and exhibits 
the Coulomb-stair case behavior as a function of $\epsilon_{d}^{}$. 
It consists of a series of plateaus of 
the width $U$ and the height $N_{d}=M$ for  $M=1$, $2$, $\ldots$, $N-1$, 
 emerging  at $-M U < \epsilon_{d}^{} < -(M - 1) U$ around the midpoint  
$\epsilon_{d,M}^{\mathrm{mid}}\equiv -(M-\frac{1}{2}) U$.  
We will use the following notation for the shifted impurity level $\xi_{d}^{}$, 
which includes the Hartree-Fock energy shift defined with respect to the half filling 
in such a way that 
\begin{align}
 \xi_{d}^{} \,\equiv\,\epsilon_{d}^{}\,+\,\frac{N-1}{2}\,U\,.
\label{eq:def_xi_d}
\end{align}
The system has an electron-hole symmetry at  $\xi_{d}^{} = 0$.  
When the tunneling couplings are switched on,  
the stair-case structure emerges for strong interactions $U \gg \Delta$.

In the SU($N$) symmetric case, 
the linear susceptibilities have two linearly independent components, i.e., 
the diagonal one $\chi_{\sigma\sigma}$ 
and the off-diagonal one  $\chi_{\sigma\sigma'}^{}$  for $\sigma \neq \sigma'.$ 
These two components determine the essential properties of quasiparticles:  
\begin{align}
T^\ast \, \equiv \,  \frac{1}{4\chi_{\sigma\sigma}^{}} \,, 
\qquad \quad 
R \,\equiv\, 1-\frac{\chi_{\sigma\sigma'}^{}}{\chi_{\sigma\sigma}^{}}
\,. 
\label{eq:Fermiparaorigin}
\end{align}
Here,  $T^\ast $  is a characteristic energy scale  of the SU($N$) Fermi liquid, 
by which the  $T$-linear specific heat of impurity electrons 
can be expressed in the form  $\mathcal{C}_\mathrm{imp}^{} 
= \frac{N\pi^2}{12} T/T^{*}$ 
\cite{YamadaYosida2,ShibaKorringa,Yoshimori}.  
The Wilson ratio $R$ corresponds to  
a dimensionless residual interaction between quasiparticles 
\cite{HewsonRPT2001};
we will also use the rescaled ratio,   
\begin{align}
\widetilde{K}\,\equiv\, (N-1)(R-1)\,,  
\label{eq:K_tilde_def}
\end{align}
which is bounded in the range  $0 \leq  \widetilde{K} \leq 1$.

\subsection{Charge and spin susceptibilities}

The charge susceptibility is given by a 
linear combination of the two-body correlations, 
\begin{align}
& \!\!\!\!\!\! 
\overline{\chi}_C^{} \,  
\equiv  \,   -  
\frac{1}{N}
\frac{\partial^2 \Omega}{\partial \epsilon_{d}^{2}}
\ = \ 
\frac{1}{N}
\sum_{\sigma_1\sigma_2} 
\chi_{\sigma_1\sigma_2}^{} 
\\
& \xrightarrow{\,\mathrm{SU}(N)\,}  \,    
\chi_{\sigma\sigma}^{}+(N-1)\chi_{\sigma\sigma'}^{} 
\ = \ 
\frac{1 -\widetilde{K}}{4T^{*}}\,
\,. \label{eq:chi_C_def}
\end{align}
Here,  
 $\Omega\equiv - T
\log \left[\mathrm{Tr}\,e^{-\mathcal{H}/T}\right]$ 
 is the free energy.

 Next, we consider the spin susceptibility for the SU($N$) symmetric case,   
 using the notation in which the internal degrees freedom are 
separated into two parts $\sigma = (m,s)$, where 
 $m=1,2,\ldots, N/2$ and $s=\uparrow, \downarrow$, 
assuming  $N$ to be even (extension to odd $N$ is straightforward). 
The Zeeman splitting is induced by an external field $b$, 
which couples to the impurity spin $s$: 
\begin{align}
\epsilon_{d,m,\uparrow}^{} = \epsilon_{d}^{}\, -\,  b \,, 
\qquad 
\epsilon_{d,m,\downarrow}^{} = \epsilon_{d}^{} \,+\,  b \,. 
\label{eq:def_b_field}
\end{align}
The magnetization $\mathcal{M}$ of the impurity spin is given by 
\begin{align}
\mathcal{M} \, \equiv & \ 
- \frac{1}{N}\frac{\partial \Omega}{\partial b} 
\,= \, \frac{1}{N}\,
\sum_{m=1}^{\frac{N}{2}} 
\left\langle 
n_{d,m\uparrow} - n_{d,m\downarrow} 
\right\rangle
\,.
\end{align}
Note that  $\Omega$  is an even function of $b$. 
The spin susceptibility 
$\overline{\chi}_S^{} \equiv -(1/N) \frac{\partial^2 \Omega}{\partial b^2}
\big|_{b=0}$ 
can be expressed in the following form,
\begin{align}
& \!\!
\overline{\chi}_S^{}\, 
 \, =   \,     
\frac{1}{N}
\sum_{m=1}^{\frac{N}{2}} 
\sum_{m'=1}^{\frac{N}{2}} 
\biggl[\,
\chi_{m\uparrow,m'\uparrow}^{}
+ \chi_{m\downarrow,m'\downarrow}^{}
\nonumber \\
& \qquad \qquad \qquad \qquad  
- \chi_{m\uparrow,m'\downarrow}^{}
- \chi_{m\downarrow,m'\uparrow}^{}
\, \biggr] 
\\
& 
\xrightarrow{\,\mathrm{SU}(N)\,}  \,   
\chi_{\sigma\sigma}^{} - \chi_{\sigma\sigma'}^{} 
\ = \ 
\frac{1}{4T^{*}}
\left(1+\frac{\widetilde{K}}{N-1}\right), 
\label{eq:chi_S_def}
\end{align}
where  $\sigma \neq \sigma'$. 

\subsection{Three-body correlation functions}

Among $N^3$ components of the three-body correlation 
$\chi_{\sigma_1\sigma_2\sigma_3}^{[3]}$, 
only three components become linearly independent in the SU($N$) symmetric case. 
They can be expressed in terms of the derivatives of  the linear susceptibilities, 
using Eqs.\ \eqref{eq:SUN_chi_3_relation_1}, \eqref{eq:SUN_chi_3_relation_2}, 
and \eqref{eq:chi_B3_def_new_appendix} 
given in  Appendix \ref{sec:3body_SUN_properties}:   
\begin{align}
\chi_{\sigma\sigma\sigma}^{[3]}
\,= & \  
\frac{1}{N}
\frac{\partial \chi_{\sigma\sigma}^{}}{\partial  \epsilon_{d}^{}} 
- \frac{N-1}{N} \,  \chi_{B}^{[3]} 
\;,
\label{eq:chi3_SUN_1}
\\
\widetilde{\chi}_{\sigma\sigma'\sigma'}^{[3]}
\,= & \  
\frac{N-1}{N}
\frac{\partial \chi_{\sigma\sigma}^{}}{\partial  \epsilon_{d}^{}} 
+ \frac{N-1}{N} \,  \chi_{B}^{[3]} 
\;,
\label{eq:chi3_SUN_2}
\\
\widetilde{\chi}_{\sigma\sigma'\sigma''}^{[3]}
\,= & \  
-\frac{N-1}{N}
\frac{\partial \chi_{\sigma\sigma}^{}}{\partial  \epsilon_{d}^{}} 
\,+\,\frac{N-1}{2}
\frac{\partial \chi_{\sigma\sigma'}^{}}{\partial  \epsilon_{d}^{}} \,-\, \frac{N-1}{N} \,  \chi_{B}^{[3]}  ,
\label{eq:chi3_SUN_3} 
\end{align}
for $\sigma \neq \sigma' \neq \sigma'' \neq \sigma$.  
Here, 
\begin{align}
\chi_{B}^{[3]}   \,\equiv & \     
\frac{\partial}{\partial b}
\left. \left(
\frac{\chi_{m\uparrow,m\uparrow}^{} - \chi_{m\downarrow,m\downarrow}^{}}{2} 
\right) \right|_{b=0}^{} 
%
\nonumber \\
=  & \ 
- \chi_{\sigma\sigma\sigma}^{[3]}  + \chi_{\sigma\sigma'\sigma'}^{[3]}  \,,
\rule{0cm}{0.6cm}
\label{eq:chi_B3_def_new}
\end{align}
and the scale factors $(N-1)$ and $(N-1)(N-2)/2$ have been introduced  
for the off-diagonal three-body components in such a way that   
\begin{align}
\widetilde{\chi}_{\sigma\sigma'\sigma'}^{[3]} \,&\equiv\,\left(N\,-\,1\right)\,\chi_{\sigma\sigma'\sigma'}^{[3]}\, , 
\label{eq:chi_II_tilde_def} \\
\widetilde{\chi}_{\sigma\sigma'\sigma''}^{[3]} \,&\equiv\,\frac{\left(N\,-\,1\right)\left(N\,-\,2\right)}{2}\,\chi_{\sigma\sigma'\sigma''}^{[3]}\, . 
\label{eq:chi_III_tilde_def} 
\end{align}
We will also use the dimensionless three-body correlations, defined by   
\begin{align}
\!\!\!\!\!\!\! 
\Theta_\mathrm{I}^{} 
\,\equiv& \ 
\frac{\sin 2\delta}{2\pi}\,
\frac{\chi_{\sigma\sigma\sigma}^{[3]}}{\chi_{\sigma\sigma}^2}\,, \qquad
\widetilde{\Theta}_\mathrm{II}^{} 
\,\equiv \ 
\frac{\sin 2\delta}{2\pi}\,
\frac{\widetilde{\chi}_{\sigma\sigma'\sigma'}^{[3]} }{\chi_{\sigma\sigma}^2}\,,
\label{eq:Theta_I_and_II_tilde_definition} \\
\widetilde{\Theta}_\mathrm{III}^{} 
\,\equiv& \  
\frac{\sin 2\delta}{2\pi}\,
\frac{\widetilde{\chi}_{\sigma\sigma'\sigma''}^{[3]}}{\chi_{\sigma\sigma}^2} 
\,. 
\label{eq:Theta_III_tilde_definition} 
\end{align}
In this work, we  calculate the right-hand side 
of Eqs.\  \eqref{eq:chi3_SUN_1}--\eqref{eq:chi3_SUN_3}
with the NRG to obtain the three-body correlations for the SU($N$) case. 
Note that,  for noninteracting electrons at $U=0$, 
 only the diagonal components  of the three-body correlation 
 $\Theta_\mathrm{I}^{0}$ and  the susceptibility $\chi_{\sigma\sigma}^{0}$ 
remain finite due to the Pauli exclusion principle,  
\begin{align}
\Theta_\mathrm{I}^{0} 
\,=\, 
\frac{-2\epsilon_{d}^{2}}{\epsilon_{d}^{2}+\Delta_{}^{2}}\,, 
\qquad \qquad 
\chi_{\sigma\sigma}^{0} 
\,=\, \frac{1}{\pi}\frac{\Delta}{\epsilon_{d}^{2}+\Delta_{}^{2}},  
\label{eq:ThetaU0}
\end{align}
and 
 $T^{*} \xrightarrow{\,U\to 0\,}\pi \Delta
 \bigl [\,1+(\epsilon_{d}^{}/\Delta)^2\,\bigr]/4$.

\begin{table}[t]
\caption{Low-energy expansion of transport coefficients 
through SU($N$) quantum dots (QD) and magnetic alloys (MA), 
described in Eqs.\ \eqref{eq:FL_cond_QD}--\eqref{eq:Lorenz_number_SUN_def} 
and  Eqs.\ \eqref{eq:FL_rho_MA}--\eqref{eq:L_MA_SUN_formula}. 
Here, $T^* \equiv 1/(4 \chi_{\sigma\sigma}^{})$ is a characteristic FL energy scale.
}
\begin{tabular}{l} 
\hline \hline
\ \ 
$
\frac{dJ}{dV} \,=\,  
\frac{Ne^2}{h} 
\left[\, 
\sin^2 \delta
- C_{T}^{} \left(\frac{\pi T}{T^*}\right)^2 
-   C_{V}^{} \left(\frac{eV}{T^*} \right)^2  
+ \cdots 
\right] 
$
\rule{0cm}{0.55cm}
\\
\ \ 
$
S_\mathrm{noise}^\mathrm{QD}
=    \frac{2Ne^2|eV| }{h} 
\left[\,
\sin^2 \delta \, \bigl(1-\sin^2 \delta\bigr) 
\, + C_{S}^{} 
\left(\frac{eV}{T^*}\right)^2
+ \cdots
\right] 
$
\rule{0cm}{0.55cm}
\\
\ \ 
$
\kappa_\mathrm{QD}^{} \,= \,   
\frac{N\pi^2 T}{3 h}
\,  \left[\,
\sin^2 \delta
\,- 
C_{\kappa}^\mathrm{QD}
\, 
\left( \frac{\pi T}{T^*}\right)^2
+ \cdots
\right] 
$
\rule{0cm}{0.55cm}
\\
\ \ 
$
L_{\mathrm{QD}}^{}
\,\equiv\,
\frac{\kappa_{\mathrm{QD}}^{}}{g\,T} 
\,= \,   
\frac{\pi^{2}}{3e^{2}}\,\left[1\,-\,
\frac{C_{L}^{\mathrm{QD}}}{\sin^{2}\delta}\,
\left(\frac{\pi T}{T^{*}}\right)^{2} \, +\, \cdots \right] 
$
\rule{0cm}{0.55cm}
\\
\hline
\ \ 
$
\varrho_\mathrm{MA}^{}
\,\equiv \, 
\frac{1}{\sigma_{\mathrm{MA}^{}}} 
\,=\, 
\frac{1}{\sigma_{\mathrm{MA}}^{\mathrm{unit}}}  
\,\left[\, 
\sin^2 \delta \,- \,  
C_\varrho^\mathrm{MA} 
\left(\frac{\pi T}{T^*}\right)^2   
+ \cdots \,\right]
$
\rule{0cm}{0.55cm}
\\
\ \ 
$
\frac{1}{\kappa_\mathrm{MA}^{}}\, =\,
\frac{3\,e^{2}}{\pi^{2}\,\sigma_\mathrm{MA}^{\mathrm{unit}}\,T}
\left[\, \sin^2 \delta - 
C_{\kappa}^\mathrm{MA} 
\,\left(\frac{\pi T}{T^*}\right)^2  + \cdots \,\right] 
$
\rule{0cm}{0.55cm}
\\
\ \ 
$
L_{\mathrm{MA}}^{}
\,\equiv\,
\frac{\kappa_{\mathrm{MA}}^{}}{\sigma_{\mathrm{MA}^{}} T} 
\,=\,\frac{\pi^{2}}{3\,e^{2}}\,\left[\,1\,-
\,
\frac{C_{L}^{\mathrm{MA}}}{\sin^2 \delta}
\,
\left( \frac{\pi T}{T^{*}}\right)^{2} \ + \  \cdots \,\right] 
$
\rule{0cm}{0.55cm}
\\
\hline
\hline
\end{tabular}
\label{tab:low_energy_expansion}
\end{table}

\begin{table*}[t]
\caption{Coefficients $C$'s for the next-leading order terms 
of SU($N$) quantum dots and magnetic alloys (MA), 
summarized in Table \ref{tab:low_energy_expansion}.
In these formulas, $W$'s represent the contributions 
determined by the phase shift $\delta$ and the rescaled Wilson ratio 
$\widetilde{K}=(N-1)(R-1)$.
Three-body correlations enter through 
$\Theta_\mathrm{I}^{} 
\equiv
\frac{\sin 2\delta}{2\pi}\,
\frac{\chi_{\sigma\sigma\sigma}^{[3]}}{\chi_{\sigma\sigma}^2}$ 
and 
$\widetilde{\Theta}_\mathrm{II}^{} 
\equiv 
\frac{\sin 2\delta}{2\pi}\,
\frac{\widetilde{\chi}_{\sigma\sigma'\sigma'}^{[3]} }{\chi_{\sigma\sigma}^2}$.  
}
\begin{tabular}{l|l} 
\hline \hline
\ $C_{T}^{} 
\ =   \, 
\frac{\pi^2}{48}
 \,\bigl[\,
W_{T}^{} 
\,+ \,
\Theta_\mathrm{I}^{}
+
\widetilde{\Theta}_\mathrm{II}^{}
 \,\bigr] \quad$
& \ \ \  
$W_{T}^{} 
\ \equiv \, 
-
\left[\,
1
+ 
\frac{2\widetilde{K}^2}{N-1}
\,\right] \cos 2 \delta $
\rule{0cm}{0.45cm}
\\
\ $
C_{V}^{} 
\ =  \, 
\frac{\pi^2}{64}
 \,\bigl[
\,
W_{V}^{} 
\,+ \,
\Theta_\mathrm{I}^{}
+
3\,
\widetilde{\Theta}_\mathrm{II}^{}
\, \bigr]$ 
& \  \ \ 
$W_{V}^{} 
\  \equiv \, 
-
\left[\,
1
+ 
\frac{5\widetilde{K}^2}{N-1}
\,\right] \cos 2 \delta $ 
\rule{0cm}{0.45cm}
\\
\ $C_{S}^{} 
\ =  \,   
\frac{\pi^2}{192} 
\left[\, W_S^{}  
-
\cos 2\delta\,
\Bigl\{
 \Theta_\mathrm{I}^{} 
+
3 \widetilde{\Theta}_\mathrm{II}^{} 
\Bigr\} 
\,\right]$ \ \ \ \ \  \ \ 
& \ \ \ 
$W_S^{} 
 \   \equiv  \, 
%
\cos 4 \delta\,
+ 
\Bigl[\,
 4+5\cos 4 \delta  + 
\frac{3}{2}\bigl(1- \cos 4\delta \bigr)\,(N-2)
\,\Bigr] 
\frac{\widetilde{K}^2}{N-1}
$
\rule{0cm}{0.45cm}
\\
\ $C_{\kappa}^\mathrm{QD} 
=  \,   
\frac{7\pi^2}{80} 
\,\bigl[\,
W_{\kappa}^\mathrm{QD} 
\,+\, 
\Theta_\mathrm{I}^{} 
+
\frac{5}{21}\, 
\widetilde{\Theta}_\mathrm{II}^{} 
\,\bigr]$ 
  &  \ \ \  
$ W_{\kappa}^\mathrm{QD} 
\,  \equiv  \,
\frac{10 - 11\cos 2\delta}{21}
 - \frac{6}{7} \frac{\widetilde{K}^2}{N-1}\, \cos 2\delta$ 
\rule{0cm}{0.45cm}
\\
\ $C_{L}^\mathrm{QD} 
=  \,   
\frac{\pi^2}{240} 
\,\bigl[\,
W_{L}^\mathrm{QD} 
\,+\, 
16\,\Theta_\mathrm{I}^{} 
\,\bigr]
\ = \ C_{\kappa}^\mathrm{QD} -C_{T}^{}  
$ 
  &  \ \ \  
$ W_{L}^\mathrm{QD} 
 \, \equiv \, 
10 - 6 \cos 2\delta -\frac{8\widetilde{K}^{2}}{N-1}\,\cos 2\delta$ 
\rule{0cm}{0.45cm}
\\
\hline
\ $C_{\varrho}^{\mathrm{MA}} 
 =   \, 
\frac{\pi^2}{48}
 \,\bigl[\,
W_\mathcal{\varrho}^{\mathrm{MA}}
\,+ \,
\Theta_\mathrm{I}^{}
+
\widetilde{\Theta}_\mathrm{II}^{}
 \,\bigr] \quad$
& \ \ \  
$W_\mathcal{\varrho}^{\mathrm{MA}}
 \equiv \, 
2+ \cos 2 \delta  
-\frac{2\widetilde{K}^2}{N-1}
\, \cos 2 \delta 
\ \, = \ 
4 \cos^2 \delta + W_{T}^{}
$
\rule{0cm}{0.45cm}
\\
\ $C_{\kappa}^\mathrm{MA} 
=  \,   
\frac{7\pi^2}{80} 
\,\bigl[\,
W_{\kappa}^\mathrm{MA} 
\,+\, 
\Theta_\mathrm{I}^{} 
+
\frac{5}{21}\, 
\widetilde{\Theta}_\mathrm{II}^{} 
\,\bigr]$ 
  &  \ \ \  
$ W_{\kappa}^\mathrm{MA} 
  \equiv  \, 
\frac{32+ 11\cos 2\delta}{21}
 - \frac{6}{7} \frac{\widetilde{K}^2}{N-1}
 \cos 2\delta 
\ \, = \  \frac{44}{21} \cos^2 \delta + W_{\kappa}^\mathrm{QD} $ 
\rule{0cm}{0.45cm}
\\
\ $C_{L}^\mathrm{MA} 
=  \,   
\frac{\pi^2}{240} 
\,\bigl[\,
W_{L}^\mathrm{MA} 
\,-\, 
16\,\Theta_\mathrm{I}^{} 
\,\bigr] 
\ = \ C_{\varrho}^\mathrm{MA} -C_{\kappa}^\mathrm{MA} 
$ \ \ 
  &  \ \ \  
$ W_{L}^\mathrm{MA} 
 \equiv \, 
-22 - 6 \cos 2\delta +\frac{8\widetilde{K}^{2}}{N-1}\,\cos 2\delta 
\ \, = \  \ -24\cos^2 \delta -W_{L}^\mathrm{QD}$ 
\rule{0cm}{0.45cm}
\\
\hline
\hline
\end{tabular}
\label{tab:C_and_W_extended}
\end{table*}

\subsection{Transport formulas for quantum dots  in SU($N$) Fermi liquid regime}
\label{subsec:Transport_coefficients_for_the_SUN_FL_liquid}

We next consider  the low-energy expansion of transport coefficients 
$dJ/dV$, $S_\mathrm{noise}^\mathrm{QD}$, $\kappa_\mathrm{QD}^{}$,  
and the Lorenz number  $L_{\mathrm{QD}}^{}$.
Specifically, for the junctions having tunneling and bias symmetries  
$\Gamma_L=\Gamma_R$ and $\mu_L=-\mu_R =eV/2$,   
these transport coefficients take the following 
form in the SU($N$) symmetric case,  
\begin{align}
&
\!\!\!
\frac{dJ}{dV} =  
\frac{Ne^2}{h} 
\left[\, 
\sin^2 \delta \, 
- C_{T}^{} \left(\frac{\pi T}{T^*}\right)^2 
-   C_{V}^{} \left(\frac{eV}{T^*} \right)^2  
+ \cdots 
\right] , 
\label{eq:FL_cond_QD} 
\\
&
\!\!\!
S_\mathrm{noise}^\mathrm{QD}
=    \frac{2Ne^2|eV| }{h} 
\left[\,
\frac{\sin^2 2\delta}{4} 
\,  + C_{S}^{} 
\left(\frac{eV}{T^*}\right)^2
+ \cdots
\right], 
\label{eq:FL_noise}
\\
&
\!\!\!
\kappa_\mathrm{QD}^{} \,= \,   
\frac{N\pi^2 T}{3 h}
\,  \left[\,
\sin^2 \delta
\,- 
C_{\kappa}^\mathrm{QD}
\, 
\left( \frac{\pi T}{T^*}\right)^2
+ \cdots
\right] ,
\label{eq:FL_kappa_QD}
\\
&
\!\!\!
L_{\mathrm{QD}}^{}\,= \,   
\frac{\pi^{2}}{3e^{2}}\,\left[\,1\,-\,
\frac{C_{L}^{\mathrm{QD}}}{\sin^{2}\delta}\,
\left(\frac{\pi T}{T^{*}}\right)^{2} \, +\, \cdots \right] \,.
\label{eq:Lorenz_number_SUN_def} 
\end{align}
The formulas for the coefficients  
$C_{T}^{}$, $C_{V}^{}$, $C_{S}^{}$,
$C_{\kappa}^\mathrm{QD}$, and 
$C_{L}^{\mathrm{QD}}$ 
of the next-leading order terms 
are summarized in  Tables \ref{tab:low_energy_expansion} 
and   \ref{tab:C_and_W_extended}.
Each of these $C$'s  can be decomposed into two parts, 
denoted as $W$'s and $\Theta$'s. 
 The $W$ part,
defined in the right column of Table \ref{tab:C_and_W_extended}, 
 represents the two-body contributions  determined by  $\widetilde{K}$ and $\delta$. 
The $\Theta$ part represents the three-body contributions 
which can be described in terms of the 
dimensionless parameters defined in  
  Eq.\ \eqref{eq:Theta_I_and_II_tilde_definition}.

\begin{figure}[b]

\leavevmode
 \centering

\includegraphics[width=0.7\linewidth]{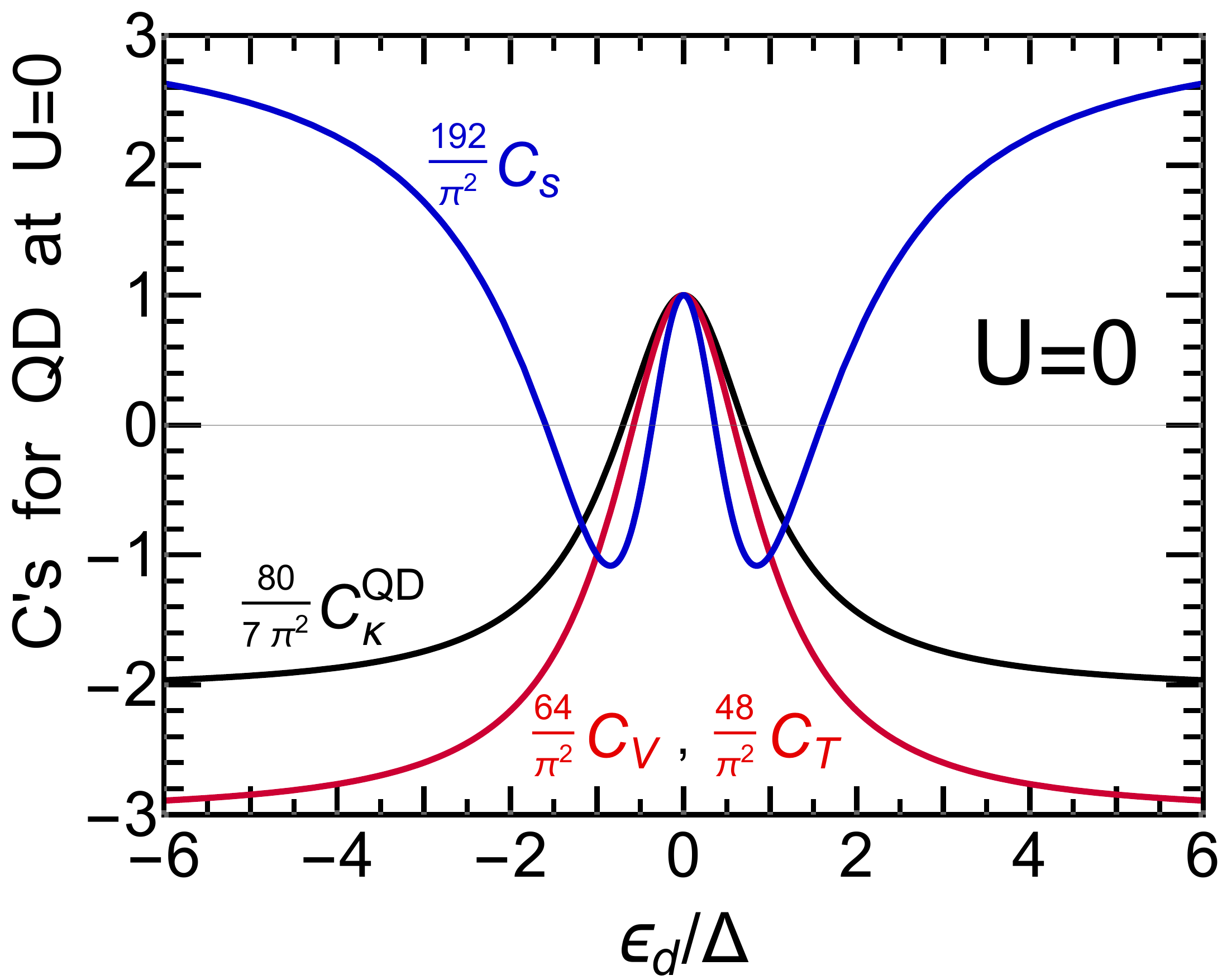}

\caption{Coefficients  $C_{V}^{}$,   
$C_{T}^{}$, 
$C_{S}^{}$, and  
$C_{\kappa}^{\mathrm{QD}}$   
 for noninteracting $U=0$ 
quantum dots  plotted  vs  $\epsilon_{d}^{}$.
}
\label{fig:C_QD_free}
\end{figure}

These transport formulas in the SU($N$) Fermi liquid regime clarify the fact that 
the next-leading order terms for the symmetric tunnel junctions 
are completely determined by {\it five\/}  parameters: 
$\delta$, $T^*$, $\widetilde{K}$,  $\Theta_\mathrm{I}^{}$, 
and $\widetilde{\Theta}_{\mathrm{II}}$. 
The three-body correlations  can be experimentally deduced   
through the measurements of the coefficients $C$'s.  
The other three-body component,    
$\widetilde{\Theta}_\mathrm{III}^{}$, 
defined with respect to  three different levels,   
couples to the tunnel and bias asymmetries, i.e.,   
 $\Gamma_L \neq \Gamma_R$ and  $\mu_L \neq - \mu_R$,  
and  contributes to the nonlinear current 
\cite{TsutsumiTerataniSakanoAO2021prb,TsutsumiSUNpaper}.
The behavior of  $C$'s depend significantly on  
the electron filling  $N_d^{}$ of the impurity levels.  
For instance, in the noninteracting case at $U=0$, 
these coefficients vary with the level position $\epsilon_{d}^{}$ 
as shown in Fig.\ \ref{fig:C_QD_free}.

 In the rest of this paper,  we will demonstrate the behavior 
of the next-leading order terms of the transport coefficients for $N=4$ and $6$. 
To this end, we calculate the correlation functions $\delta$,  
$\chi_{\sigma_1\sigma_2}^{}$, and  $\chi_{\sigma_1\sigma_2\sigma_3}^{[3]}$  
using the NRG approach \cite{HewsonOguriMeyer, TerataniSakanoOguri2020},  
with parameter settings described in Appendix \ref{sec:brief_exp_NRG}.
Specifically,  Eqs.\ \eqref{eq:chi3_SUN_1}--\eqref{eq:chi_B3_def_new} 
are used for obtaining the three-body correlations. 
We have reported part of the results for coefficient $C_{V}^{}$ 
 in a previous paper,  studying 
the role of bias and tunneling asymmetries 
on the nonlinear terms of $dJ/dV$ at $T=0$ \cite{TsutsumiSUNpaper}.  
In this paper,  we provide 
a comprehensive view of  
the three-body Fermi liquid effect through 
a systematic analysis of the next-leading order terms 
$C_{S}^{}$, $C_{T}^{}$, and  
$C_{\kappa}^\mathrm{QD}$ for quantum dots,  
and through the related coefficients for magnetic alloys,
 $C_{\varrho}^\mathrm{MA}$, 
and $C_{\kappa}^\mathrm{MA}$.  
In order to quickly grasp the underlying physics derived from quasiparticle properties 
in the SU(4) and SU(6) cases, 
we provide a brief review of the key characteristics of the two-body correlation functions 
in Appendix \ref{sec:NRG_results_2body_functions_for_sun},
 extending the interaction range up to $U/(\pi \Delta) = 6.0$. 
Additionally, we also include some new results for the renormalized 
impurity level  $\widetilde{\epsilon}_{d\sigma}^{}$ there.

\section{Three-body correlations in the SU(4) $\&$ SU(6) Anderson impurity}
\label{sec:NRG_results_three_body}

In this section, we discuss the behavior of charge and spin susceptibilities  
defined in Eqs.\ \eqref{eq:chi_C_def} and \eqref{eq:chi_S_def}.
In particular, we focus on the derivatives 
 $\partial \overline{\chi}_{C}^{}/\partial\epsilon_d^{}$ and 
 $\partial\overline{\chi}_{S}^{}/\partial\epsilon_d^{}$,   
which can also be expressed in terms of the three-body correlation functions:  
\begin{align}
\frac{\partial \overline{\chi}_{C}^{}}{\partial \epsilon_{d}^{}}\,
=&\ \chi_{\sigma\sigma\sigma}^{[3]}
\,+\,3\,\widetilde{\chi}_{\sigma\sigma'\sigma'}^{[3]}
\,+\,2\,\widetilde{\chi}_{\sigma\sigma'\sigma''}^{[3]}\;,
\label{eq:derivative_charge_susceptibility_SUN}  
\\
\frac{\partial \overline{\chi}_{S}^{}}{\partial \epsilon_{d}^{}}\,
=&\ \chi_{\sigma\sigma\sigma}^{[3]}
\,+\,\frac{N-3}{N-1}\,\,\widetilde{\chi}_{\sigma\sigma'\sigma'}^{[3]}
\,-\,\frac{2}{N-1}\,\,\widetilde{\chi}_{\sigma\sigma'\sigma''}^{[3]}\,.
\label{eq:derivative_spin_susceptibility_SUN}  
\end{align}
The NRG results reveal the fact that these derivatives in the left-hand side
are suppressed in the strong-coupling region $|\xi_d^{}| \lesssim (N-1)U/2$ 
for large $U$.  
This implies that  the linear combinations of the three-body correlations 
in left-hand side of 
Eqs.\ \eqref{eq:derivative_charge_susceptibility_SUN} and  
\eqref{eq:derivative_spin_susceptibility_SUN} approach zero, 
reducing the number of independent components of 
the three-body correlation functions 
$\chi_{\sigma_1\sigma_2\sigma_3}^{[3]}$,   
 as demonstrated below.

\begin{figure}[t]

 \leavevmode
 \centering

\includegraphics[width=0.47\linewidth]{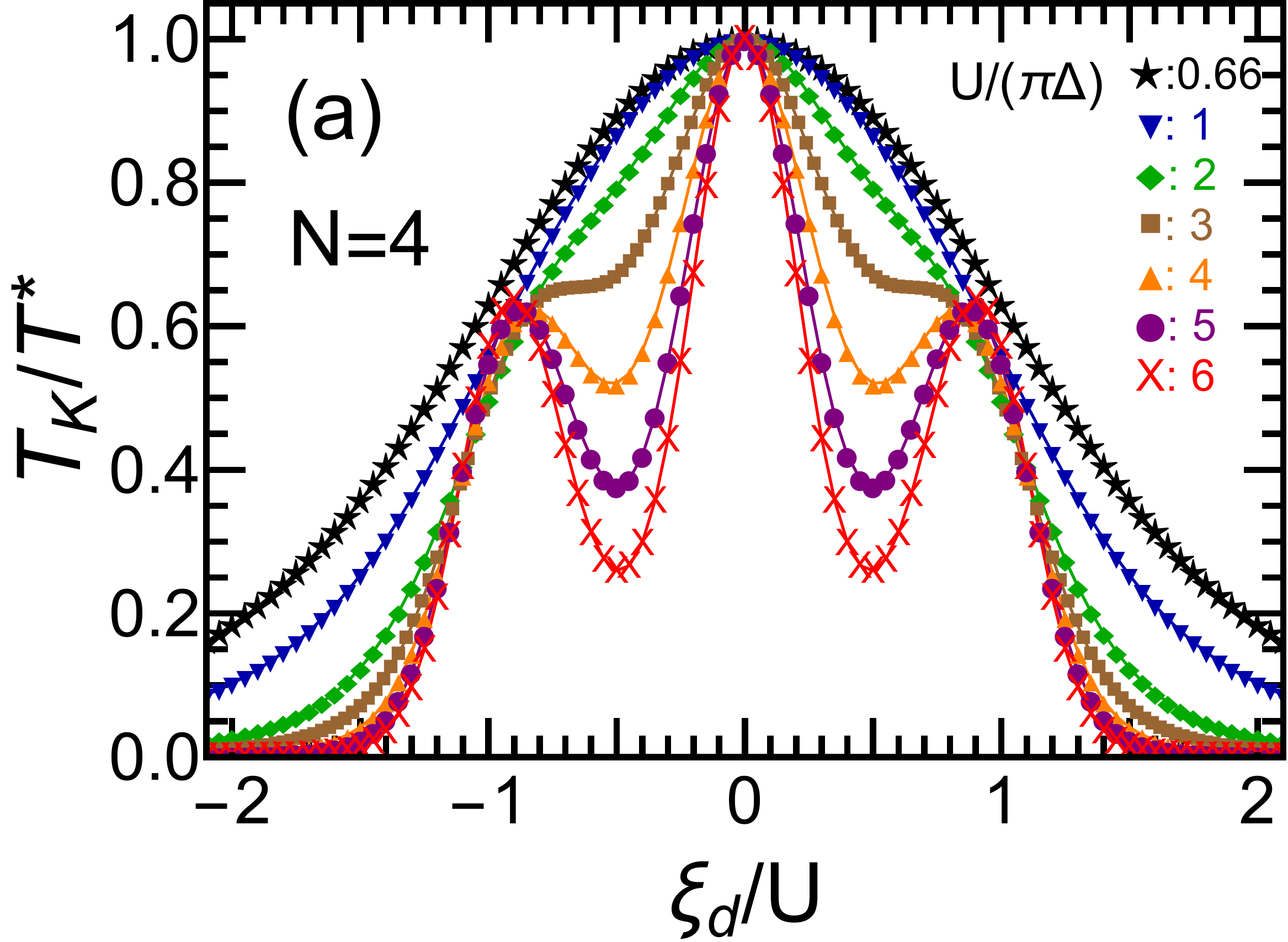}
 \hspace{0.01\linewidth} 
\includegraphics[width=0.47\linewidth]{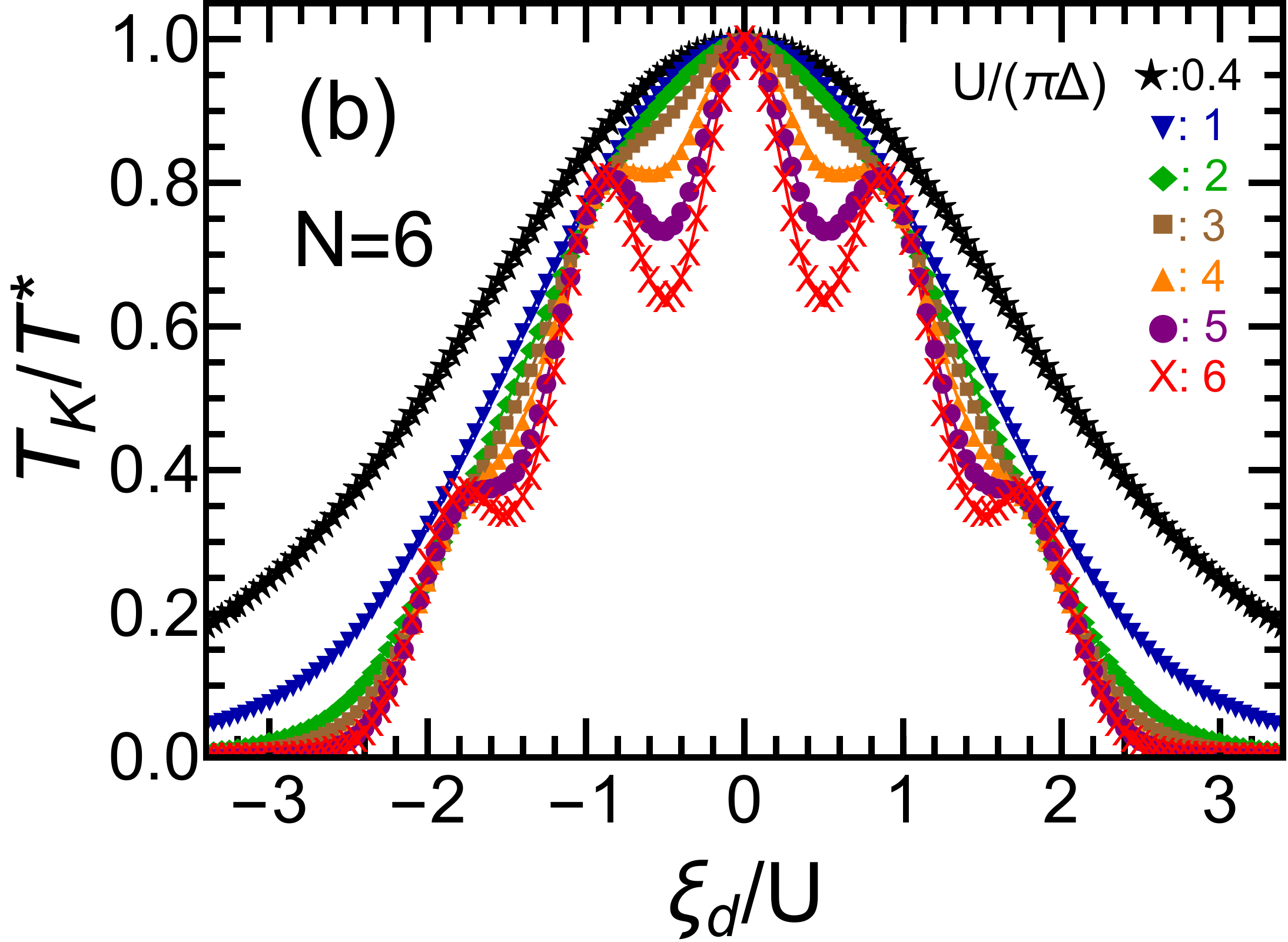}
\\
\includegraphics[width=0.47\linewidth]{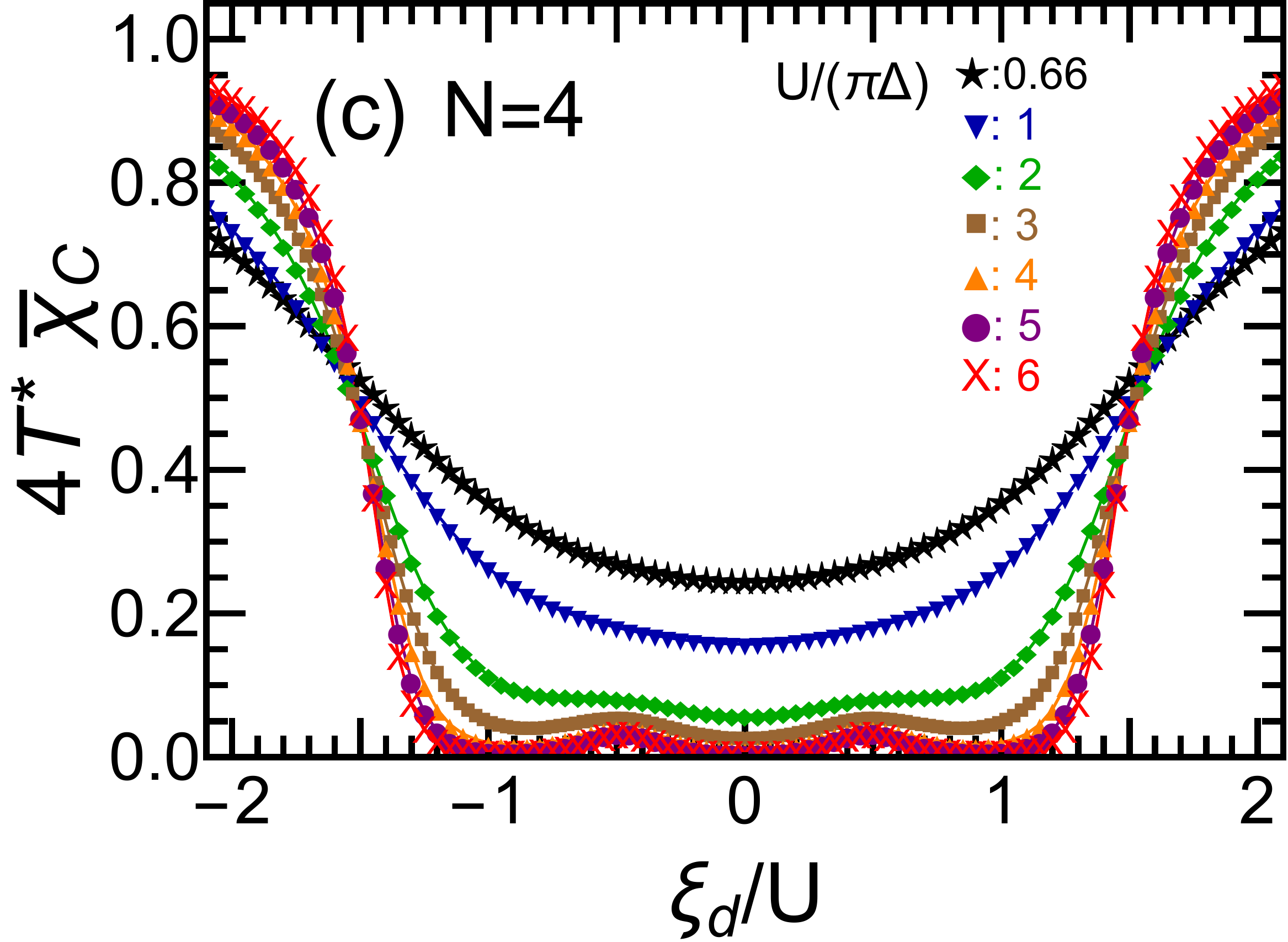}
 \hspace{0.01\linewidth} 
\includegraphics[width=0.47\linewidth]{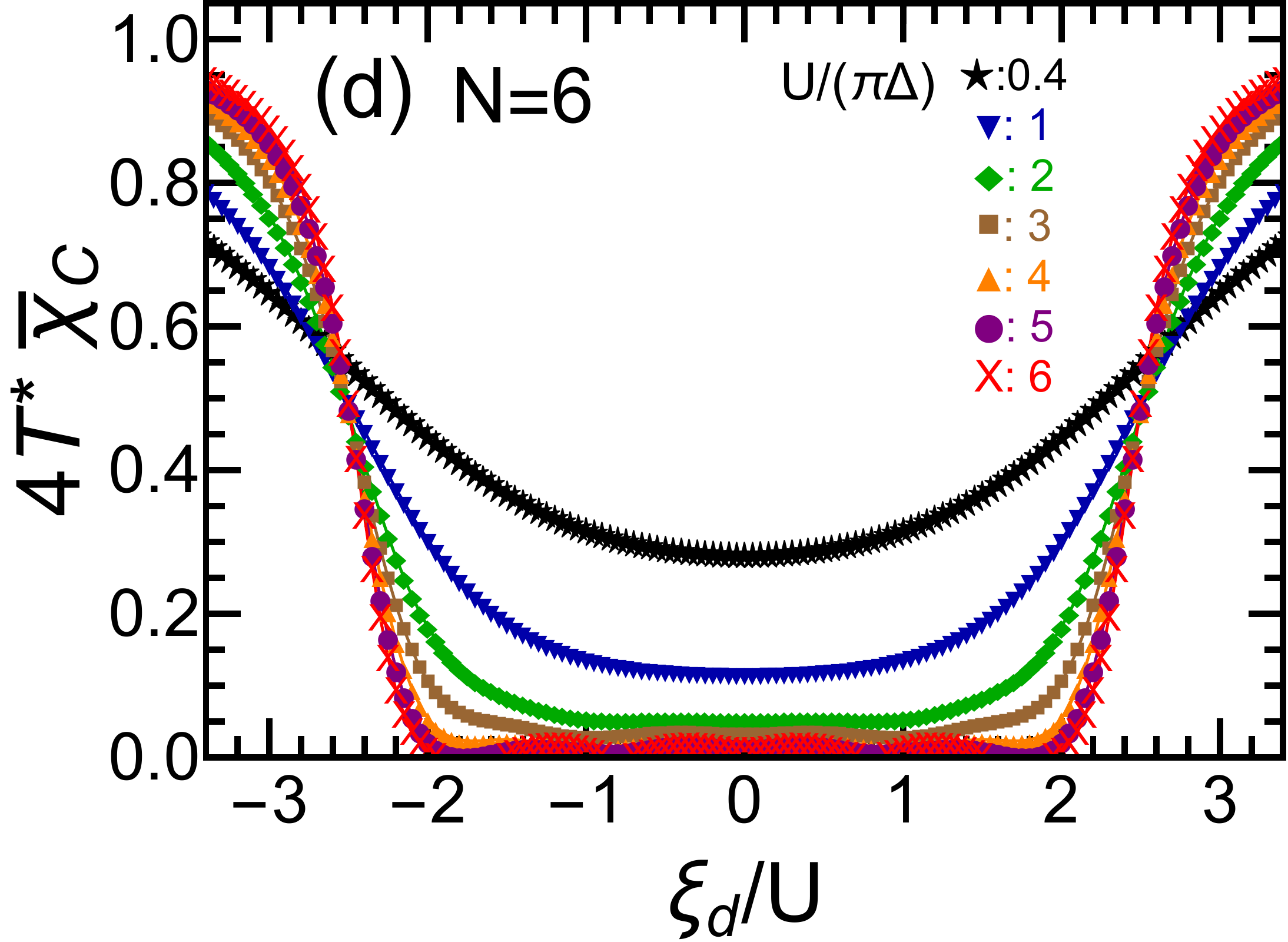}
\\
\includegraphics[width=0.47\linewidth]{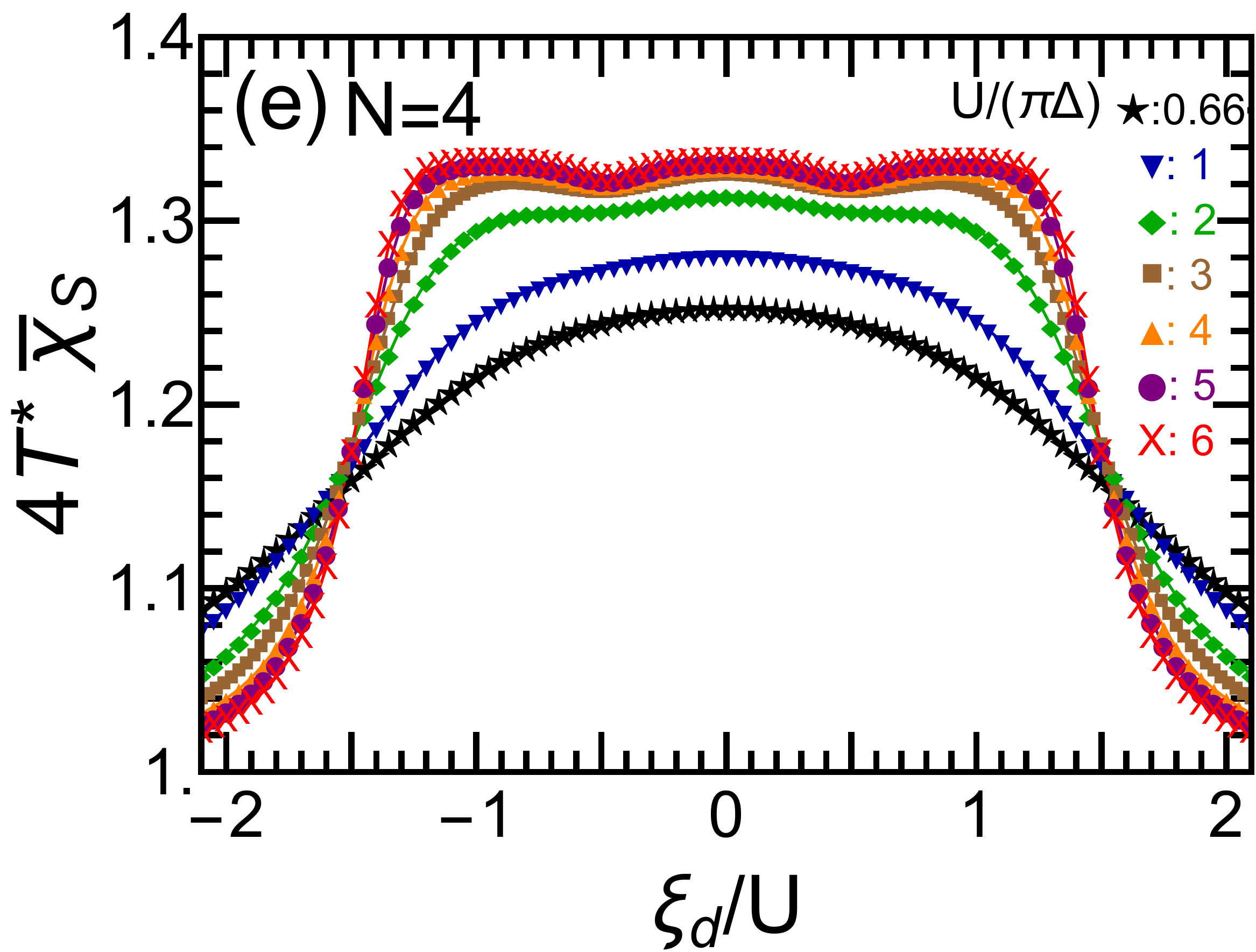}
 \hspace{0.01\linewidth} 
\includegraphics[width=0.47\linewidth]{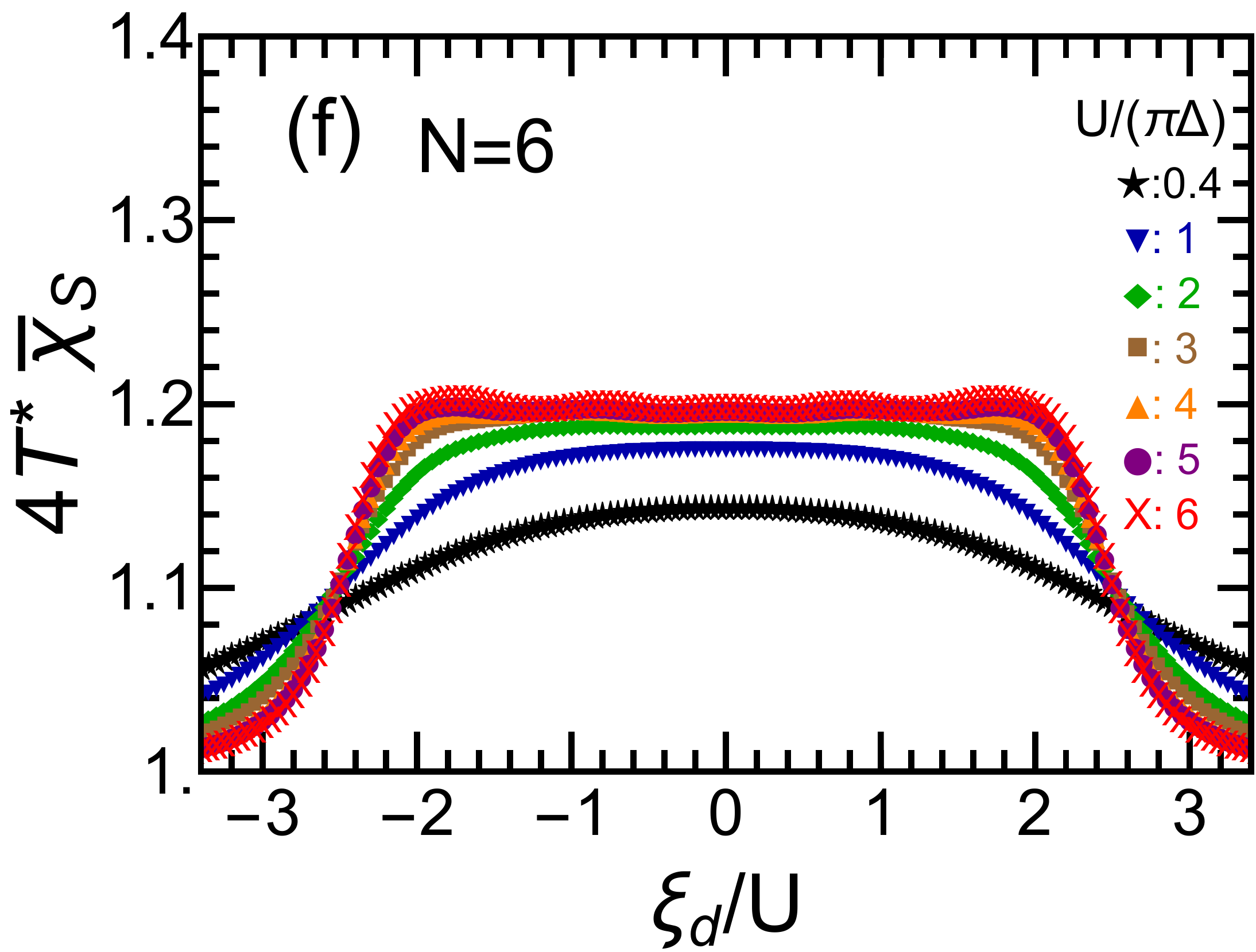}
\caption{Inverse energy scale  $1/T^{*}$, 
charge susceptibility   $\overline{\chi}_{C}^{}$ and 
spin  $\overline{\chi}_{S}^{}$ susceptibilities are plotted 
 vs  $\xi_{d}^{}$  ($= \epsilon_{d}^{}+U/2$) for $N=4$ and $6$.
Here, $4T^{*}$ ($=1/\chi_{\sigma \sigma}^{}$), 
and  $T_{K}^{} \equiv T^{*}\big|_{\xi_d^{}=0}^{}$.  
Interaction strengths are chosen such that,  
for $N=4$, 
  $U/(\pi\Delta) = 2/3(\star)$, $1(\blacktriangledown)$, $2(\blacklozenge)$, 
$3(\blacksquare)$, $4(\blacktriangle)$, $5(\bullet)$, and $ 6(\times)$, 
at which  $T_{K}^{}/(\pi \Delta)= 0.20$, $0.18$, $0.13$, 
$0.092$, $0.063$, $0.041$, and $0.026$, respectively. 
For $N=6$,   
  $U/(\pi\Delta) = 2/5(\star)$, $1(\blacktriangledown)$, $2(\blacklozenge)$, 
$3(\blacksquare)$, $4(\blacktriangle)$, $5(\bullet)$, and $ 6(\times)$, 
at which  $T_{K}^{}/(\pi \Delta)= 0.22$, $0.19$, $0.15$, $0.13$, $0.10$,
 $0.085$, and $0.068$, respectively.
} 
\label{fig:Tstar-ChiC-ChiS-N4-6}
\end{figure}

\begin{figure}[t]

 \leavevmode
 \centering

\includegraphics[width=0.47\linewidth]{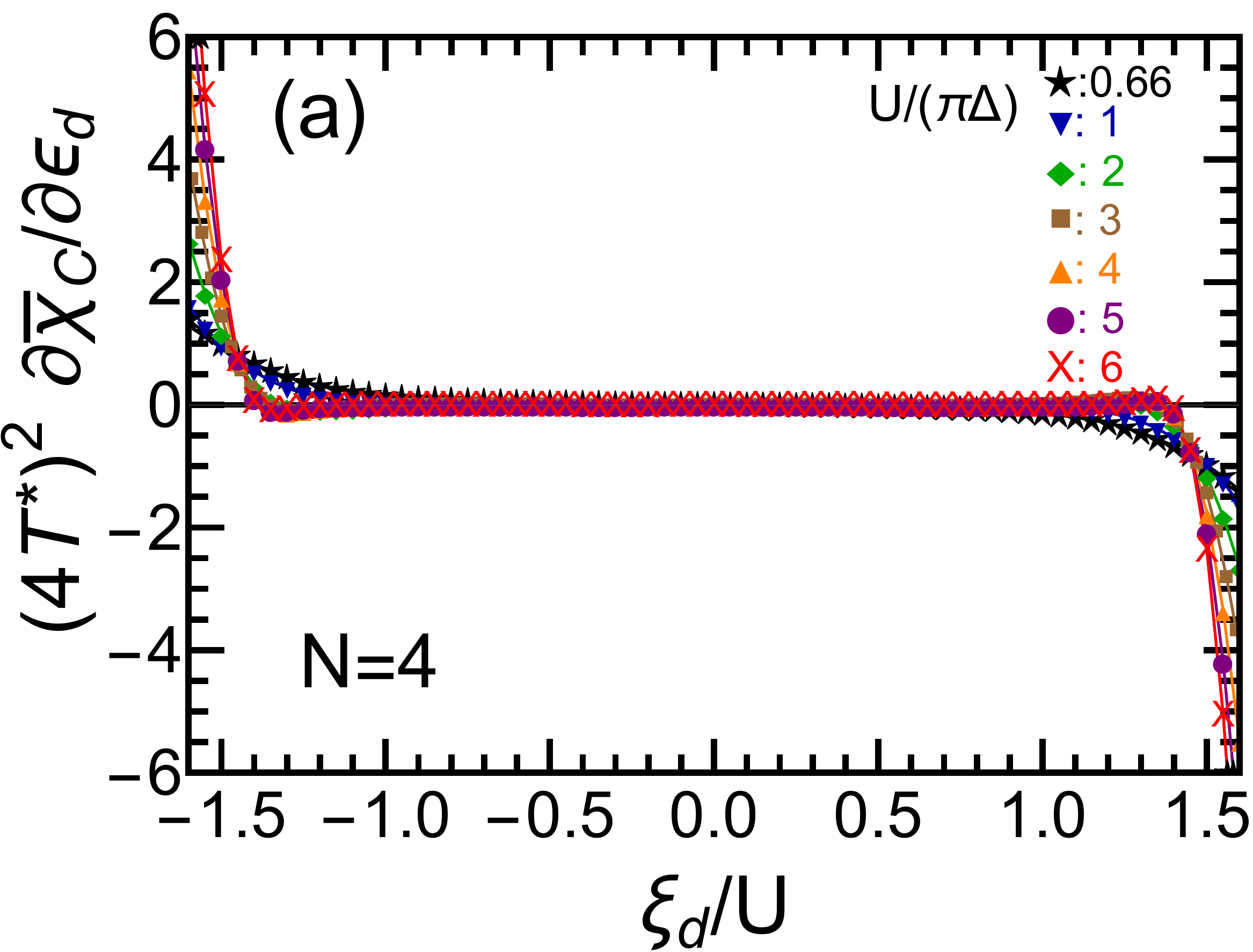}
 \hspace{0.01\linewidth} 
\includegraphics[width=0.47\linewidth]{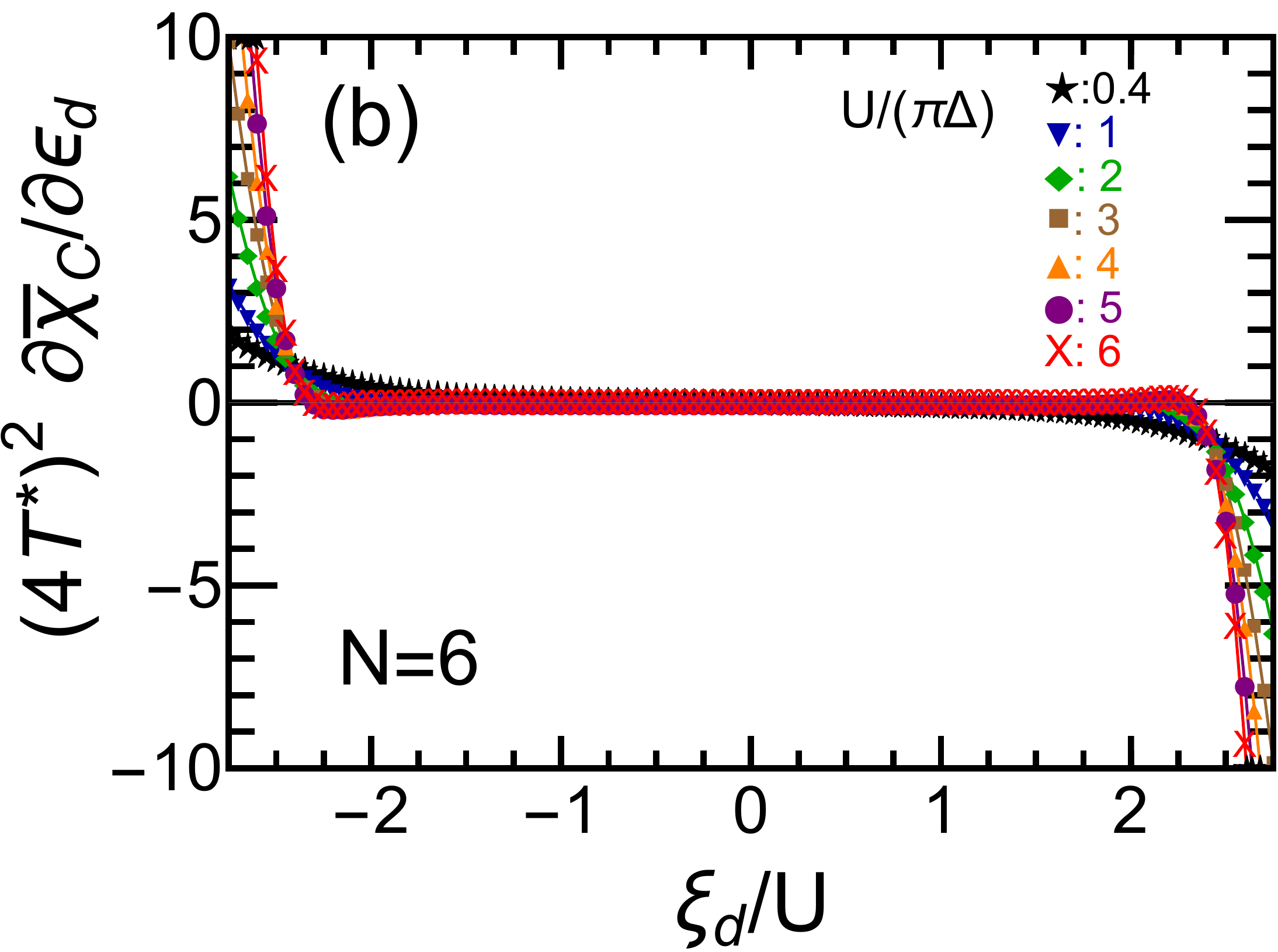}
\\
\includegraphics[width=0.47\linewidth]{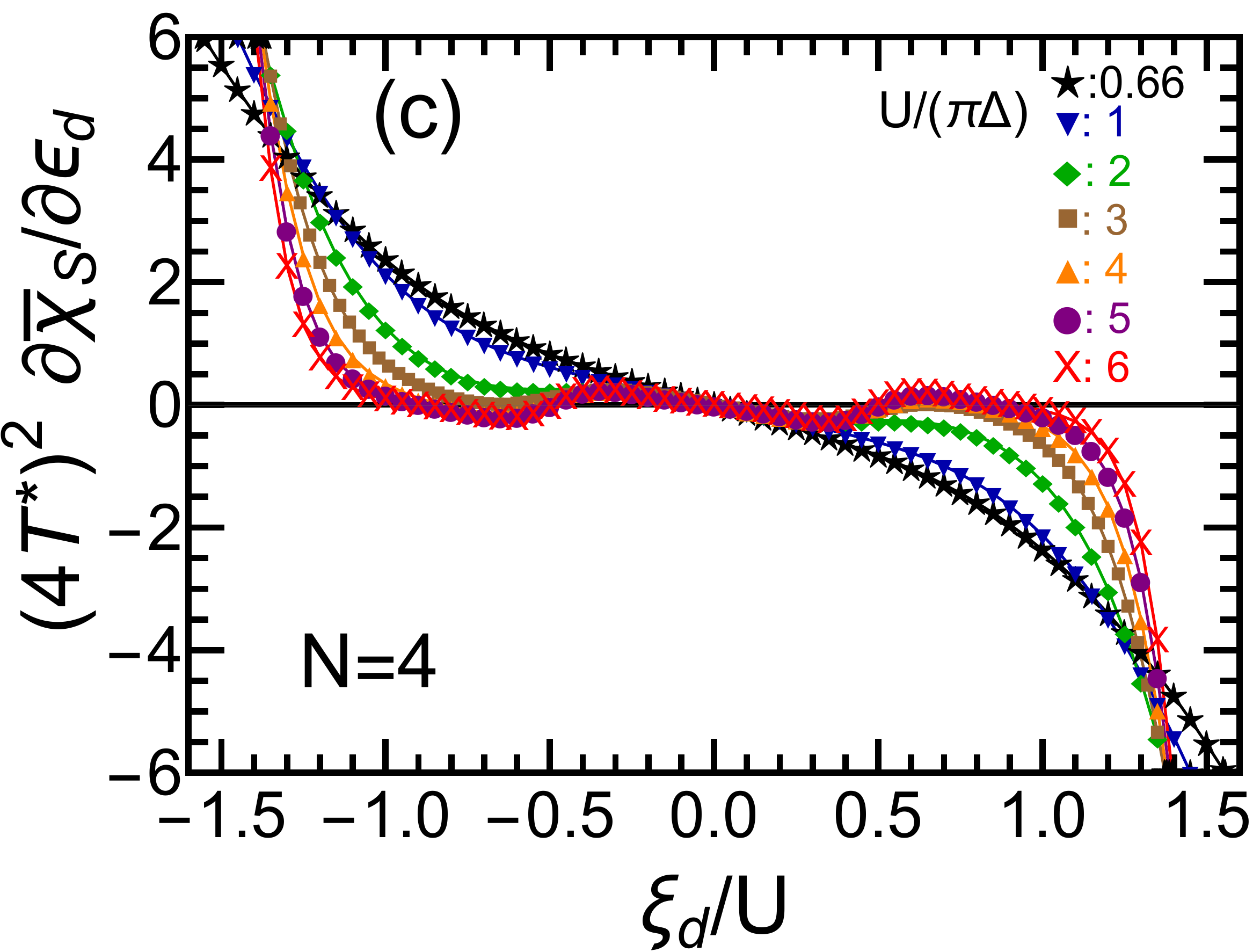}
 \hspace{0.01\linewidth} 
\includegraphics[width=0.47\linewidth]{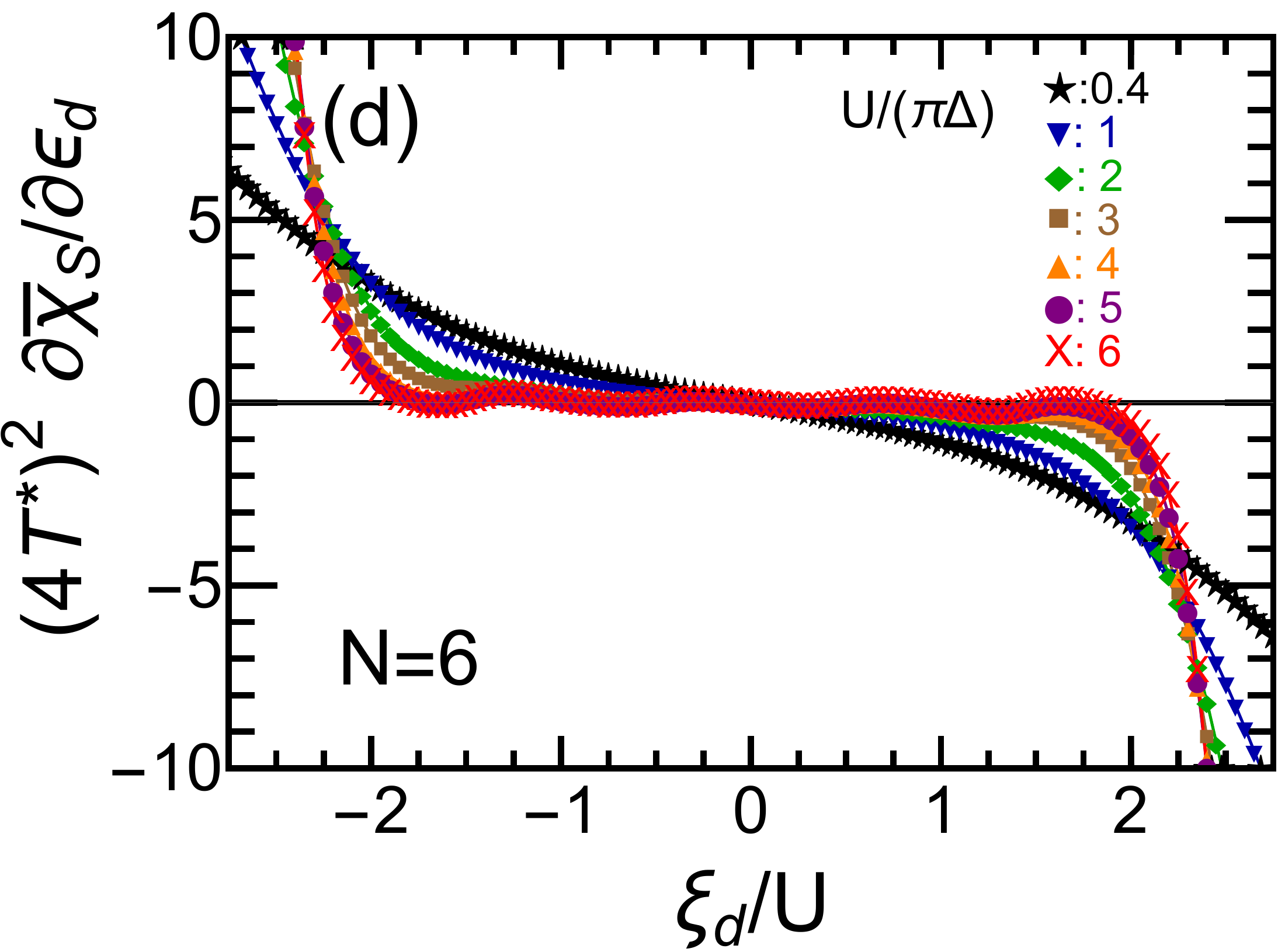}
\\
\hspace{-0.034\linewidth}
\includegraphics[width=0.498\linewidth]{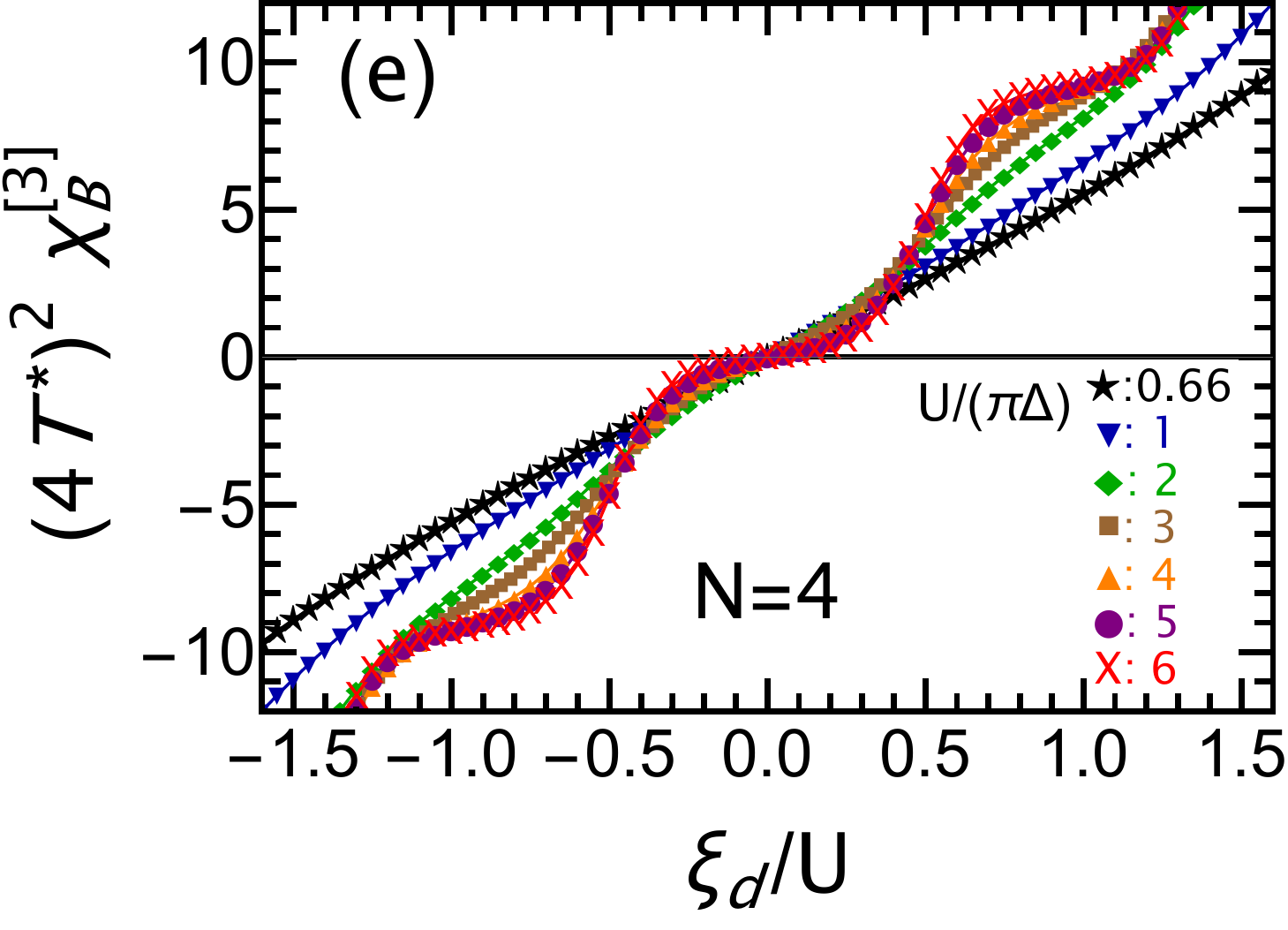}
 \hspace{0.0005\linewidth} 
\includegraphics[width=0.490\linewidth]{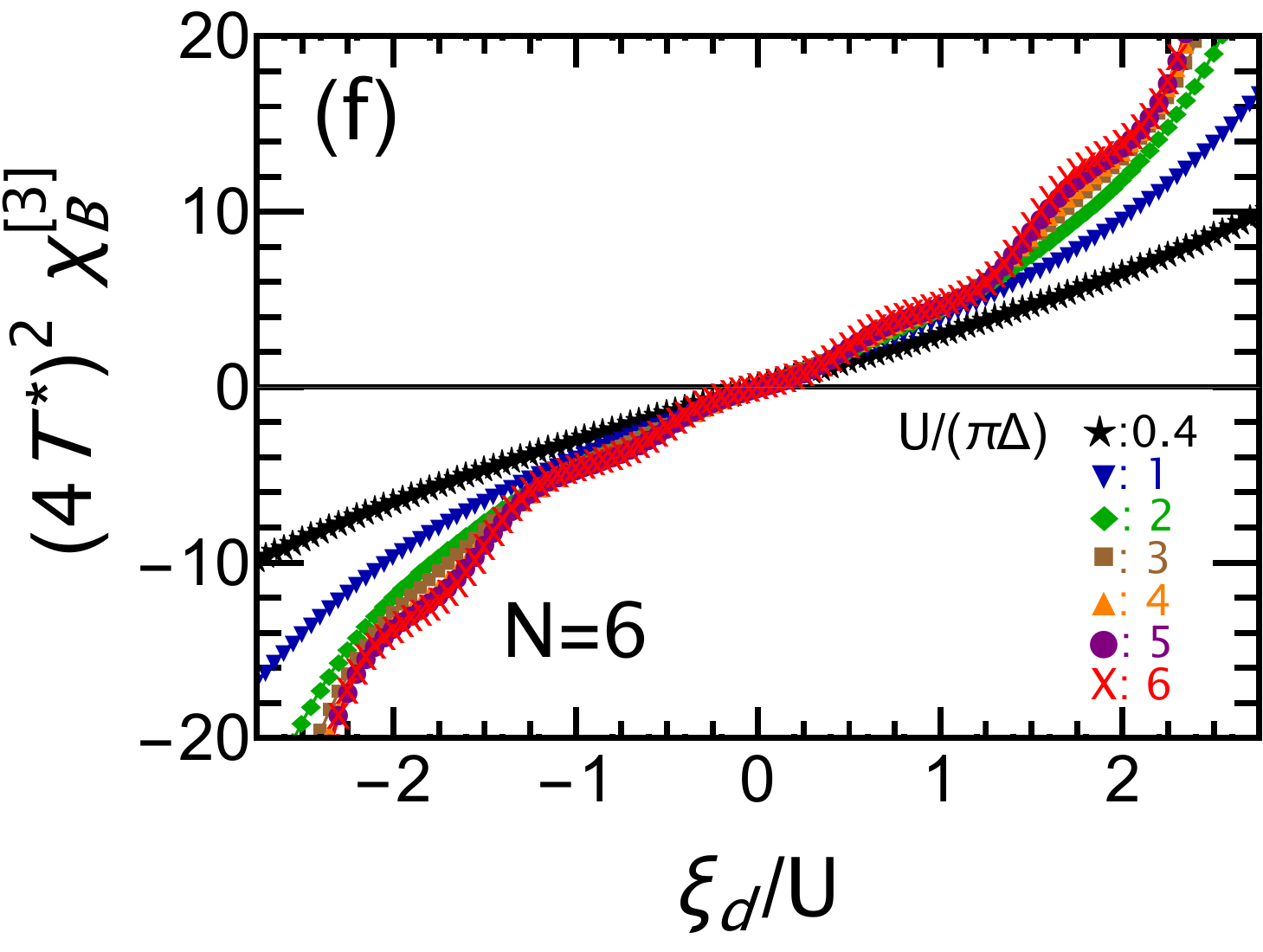}
 \hspace{0.0005\linewidth} 
\caption{$(4T^{*})^{2}\partial  \overline{\chi}_C^{}/\partial \epsilon_{d}$, 
$(4T^{*})^{2}\partial \overline{\chi}_S^{}/\partial \epsilon_{d}$, 
and $(4T^{*})^{2}\chi_{B}^{[3]}$ are plotted vs  $\xi_{d}^{}$ 
for $N=4$ and $6$.
Interaction strengths are chosen such that, 
 for $N=4$, 
 $U/(\pi\Delta) = 2/3(\star)$, $1(\blacktriangledown)$, $2(\blacklozenge)$, 
$3(\blacksquare)$,  $4(\blacktriangle)$, $5(\bullet)$,  $6(\times)$.
 For $N=6$,  
 $U/(\pi\Delta) = 2/5(\star)$, $1(\blacktriangledown)$, $2(\blacklozenge)$, 
$3(\blacksquare)$,  $4(\blacktriangle)$, $5(\bullet)$, $6(\times)$. 
}
\label{fig:DChiC_DChiS_chi3B_N4-6_u6}	
\end{figure}

\subsection{Charge and spin susceptibilities:
$\overline{\chi}_C^{}$ \&  $\overline{\chi}_S^{}$ }

\label{subsec:charge_spin_susceptibilities_N4-6}

One of the most fundamental quantities that play a central role 
in the low-energy physics of quantum impurities   
is the characteristic energy scale 
  $T^{*} \equiv 1/(4\chi_{\sigma\sigma}^{})$, 
 defined in Eq.\ \eqref{eq:Fermiparaorigin} 
as an inverse of the diagonal susceptibility  $N \chi_{\sigma\sigma}^{} 
= \overline{\chi}_C^{}+(N-1)\overline{\chi}_S^{}$. 
We will use, in the following discussions, the Kondo temperature  
$T_{K}^{} \equiv T^{*}\big|_{\xi_d^{}=0}^{}$, 
defined as the value of $T^{*}$ 
at the electron-hole symmetric point $\xi_d^{}=0$.

The NRG results for  $1/T^{*}$ in the SU(4) and SU(6) cases 
 are plotted vs $\xi_d^{}$ in Figs.\  \ref{fig:Tstar-ChiC-ChiS-N4-6}(a) 
and \ref{fig:Tstar-ChiC-ChiS-N4-6}(b), respectively, 
by multiplying them by $T_K^{}$.   
We see that  $1/T^{*}$ has $N-1$ local maxima for strong interactions,
at integer-filling points, i.e.,  $\xi_{d}^{}\simeq 0, \pm U, \ldots, \pm (N-2)U/2$, 
reflecting the oscillatory behavior of the wave function renormalization 
 factor $z$ ($=\rho_{d\sigma}^{}/\chi_{\sigma\sigma}^{}$) 
 described in Appendix \ref{sec:NRG_results_2body_functions_for_sun}.
At  $|\xi_{d}^{}| \gg (N-1) U/2$,   
the energy scale $T^{*}$ approaches the non-interacting value, 
 $T_{K}^{}/T^{*} \xrightarrow{\,|\xi_{d}^{}| \gg (N-1)U/2\,}
 \Delta^2/\xi_d^{2}$, 
as the electron filling of the impurity levels approaches $N_d^{} \simeq 0$ or $N$.

The charge susceptibilities for $N=4$ and $6$ 
are plotted in Figs.\ \ref{fig:Tstar-ChiC-ChiS-N4-6}(c) 
and \ref{fig:Tstar-ChiC-ChiS-N4-6}(d), 
using $T^*$ as a normalization factor, i.e.,  
 $4 T^{\ast}\overline{\chi}_C^{}= 1 -\widetilde{K}$ 
from Eq.\ \eqref{eq:chi_C_def}.  
Therefore, the normalized value $4T^{\ast}\overline{\chi}_C^{}$ 
is determined by the rescaled Wilson ratio $\widetilde{K}$,  
 described in Appendix \ref{sec:NRG_results_2body_functions_for_sun}. 
As $U$ increases,  $4 T^{\ast}\overline{\chi}_C^{}$ decreases 
in a wide region of the impurity level  $|\xi_d^{}| \lesssim (N-1)U/2$,  
where the impurity levels are partially filled $1\lesssim N_d^{} \lesssim N-1$. 
In this filling range, the charge susceptibility is significantly suppressed 
by the Coulomb repulsion, 
and it vanishes 
$4 T^{\ast}\overline{\chi}_C^{} \xrightarrow{\,U \gg \pi \Delta\,} 0$ 
in the strong interaction limit.
Outside this region, i.e., at  $|\xi_d^{}| \gg (N-1)U/2$,  
the charge susceptibility approaches the noninteracting value 
$4T_{}^{*}\overline{\chi}_{C}^{} \xrightarrow{\,|\xi_{d}^{}|\to\infty\,} 1$ 
as the filling of the impurity levels approaches $N_d^{} \simeq 0$ or $N$.

Figures \ref{fig:Tstar-ChiC-ChiS-N4-6}(e) and \ref{fig:Tstar-ChiC-ChiS-N4-6}(f) 
show the spin susceptibilities for $N=4$ and $6$, which are 
normalized with the same scaling factor,
 i.e.,  $4T^{*}\overline{\chi}_{S}^{} = 1 + \widetilde{K}/(N-1)$  
[see  Eq.\ \eqref{eq:chi_S_def}].  
As the interaction $U$ increases,  
$4T^{*}\overline{\chi}_{S}^{}$ increases from the non-interacting value $1$. 
In the strong interaction limit, 
it approaches  the value $4T^{*}\overline{\chi}_{S}^{}  \to N/(N-1)$, i.e.,   
$4/3$ for $N=4$ and $6/5$ for $N=6$, 
and exhibits a wide plateau structure  
in the strong-coupling region  $|\xi_{d}^{}|\lesssim (N-1)U/2$.  
At  $|\xi_{d}^{}| \gg  (N-1)U/2$,  
where the occupation number approaches  $N_d^{} \simeq 0$ or  $N$, 
the spin susceptibility also approaches the noninteracting value 
$4T_{}^{*}\overline{\chi}_{S}^{} \xrightarrow{\,|\xi_{d}^{}|\to\infty\,} 1$.

\subsection{Derivative of $\overline{\chi}_C^{}$ 
and  $\overline{\chi}_S^{}$ with respect to $\epsilon_{d}^{}$}
\label{subsec:Derivatives_of_chi_C_and_of_chi_S_N4-6}

The three-body correlation functions can be obtained from 
the derivatives of  $\chi_{\sigma_1\sigma_2}^{}$ 
with respect to $\epsilon_d^{}$ and $b$, 
 using Eqs.\ \eqref{eq:chi3_SUN_1}--\eqref{eq:chi3_SUN_3}.   
In particular, $\partial \chi_{\sigma_1\sigma_2}^{}/\partial \epsilon_d^{}$
can be rewritten in terms of  the derivatives of the charge and spin susceptibilities, as 
\begin{align}
\frac{\partial \chi_{\sigma\sigma}^{}}{\partial  \epsilon_{d}^{}} 
\,=&\ 
\frac{1}{N}
\frac{\partial \overline{\chi}_{C}^{}}{\partial \epsilon_{d}^{}}
\,+\,
\frac{N-1}{N}
\frac{\partial \overline{\chi}_{S}^{}}{\partial \epsilon_{d}^{}}
\,,
\label{eq:derivative_chi11_SUN_add}  
\\
(N-1)
\frac{\partial \chi_{\sigma\sigma'}^{}}{\partial  \epsilon_{d}^{}} 
\,=&\ 
\frac{N-1}{N}
\frac{\partial \overline{\chi}_{C}}{\partial \epsilon_{d}^{}}
\,-\,
\frac{N-1}{N}
\frac{\partial \overline{\chi}_{S}}{\partial \epsilon_{d}^{}}
\,,
\label{eq:derivative_chi12_SUN_add}  
\end{align}
for $\sigma \neq \sigma'$.
Figures  \ref{fig:DChiC_DChiS_chi3B_N4-6_u6}(a) 
and \ref{fig:DChiC_DChiS_chi3B_N4-6_u6}(b) show the 
NRG results for $\partial \overline{\chi}_C^{}/\partial  \epsilon_{d}^{}$ 
for $N=4$ and $6$, respectively. 
Similarly, the derivatives of the spin susceptibility  
 $\partial \overline{\chi}_S^{}/\partial  \epsilon_{d}^{}$ 
for $N=4$ and $6$ are 
 plotted  in  Figs.\ \ref{fig:DChiC_DChiS_chi3B_N4-6_u6} (c) 
and \ref{fig:DChiC_DChiS_chi3B_N4-6_u6}(d). 
Note that in these figures, the derivatives have been multiplied by 
a factor of  $(4T^{*})^{2}$  to make them dimensionless.
These derivatives are significantly suppressed 
in the strong-coupling region  $|\xi_d^{}| \lesssim (N-1)U/2$ for large  $U$.
More specifically, 
$|\partial \overline{\chi}_C^{}/\partial \epsilon_{d}^{}| 
  \ll |\partial \overline{\chi}_S^{}/\partial \epsilon_{d}^{}| \ll 1/(4T^{*})^{2}$: 
 the derivative of spin susceptibility becomes much smaller than $1/(4T^{*})^{2}$  
while it is still larger than the derivative of charge susceptibility.
Therefore, both  $\partial \chi_{\sigma\sigma}/\partial \epsilon_{d}^{}$ 
and $\partial \chi_{\sigma\sigma'}/\partial \epsilon_{d}^{}$   
are suppressed in a wide range of electron fillings 
$1 \lesssim N_d^{} \lesssim N-1$ for large $U$.

We next consider $\chi_{B}^{[3]}$,  defined 
in Eq.\ \eqref{eq:chi_B3_def_new} as a derivative of a  linear combination of 
the two different diagonal susceptibilities with respect to 
the magnetic fields $b$.   
Figures \ref{fig:DChiC_DChiS_chi3B_N4-6_u6}(e) 
and \ref{fig:DChiC_DChiS_chi3B_N4-6_u6}(f) 
show  $(4T^{*})^{2} \chi_{B}^{[3]}$ for $N=4$ and $6$,  
respectively.
We see that $(4T^{*})^{2}\chi_{B}^{[3]}$ 
exhibits a staircase structure with a flat plateau emerging 
around the integer filling points 
  $\xi_{d}^{}=0$, $\pm U$, $\pm 2U$, $\ldots$, $(N-2)U/2$
for large  $U$.  
The magnitude $\bigl|\chi_{B}^{[3]}\bigr|$ becomes much larger 
than the derivative of the charge and spin susceptibilities, 
$|\partial \overline{\chi}_C^{}/\partial \epsilon_{d}^{}|$ and 
$|\partial \overline{\chi}_S^{}/\partial \epsilon_{d}^{}|$.
Therefore,  $\chi_{B}^{[3]}$ dominates the terms 
in the right-hand side of Eqs.\ \eqref{eq:chi3_SUN_1}--\eqref{eq:chi3_SUN_3} 
in the strong-coupling region  $|\xi_{d}^{}| \lesssim (N-1)U/2$, 
and the three independent components of three-body correlations 
 approach one another in such a way that   
\begin{align}
  \chi_{\sigma\sigma\sigma}^{[3]}
\,\simeq\,
-\,\widetilde{\chi}_{\sigma\sigma'\sigma'}^{[3]}
\,\simeq\,\widetilde{\chi}_{\sigma\sigma'\sigma''}^{[3]}
\ \simeq \ 
-\frac{N-1}{N}\,\chi_{B}^{[3]}\,  .
  \label{eq:three_cody_cor_relations_strong_U_simplified}
\end{align}
This means that the three-body correlations are described 
by a single parameter $\chi_{B}^{[3]}$  
in  a wide filling range  $1\lesssim N_d^{} \lesssim N-1$ 
 for large $U$.

\begin{figure}[t]

 \leavevmode
 \centering

\includegraphics[width=0.47\linewidth]{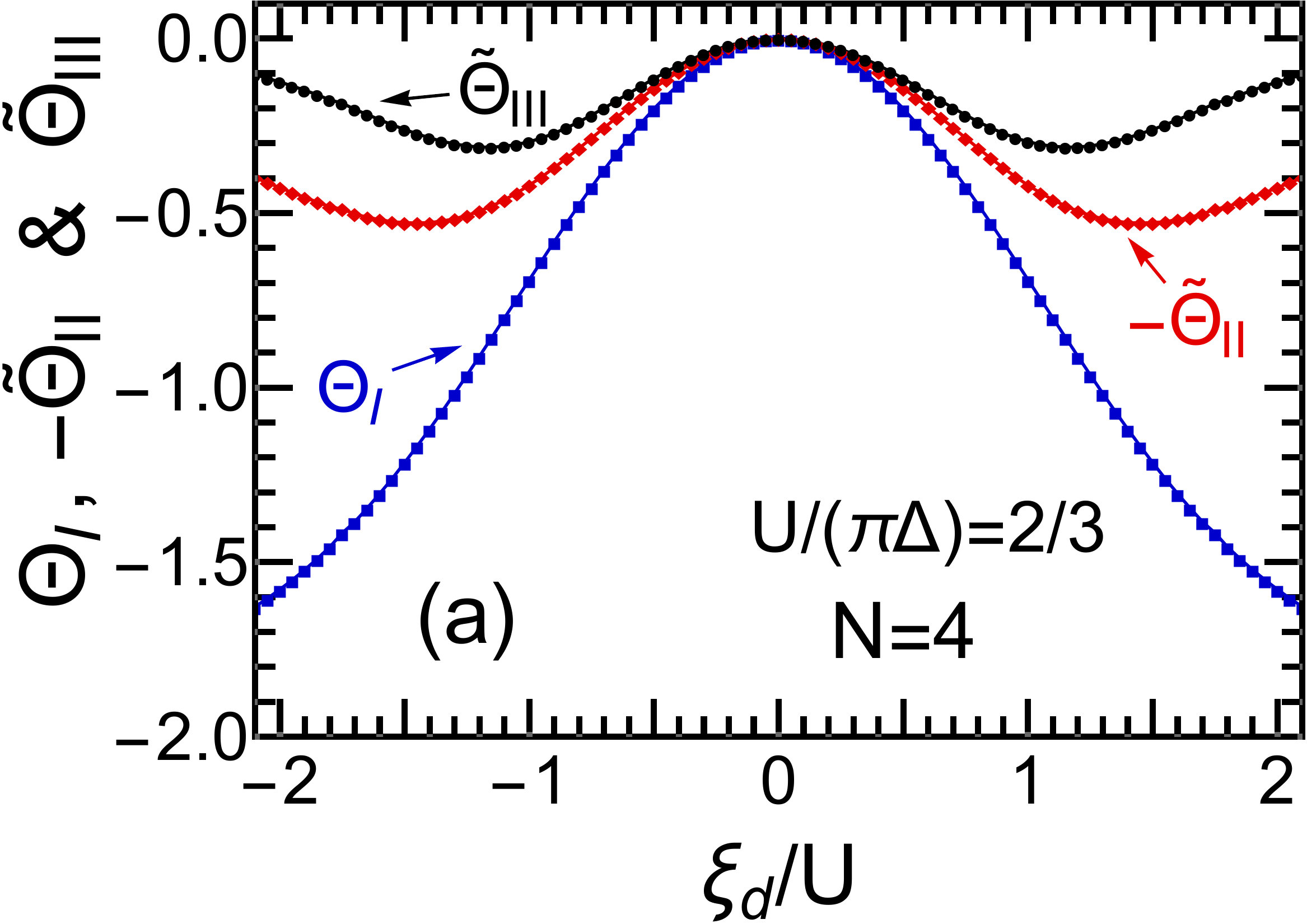}
 \hspace{0.01\linewidth} 
\includegraphics[width=0.47\linewidth]{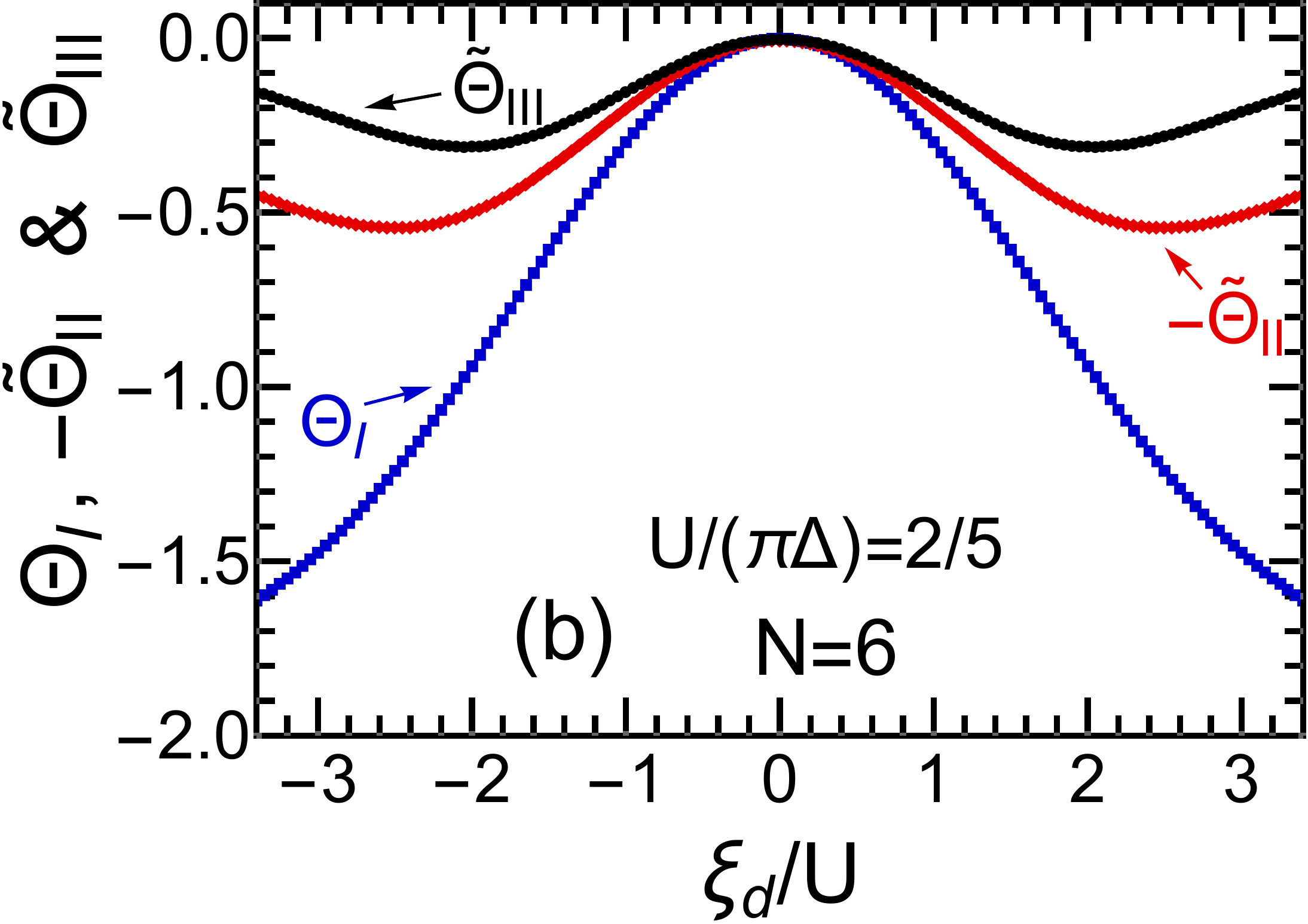}
\\
\includegraphics[width=0.47\linewidth]{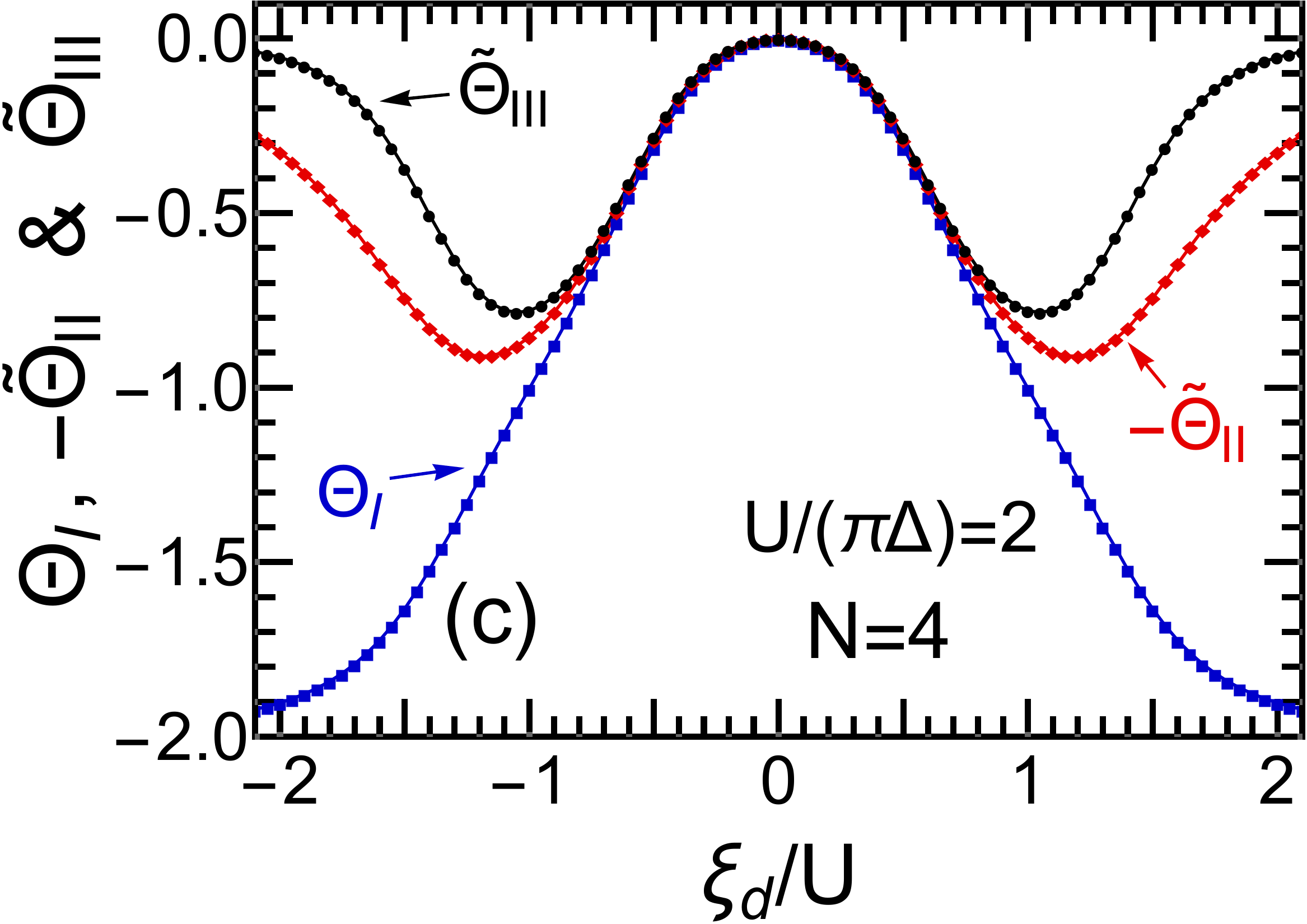}
 \hspace{0.01\linewidth} 
\includegraphics[width=0.47\linewidth]{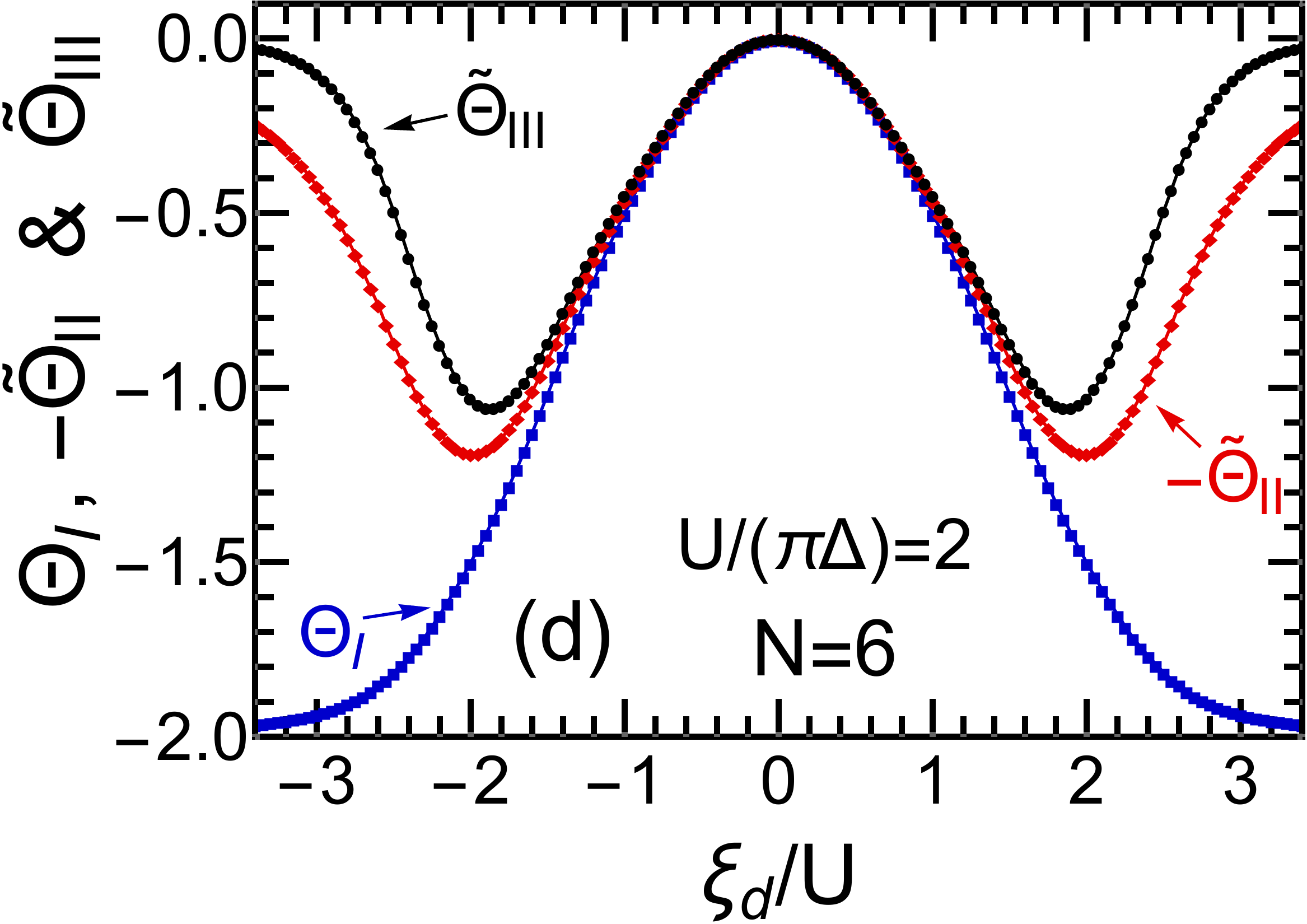}
\\
\includegraphics[width=0.47\linewidth]{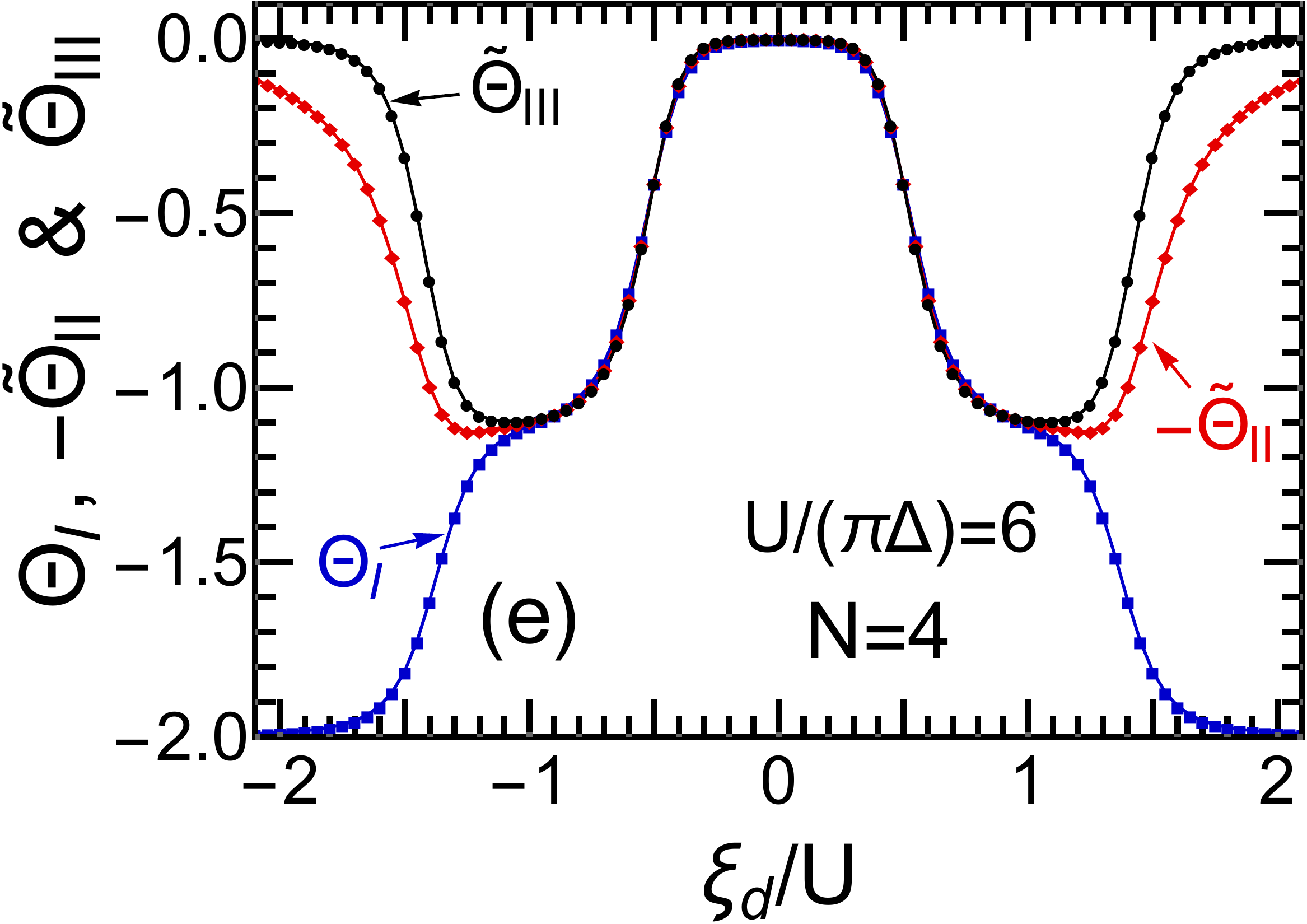}
 \hspace{0.01\linewidth} 
\includegraphics[width=0.47\linewidth]{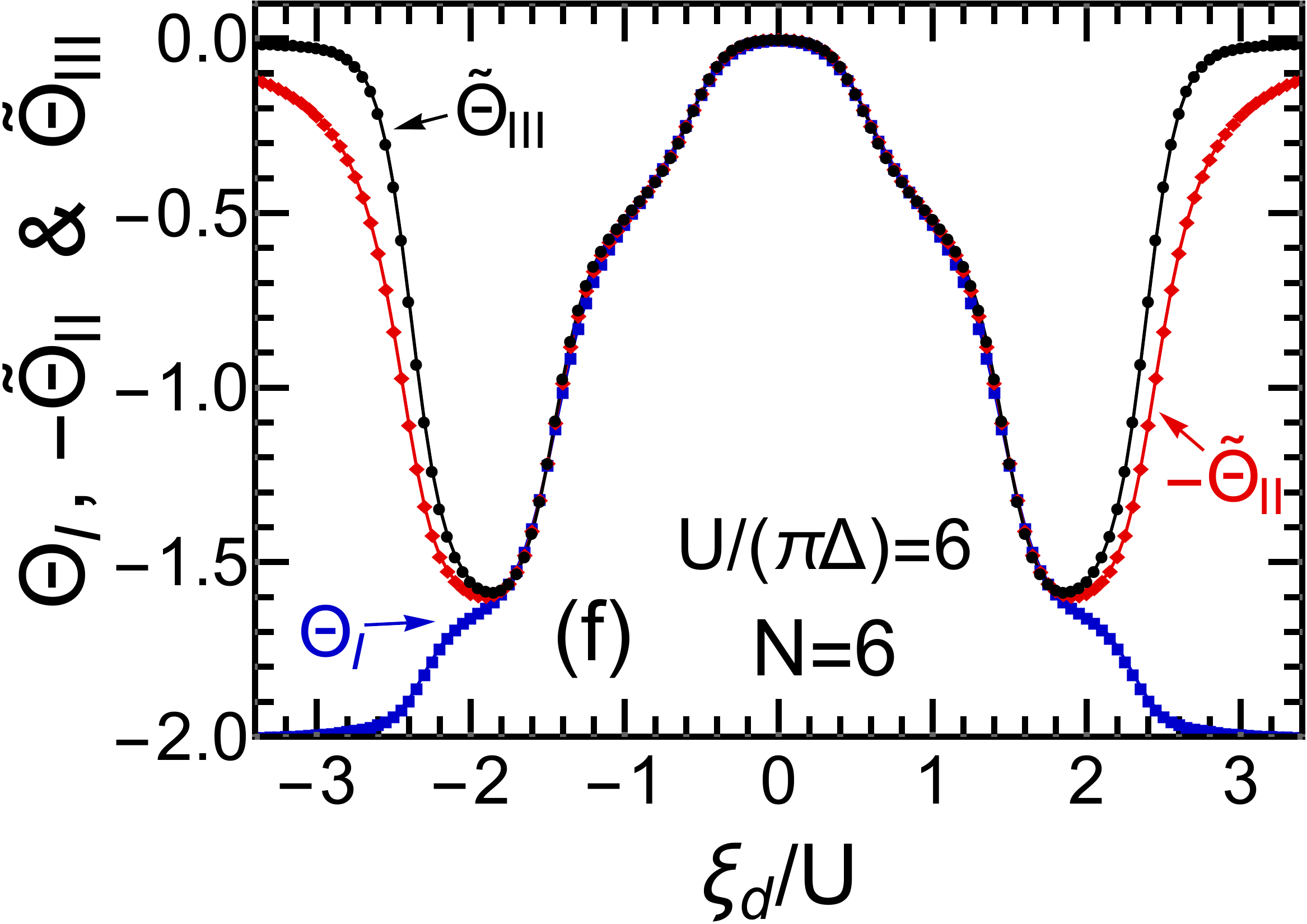}
\caption{Dimensionless three-body corrections 
$\Theta_{\mathrm{I}}^{}(\blacksquare)$ , 
$-\widetilde{\Theta}_{\mathrm{II}}^{}(\blacklozenge)$,  
and $\widetilde{\Theta}_{\mathrm{III}^{}}(\bullet)$ are plotted 
vs $\xi_{d}^{}$,
 for interaction strengths   
 $U/(\pi\Delta) = 
2/3$ (a), $2/5$ (b), 
$2$ (c,d), and $6$ (e,f).
}
\label{fig:3body_correlation_N4-6_some_U}
\end{figure}

\subsection{Three-body correlations for $N=4$ and $6$}
\label{subsec:Three_body_cor_N4}

We have calculated  the three-body correlation functions 
 $\chi_{\sigma\sigma\sigma}^{[3]}$, 
 $\widetilde{\chi}_{\sigma\sigma'\sigma'}^{[3]}$ and 
 $\widetilde{\chi}_{\sigma\sigma'\sigma''}^{[3]}$ 
for $\sigma \neq \sigma' \neq\sigma'' \neq \sigma$, 
using  Eqs.\ \eqref{eq:chi3_SUN_1}--\eqref{eq:chi3_SUN_3}. 
Figure \ref{fig:3body_correlation_N4-6_some_U} shows 
 the dimensionless three-body correlations   
$\Theta_{\mathrm{I}}^{}$, $\widetilde{\Theta}_{\mathrm{II}}^{}$ 
and $\widetilde{\Theta}_{\mathrm{III}}^{}$, 
defined in Eqs.\  \eqref{eq:Theta_I_and_II_tilde_definition} 
and \eqref{eq:Theta_III_tilde_definition}.  
The left and right panels describe the NRG results in the SU(4) and SU(6) cases, 
respectively, and three different 
interaction strengths (from weak to strong)   
are chosen for the top, middle, and bottom panels.

All components of the three-body correlation functions vanish, 
$\Theta_{\mathrm{I}}^{}=\widetilde{\Theta}_{\mathrm{II}}^{}
=\widetilde{\Theta}_{\mathrm{III}}^{}=0$,  
at the electron-hole symmetric point $\xi_{d}^{} = 0$,
and evolve as $\xi_d^{}$ deviates from this point.
Among the three independent components, 
the intra-level component 
 $\Theta_{\mathrm{I}}^{}$ has the largest magnitude, 
and exhibits plateau structures for large $U$  
at integer filling points $N_d^{} \simeq 1$ , $2$, $\ldots$, $N-1$.  
The other components, 
 $\widetilde{\Theta}_{\mathrm{II}}^{}$ 
and $\widetilde{\Theta}_{\mathrm{III}}^{}$,  involve inter-level correlations   
and evolve as  the Coulomb interaction $U$ increases.  
In particular, the correlation between the three different levels 
 $\widetilde{\Theta}_{\mathrm{III}}^{}$ becomes the weakest. 
In the limit of $|\xi_{d}^{}|\to \infty$, the diagonal component 
$\Theta_{\mathrm{I}}^{}$ approaches the noninteracting value 
while the other two vanish:
\begin{align}
\Theta_{\mathrm{I}}^{} 
\,\xrightarrow{|\xi_d^{}|\to \infty\,}\, 
-2 ,  
\quad
\widetilde{\Theta}_{\mathrm{II}}^{}\, 
 \xrightarrow{|\xi_d^{}|\to \infty\,}
\, 0,
\quad
\widetilde{\Theta}_{\mathrm{III}}^{}\, 
 \xrightarrow{|\xi_d^{}|\to \infty\,}
\, 0.
\label{eq:Theta_I_empty_limit}
\end{align}

Figures \ref{fig:3body_correlation_N4-6_some_U}(e) and  
 \ref{fig:3body_correlation_N4-6_some_U}(f) clearly demonstrate   
the relation  Eq.\ \eqref{eq:three_cody_cor_relations_strong_U_simplified}, 
which holds at  $|\xi_d^{}| \lesssim (N-1)U/2$ for large $U$:  
\begin{align}
\Theta_{\mathrm{I}}^{}
\,\simeq\,-\widetilde{\Theta}_{\mathrm{II}}^{}
\,\simeq\,\widetilde{\Theta}_{\mathrm{III}}^{}\,. 
\label{eq:Theta_I_II_III_relations_strong_U_simplified}
\end{align}
It means that all the three-body components are determined 
by a single parameter $\chi_{B}^{[3]}$ in the strong-coupling region  for large $U$, 
as mentioned.  
The dimensionless three-body correlation functions clearly exhibit the plateau structure 
around the integer filling points 
 $\xi_{d}^{}= 0$, $\pm U$, $\pm2U$, $\ldots$,  $\pm (N-2)U/2$,
 where the SU($N$) Kondo effect occurs.
These structures reflect the behavior of  $\chi_{B}^{[3]}$,  
 described in Figs.\ \ref{fig:DChiC_DChiS_chi3B_N4-6_u6}(e) 
and \ref{fig:DChiC_DChiS_chi3B_N4-6_u6}(f),   
and evolve as the interaction strength $U$ increases. 
We see, in Figs.\ \ \ref{fig:3body_correlation_N4-6_some_U}(e) and  
 \ref{fig:3body_correlation_N4-6_some_U}(f),  
that the plateaus appear much clearer for $N=4$ than  $N=6$,  
for the same interaction strength $U/(\pi \Delta) =6.0$.  

We will examine, in the subsequent sections,   
how these three-body correlation functions 
affect the next-leading order terms of the transport coefficients
 in the low-energy  Fermi liquid regime away from half filling.

\section{Nonlinear current noise of \\ SU(4)  $\&$ SU(6) quantum dots}
\label{sec:NRG_CvCs}

In this section,  we discuss  
the nonlinear terms of the steady current $J$ 
and the current noise  $S_\mathrm{noise}^{\mathrm{QD}}$, 
specifically, the order $(eV)^3$ term   
for symmetric junctions, i.e.,  
$\Gamma_L^{}=\Gamma_R^{}$ and $\mu_L^{}=-\mu_R^{}= {eV}/{2}$. 
To provide a comprehensive view of the low-bias behavior of these terms, 
 we begin with a brief review on  
the previous results for the coefficient $C_{V}^{}$ 
of the nonlinear conductance \cite{TsutsumiSUNpaper},  
 extending slightly the interaction range up to $U/(\pi\Delta)=6.0$.
We will then discuss the results for  $C_{S}^{}$,  i.e., 
the order $|eV|^3$ term of nonlinear noise.

\subsection{$C_{V}^{}$: order  $(eV)^2$ term of $dJ/dV$}

The leading-order term of the conductance, at $T=0$, 
is determined by 
the transmission probability $\sin^2 \delta$,  i.e., 
 the first term in the right-hand side of Eq.\ \eqref{eq:FL_cond_QD}.
It exhibits the well-known Kondo plateaus, 
which develop in the  strong-coupling region,  
as shown in Figs.\ \ref{fig:FL-parameters-N4-U-from-2-over-3-to-6}(b)
and \ref{fig:FL-parameters-N6-U-from-2-over-5-to-6}(b) 
 in Appendix \ref{sec:NRG_results_2body_functions_for_sun}.

\begin{figure}[t]

 \leavevmode
  \centering
 
\includegraphics[width=0.47\linewidth]{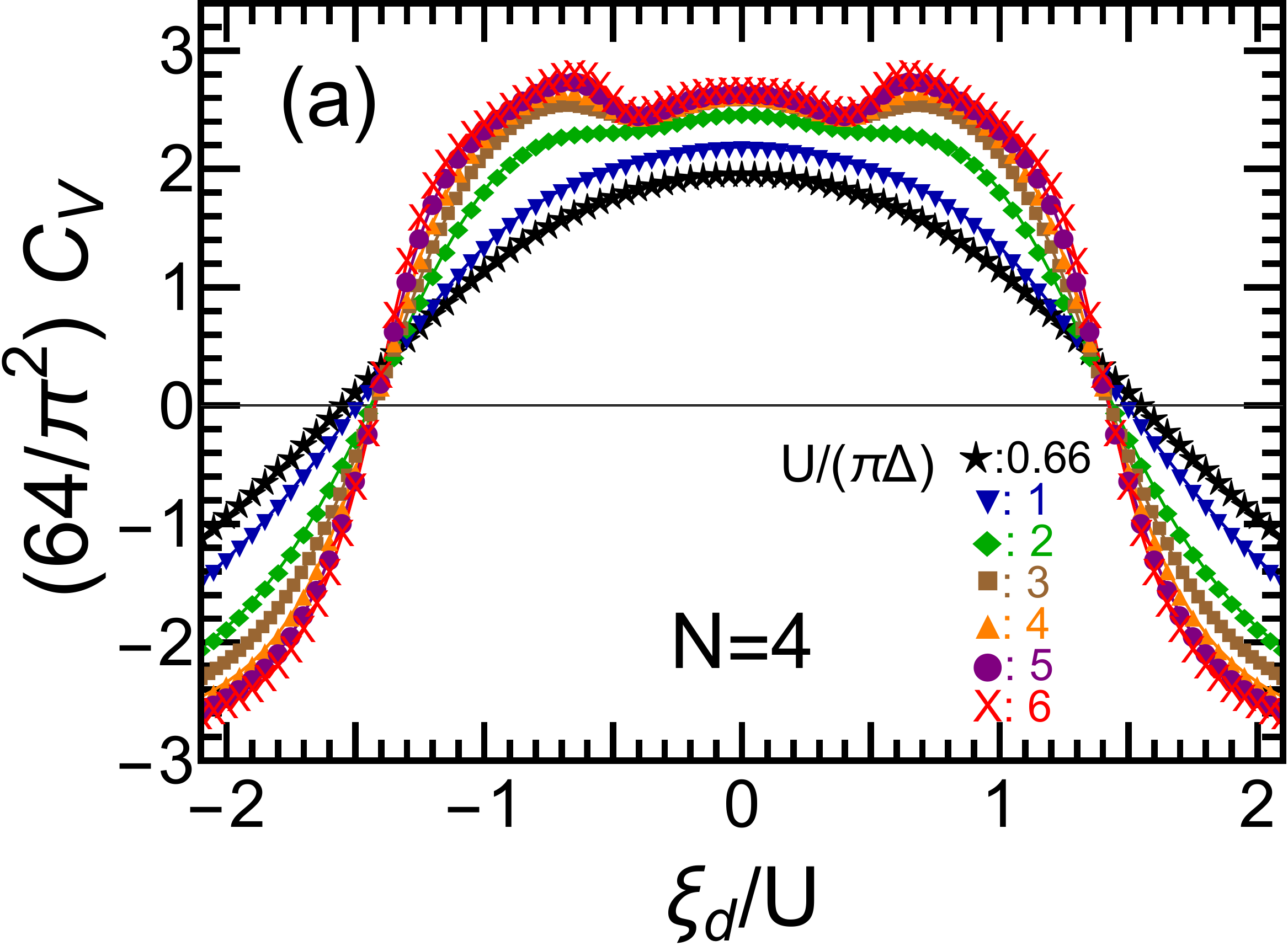}
  \hspace{0.01\linewidth} 
\includegraphics[width=0.47\linewidth]{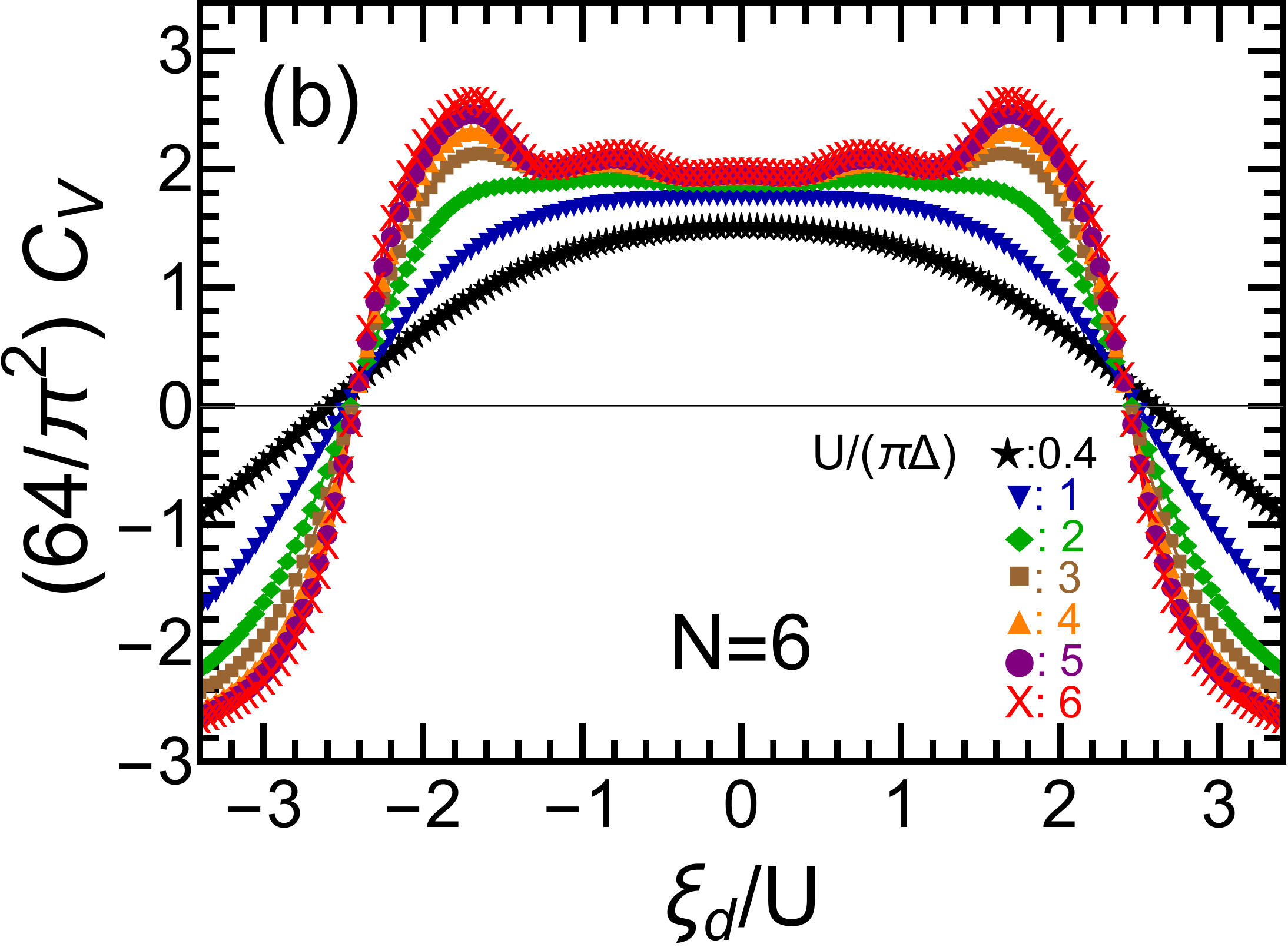}
 \\
\includegraphics[width=0.47\linewidth]{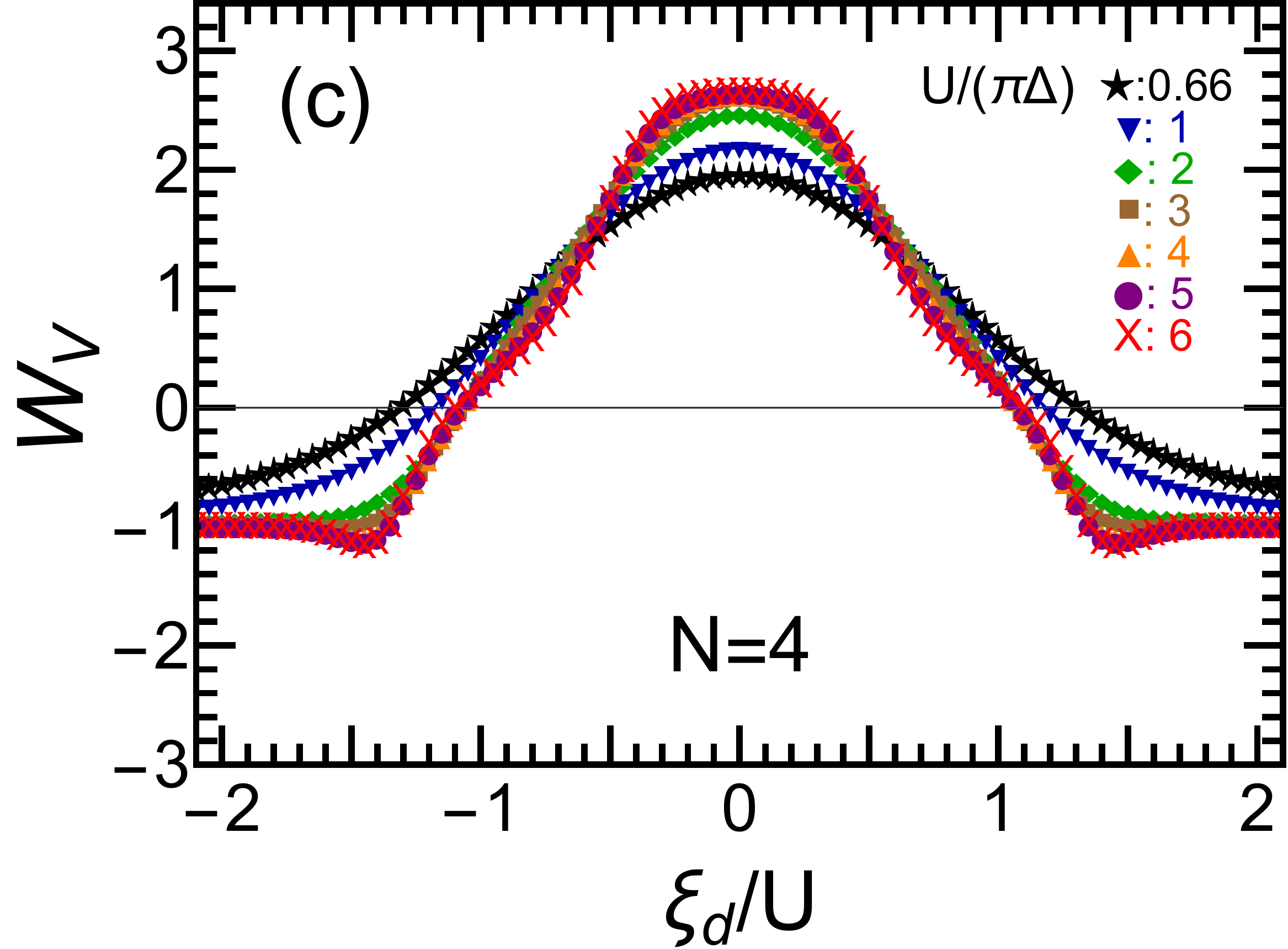}
  \hspace{0.01\linewidth} 
\includegraphics[width=0.47\linewidth]{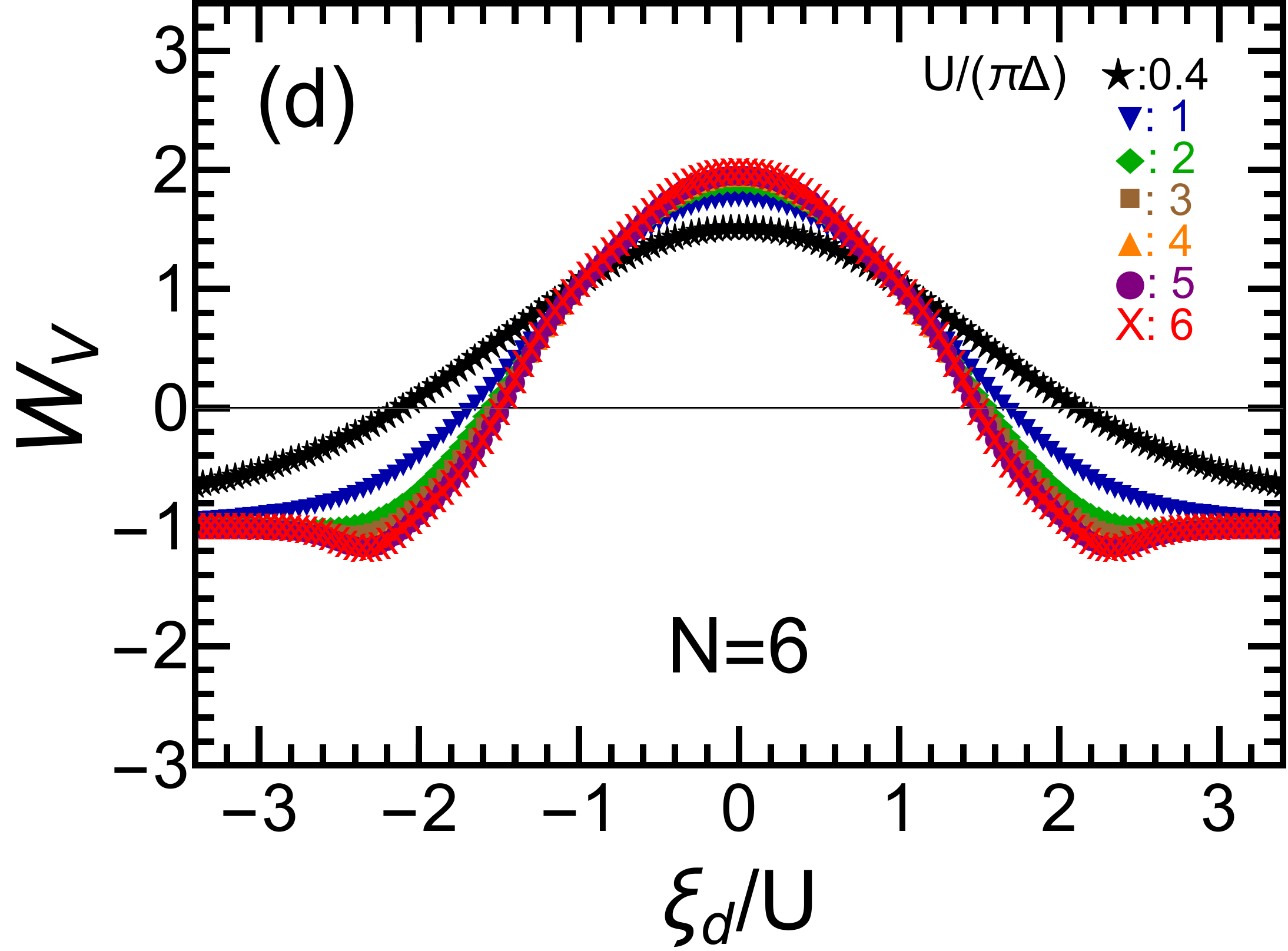}
 \\
\includegraphics[width=0.47\linewidth]{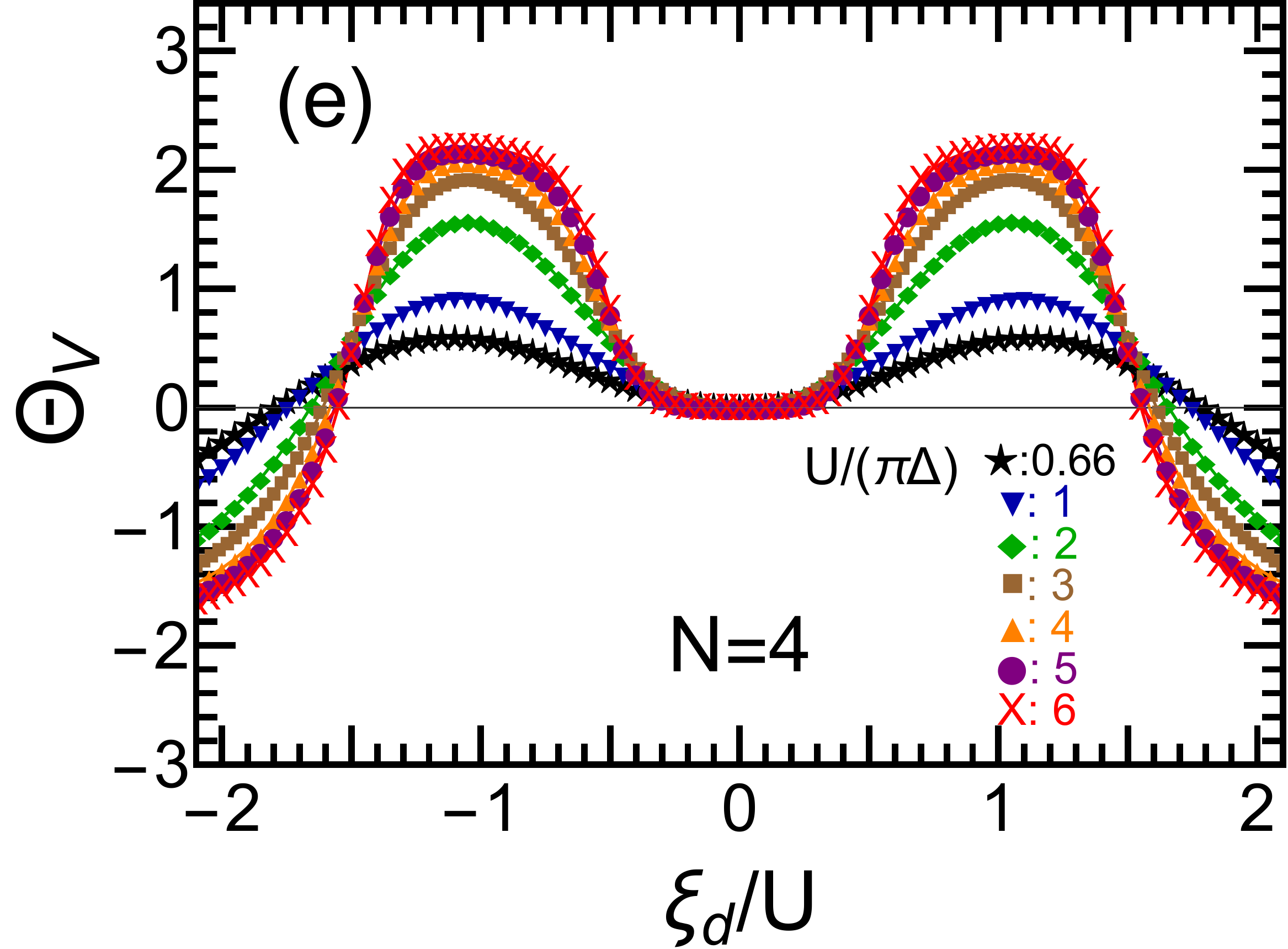}
  \hspace{0.01\linewidth} 
\includegraphics[width=0.47\linewidth]{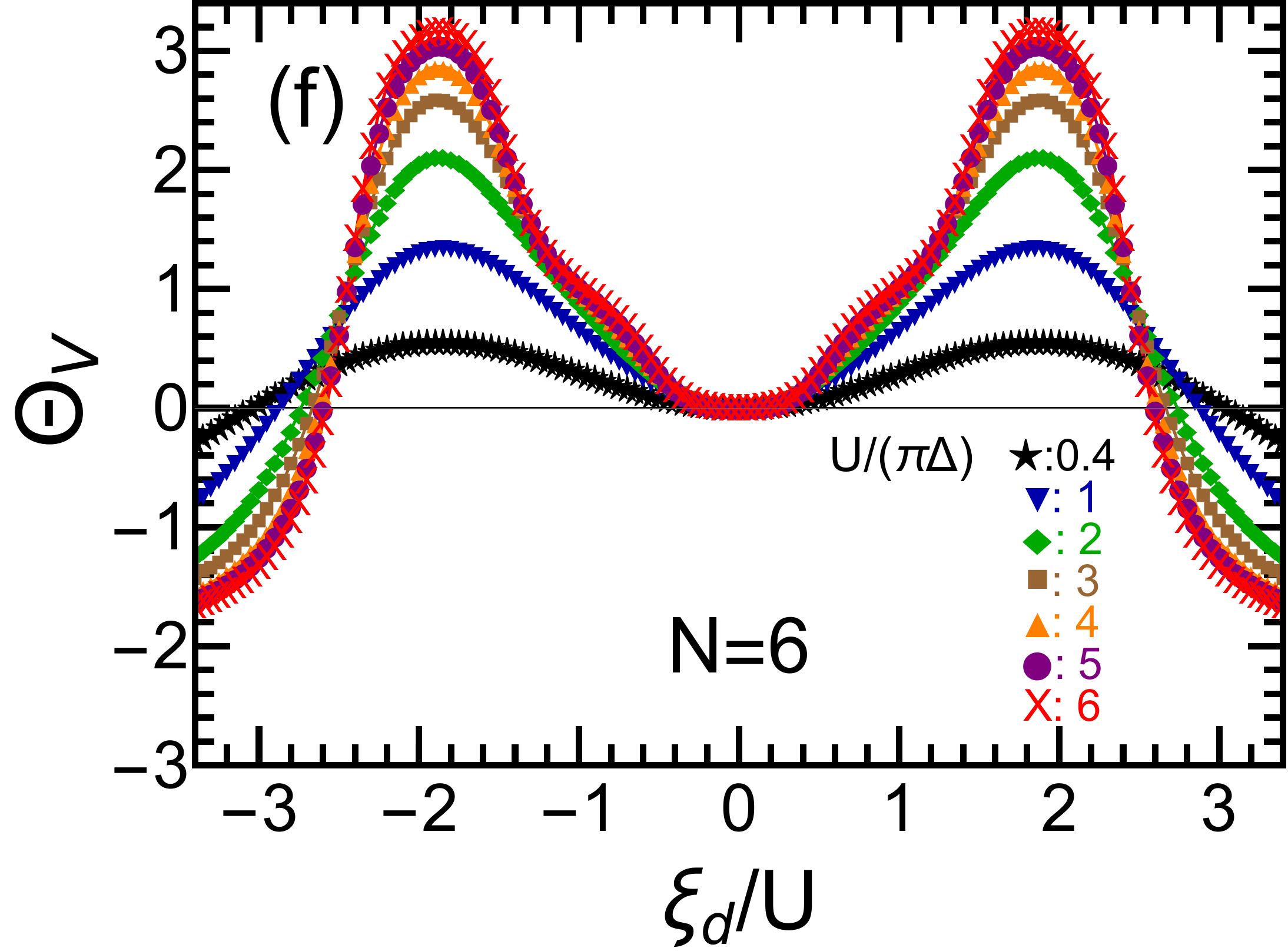}
\caption{$\xi_d^{}$ dependence of 
 $C_{V}^{}=(\pi^2/64) (W_{V}^{}+\Theta_{V}^{})$,  
two-body part $W_{V}^{}$, and three-body part 
$\Theta_{V}^{} \equiv \Theta_{\mathrm{I}}^{}
+3\,\widetilde{\Theta}_{\mathrm{II}}^{}$. 
Left panels: $N=4$, for 
$U/(\pi\Delta) = 2/3(\star)$, $1(\blacktriangledown)$, $2(\blacklozenge)$, $3(\blacksquare)$, $4(\blacktriangle)$, $5(\bullet)$, $6(\times)$.
Right panels: $N=6$, for 
$U/(\pi\Delta) = 2/5(\star)$, $1(\blacktriangledown)$, $2(\blacklozenge)$, 
$3(\blacksquare)$, $4(\blacktriangle)$, $5(\bullet)$, $6(\times)$.} 
\label{fig:CV-QD-N4-6-U}
\end{figure}

The next-leading order term $C_{V}^{}$ of  the nonlinear conductance 
 can be decomposed into the two-body part  $W_{V}^{}$, 
defined in Table \ref{tab:C_and_W_extended},
 and  the three-body part $\Theta_{V}^{}$, as     
\begin{align}
C_{V}^{} 
\, =  \, 
\frac{\pi^2}{64}
 \,\bigl(W_{V}^{} + \Theta_{V}^{}
\bigr) , \qquad
\Theta_{V}^{} \,\equiv \, 
\Theta_{\mathrm{I}}^{} +  3\,\widetilde{\Theta}_{\mathrm{II}}^{} .
\label{eq:def_Theta_V}
\end{align}
These coefficients  
 $C_{V}^{}$, $W_{V}^{}$ and $\Theta_{V}^{}$ 
are plotted vs $\xi_{d}^{}$ 
in Figs.\ \ref{fig:CV-QD-N4-6-U}(a)--\ref{fig:CV-QD-N4-6-U}(f) 
for $N=4$ and $6$.

The two-body part $W_{V}^{}$ dominates 
 the next-leading order term 
near the electron-hole symmetric point $\xi_{d}^{}= 0$, 
where $\delta = \pi/2$ and 
 the three-body part $\Theta_{V}^{}$ vanishes:   
\begin{align}
 W_{V}^{}\,\xrightarrow{\, \xi_d^{}=0}\, 
 1 + \frac{5 \widetilde{K}^2}{N-1} , 
\qquad 
\Theta_{V}^{} \xrightarrow{\,\xi_d^{}=0\,} 0.
\end{align}
Therefore,  in the strong interaction limit,  the peak at $\xi_d^{}=0$ reaches   
 $(64/\pi^2) C_{V}^{}\xrightarrow{\, \xi_d^{}=0\,\& U\to \infty} 8/3$ 
and $2$ for $N=4$ and $6$, respectively.
Note that the rescaled Wilson ratio approaches  $\widetilde{K} \simeq 1$
in a wide region of $|\xi_d^{}| \lesssim (N-1)U/2$ for large $U$. 
Outside this region, the two-body and three-body parts 
of $ C_{V}^{}$ approach to the noninteracting values, 
\begin{align}
W_{V}^{}
\,\xrightarrow{\, |\xi_d^{}| \to \infty}\, -1, 
\qquad
\Theta_{V}^{}
\,\xrightarrow{\, |\xi_d^{}| \to \infty}\, -2,  
\end{align}
since  $\cos 2\delta \xrightarrow{\, |\xi_d^{}| \to \infty} 1$, 
$\widetilde{K} \xrightarrow{\, |\xi_d^{}| \to \infty} 0$, and 
the three-body contributions approach the values given 
in Eq.\ \eqref{eq:Theta_I_empty_limit}.

The SU($N$) Kondo effect occurs in the strong-coupling region 
 $|\xi_{d}^{}| \lesssim (N-1)U/2$ at the integer filling points.  
The plateau structure evolves as $U$ increases, 
especially in the three-body part $\Theta_{V}^{}$, 
as seen in Figs.\ \ref{fig:CV-QD-N4-6-U}(e) and \ref{fig:CV-QD-N4-6-U}(f).   
In the strong-coupling region,    
it can be expressed in the form $\Theta_{V}^{} \simeq -2\Theta_\mathrm{I}^{}$ 
due to  Eq.\ \eqref{eq:Theta_I_II_III_relations_strong_U_simplified}. 
The two-body part $W_{V}^{}$, 
shown in Figs.\ \ref{fig:CV-QD-N4-6-U}(c) and \ref{fig:CV-QD-N4-6-U}(d) 
for $U/(\pi \Delta) \leq 6.0$, 
 does not exhibit a clear plateau other than the one appearing near half filling.  
For $N=4$, this is because the factor $\cos 2\delta$ 
for $W_{V}^{}$ vanishes at $\delta =\pi/4$ and $3\pi/4$, 
i.e., at the quarter and three-quarters filling points.  
As a sum of $W_{V}^{}$ and $\Theta_{V}^{}$, 
the coefficient $C_{V}^{}$ exhibits a wide and rather flat structure   
in the region of $|\xi_{d}^{}| \lesssim (N-1)U/2$ for large $U$,  
seen in Figs.\ \ref{fig:CV-QD-N4-6-U}(a) and \ref{fig:CV-QD-N4-6-U}(b).  
There also emerge some weak local maxima in this flat structure 
at the integer filling points 
 $\xi_{d}^{} = 0$, $ \pm U$, $ \pm 2U$, $\ldots$,  $\pm (N-2)U/2$. 
In particular, the peak is most pronounced 
 at  $\xi_{d}^{} \simeq \pm (N-2)U/2$, 
where  $N_d^{}\simeq 1$ or  $N-1$.

\begin{figure}[t]

 \leavevmode
 \centering

\includegraphics[width=0.47\linewidth]{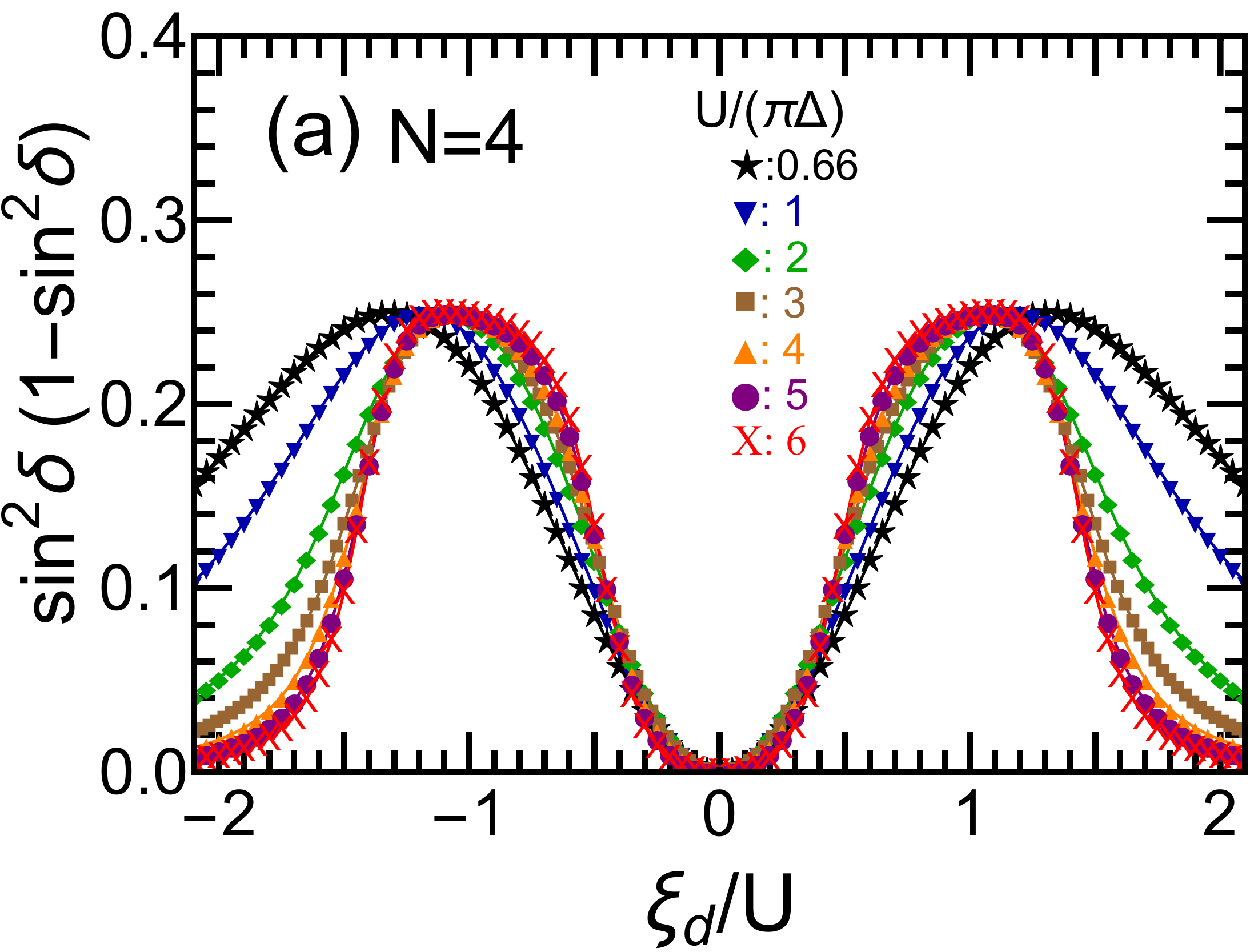}
 \hspace{0.01\linewidth} 
\includegraphics[width=0.47\linewidth]{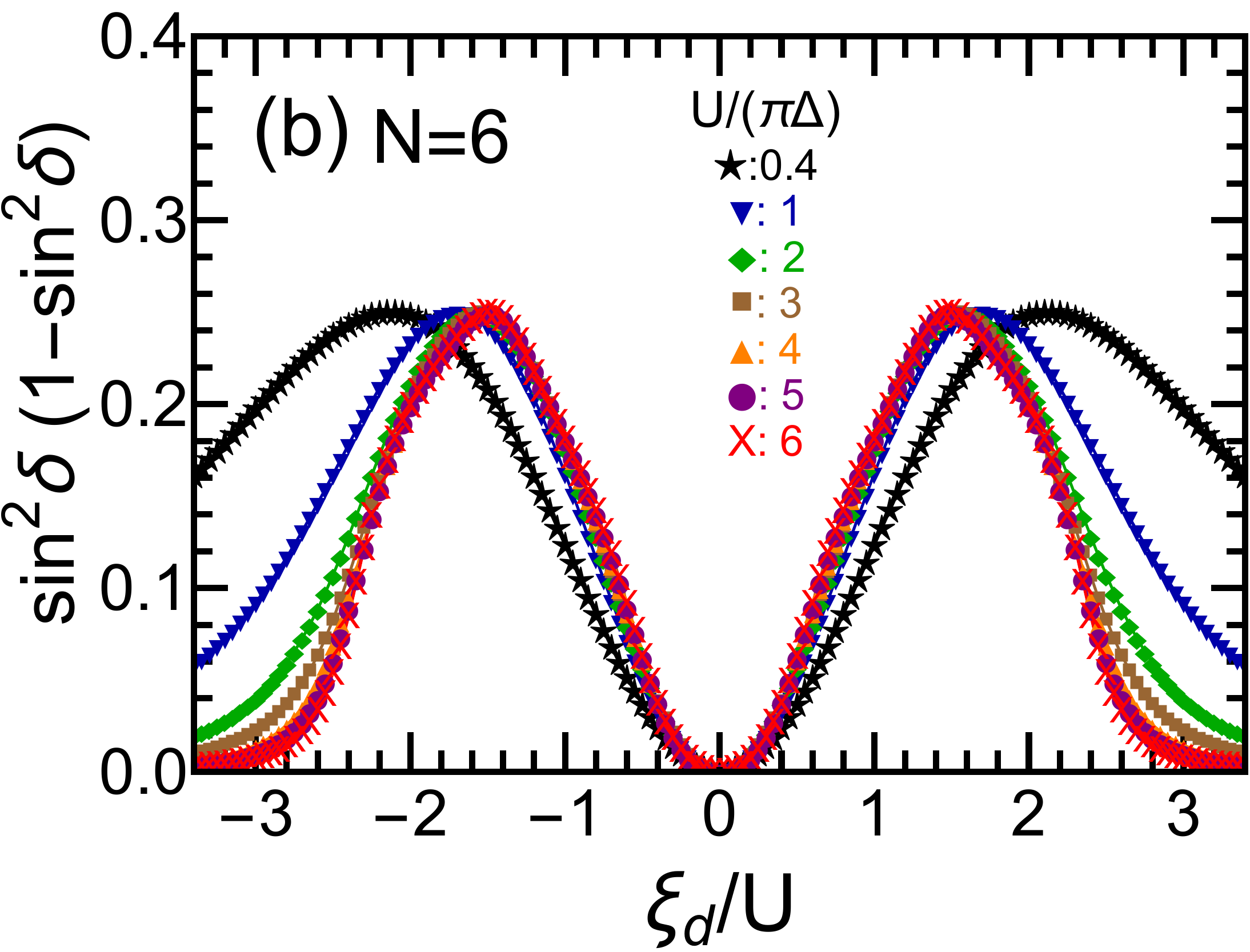}
\caption{Linear part,  
$\sin^{2}\delta \, (1-\sin^{2}\delta)$,  
of the noise $S_\mathrm{noise}^\mathrm{QD}$.
Interaction strengths are chosen for (a) $N=4$ to be 
$U/(\pi\Delta) = 2/3 (\star)$, $1 (\blacktriangledown)$, 
$2 (\blacklozenge)$, $3 (\blacksquare)$,
 $4 (\blacktriangle)$, $5 (\bullet)$, $6 (\times)$. 
For (b) $N=6$,  
$U/(\pi\Delta) = 2/5 (\star)$, $1 (\blacktriangledown)$, 
$2 (\blacklozenge)$, $3 (\blacksquare)$,
 $4 (\blacktriangle)$, $5 (\bullet)$, $6 (\times)$. 
} 
\label{fig:LinearNoise}
\end{figure}

\subsection{Order $|eV|$ and  Order $|eV|^3$ terms 
of $S_\mathrm{noise}^\mathrm{QD}$}
\label{subsec:CS_N4-6}

We next consider the current noise $S_\mathrm{noise}^\mathrm{QD}$,  
the low-energy asymptotic form of which is given in Eq.\ \eqref{eq:FL_noise} 
and Table  \ref{tab:C_and_W_extended}.
The leading-order term, 
$\sin^{2} \delta\, (1-\sin^{2} \delta)=(1-\cos 4\delta)/8$, 
corresponds to the linear-response noise,    
the NRG results for which  are shown as a function of $\xi_d^{}$ 
in Figs.\ \ref{fig:LinearNoise}(a) and \ref{fig:LinearNoise}(b),  
 for $N=4$ and $6$, respectively.  
The linear noise is maximized at the points where 
the phase shift reaches $\delta=\pi/4$ and $3\pi/4$.   
It occurs at the integer filling points $N_d^{}=1$ and $3$ for $N=4$,  
at which the SU(4) Kondo effect makes the peaks wide and flat. 
In contrast, for $N=6$, the peak emerges   
at the half-integer filling points $N_d^{}=3/2$ and  $9/2$. 
More generally, 
the peak of the linear noise forms a flat plateau structure 
for $N\equiv 0$ (mod $4$),
whereas the peak becomes round for $N\equiv 2$ (mod $4$) 
due to the fluctuations occurring between two adjacent integer filling states.  
For large $U$, the quarter and three-quarters fillings 
 occur near  $|\xi_d^{}|/U \simeq N/4$,
at which the ground state is highly correlated for multilevel systems of $N\geq 4$.   
In contrast, these fillings occur at $\xi_d^{} \simeq \pm U/2$ for SU(2) quantum dots, 
where the electron correlation becomes less important 
due to the valence fluctuations (see Appendix \ref{sec:NoiseN2}).

\begin{figure}[t]

\leavevmode
 \centering

\includegraphics[width=0.47\linewidth]{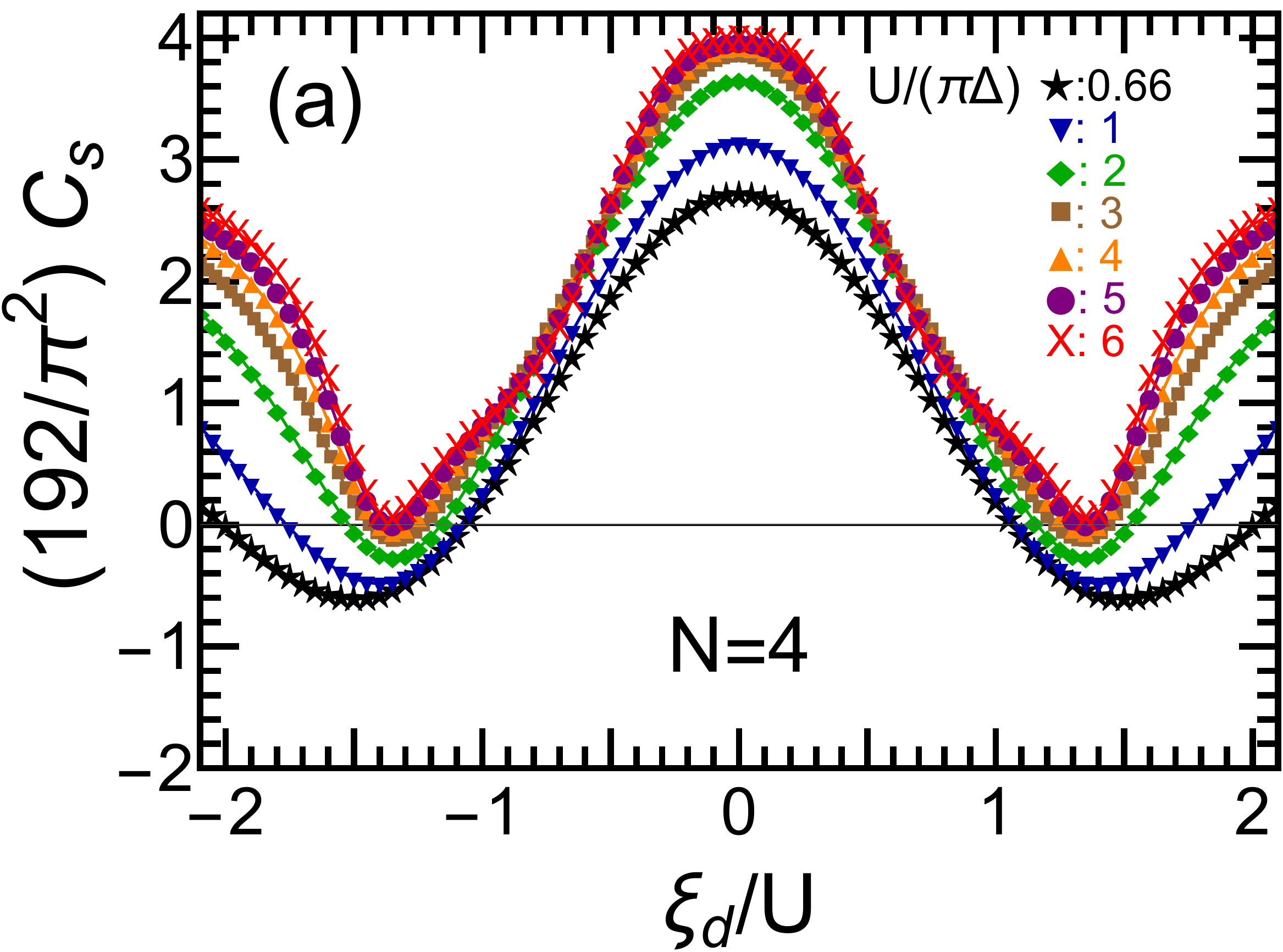}
 \hspace{0.01\linewidth} 
\includegraphics[width=0.47\linewidth]{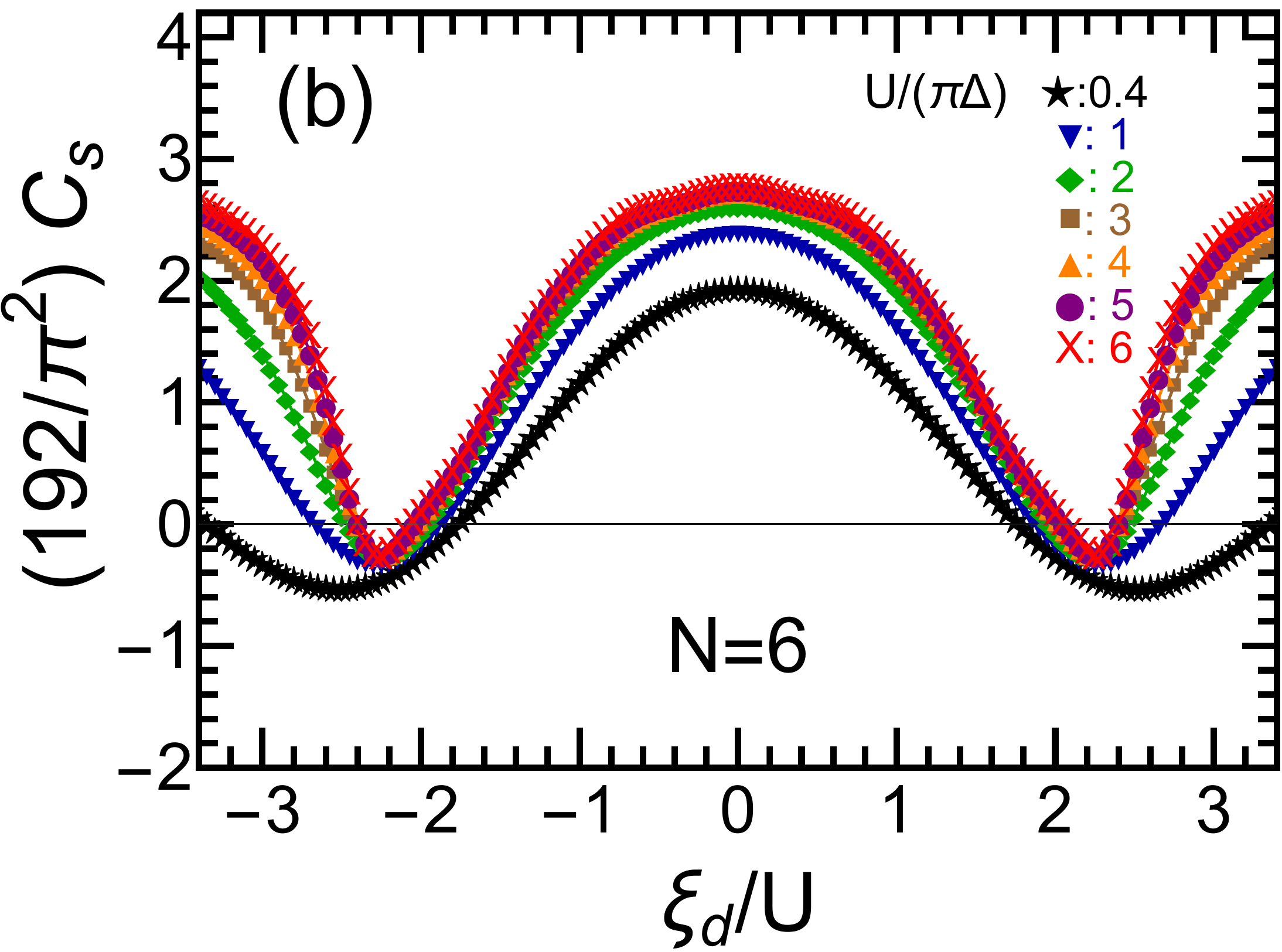}
\\
\includegraphics[width=0.47\linewidth]{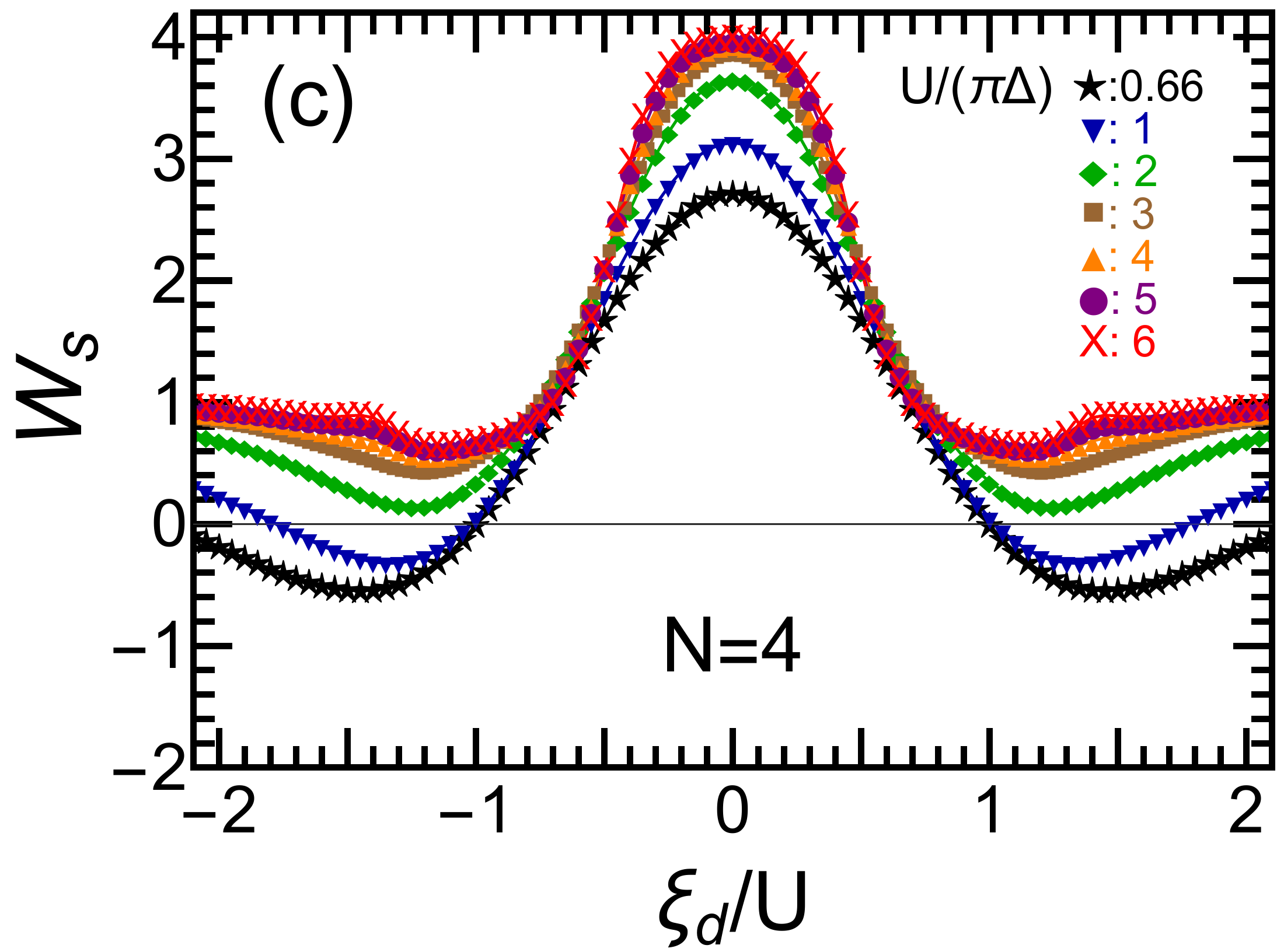}
 \hspace{0.01\linewidth} 
\includegraphics[width=0.47\linewidth]{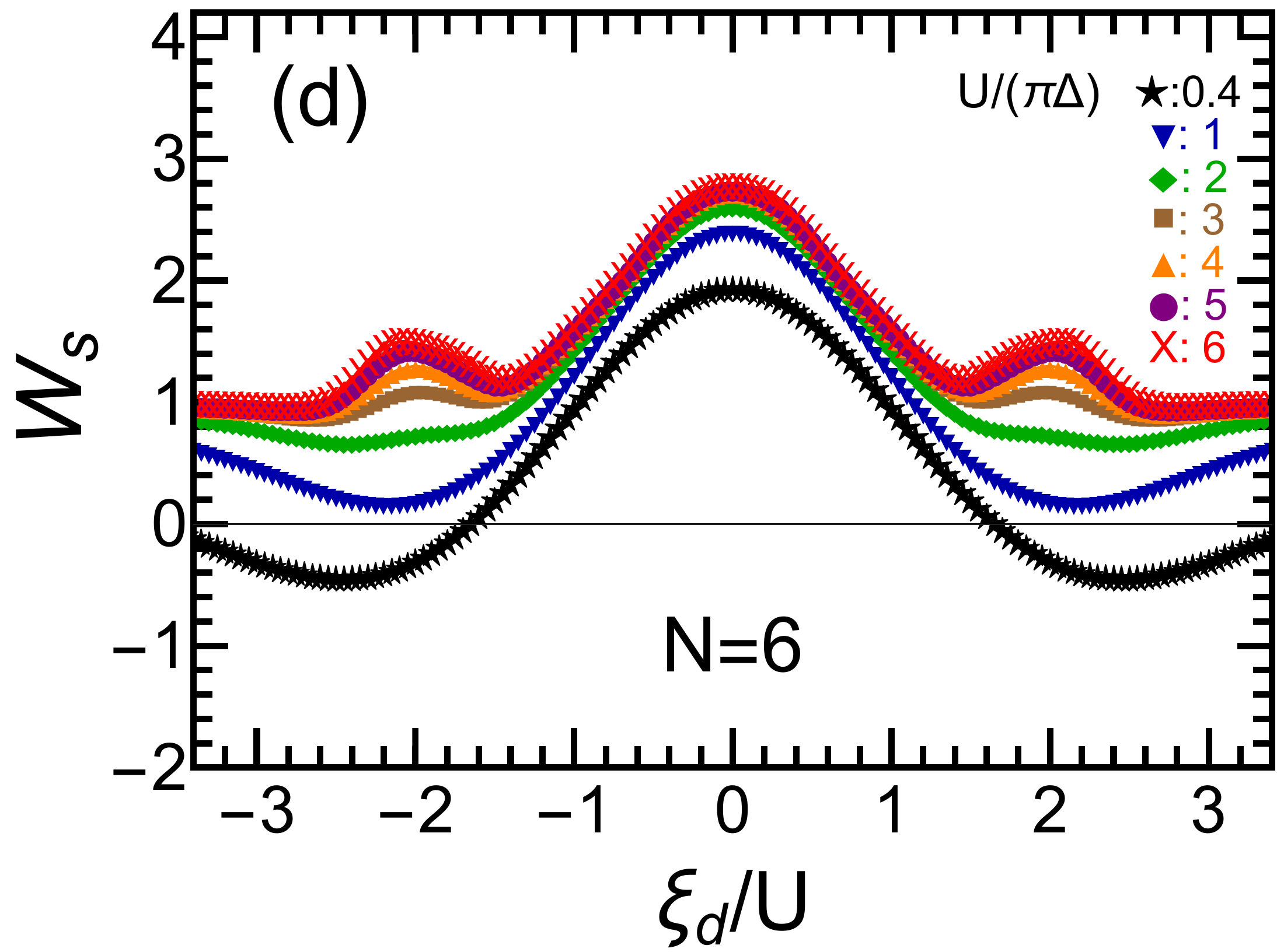}
\\
\includegraphics[width=0.47\linewidth]{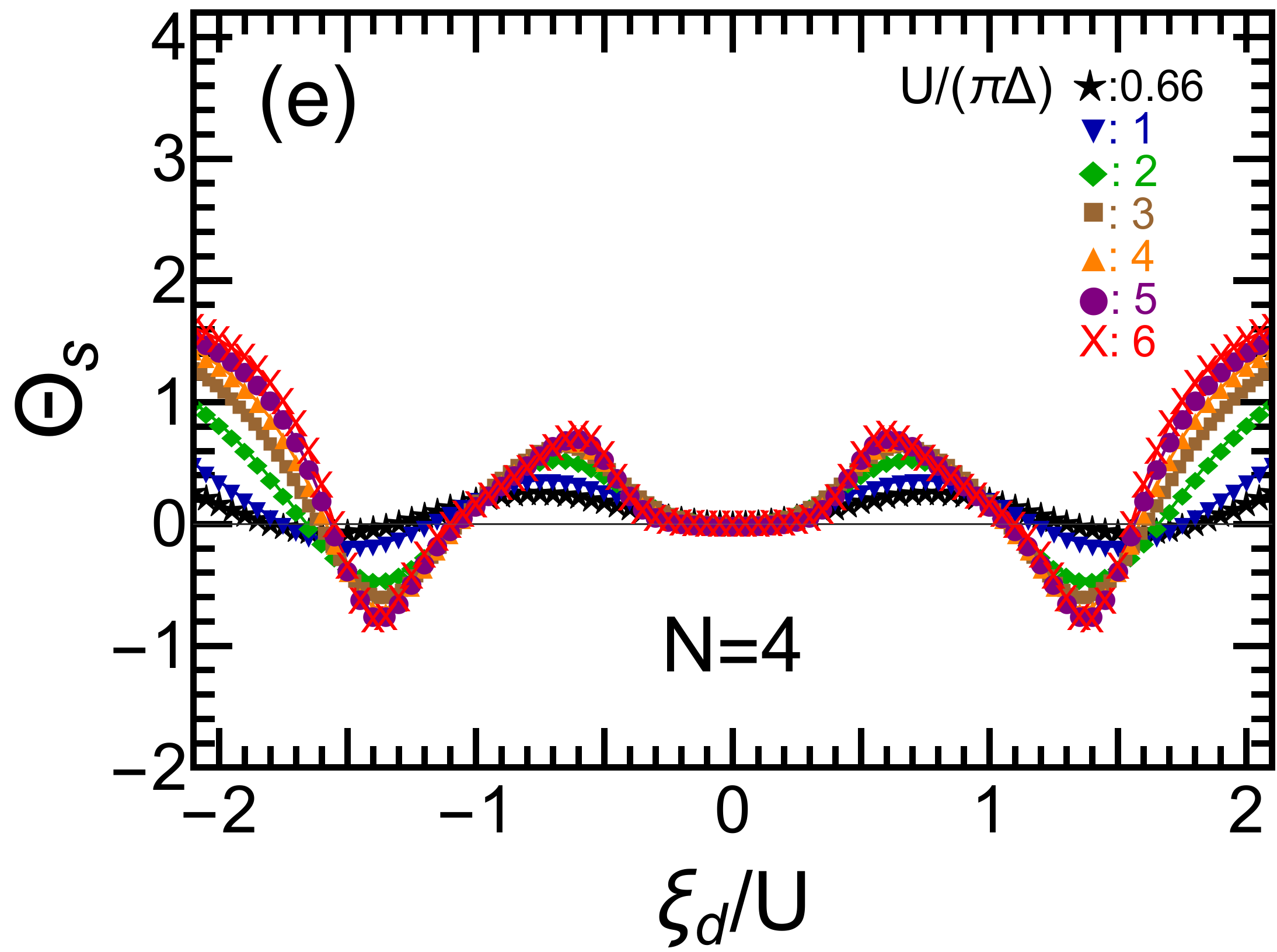}
 \hspace{0.01\linewidth} 
\includegraphics[width=0.47\linewidth]{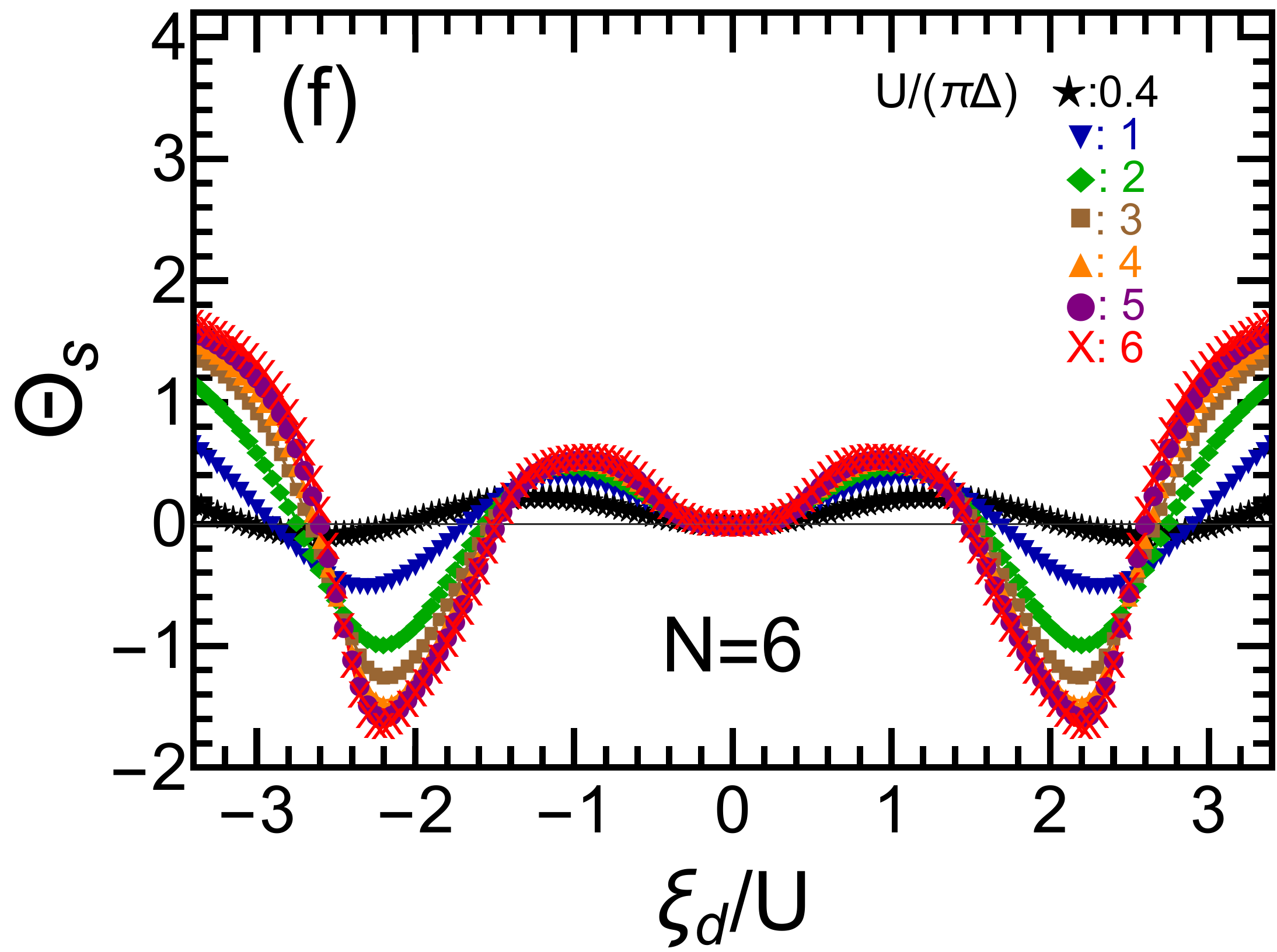}

\caption{
$\xi_d^{}$ dependence of 
 $C_{S}^{}=(\pi^2/192) (W_{S}^{}+\Theta_{S}^{})$,  
two-body part $W_{S}^{}$, and three-body part 
$\Theta_{S}^{} \equiv - \bigl(\Theta_\mathrm{I}^{} 
+ 3 \,\widetilde{\Theta}_{\mathrm{II}}^{} 
\bigr)  \cos 2\delta$. 
Left panels: $N=4$, for 
$U/(\pi\Delta) = 2/3(\star)$, $1(\blacktriangledown)$, $2(\blacklozenge)$, $3(\blacksquare)$, $4(\blacktriangle)$, $5(\bullet)$, $6(\times)$.
Right panels: $N=6$, for 
$U/(\pi\Delta) = 2/5(\star)$, $1(\blacktriangledown)$, $2(\blacklozenge)$, 
$3(\blacksquare)$, $4(\blacktriangle)$, $5(\bullet)$, $6(\times)$.
}

\label{fig:CS-QD-N4-6-U}

\end{figure}

The coefficient $C_{S}^{}$ for 
the order  $|eV|^{3}$ term of current noise $S_{\mathrm{noise}}^{\mathrm{QD}}$
can also be decomposed into the two-body $W_{S}^{}$ and 
 three-body $\Theta_{S}^{}$ parts, 
as shown in Table \ref{tab:C_and_W_extended}:
\begin{align}
\!\!
C_{S}^{} 
=    
\frac{\pi^2}{192}
\bigl(
W_{S}^{} 
+ 
\Theta_{S}^{}
\bigr), 
\quad 
\Theta_{S}^{} 
 \equiv  - \bigl(\Theta_\mathrm{I}^{} + 3 \widetilde{\Theta}_{\mathrm{II}} 
\bigr)  \cos 2\delta .
\label{eq:def_Theta_S}
\end{align}
The behavior of  three-body part  $\Theta_{S}^{}$ can be deduced from 
the one for $dJ/dV$, 
i.e., $\Theta_{V}^{}= 
\Theta_\mathrm{I}^{} + 3 \widetilde{\Theta}_{\mathrm{II}}$ 
shown in Figs.\ \ref{fig:CV-QD-N4-6-U}(e) and \ref{fig:CV-QD-N4-6-U}(f), 
by multiplying them by a factor of ``$-\cos 2\delta$" which induces the  modulations.   
One of the most distinctive features of  $C_{S}^{}$, compared to   
 the other coefficients $C$'s listed in Table \ref{tab:C_and_W_extended},
is that it depends upon the higher harmonics $\cos 4\delta$ and $\sin 4\delta$  
with respect to the phase shift,  
which enter through not only through $W_{S}^{}$ but $\Theta_{S}$: 
note that $\Theta_\mathrm{I}^{}$ and $\widetilde{\Theta}_{\mathrm{II}}^{}$  
defined in Eq.\ \eqref{eq:Theta_I_and_II_tilde_definition}  
are proportional to the factor $\sin 2\delta$. 
As  $\xi_d^{}$ varies, these higher harmonics evolve continuously 
in the range $0 \leq 4\delta \leq 4\pi$,  
simultaneously with the electron filling  $0 \leq N_d^{} \leq N$. 
Figures \ref{fig:CS-QD-N4-6-U}(a)--\ref{fig:CS-QD-N4-6-U}(f) 
show results for $C_{S}^{}$, $W_{S}^{}$, and  $\Theta_{S}^{}$ 
for $N=4$ and $6$.

Each curve for $C_{S}^{}$,  
shown in Figs.\ \ref{fig:CS-QD-N4-6-U}(a) and \ref{fig:CS-QD-N4-6-U}(b),  
exhibits two valleys situated  at the valence fluctuation regions 
$\xi_d^{} \simeq \pm (N-1)U/2$ for large $U$. 
These valleys rise all around as $U$ increases. 
Notably, for $N=4$, the bottom value of $C_{S}^{}$ 
turns positive  for large interactions $U/(\pi\Delta) \gtrsim 5$, 
while for $N=6$, it remains negative 
even for the largest interaction $U/(\pi\Delta)=6$ 
examined in this study.

Near the electron-hole symmetric point $\xi_{d}^{}= 0$,  
where  $\delta = \pi/2$ and  $N_d^{}=N/2$,  
the two-body part $W_{S}^{}$ dominates the nonlinear noise coefficient $C_{S}^{}$  
since the three-body correlations disappear around this point: 
\begin{align}
 W_{S}^{}\,\xrightarrow{\, \xi_d^{}=0\,}\, 
 1 + \frac{9 \widetilde{K}^2}{N-1} \,, 
\qquad 
\Theta_{S}^{} \xrightarrow{\,\xi_d^{}=0\,} 0. 
\label{eq:Cs_half-filling}
\end{align}
In particular,  in the limit of  $U \to \infty$, 
the rescaled Wilson ratio approaches  the saturation value $\widetilde{K} \to 1$. 
Hence, the coefficient for the order $|eV|^3$ term 
 reaches  $(192/\pi^2) C_{S}^{}\xrightarrow{\xi_d^{}=0\,\& \,U\to \infty} 4$ 
for $N=4$,  and it reaches $14/5$ for $N=6$:  
the  height of this ridge decreases as $N$ increases.

In contrast, in the opposite limit $|\xi_{d}^{}| \to \infty$, 
both the two-body and the three-body parts approach the noninteracting values  
\begin{align}
W_{S}^{}
\,\xrightarrow{\, |\xi_d^{}| \to \infty}\, 1, 
\qquad
\Theta_{S}^{}
\,\xrightarrow{\, |\xi_d^{}| \to \infty}\, 2, 
\label{eq:Cs_empty}
\end{align}
since 
 $\cos 2\delta \to 1$, $\widetilde{K} \to 0$, and 
$\Theta_{\mathrm{I}}^{}\to  -2$ in this limit.
Therefore, $W_{S}^{}$ and $\Theta_{S}^{}$  
 contribute comparably to $C_{S}^{}$ 
in the region  $|\xi_d^{}| \gtrsim (N-1)U/2$, 
where the impurity levels are either 
almost empty $N_d^{}\simeq 0$ or 
fully occupied $N_d^{}\simeq N$. 
In particular,  as seen in   
Figs.\ \ref{fig:CS-QD-N4-6-U}(c) and \ref{fig:CS-QD-N4-6-U}(d),  
 the two-body part approaches 
the saturation value $W_{S}^{}\to 1$ 
already at $|\xi_d^{}|\simeq (N-1)U/2$ 
for large interactions $U/(\pi \Delta)\gtrsim 3$.

In the strong-coupling region $|\xi_d^{}|\lesssim (N-1)U/2$ for large $U$, 
the behavior of two-body part $W_{S}^{}$ 
is determined by the higher-harmonic $\cos 4\delta$ term, as   
\begin{align}
W_S^{} 
 \,   \simeq  \, 
\frac{1}{2} \left( 3+\frac{5}{N-1}\right)
\,+\,
\frac{1}{2} \left(\frac{13}{N-1} -1\right) \cos 4 \delta
\,,
%
\label{eq:Ws_largeU}
\end{align}
since  the Wilson ratio is locked at $\widetilde{K} \simeq 1.0$ in this region.
 As seen in Figs.\ \ref{fig:CS-QD-N4-6-U}(c) 
and \ref{fig:CS-QD-N4-6-U}(d),    
the two-body part $W_S^{}$ has local minima 
 at quarter and three-quarters filling points,
which occur at  $\delta=\pi/4$ and $3\pi/4$, 
or  $|\xi_{d}^{}|/U\simeq  N/4$ for large $U$.
At these local minima, $W_S^{}$ take  a positive value for $N\geq 4$, 
as it can be deduced from Eq.\ \eqref{eq:Ws_largeU}. 
This is in contrast to the SU(2) case   
where the local minima of $W_S^{}$ take a negative value, 
as shown in Appendix \ref{sec:NoiseN2}.
In the strong-coupling region $|\xi_d^{}|\lesssim (N-1)U/2$,
 the three-body part  takes the following form,  for large $U$, 
\begin{align}
 \Theta_{S}^{}\, \simeq 
\, \frac{\chi_{\sigma\sigma\sigma}^{[3]}}{2 \pi \,\chi_{\sigma\sigma}^{2}} 
\,\sin 4\delta \,, 
\label{eq:ThetaS_StrongCoupl}
\end{align}
due to the property described 
in Eq.\ \eqref{eq:Theta_I_II_III_relations_strong_U_simplified}. 
Equation \eqref{eq:ThetaS_StrongCoupl} clearly shows that 
the three-body part  vanishes, $\Theta_S^{}=0$,  
at quarter $\delta=\pi/4$ and three-quarters $\delta= 3\pi/4$ fillings.  
Therefore, at these filling points,  $\Theta_S^{}$ 
does not exhibit  the plateau structures for $N=4$ in Fig.\ \ref{fig:CS-QD-N4-6-U}(e),
 despite the fact that 
$\Theta_V^{}$ clearly exhibits the plateau structures 
as seen in Fig.\ \ref{fig:CV-QD-N4-6-U}(e). 
 In contrast,
 for $N=6$, $\Theta_S^{}$ shows 
the clear plateau structures in Fig.\ \ref{fig:CS-QD-N4-6-U}(f) 
at the fillings of $N_d^{}= 2$ and $4$.
The three-body parts  $\Theta_S^{}$ for both $N=4$ and $6$ cases 
also have pronounced local minima near the valence fluctuation regions 
$\xi_d^{}\simeq \pm (N-1)U/2$, 
which cause the valley structure appearing in $C_S^{}$.  
A similar valley structure also emerges in  $C_S^{}$ for $N=2$, 
as shown in Appendix \ref{sec:NoiseN2}.
However, it stems from the two-body correlations  $W_S^{}$, 
instead of $\Theta_S^{}$,  in the SU($2$) case.

\section{Thermoelectric transport of \\ SU(4)  $\&$ SU(6) quantum dots}
\label{sec:NRG_CtCkappa}

We next consider the order  $T^{2}$ term 
of the linear conductance $g=dJ/dV\big |_{eV=0}^{}$ 
 and the order $T^3$ term of thermal conductance $\kappa_\mathrm{QD}^{}$ of 
SU($N$) quantum dots.

\subsection{$C_{T}^{}$: order  $T^2$ term of $dJ/dV$}

\begin{figure}[t]

\leavevmode
 \centering

\includegraphics[width=0.47\linewidth]{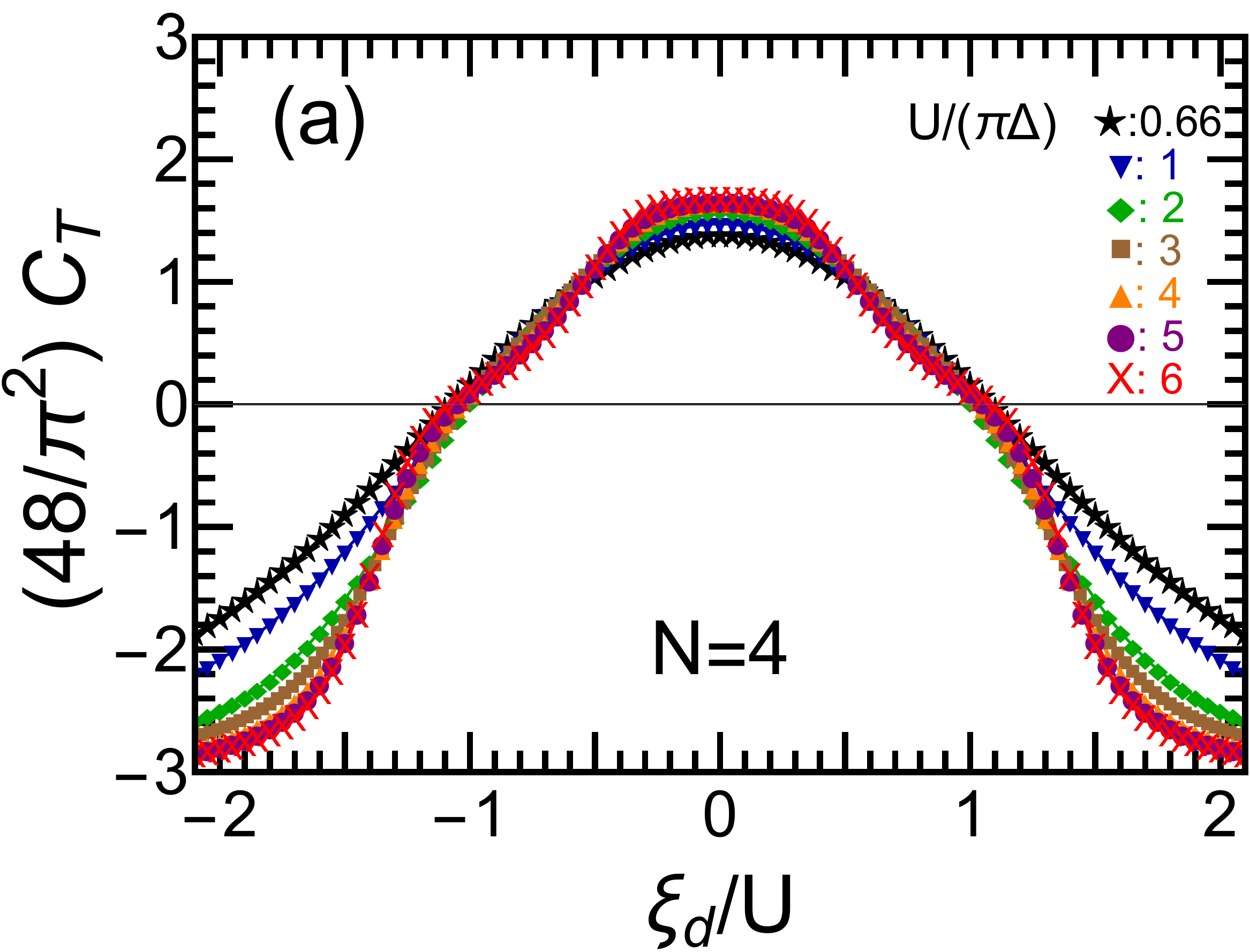}
 \hspace{0.01\linewidth} 
\includegraphics[width=0.47\linewidth]{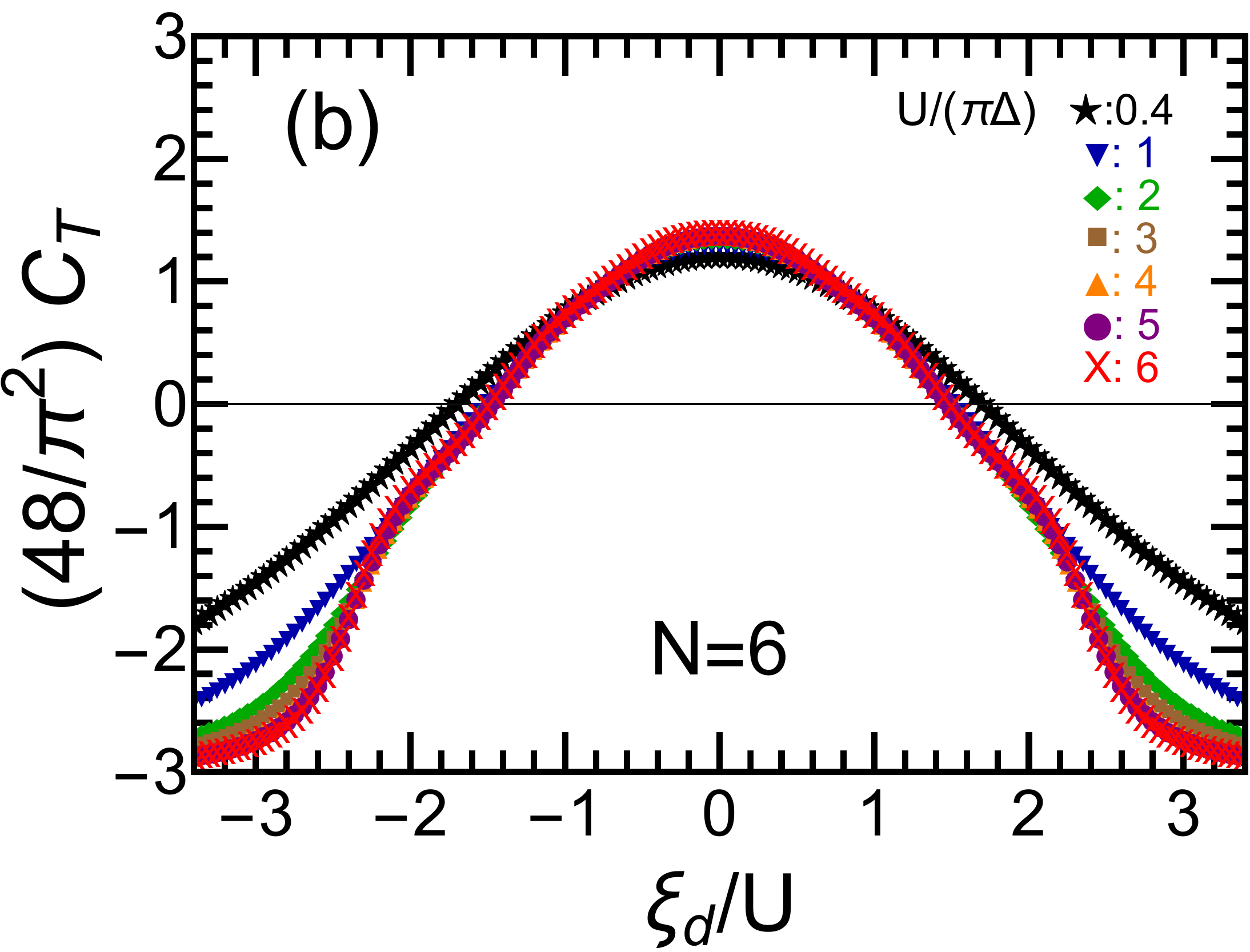}
\\
\includegraphics[width=0.47\linewidth]{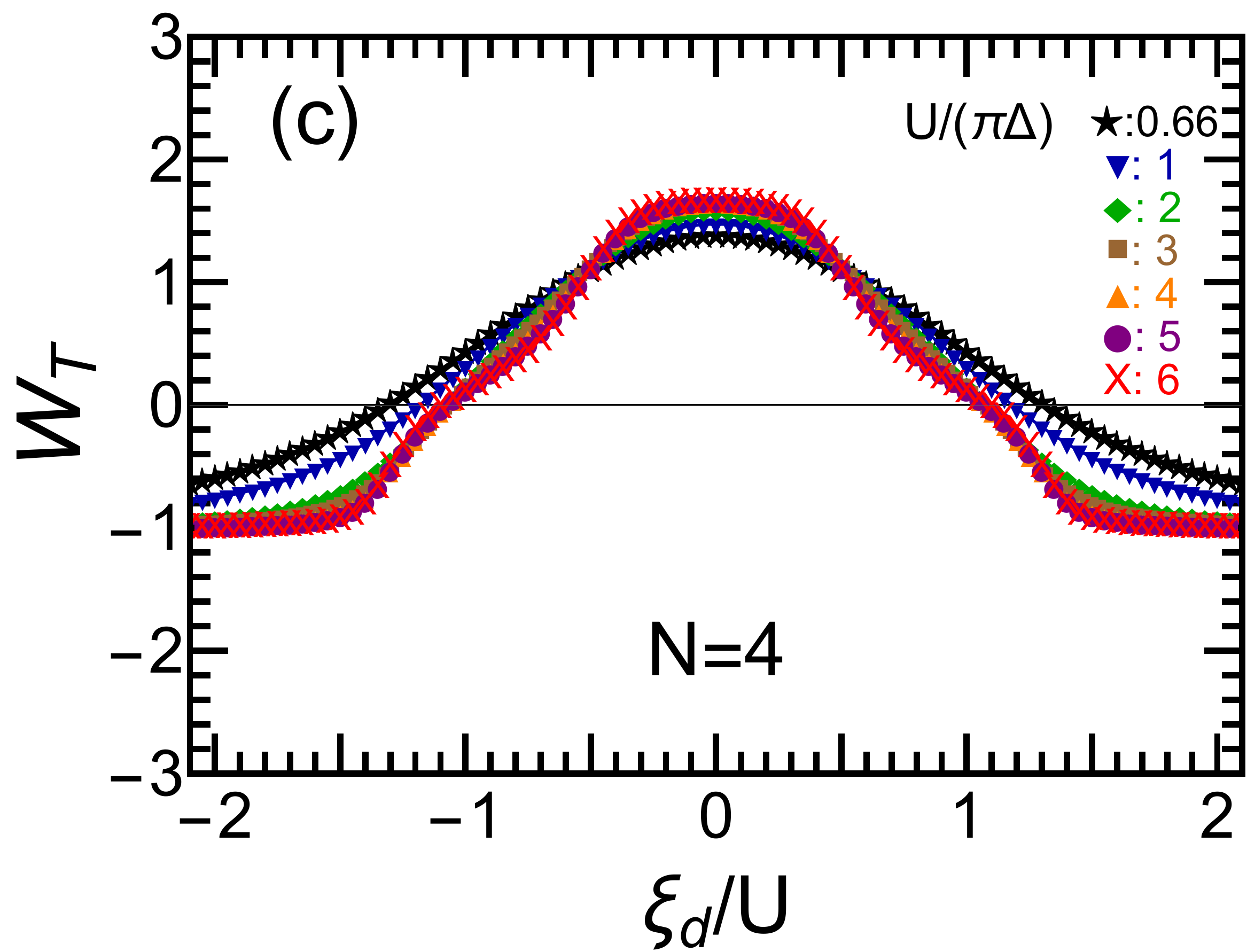}
 \hspace{0.01\linewidth} 
\includegraphics[width=0.47\linewidth]{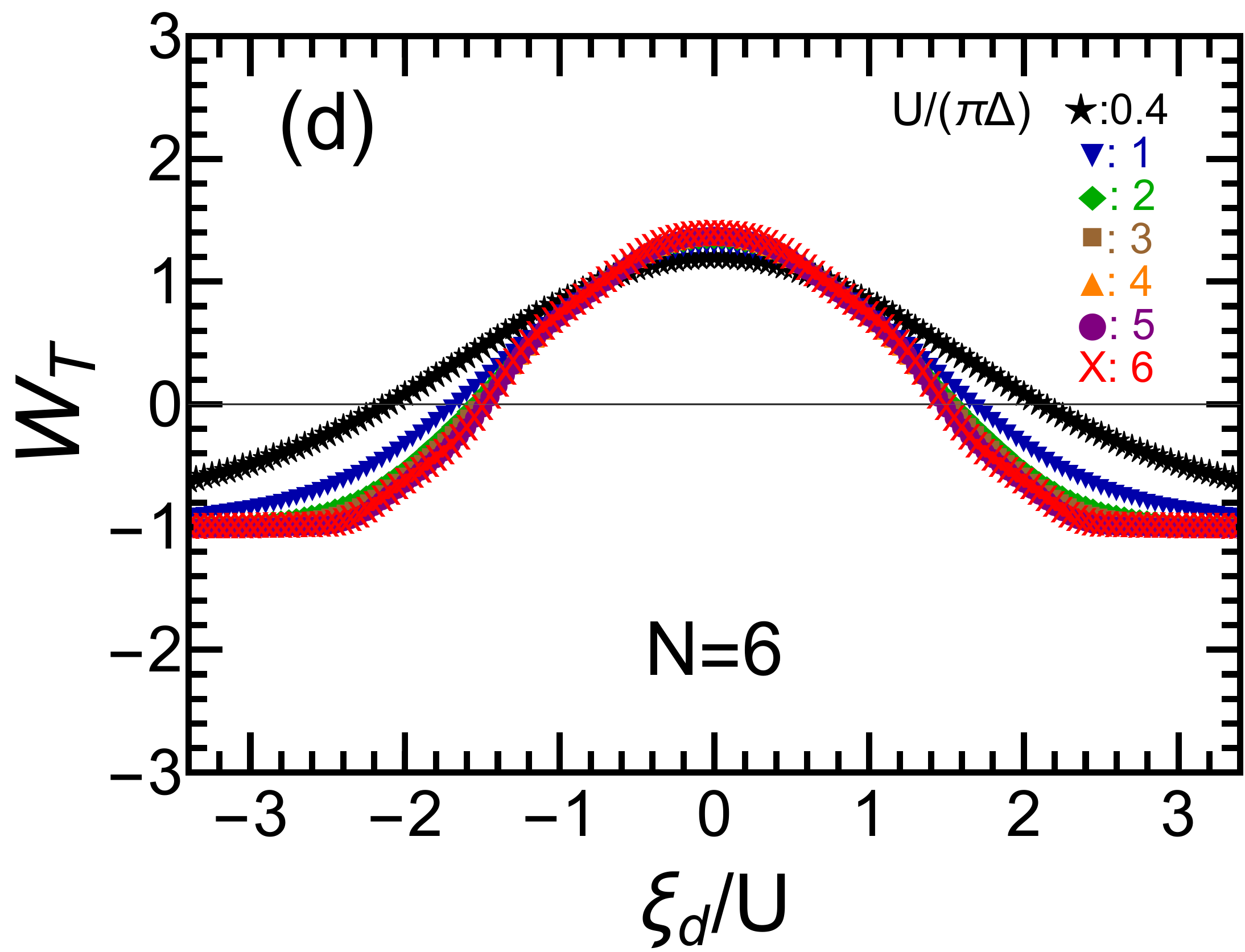}
\\
\includegraphics[width=0.47\linewidth]{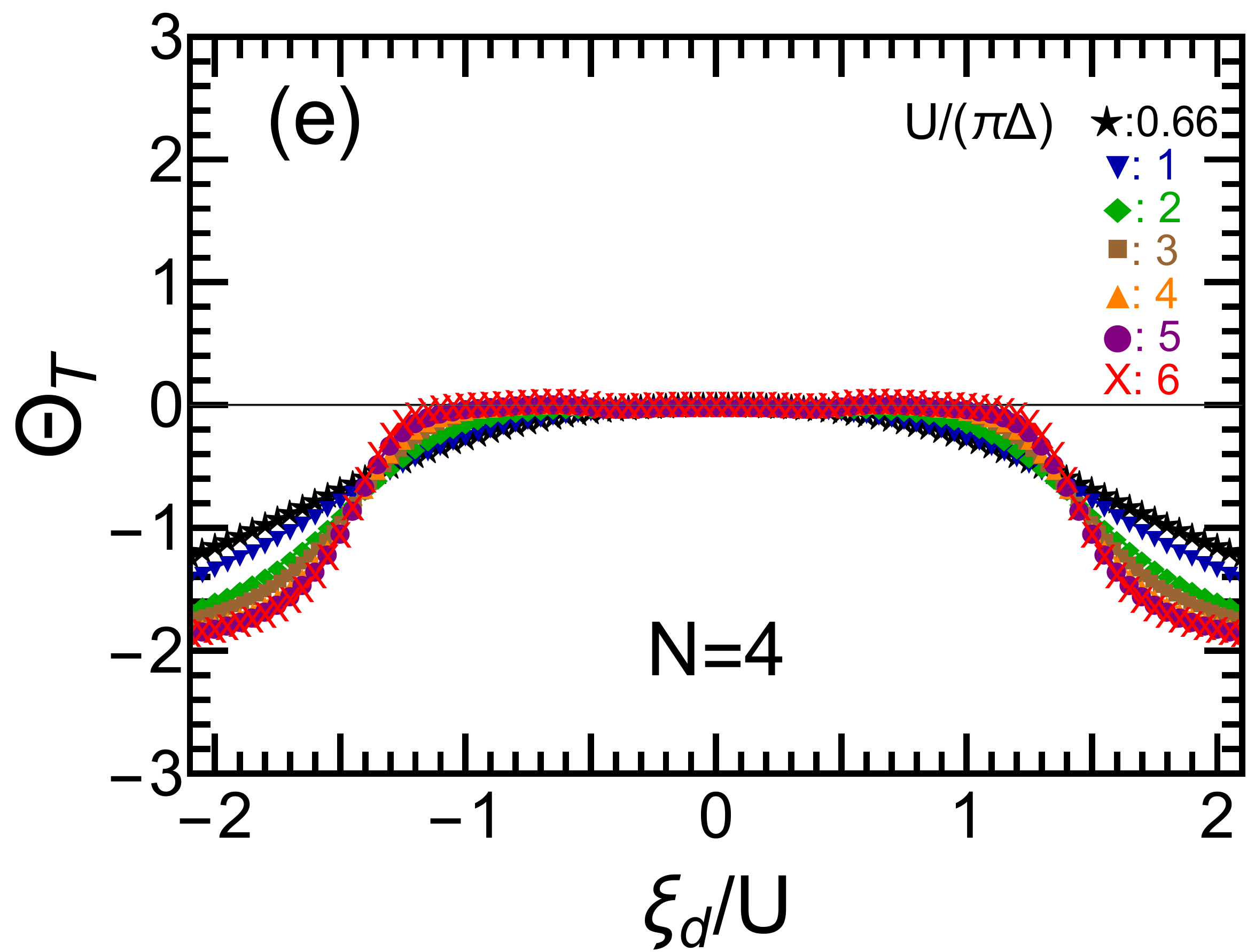}
 \hspace{0.01\linewidth} 
\includegraphics[width=0.47\linewidth]{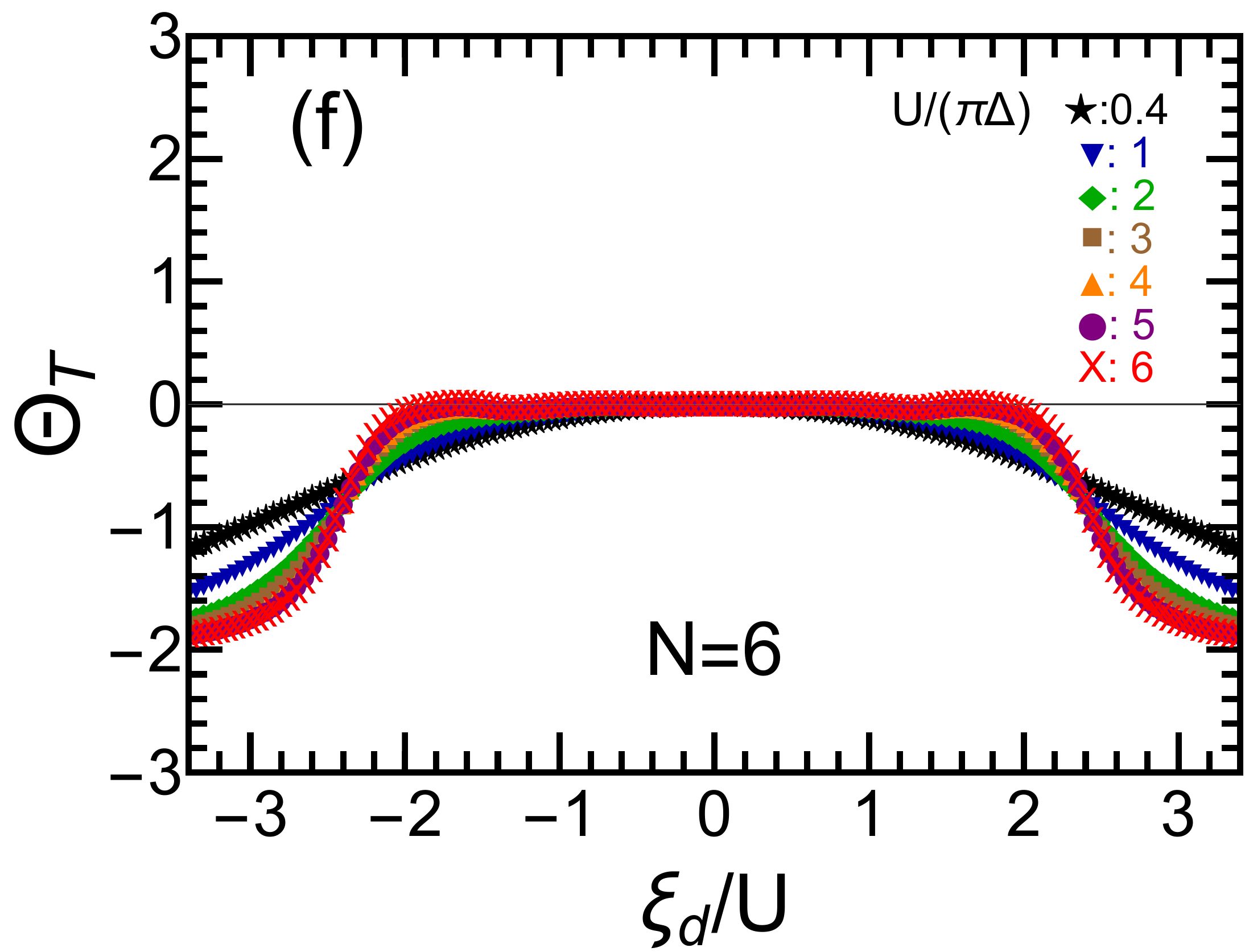}
\caption{$\xi_d^{}$ dependence of 
 $C_{T}^{}=(\pi^2/48) (W_{T}^{}+\Theta_{T}^{})$,  
two-body part $W_{T}^{}$, and three-body part 
$\Theta_{T}^{} \equiv \Theta_{\mathrm{I}}^{}
+3\,\widetilde{\Theta}_{\mathrm{II}}^{}$. 
Left panels: $N=4$, for 
$U/(\pi\Delta) = 2/3(\star)$, $1(\blacktriangledown)$, $2(\blacklozenge)$, $3(\blacksquare)$, $4(\blacktriangle)$, $5(\bullet)$, $6(\times)$.
Right panels: $N=6$, for 
$U/(\pi\Delta) = 2/5(\star)$, $1(\blacktriangledown)$, $2(\blacklozenge)$, 
$3(\blacksquare)$, $4(\blacktriangle)$, $5(\bullet)$, $6(\times)$.} 
\label{fig:CT-QD-N4-6-U}
\end{figure}

The coefficient $C_{T}^{}$ for the order  $T^{2}$ conductance, 
 defined  in Table \ref{tab:C_and_W_extended}, 
 also consists of two-body parts  $W_{T}^{}$ and three-body  
part $\Theta_{T}^{}$:    
\begin{align}
C_{T}^{} 
\, =  \, 
\frac{\pi^2}{48}
 \,\bigl(W_{T}^{} + \Theta_{T}^{}
\bigr)\, , \qquad
\Theta_{T}^{} \,\equiv \, \Theta_{\mathrm{I}}^{} 
+ \widetilde{\Theta}_{\mathrm{II}}^{} 
\,.
\end{align}
In particular, 
the  three-body part $\Theta_{T}^{}$ 
is solely determined by the derivatives of the charge and spin susceptibilities, 
given in Eqs.\ \eqref{eq:derivative_charge_susceptibility_SUN} 
and \eqref{eq:derivative_spin_susceptibility_SUN}, 
and does not depend on   $\chi_{B}^{[3]}$: 
\begin{align}
\Theta_{T}^{} \,=\, 
\frac{(4T^*)^2}{N}
\left[\,
\frac{\partial \overline{\chi}_{C}^{}}{\partial \epsilon_{d}^{}}\,
+(N-1)\frac{\partial \overline{\chi}_{S}^{}}{\partial \epsilon_{d}^{}}\,
\,\right] 
\frac{\sin 2\delta}{2 \pi} \,.  
\label{eq:ThetaT_vs_derivativeCHI}
\end{align}
This is a quite distinct characteristics of $C_{T}^{}$ from 
 the next-leading order terms of the other transport coefficients.
Figures \ref{fig:CT-QD-N4-6-U}(a)--\ref{fig:CT-QD-N4-6-U}(f) 
show the NRG results for 
$C_{T}^{}$, $W_{T}^{}$, and  $\Theta_{T}^{}$.

We see  in Figs.\ \ref{fig:CT-QD-N4-6-U}(e) and \ref{fig:CT-QD-N4-6-U}(f) 
that the three-body contribution almost vanishes, $\Theta_{T}^{} \simeq 0.0$, 
in the wide strong-coupling region $|\xi_{d}^{}|\lesssim (N-1)U/2$, 
in which the occupation number varies with $\xi_{d}^{}$,  
in the range of $1 \lesssim N_d^{} \lesssim N-1$.   
This is because  the magnitudes of the derivatives 
  $\partial \overline{\chi}_{C}^{}/\partial \epsilon_{d}^{}$ 
and  $\partial \overline{\chi}_{S}^{}/\partial \epsilon_{d}^{}$,  
appearing in the right-hand side of  Eq.\ \eqref{eq:ThetaT_vs_derivativeCHI},  
are significantly suppressed by the Coulomb repulsion in this region, 
as demonstrated in Figs.\ 
\ref{fig:DChiC_DChiS_chi3B_N4-6_u6}(a)--\ref{fig:DChiC_DChiS_chi3B_N4-6_u6}(d).

Therefore, 
the two-body part  $W_{T}^{}$ dominates $C_{T}^{}$ 
in the strong-coupling region  $|\xi_{d}^{}|\lesssim (N-1)U/2$,  
and it takes the following form for large $U$, 
\begin{align}
  W_{T}^{} \simeq
 \left[\,1+ \frac{2}{N-1}\,\right]\,(2\sin^2 \delta -1) , 
\qquad 
 \Theta_{T}^{} \simeq 0, 
\label{eq:WT_strong_NRG}
\end{align}
as the rescaled Wilson ratio reaches the saturation value  $\widetilde{K} \to 1$. 
Hence, the plateau structure of $C_{T}^{}$ is determined by 
the  $\sin^2 \delta$ term of $W_{T}^{}$ in Eq.\ \eqref{eq:WT_strong_NRG}.
In particular, the plateau around the half filling point  $|\xi_d^{}| \lesssim U/2$ 
reaches the height of 
 $(48/\pi^2) C_{T}^{}\xrightarrow{U\to \infty\,} 5/3$ 
and $7/5$ for  $N=4$ and $6$, respectively, 
since  $\delta \simeq \pi/2$  in this region. 
The order $T^2$ conductance,  $C_{T}^{}$, vanishes 
 at  $|\xi_{d}^{}|/U\simeq N/4$, 
more specifically at the quarter and the three-quarter fillings where 
the phase shift reaches $\delta=\pi/4$ or $3\pi/4$. 
The zero points of $C_{T}^{}$ emerge at integer fillings 
in the case  of $N \equiv 0$  (mod $4$) at which the SU($N$) 
Kondo effect is occurring,  
whereas for $N \equiv 2$  (mod $4$) the zeros emerge at half-integer filings  
in between the two adjacent Kondo states. 
This explains the reason why   
the SU(4) Kondo state at quarter filling exhibits  universal $T/T^*$-scaling behavior, 
which shows the $(T/T^*)^4$ dependence at low temperatures 
instead of the $(T/T^*)^2$ dependence \cite{TerataniPRB2020}.

In the valence fluctuation and empty (or fully-occupied) orbital regimes, 
which spread over the regions of  $|\xi_{d}^{}| \gtrsim (N-1)U/2$, 
 the three-body part $\Theta_{T}^{}$ becomes  
comparable to the two-body part $W_{T}^{}$. 
Both of these parts approach the noninteracting values 
in the limit of  $|\xi_{d}^{}| \to \infty$:  
\begin{align}
W_{T}^{}
\,\xrightarrow{\, |\xi_d^{}| \to \infty}\, -1, 
\qquad
\Theta_{T}^{}
\,\xrightarrow{\, |\xi_d^{}| \to \infty}\, -2.  
\end{align}

\begin{figure}[t]

\leavevmode
 \centering

\includegraphics[width=0.47\linewidth]{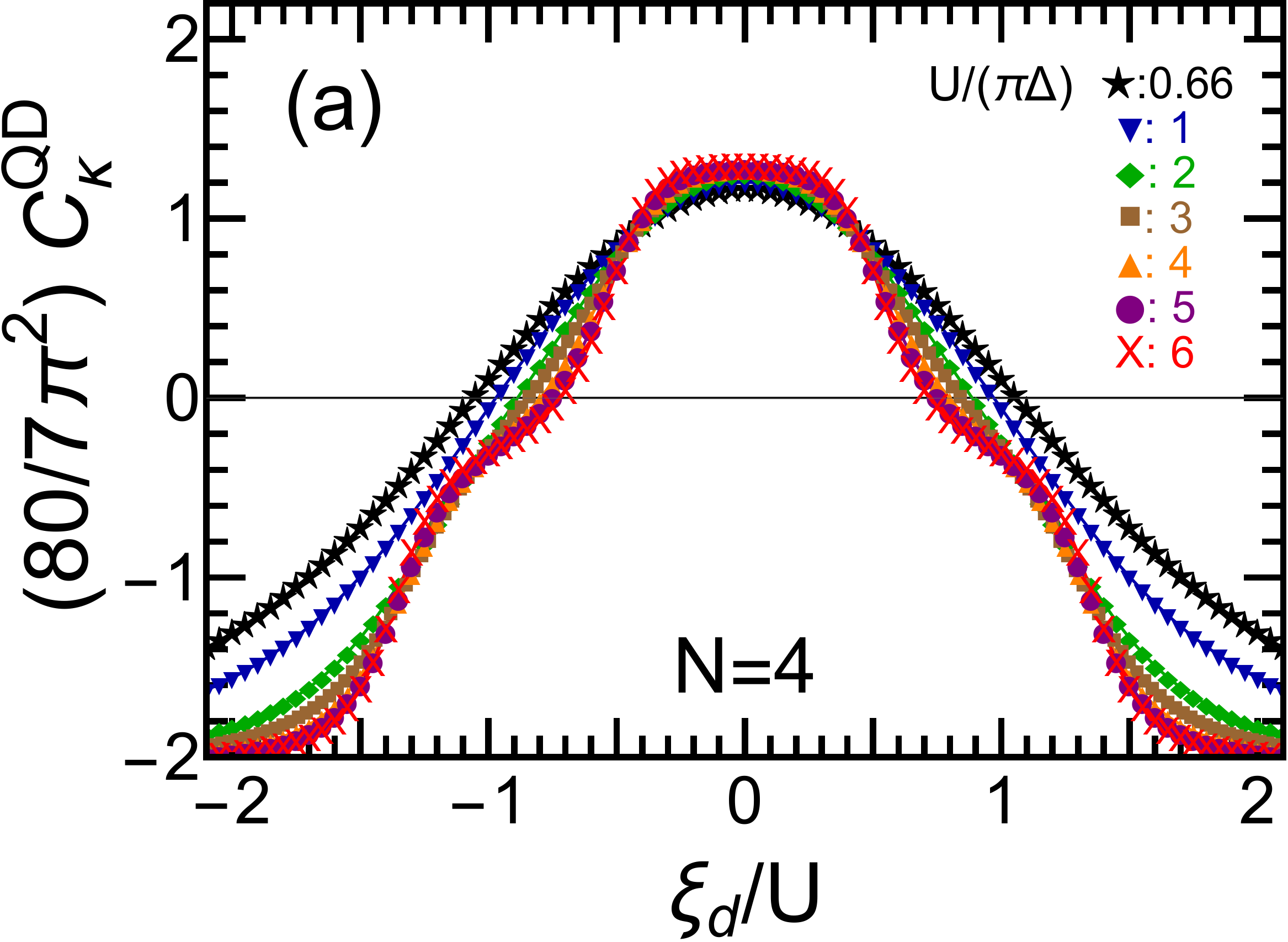}
 \hspace{0.01\linewidth} 
\includegraphics[width=0.47\linewidth]{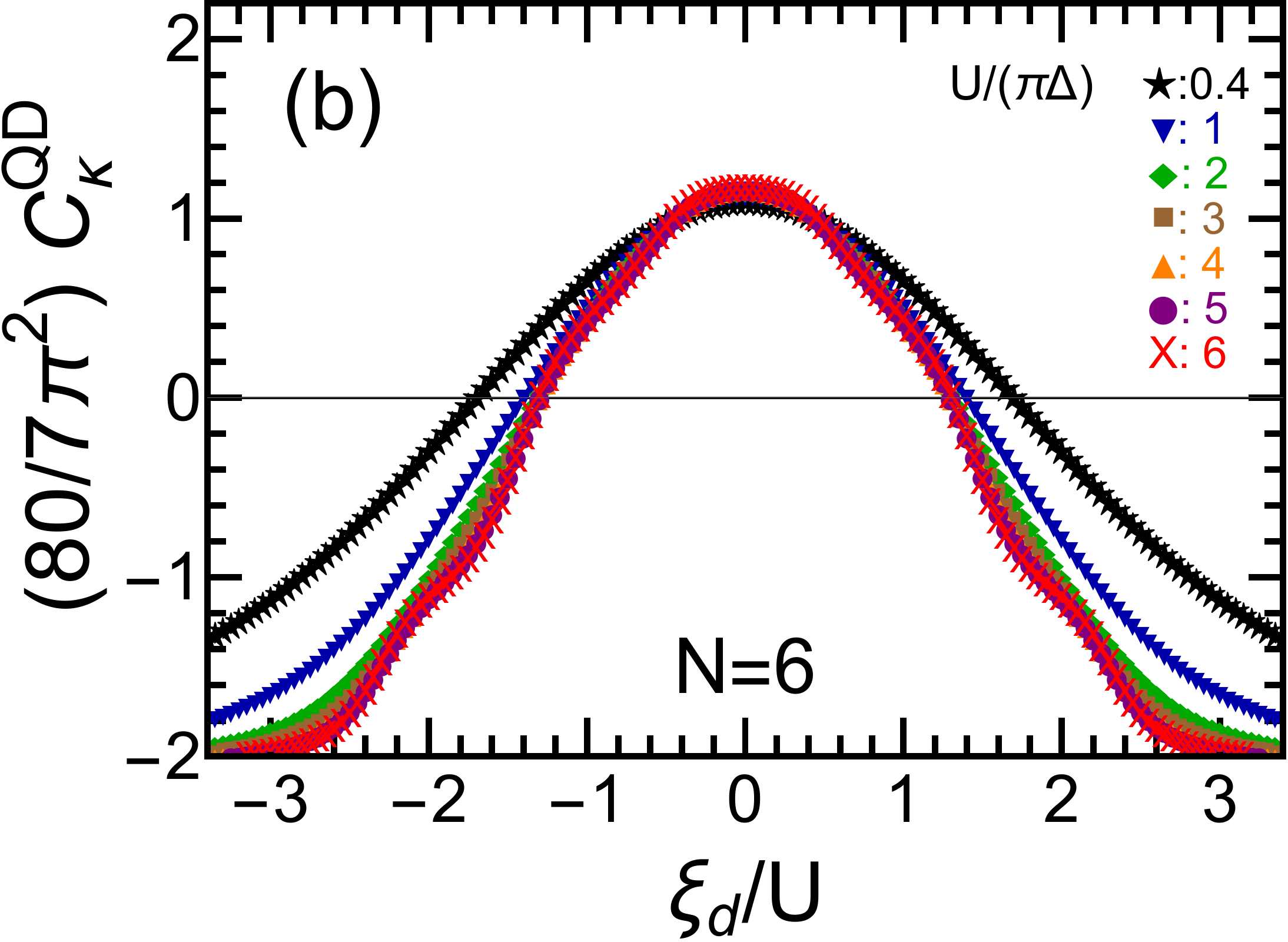}
\\
\includegraphics[width=0.47\linewidth]{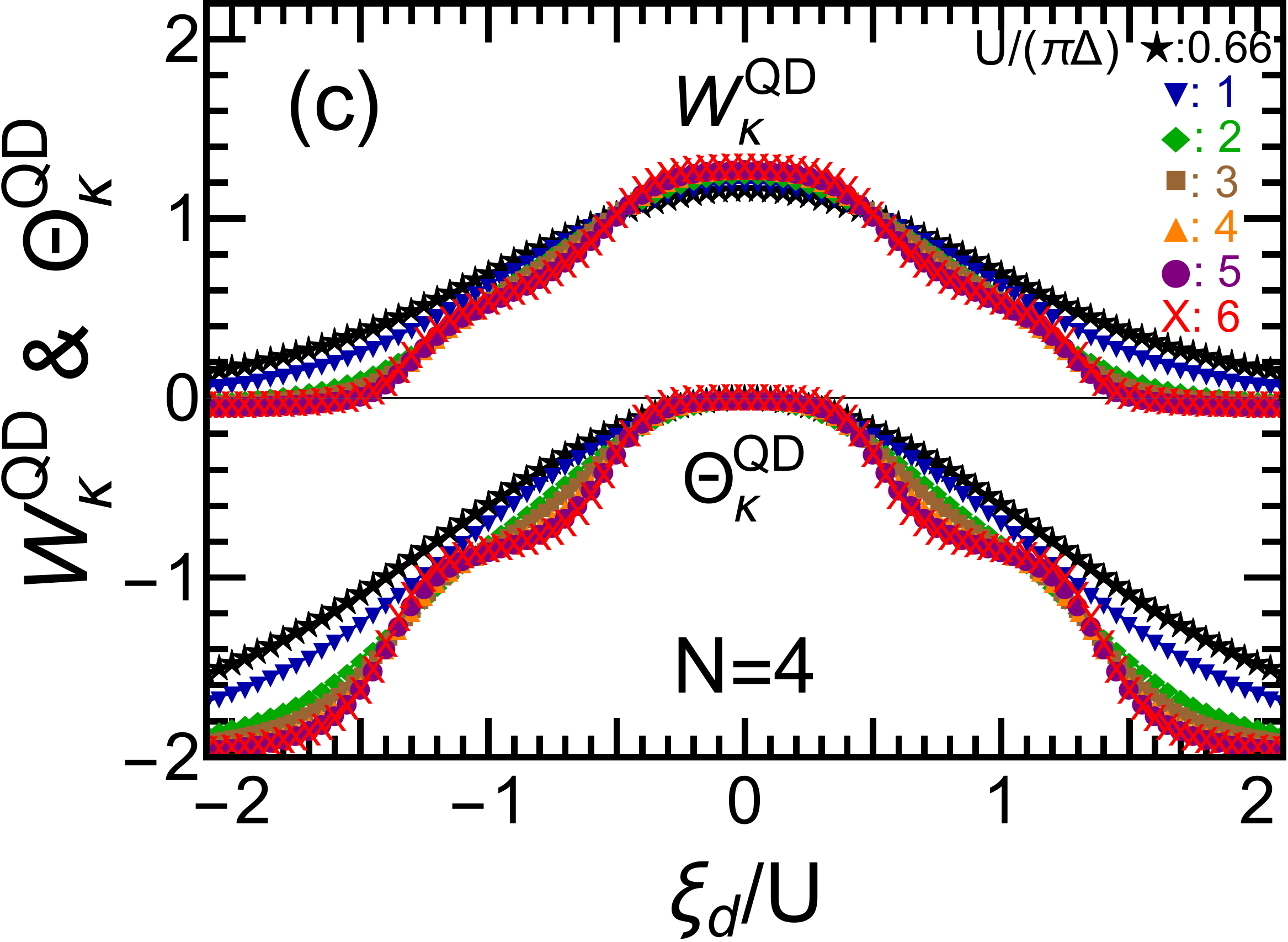}
 \hspace{0.01\linewidth} 
\includegraphics[width=0.47\linewidth]{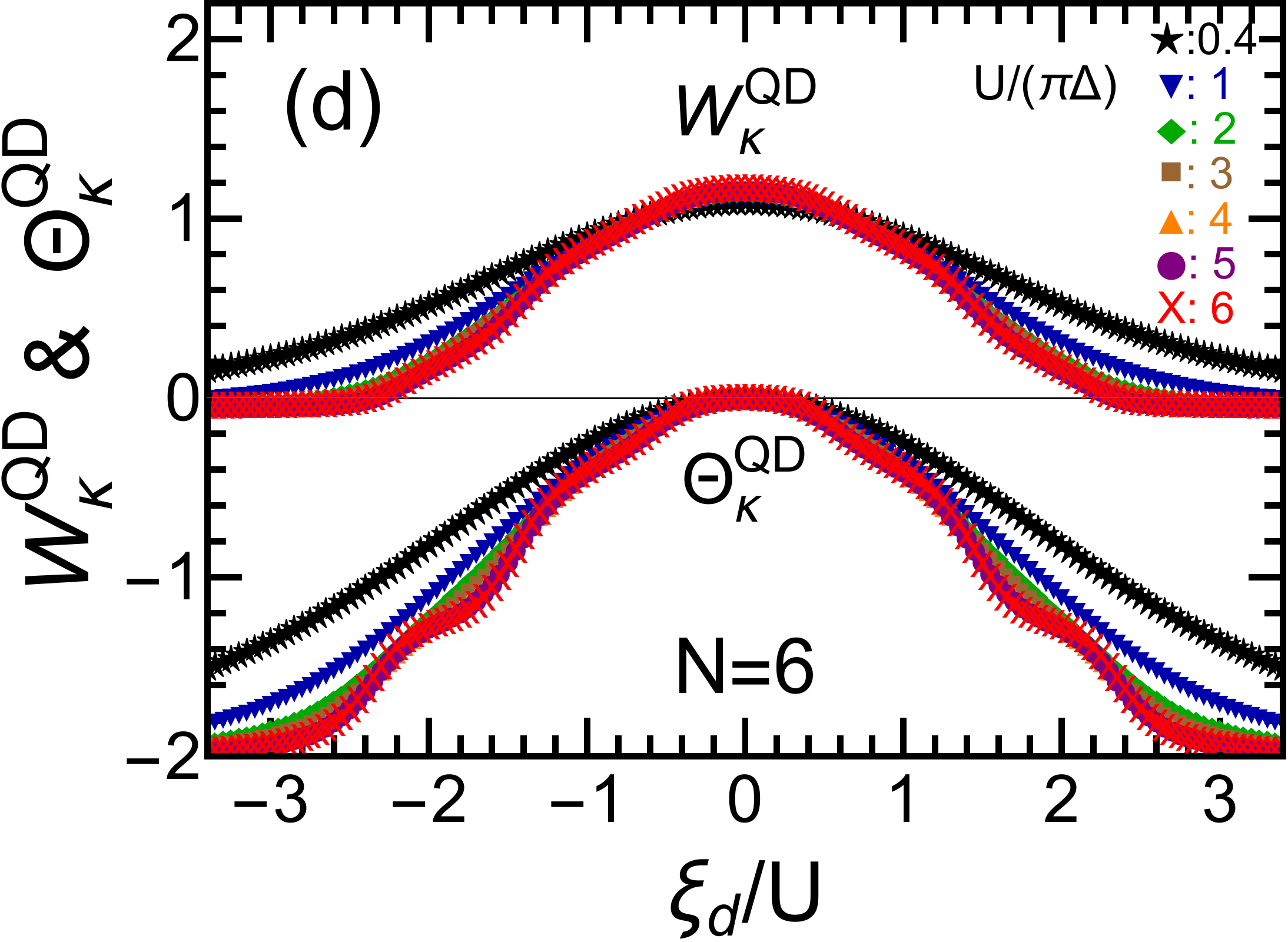}
\caption{$\xi_d^{}$ dependence of 
 $C_{\kappa}^{\mathrm{QD}}=(7\pi^2/80) (W_{\kappa}^{\mathrm{QD}}
+\Theta_{\kappa}^{\mathrm{QD}})$,  
two-body part $W_{\kappa}^{\mathrm{QD}}$, and three-body part 
$\Theta_{\kappa}^{\mathrm{QD}} 
 = \Theta_\mathrm{I}^{} +
\frac{5}{21}\widetilde{\Theta}_{\mathrm{II}}^{}$.
Left panels: $N=4$, for 
$U/(\pi\Delta) = 2/3(\star)$, $1(\blacktriangledown)$, $2(\blacklozenge)$, $3(\blacksquare)$, $4(\blacktriangle)$, $5(\bullet)$, $6(\times)$.
Right panels: $N=6$, for 
$U/(\pi\Delta) = 2/5(\star)$, $1(\blacktriangledown)$, $2(\blacklozenge)$, 
$3(\blacksquare)$, $4(\blacktriangle)$, $5(\bullet)$, $6(\times)$.} 
\label{fig:Ckappa-QD-N4-6-U}
\end{figure}

\subsection{$C_{\kappa}^{\mathrm{QD}}$: order  $T^3$ term 
of $\kappa_\mathrm{QD}^{}$ }
\label{subsec:Ckappa_QD_N4-6}

The coefficient  $C_{\kappa}^{\mathrm{QD}}$ 
for the order $T^{3}$ term of the thermal conductance  $\kappa_\mathrm{QD}^{}$  
can also be decomposed  into two parts,  
$W_{\kappa}^{\mathrm{QD}}$ and $\Theta_{\kappa}^{\mathrm{QD}}$, 
as shown in Table \ref{tab:C_and_W_extended}: 
\begin{align}
C_{\kappa}^{\mathrm{QD}} 
\, =  \, 
\frac{7\pi^2}{80}
 \,\bigl(W_{\kappa}^{\mathrm{QD}} + \Theta_{\kappa}^{\mathrm{QD}}
\bigr) , \qquad
\Theta_{\kappa}^{\mathrm{QD}} 
\, \equiv\, \Theta_\mathrm{I}^{}  +
\frac{5}{21}\widetilde{\Theta}_{\mathrm{II}}^{} .
\end{align}
In contrast to $\Theta_{T}^{}$ of the conductance  
given in Eq.\ \eqref{eq:ThetaT_vs_derivativeCHI}, 
the three-body part $\Theta_{\kappa}^{\mathrm{QD}}$ 
 of the thermal conductance depends on 
 $\chi_{B}^{[3]}$ as well as 
$\partial \overline{\chi}_{C}^{}/\partial \epsilon_{d}^{}$ and 
$\partial \overline{\chi}_{S}^{}/\partial \epsilon_{d}^{}$. 
The contribution of $\chi_{B}^{[3]}$ enters 
through  $\chi_{\sigma\sigma\sigma}^{[3]}$ and
 $\widetilde{\chi}_{\sigma\sigma'\sigma'}^{[3]}$, 
 described in  Eqs.\ \eqref{eq:chi3_SUN_1} and \eqref{eq:chi3_SUN_2}.
This component $\chi_{B}^{[3]}$ yields the plateau structure 
in $\Theta_{\kappa}^{\mathrm{QD}}$,     
which reflects the staircase behavior seen 
in Figs.\ \ref{fig:DChiC_DChiS_chi3B_N4-6_u6}(e) 
and \ref{fig:DChiC_DChiS_chi3B_N4-6_u6}(f). 
Figure \ref{fig:Ckappa-QD-N4-6-U} shows the NRG results for 
$C_{\kappa}^{\mathrm{QD}}$,
$W_{\kappa}^{\mathrm{QD}}$, and 
$\Theta_{\kappa}^{\mathrm{QD}}$ in the SU(4) and SU(6) cases.

Near half filling in the region of $|\xi_d^{}| \lesssim U/2$,  
the two-body part $W_{\kappa}^{\mathrm{QD}}$ dominates 
  $C_{\kappa}^{\mathrm{QD}}$ as 
the three-body part $\Theta_{\kappa}^{\mathrm{QD}}$ disappears 
near half filling $\xi_{d}^{}= 0$, i.e.,  $\delta = \pi/2$:   
\begin{align}
 W_{\kappa}^{\mathrm{QD}} \xrightarrow{\, \xi_d^{}=0\,}\, 
 1+ \frac{6 \widetilde{K}^2}{7(N-1)}\,, 
\qquad 
\Theta_{\kappa}^{\mathrm{QD}} \xrightarrow{\,\xi_d^{}=0\,} 0\,.
\end{align}
In particular, in the strong interaction limit $U \to \infty$,  
 the coefficient   $C_{\kappa}^{\mathrm{QD}}$ 
for $N=4$ and $6$ approach the values of 
 $[80/(7\pi^2)] C_{\kappa}^{\mathrm{QD}} 
\xrightarrow{\, \xi_d^{}=0\,\&\, U\to \infty\,} 9/7$ 
and $41/35$, respectively.

The rescaled Wilson ratio takes the values very close to 
 the saturation value $\widetilde{K} \to 1$ for large $U$  
in the strong-coupling region of $|\xi_{d}^{}|\lesssim (N-1)U/2$,
 as described in Appendix \ref{sec:NRG_results_2body_functions_for_sun}. 
Similarly, in this region, 
the three-body correlations show the property described  in  
  Eq.\ \eqref{eq:Theta_I_II_III_relations_strong_U_simplified}, 
and thus 
  $W_{\kappa}^{\mathrm{QD}}$ and 
  $\Theta_{\kappa}^{\mathrm{QD}}$ 
 take the form
\begin{align}
 W_{\kappa}^{\mathrm{QD}}
\,\simeq& \ 
\frac{1}{21}
\left[ 
10 + \left(11  + \frac{18}{N-1}\right)\left(2\sin^2\delta -1\right)\right]  ,
\nonumber 
\\
 \Theta_{\kappa}^{\mathrm{QD}}
\,\simeq& \  \frac{16}{21}\, \Theta_\mathrm{I}^{}\,. 
\end{align}
Therefore, the plateau structure of  $W_{\kappa}^{\mathrm{QD}}$ 
is determined by the $\sin^2 \delta$ term appearing in the right-hand side,  
while the structure of  $\Theta_{\kappa}^{\mathrm{QD}}$ 
reflects the behavior of  $\Theta_\mathrm{I}^{}$  
shown in Fig.\ \ref{fig:3body_correlation_N4-6_some_U}(e) 
and \ref{fig:3body_correlation_N4-6_some_U}(f). 
 The coefficient $C_{\kappa}^{\mathrm{QD}}$  changes sign 
in the strong-coupling region at two points of $\xi_d^{}$, 
which are incommensurate with the occupation number $N_d^{}$
 in contrast to $C_{T}^{}$ that changes sign at $1/4$ and $3/4$ fillings.

In the opposite limit $|\xi_{d}^{}| \to \infty$, 
both the two-body and three-body parts 
approach the noninteracting values, 
\begin{align}
W_{\kappa}^{\mathrm{QD}} 
\xrightarrow{\, |\xi_d^{}| \to \infty\,}\, -\frac{1}{21}, 
\qquad
\Theta_{\kappa}^{\mathrm{QD}} 
\xrightarrow{\, |\xi_d^{}| \to \infty\,}\, -2.
\end{align}

\begin{figure}[t]

\leavevmode
 \centering

\includegraphics[width=0.47\linewidth]{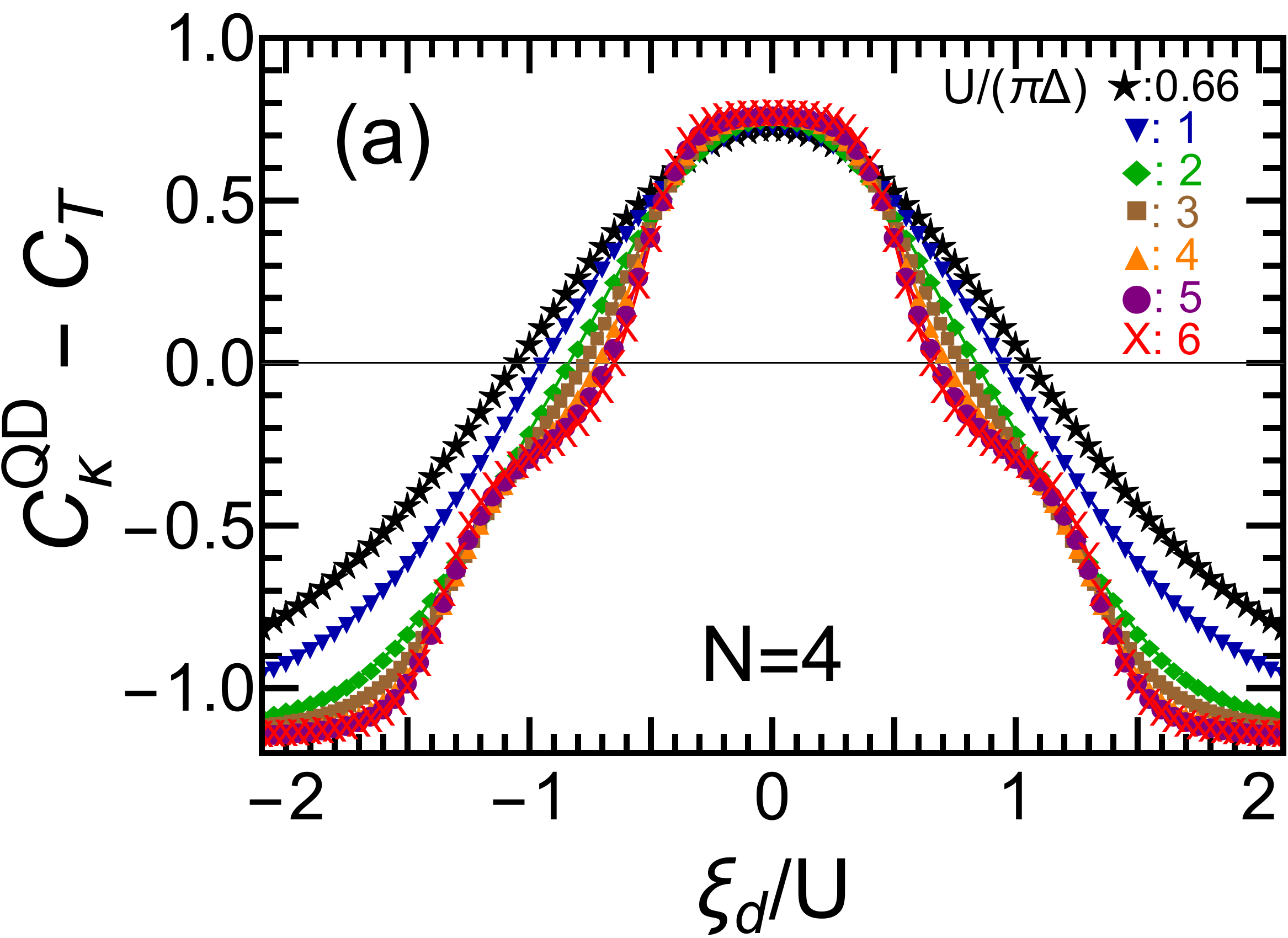}
 \hspace{0.01\linewidth} 
\includegraphics[width=0.47\linewidth]{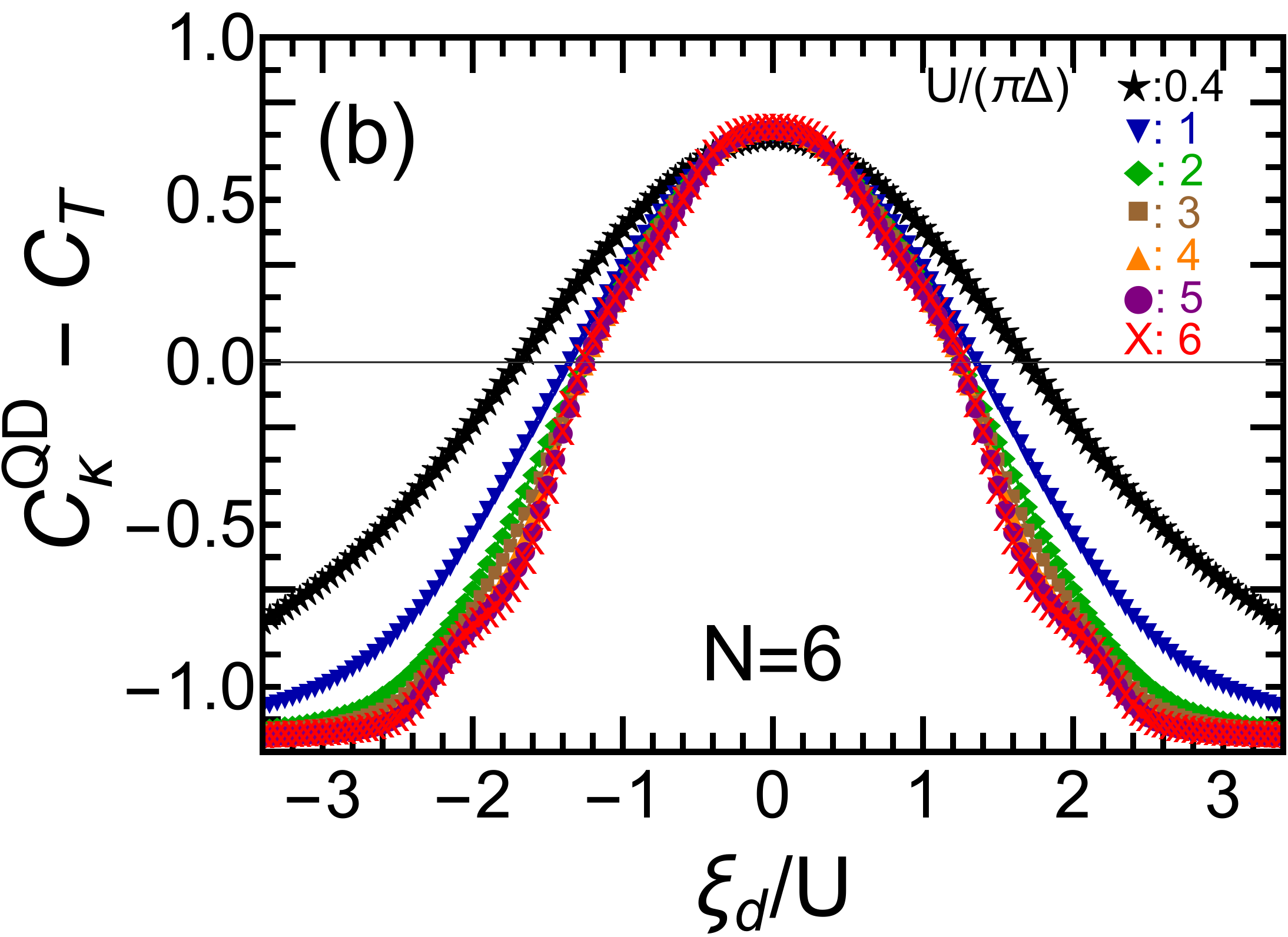}
\caption{Order $T^{2}$ term  
$C_{L}^{\mathrm{QD}}=C_{\kappa}^{\mathrm{QD}}- C_{T}^{}$ 
of Lorenz number $L_{\mathrm{QD}}$ 
is plotted vs $\xi_{d}^{}$. 
Left panel: $N=4$,  for $U/(\pi\Delta) = 2/3(\star)$, 
$1(\blacktriangledown)$, $2(\blacklozenge)$, $3(\blacksquare)$, $4(\blacktriangle)$, 
$5(\bullet)$, $6(\times)$.
Right panel: $N=6$,  
for $U/(\pi\Delta) = 2/5(\star)$, $1(\blacktriangledown)$, $2(\blacklozenge)$, 
$3(\blacksquare)$,  $4(\blacktriangle)$, $5(\bullet)$, $6(\times)$.}
\label{fig:CL-QD-N4-6-U}

\end{figure}

\subsection{$C_{L}^{\mathrm{QD}}$: order  $T^2$ term of  $L_{\mathrm{QD}}$}

The Lorenz number 
$L_{\mathrm{QD}} \equiv \kappa_{\mathrm{QD}}^{}/( g\, T)$ 
 for quantum dots  is defined as 
the ratio of the thermal conductance $\kappa_{\mathrm{QD}}^{}/T$ 
to the electrical conductance $g$.  
It takes the universal Wiedemann-Franz value at zero temperature:  
$L_{\mathrm{QD}} \xrightarrow{T\to 0}\pi^{2}/(3e^{2})$. 
However, as the temperature rises,  
it deviates from the universal value,   
showing the $T^2$ dependence described 
 in Tables \ref{tab:low_energy_expansion} and  
 \ref{tab:C_and_W_extended}.
The coefficient for the order $T^2$ term 
is given by
$C_{L}^{\mathrm{QD}}=C_{\kappa}^{\mathrm{QD}}-C_{T}^{}$, 
as  a difference between 
the next-leading order terms of  $\kappa_{\mathrm{QD}}^{}/T$ and  $g$.

Figure \ref{fig:CL-QD-N4-6-U}(a) and \ref{fig:CL-QD-N4-6-U}(b) show the 
NRG results for  $C_{L}^{\mathrm{QD}}$  in the SU(4) and SU(6) cases,   
respectively. 
Near half filling $|\xi_d^{}| \lesssim U/2$, 
the coefficient  $C_{L}^{\mathrm{QD}}$ is determined 
by the two-body part $W_{L}^{\mathrm{QD}}$ as 
the three-body part $\Theta_{L}^{\mathrm{QD}} 
\equiv 16 \,\Theta_\mathrm{I}^{}$  vanishes at  $\xi_d^{} = 0$, 
as seen in Fig.\ \ref{fig:3body_correlation_N4-6_some_U}(e) 
and \ref{fig:3body_correlation_N4-6_some_U}(f):
\begin{align} 
C_{L}^{\mathrm{QD}}
\xrightarrow{\, \xi_d^{}=0\,}\, 
\frac{\pi^2}{30}\left(  2+  \frac{\widetilde{K}^2}{N-1} \right)  \ >0 \,. 
\end{align} 
It takes a positive value and reaches 
$C_{L}^{\mathrm{QD}}
\xrightarrow{\xi_d^{}=0\, \&\,U=\infty}0,767\cdots$ 
 for $N=4$, and $0.723\cdots$ for $N=6$,   
in the limit of $U \to \infty$ where  $\widetilde{K} \to 1$.

In the strong-coupling region $|\xi_d^{}| \lesssim (N-1)U/2$, 
the coefficient $C_{L}^{\mathrm{QD}}$ exhibits the plateau structures for large $U$,
reflecting  the corresponding structures 
of  $C_{T}^{}$ and $C_{\kappa}^{\mathrm{QD}}$. 
The coefficient $C_{L}^{\mathrm{QD}}$ 
changes sign at the points where the order $T^2$ terms 
of $\kappa_{\mathrm{QD}}^{}/T$ and $g$ coincide, i.e.,
 $C_{\kappa}^{\mathrm{QD}}=C_{T}^{}$.

In the other regions at  $|\xi_d^{}| \gtrsim (N-1)U/2$, 
 the impurity levels approach  
the empty state $N_d^{}\simeq 0$ or 
fully occupied $N_d^{}\simeq N$ state.  
In particular, in the limit of  $|\xi_{d}^{}| \to \infty$, 
both the two-body and three-body parts 
 approach the noninteracting values,  
$W_{L}^{\mathrm{QD}}\xrightarrow{|\xi_d^{}| \to \infty} 4$ 
and  
$16 \Theta_\mathrm{I}^{} \xrightarrow{|\xi_d^{}| \to \infty} -32$,  
and thus  
 \begin{align}
C_{L}^{\mathrm{QD}}
\xrightarrow{\, |\xi_d^{}| \to \infty\,}\, 
 -\frac{7\pi^2}{60}  \  =  \ 
 - 1.151 \cdots \ \  < 0\,.
\end{align}

\section{Fermi liquid description for \\  SU($N$) symmetric magnetic alloys}
\label{sec:transport_dilute_alloys_SUN}

We have discussed, in the previous sections, 
low-energy transport properties of  SU($N$) quantum dots,  
by extending the Fermi liquid description to the next-leading 
order terms which contribute to the transport 
 at finite temperatures or at finite bias voltages.
Our formulation is applicable to a wide class of Kondo systems  
other than quantum dots,  particularly to dilute magnetic alloys (MA) 
composed of $3d$, $4f$, or $5f$ electrons \cite{HewsonBook}. 
In this and the next sections, 
we apply this formulation to dilute magnetic alloys away from half filling, 
taking into account exactly the order $\omega^2$ and $T^2$ 
energy shifts of quasiparticles that enter through the real part 
of the self-energy $\Sigma_{\sigma}^r(\omega)$ 
 given in Appendix \ref{sec:Low_energy_asymptotic_form_A}.

 The thermoelectric transport coefficients of magnetic alloys  
in the linear-response regime can be derived from the function   
 $\mathcal{L}_{n,\sigma}^\mathrm{MA}$ for $n=0$, $1$, and $2$, 
defined by \cite{CostiThermo}
\begin{align}
\mathcal{L}_{n,\sigma}^\mathrm{MA} 
= 
\int_{-\infty}^{\infty}  
d\omega\, 
\frac{\omega^n}{\pi \Delta A_{\sigma}(\omega)}
\left( -\frac{\partial f(\omega)}{\partial \omega}\right)  \, .  
\label{eq:L_thermal}
\end{align}
Here,  the inverse spectral function $1/A_{\sigma}(\omega)$ in the integrand 
represents the relaxation time of conduction electrons, 
which depends on $T$ as well as $\omega$. 
The low-temperature expansion of $\mathcal{L}_{n,\sigma}^\mathrm{MA}$ 
can be deduced from the exact low-energy asymptotic  
of $1/A_{\sigma}(\omega)$, given 
in Appendix \ref{sec:Low_energy_asymptotic_form_A}.
We have presented the expansion formulas 
for the standard $N=2$ Anderson impurity model 
in a previous paper \cite{ao2017_3_PRB}.
In this work, we extend the formulation to multi-level impurities  
in a general form without assuming the SU($N$) symmetry.
Details of the derivation are given 
 in Appendix \ref{sec:thermoelectric_transport_general_MA}.

In the following, we consider the  behavior of the next-leading order terms of 
  electrical conductivity $\sigma_\mathrm{MA}^{}$ 
 and thermal conductivity  $\kappa_\mathrm{MA}^{}$ 
of magnetic alloys in the  SU($N$) symmetric case, 
where  $\epsilon_{d\sigma}^{}\equiv \epsilon_{d}^{}$ 
for all $\sigma$, and $U_{\sigma\sigma'}^{} \equiv U$  
for all  $\sigma$ and $\sigma'$. 
 In this case, the formulas given 
 in Appendix \ref{sec:thermoelectric_transport_general_MA} 
are simplified, 
and as a result,  the electrical resistivity 
$\varrho_\mathrm{MA}^{}=1/\sigma_\mathrm{MA}^{}$, 
the thermal resistivity  $1/\kappa_\mathrm{MA}^{}$,  
and the Lorenz number $L_{\mathrm{MA}}=
\kappa_{\mathrm{MA}}^{}/(\sigma_{\mathrm{MA}}^{}\,T)$  
can be expressed in the form,  
\begin{align}
\!\! 
& 
\varrho_\mathrm{MA}^{}
\,=\, 
\frac{1}{\sigma_{\mathrm{MA}}^{\mathrm{unit}}}  
\,\left[\, 
\sin^2 \delta \,- \,  
C_\varrho^\mathrm{MA} 
\left(\frac{\pi T}{T^*}\right)^2   
+ \cdots \,\right]
\,, \label{eq:FL_rho_MA}
\\
& \!\! 
\frac{1}{\kappa_\mathrm{MA}^{}}\, =\,
\frac{3\,e^{2}}{\pi^{2}\,\sigma_\mathrm{MA}^{\mathrm{unit}}}\frac{1}{T}
\left[\, \sin^2 \delta - 
C_{\kappa}^\mathrm{MA} 
\,\left(\frac{\pi T}{T^*}\right)^2  + \cdots \,\right] , 
\label{eq:FL_kappa_inv_MA}
\\
& \!\! 
L_{\mathrm{MA}}^{}
\,=\,\frac{\pi^{2}}{3\,e^{2}}\,\left[\,1\,-
\,
\frac{C_{L}^{\mathrm{MA}}}{\sin^2 \delta}
\,
\left( \frac{\pi T}{T^{*}}\right)^{2} \ + \  \cdots \,\right] \,.
\label{eq:L_MA_SUN_formula}
\end{align}
Here,  $\sigma_\mathrm{MA}^{\mathrm{unit}}$ is the unitary-limit value 
of electrical conductivity.  
The explicit expressions of the dimensionless coefficients 
$C_{\varrho}^\mathrm{MA}$, $C_{\kappa}^\mathrm{MA}$, and 
$C_{L}^{\mathrm{MA}}$ are listed in Table \ref{tab:C_and_W_extended}. 
These coefficients  $C$'s  for magnetic alloys  
can also be decomposed into the two-body $W$ part 
and the three-body $\Theta$ part, as those for quantum dots.
 Note that the following relations hold between the coefficients 
for magnetic alloys and quantum dots:
\begin{align}
C_{\varrho}^\mathrm{MA}\,= & \ \, \frac{\pi^2}{12} \,\cos^2 \delta + C_{T}^{}
\,,
\label{eq:Crho_CT_relation}
\\
C_{\kappa}^\mathrm{MA} \,=& \ \,
\frac{11\pi^2}{60} \,\cos^2 \delta + C_{\kappa}^\mathrm{QD} \,,
\label{eq:CkappaMA_CkappaQD_relation}
\\
C_{L}^\mathrm{MA}\, =& \   
-\frac{\pi^2}{10}\,\cos^2 \delta -C_{L}^\mathrm{QD} 
\label{eq:CLMA_CLQD_relation}
\,.
\end{align}
In particular,  the $\cos^2 \delta$ term appearing in the right-hand side  
vanishes at half filling, i.e.,  $\delta = \pi/2$. 
In this case the coefficients for magnetic alloys in the left-hand side 
coincide with their quantum-dot counterparts in the right-hand side, 
except for the signs of $C_{L}^\mathrm{MA}$ and $C_{L}^\mathrm{QD}$.
The behavior of these coefficients for magnetic alloys also 
 reflects the properties of low-lying energy states and significantly 
 depends on the occupation number $N_d^{}$ and the interaction strength $U$.

\section{Thermoelectric transport of \\  SU(4) $\&$ SU(6) magnetic alloys}
\label{sec:NRG_MA}

We consider here  the next-leading order terms of 
 the  electrical resistivity $\varrho_{\mathrm{MA}}^{}$ and 
 thermal resistivity $1/\kappa_{\mathrm{MA}}^{}$
of SU($N$) symmetric magnetic alloys 
for $N=4$ and $6$.
For comparison, we also provide  
the NRG results for these transport coefficients 
in the SU(2) case  in Appendix \ref{sec:NoiseN2},
and  the analytic expressions of $C$'s  
for noninteracting magnetic alloys in Appendix \ref{sec:thermo_MA_U0}.

\subsection{$C_{\varrho}^{\mathrm{MA}}$: order  $T^2$ term 
of $\varrho_{\mathrm{MA}}$}

\begin{figure}[t]

\leavevmode
 \centering

\includegraphics[width=0.47\linewidth]{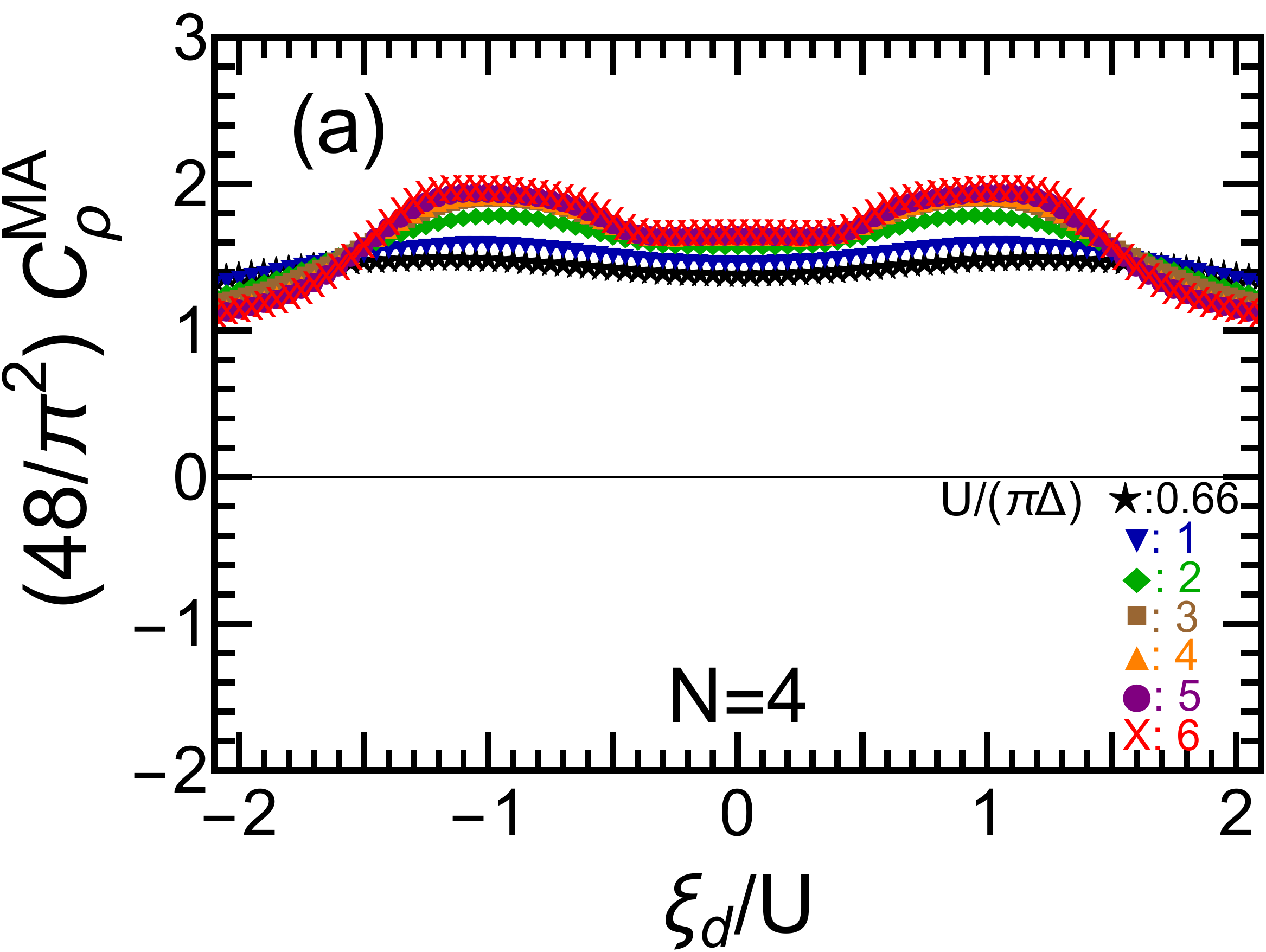}
 \hspace{0.01\linewidth} 
\includegraphics[width=0.47\linewidth]{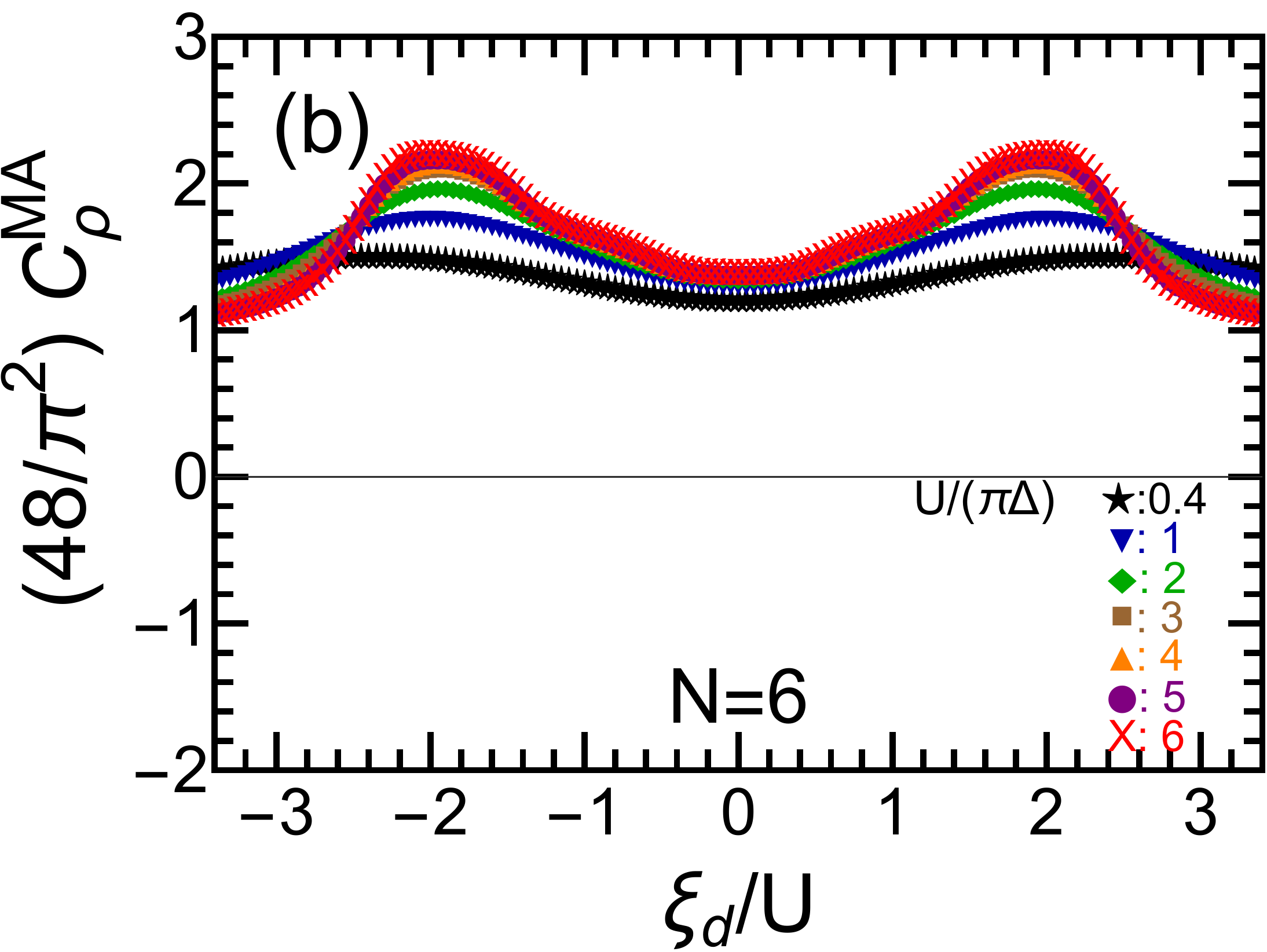}
\\
\includegraphics[width=0.47\linewidth]{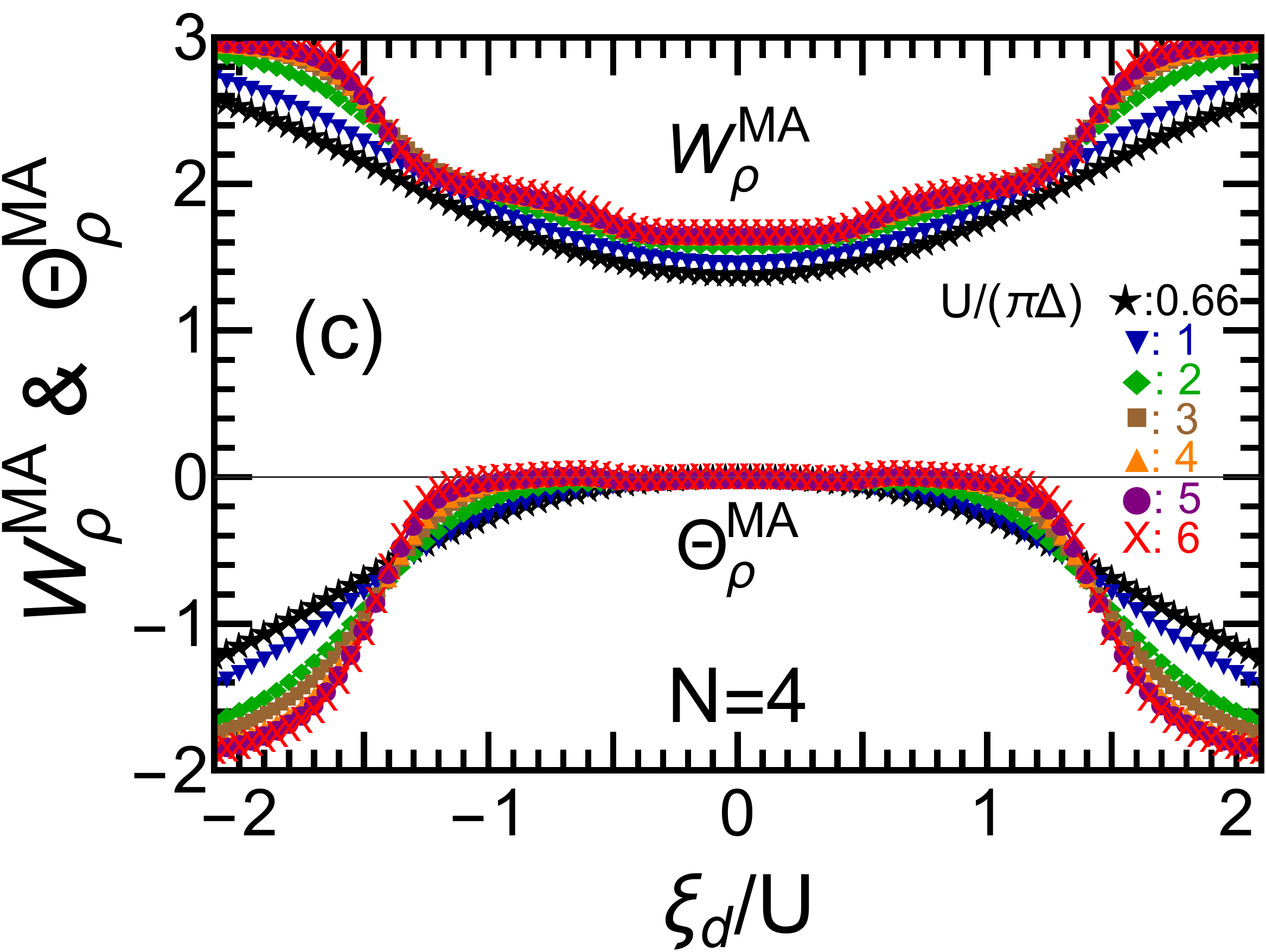}
 \hspace{0.01\linewidth} 
\includegraphics[width=0.47\linewidth]{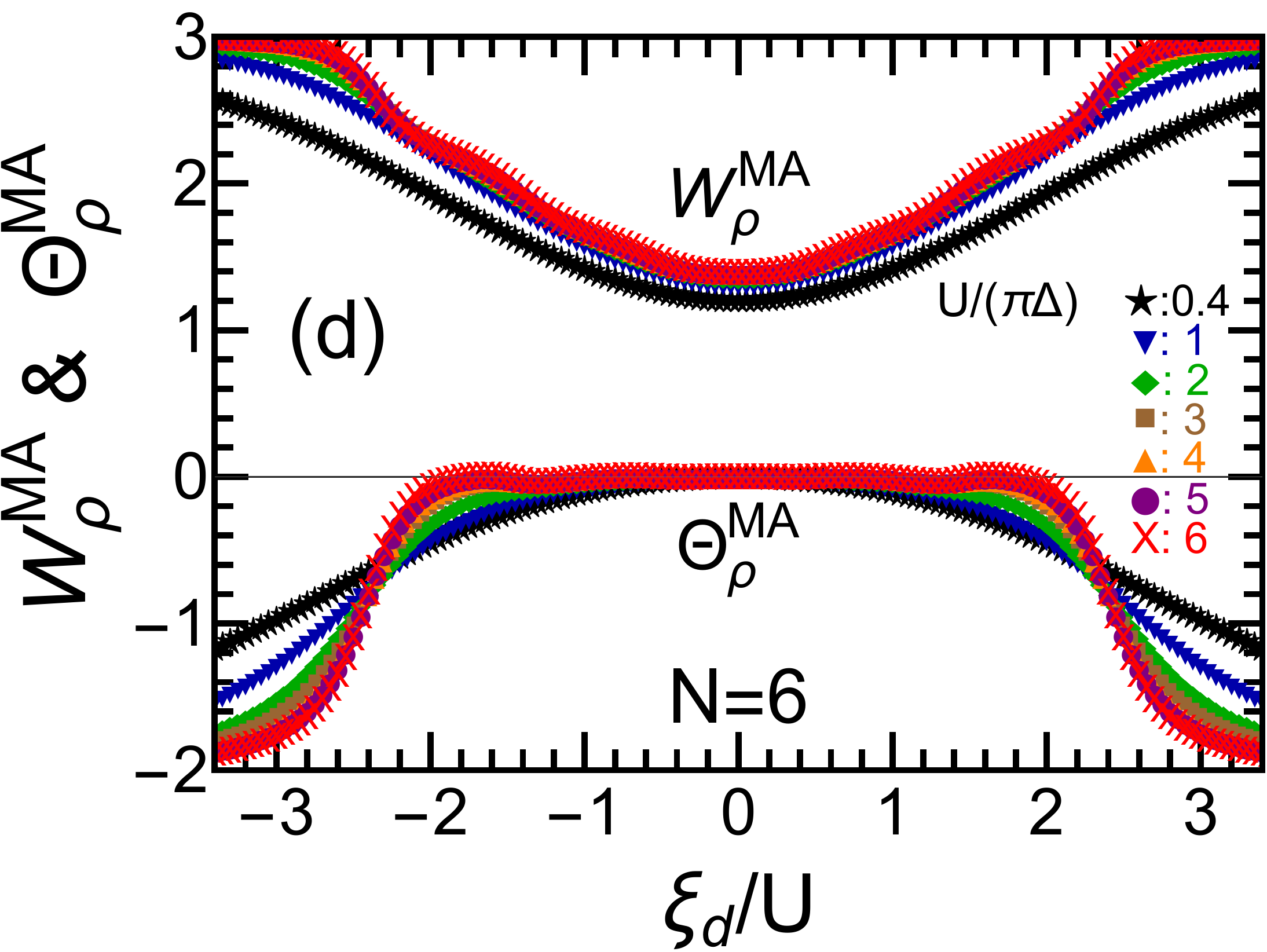}

\caption{$\xi_d^{}$ dependence of 
 $C_{\varrho}^{\mathrm{MA}}=(\pi^2/48) (W_{\varrho}^{\mathrm{MA}}
+\Theta_{\varrho}^{\mathrm{MA}})$,  
two-body part $W_{\varrho}^{\mathrm{MA}}$, and three-body part 
$\Theta_{\varrho}^{\mathrm{MA}}
 = \Theta_\mathrm{I}^{} + \widetilde{\Theta}_{\mathrm{II}}^{}$. 
Left panels: $N=4$, for 
$U/(\pi\Delta) = 2/3(\star)$, $1(\blacktriangledown)$, $2(\blacklozenge)$, $3(\blacksquare)$, $4(\blacktriangle)$, $5(\bullet)$, $6(\times)$.
Right panels: $N=6$, for 
$U/(\pi\Delta) = 2/5(\star)$, $1(\blacktriangledown)$, $2(\blacklozenge)$, 
$3(\blacksquare)$, $4(\blacktriangle)$, $5(\bullet)$, $6(\times)$.} 

\label{fig:Crho-MA-N4-6-U}
\end{figure}

The coefficient $C_{\varrho}^{\mathrm{MA}}$ for the order $T^2$  resistivity,
is defined in Table \ref{tab:C_and_W_extended}. 
It  consists of two-body part $W_{\varrho}^{\mathrm{MA}}$ and 
three-body part $\Theta_{\varrho}^{\mathrm{MA}}$:
\begin{align}
\!\!\! 
C_{\varrho}^{\mathrm{MA}} 
 =   
\frac{\pi^2}{48}
 \,
\left(
W_{\varrho}^{\mathrm{MA}} + \Theta_{\varrho}^{\mathrm{MA}}
\right),
\qquad 
\Theta_{\varrho}^{\mathrm{MA}} 
 \equiv  \Theta_\mathrm{I}^{} +
\widetilde{\Theta}_{\mathrm{II}}^{}.
\end{align} 
Note that  
 $\Theta_{\varrho}^{\mathrm{MA}} \equiv  \Theta_{T}^{}$, 
 i.e.,  the three-body part for the $T^2$ resistivity of magnetic alloys 
is identical to the one for the $T^2$ conductance of quantum dots.  
Therefore, $\Theta_{\varrho}^{\mathrm{MA}}$ does not 
depend on $\chi_{B}^{[3]}$ and 
is determined by the derivative of the charge and spin susceptibilities
through  Eq.\ \eqref{eq:ThetaT_vs_derivativeCHI}.
Figures \ref{fig:Crho-MA-N4-6-U}(a)--\ref{fig:Crho-MA-N4-6-U}(d) 
show the NRG results for 
 $C_{\varrho}^{\mathrm{MA}}$,  $W_{\varrho}^{\mathrm{MA}}$,
 and $\Theta_{\varrho}^{\mathrm{MA}}$ 
 for both the SU(4) and SU(6) symmetric cases.

The coefficient $C_{\varrho}^{\mathrm{MA}}$ is positive 
and is less sensitive to the impurity level position $\xi_d^{}$ 
as compared to the quantum-dot counterpart  $C_T^{}$ 
shown in Fig.\ \ref{fig:CT-QD-N4-6-U}. 
 This difference is caused by  the first term,  $(\pi^2/12) \cos^2\delta$, 
appearing in the right-hand side of Eq.\ \eqref{eq:Crho_CT_relation}.
 In particular, in the noninteracting case, it becomes a constant, 
  $(48/\pi^2) C_{\varrho}^{\mathrm{MA}}
\xrightarrow{U=0} 1$, independent of  the level position,   
as all effects due to  $\epsilon_d^{}$   
are absorbed into the characteristic energy $T^*$ 
for $U=0$ (see Appendix \ref{sec:thermo_MA_U0}).   
Therefore, it is the strong electron correlation 
that makes $C_{\varrho}^{\mathrm{MA}}$, 
shown in Figs.\ \ref{fig:Crho-MA-N4-6-U}(a) and \ref{fig:Crho-MA-N4-6-U}(b),   
deviates from the constant value.

The three-body part almost vanishes,  
$\Theta_{\varrho}^{\mathrm{MA}} \simeq 0$, 
for large $U$ 
in the strong-coupling region $|\xi_{d}^{}|\lesssim (N-1)U/2$,
as mentioned for $\Theta_{T}^{}$ 
in  Eq.\ \eqref{eq:ThetaT_vs_derivativeCHI}. 
This is caused by the fact that the magnitudes of the derivatives, 
  $\partial \overline{\chi}_{C}^{}/\partial \epsilon_{d}^{}$ 
and  $\partial \overline{\chi}_{S}^{}/\partial \epsilon_{d}^{}$, 
are significantly suppressed by the Coulomb repulsion 
in the wide range of electron filling $1 \lesssim N_d^{} \lesssim N-1$, 
as demonstrated in Figs.\ 
\ref{fig:DChiC_DChiS_chi3B_N4-6_u6}(a)--\ref{fig:DChiC_DChiS_chi3B_N4-6_u6}(d).
Therefore, in this region $|\xi_{d}^{}|\lesssim (N-1)U/2$,   
the coefficient  $C_{\varrho}^{\mathrm{MA}}$
is determined solely by the two-body part $W_{\varrho}^{\mathrm{MA}}$, 
which takes the form   
\begin{align}
 C_{\varrho}^{\mathrm{MA}}
\,\simeq\, 
\frac{\pi^2}{48}\left[ 
2 + \left(1- \frac{2}{N-1}\right)  \left(1-2\sin^2 \delta\right)\right]  \, > \,0,
\label{eq:CrhoMA_strong}
\end{align}
as the rescaled Wilson ratio is saturated to $\widetilde{K} \simeq 1$.
Hence, the plateaus emerging around the integer filling points 
 $\delta/\pi = 1$, $2$, $\ldots$, $N-1$, 
reflect the structures of the Kondo ridge occurring 
for the  transmission probability $\sin^2 \delta$, 
seen in Figs.\ 
\ref{fig:FL-parameters-N4-U-from-2-over-3-to-6}(b) and
\ref{fig:FL-parameters-N6-U-from-2-over-5-to-6}(b) 
in Appendix \ref{sec:NRG_results_2body_functions_for_sun}.
Among these plateaus,  the one at half filling, where $\delta=\pi/2$,  
takes the smallest value:  
 $(48/\pi^2) C_{\varrho}^{\mathrm{MA}}
 \xrightarrow{\, \xi_d^{}=0\,\&\, U\to \infty} 5/3$ 
 and $7/5$  for $N=4$ and $6$,  respectively.   
As $\xi_d^{}$ moves away from half filling,  
the coefficient $C_{\varrho}^{\mathrm{MA}}$ 
increases in the region of $|\xi_{d}^{}|\lesssim (N-1)U/2$.
Equation \eqref{eq:CrhoMA_strong} also indicates that   
the plateau becomes highest at the electron fillings 
of $N_d^{} \simeq 1$ and $N-1$:  $(48/\pi^2) C_{\varrho}^{\mathrm{MA}}
\xrightarrow{U\to \infty, \, \delta =\pi/N} 2$ and $23/10$  for $N=4$  and $6$,  
respectively.
 The NRG results for  $C_{\varrho}^{\mathrm{MA}}$,
shown in Figs.\ \ref{fig:Crho-MA-N4-6-U}(a) and \ref{fig:Crho-MA-N4-6-U}(b),  
clearly demonstrate these behaviors, 
which are quite different form the behaviors of $C_{T}^{}$ of quantum dots.  
Equation \eqref{eq:CrhoMA_strong}  also shows   
that,  in the SU(2) symmetric case where $N=2$,  
the coefficient $C_{\varrho}^{\mathrm{MA}}$ 
takes the maximum value at half filling,  
as demonstrated also in Appendix \ref{sec:NoiseN2}.

In contrast,  at  $|\xi_{d}^{}| \gtrsim (N-1)U/2$,  
the occupation number approaches 
the empty $N_d^{} \simeq 0$ or the  fully occupied $N_d^{} \simeq N$ states 
as the impurity level moves further away from the Fermi level.  
In this region,  both the two-body  
 $W_{\varrho}^{\mathrm{MA}}$
and three-body 
 $\Theta_{\varrho}^{\mathrm{MA}}$
parts give comparable contributions to  $C_{\varrho}^{\mathrm{MA}}$,
and approach the noninteracting values: 
\begin{align}
W_{\varrho}^{\mathrm{MA}}
\,\xrightarrow{\, |\xi_d^{}| \to \infty}\, 3, 
\qquad
\Theta_{\varrho}^{\mathrm{MA}}
\,\xrightarrow{\, |\xi_d^{}| \to \infty}\, -2,   
\end{align}
and thus  $(48/\pi^2)C_{\varrho}^{\mathrm{MA}}
\,\xrightarrow{\, |\xi_d^{}| \to \infty}1$.

\subsection{$C_{\kappa}^{\mathrm{MA}}$: order  $T^3$ term 
of $\kappa_\mathrm{MA}^{}$ }

\begin{figure}[t]

\leavevmode
 \centering

\includegraphics[width=0.47\linewidth]{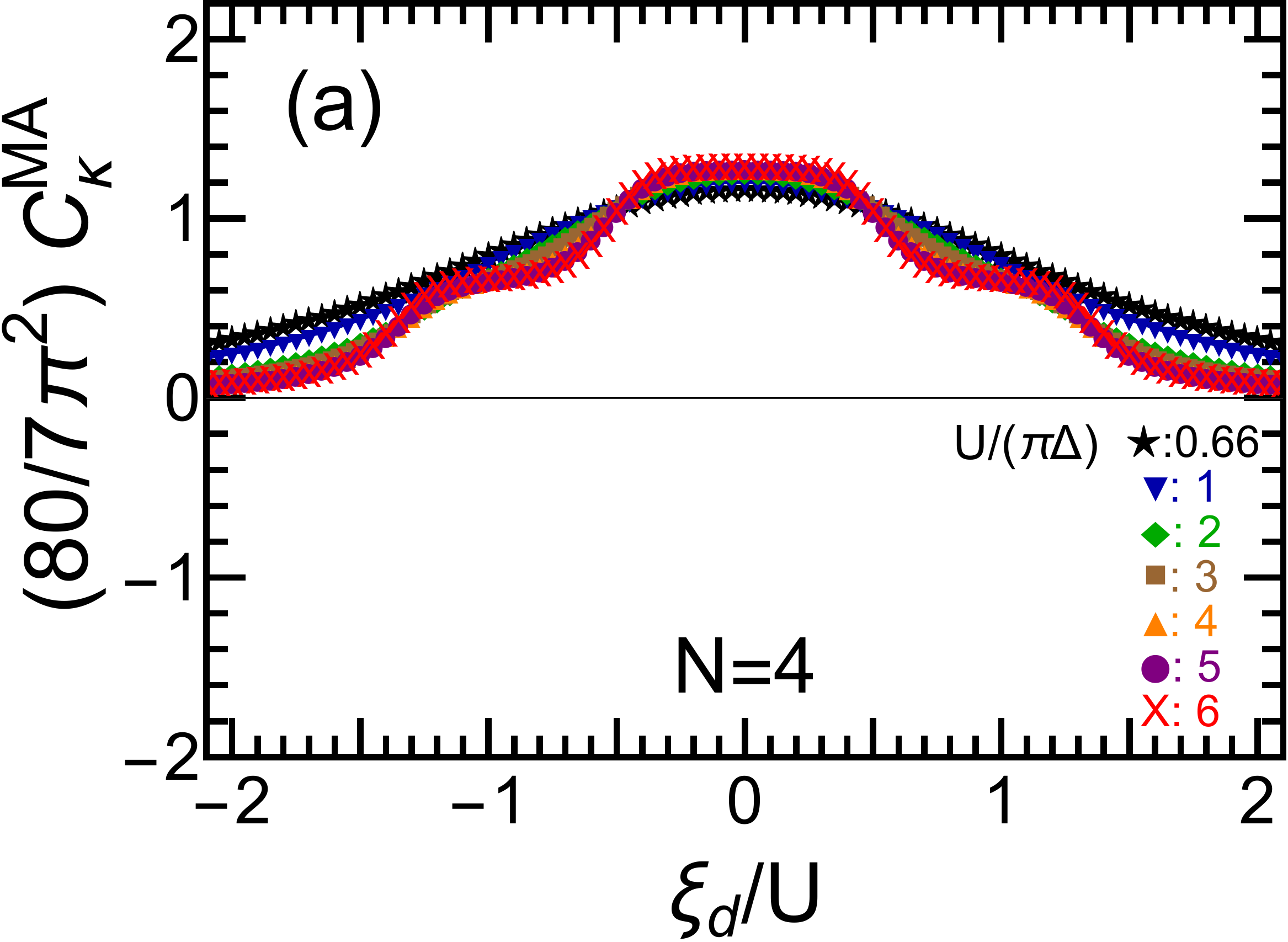}
 \hspace{0.01\linewidth} 
\includegraphics[width=0.47\linewidth]{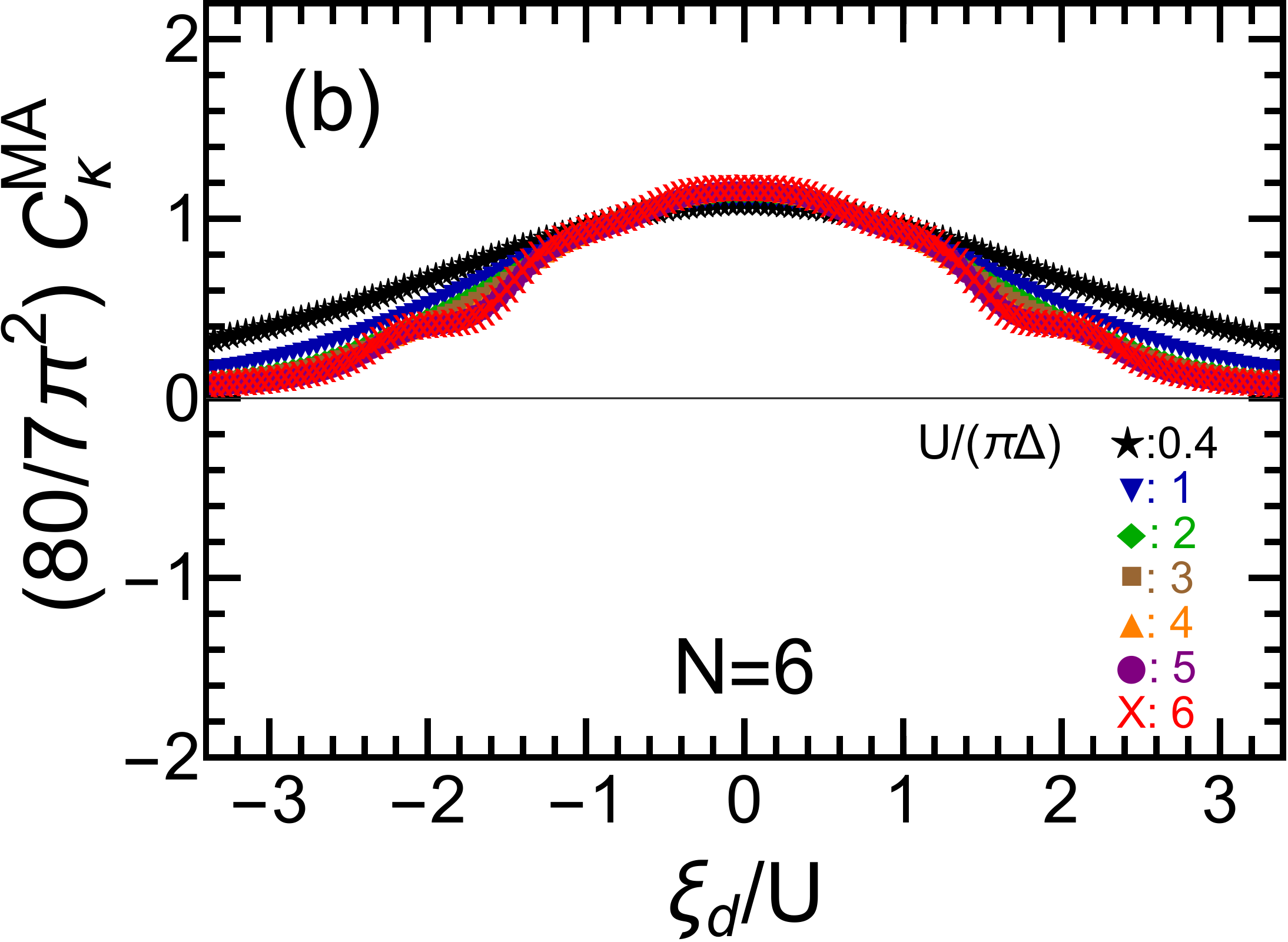}
\\
\includegraphics[width=0.47\linewidth]{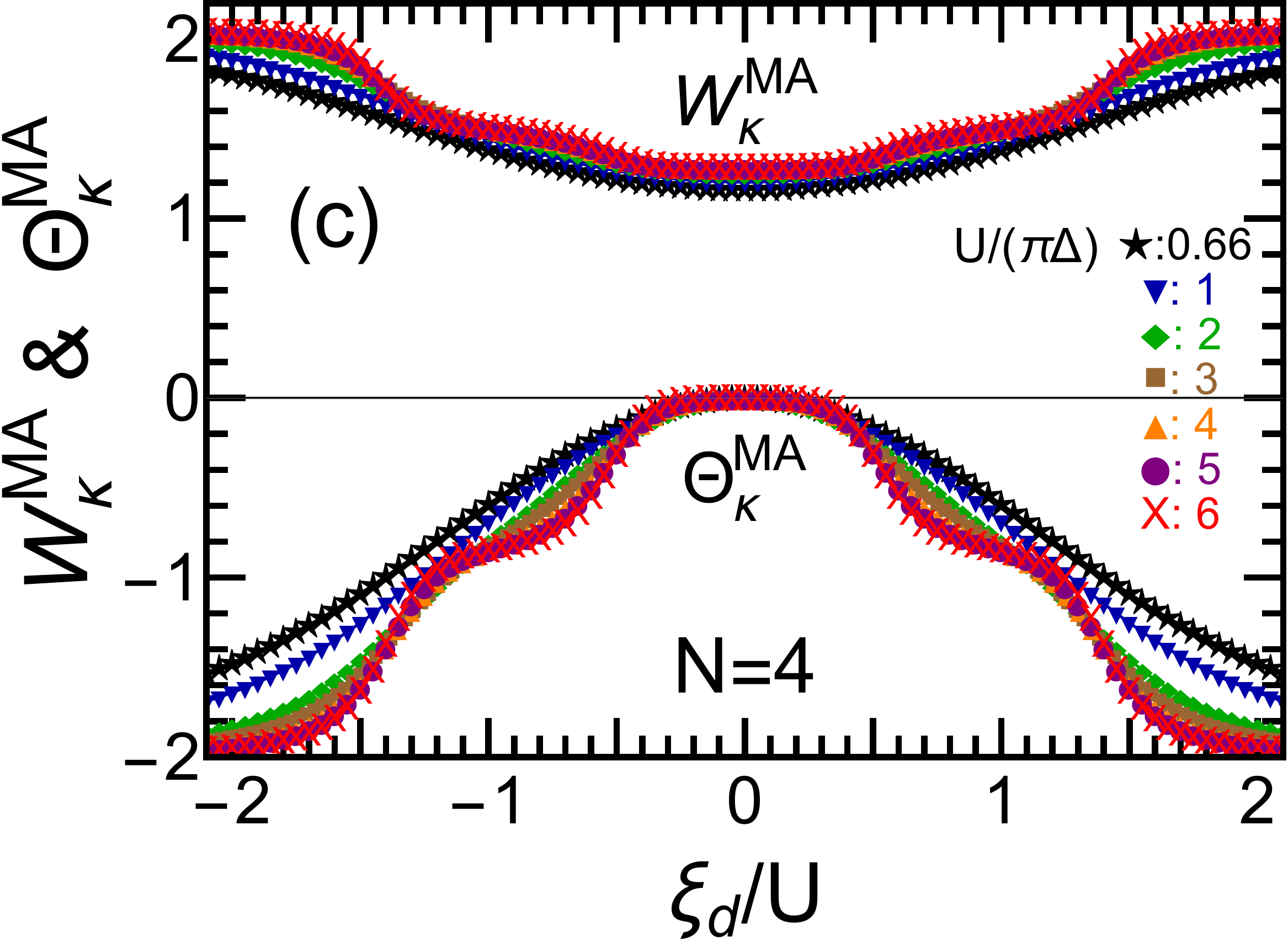}
 \hspace{0.01\linewidth} 
\includegraphics[width=0.47\linewidth]{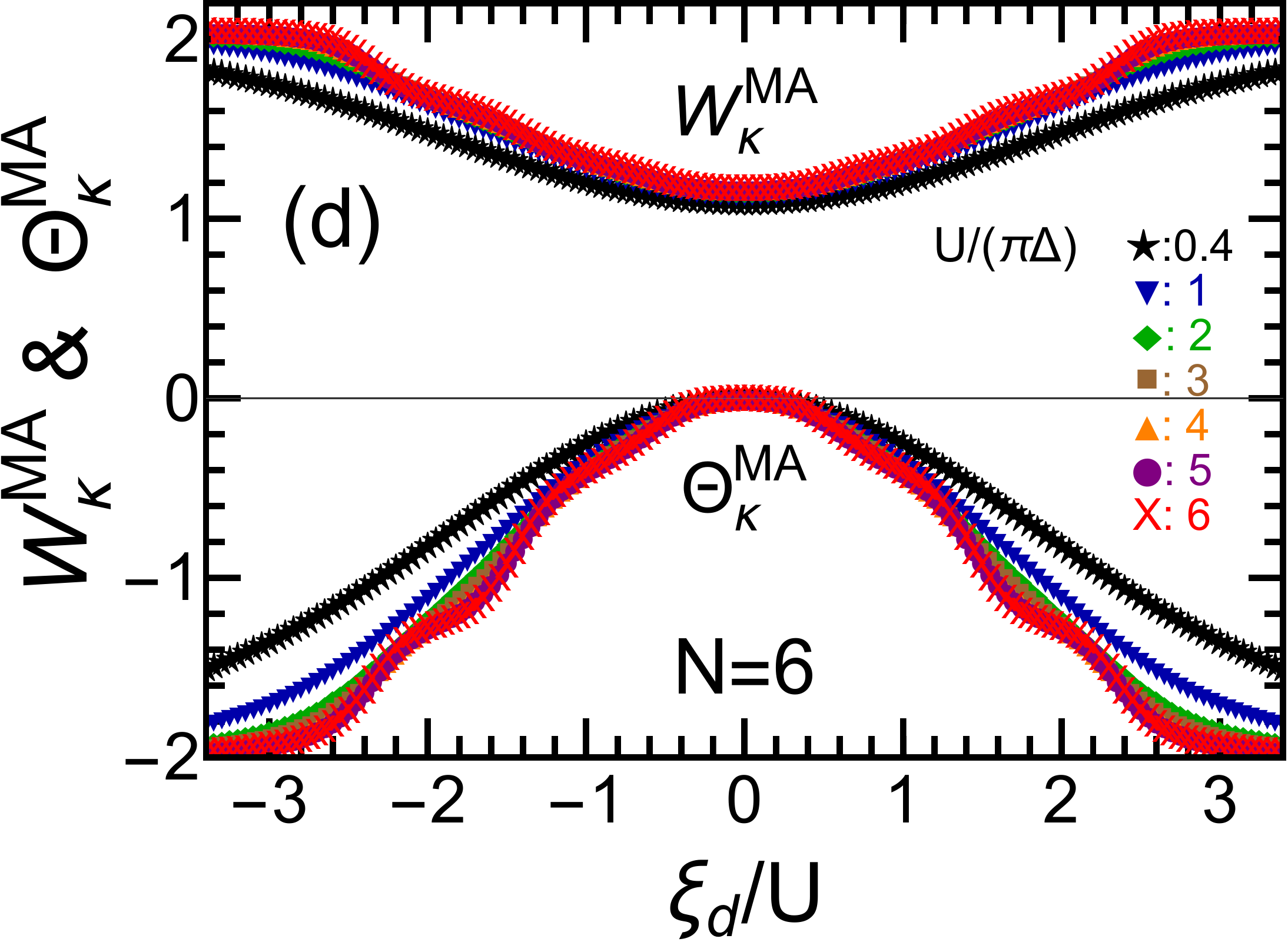}
\caption{$\xi_d^{}$ dependence of 
 $C_{\kappa}^{\mathrm{MA}}=(7\pi^2/80) (W_{\kappa}^{\mathrm{MA}}
+\Theta_{\kappa}^{\mathrm{MA}})$,  
two-body part $W_{\kappa}^{\mathrm{MA}}$, and three-body part 
$\Theta_{\kappa}^{\mathrm{MA}} 
 = \Theta_\mathrm{I}^{} +
\frac{5}{21}\widetilde{\Theta}_{\mathrm{II}}^{}$.
Left panels: $N=4$, for 
$U/(\pi\Delta) = 2/3(\star)$, $1(\blacktriangledown)$, $2(\blacklozenge)$, $3(\blacksquare)$, $4(\blacktriangle)$, $5(\bullet)$, $6(\times)$.
Right panels: $N=6$, for 
$U/(\pi\Delta) = 2/5(\star)$, $1(\blacktriangledown)$, $2(\blacklozenge)$, 
$3(\blacksquare)$, $4(\blacktriangle)$, $5(\bullet)$, $6(\times)$.} 
\label{fig:Ckappa-MA-N4-6-U}
\end{figure}

We next consider the order $T^3$ term of 
 thermal conductivity $\kappa_{\mathrm{MA}}$ 
of the SU($N$) symmetric  magnetic alloys. 
The coefficient  $C_{\kappa}^{\mathrm{MA}}$, 
defined  in Tables 
\ref{tab:low_energy_expansion} and \ref{tab:C_and_W_extended},  
consists of two-body $W_{\kappa}^{\mathrm{MA}}$ 
and  three-body $\Theta_{\kappa}^{\mathrm{MA}}$ parts:  
\begin{align}
C_{\kappa}^{\mathrm{MA}} 
\, =  \, 
\frac{7\pi^2}{80}
 \,\bigl(W_{\kappa}^{\mathrm{MA}} + \Theta_{\kappa}^{\mathrm{MA}}
\bigr) , \qquad
\Theta_{\mathrm{\kappa}}^{\mathrm{MA}} 
\, \equiv \, \Theta_\mathrm{I}^{} +
\frac{5}{21}\widetilde{\Theta}_{\mathrm{II}}^{} .
\end{align} 
Note that  $\Theta_{\mathrm{\kappa}}^{\mathrm{MA}} 
 \equiv \Theta_{\mathrm{\kappa}}^{\mathrm{QD}}$, 
i.e., the three-body part for 
 $C_{\kappa}^{\mathrm{MA}}$ of magnetic alloys 
is identical to the one for the $T^3$ thermal conductance of quantum dots.  
The NRG results for these coefficients 
 $C_{\kappa}^{\mathrm{MA}}$,  $W_{\kappa}^{\mathrm{MA}}$,
 and $\Theta_{\kappa}^{\mathrm{MA}}$ are shown  
in Figs.\ \ref{fig:Ckappa-MA-N4-6-U}(a)--\ref{fig:Ckappa-MA-N4-6-U}(d) 
 for $N=4$ and $6$.

The coefficient $C_{\kappa}^{\mathrm{MA}}$ for magnetic alloys 
 is less sensitive to  $\xi_d^{}$ 
as compared to  $C_{\kappa}^{\mathrm{QD}}$ for 
 for quantum dots. 
 This difference  is caused 
by the contribution of the first term,  $(11\pi^2/60) \cos^2\delta$, 
in the right-hand side of Eq.\ \eqref{eq:CkappaMA_CkappaQD_relation}. 
The coefficient $C_{\kappa}^{\mathrm{MA}}$ 
is positive and has a broad peak at $\xi_{d}^{}= 0$, 
the height of which is determined by 
 the two-body contribution $W_{\kappa}^{\mathrm{MA}}$
as three-body part $\Theta_{\kappa}^{\mathrm{MA}}$ vanishes  
in the electron-hole symmetric case: 
\begin{align}
 W_{\kappa}^{\mathrm{MA}}\,\xrightarrow{\, \xi_d^{}=0}\, 
 1 + \frac{6 \widetilde{K}^2}{7(N-1)} \,, 
\qquad 
\Theta_{\kappa}^{\mathrm{MA}} \xrightarrow{\,\xi_d^{}=0\,} 0\,.
\end{align}
In the limit of $U \to \infty$,   
it reaches the maximum possible value, $[80/(7\pi^2)] C_{\kappa}^{\mathrm{MA}} 
\xrightarrow{\, \xi_d^{}=0\,\&\, U\to \infty} 9/7$ 
and $41/35$  for $N=4$ and $6$, respectively.

In the strong-coupling region  $|\xi_{d}^{}|\lesssim (N-1)U/2$, 
the coefficient  $C_{\kappa}^{\mathrm{MA}}$ takes the following form 
for large $U$,  
\begin{align}
 C_{\kappa}^{\mathrm{MA}}\,\simeq\, 
\frac{\pi^2}{240}\left[
 32 + \left(11- \frac{18}{N-1}\right)  \,
\cos 2 \delta  
\,+\,16\, \Theta_\mathrm{I}^{} \right].  
\label{eq:CkappaMA_strong}
\end{align}
This is because, in this region,  
 the rescaled Wilson ratio is almost saturated $\widetilde{K} \simeq 1$ 
  (see Appendix \ref{sec:NRG_results_2body_functions_for_sun}) and  
 the three-body part is parameterized by a single component, 
 $\Theta_{\kappa}^{\mathrm{MA}}\simeq \frac{16}{21}\, 
\Theta_\mathrm{I}^{}$,  
due to the property described in  
  Eq.\ \eqref{eq:Theta_I_II_III_relations_strong_U_simplified}. 
 The plateau structures of  $C_{\kappa}^{\mathrm{MA}}$, appearing   
in Figs.\ \ref{fig:Ckappa-MA-N4-6-U}(a) and \ref{fig:Ckappa-MA-N4-6-U}(b) 
around the integer filling points 
 $N_d^{} = 1$, $2$, $\ldots$, $N-1$, 
are determined by both the $\cos 2 \delta$ term in  $W_{\kappa}^{\mathrm{MA}}$ 
and the plateaus occurring in $\Theta_{\kappa}^{\mathrm{MA}}$ 
 seen in Figs.\ \ref{fig:Ckappa-MA-N4-6-U}(c) and \ref{fig:Ckappa-MA-N4-6-U}(d). 

In contrast, at  $|\xi_{d}^{}| \gtrsim (N-1)U/2$,  
the electron filling  approaches the empty 
$N_d^{} \simeq 0$ or the fully occupied $N_d^{} \simeq N$ states   
as the impurity level goes very far away from the half filled point.
Therefore,  the order $T^3$ thermal conductivity for magnetic alloys 
also approaches  the noninteracting value in this limit:  
\begin{align}
W_{\kappa}^{\mathrm{MA}}
\,\xrightarrow{\, |\xi_d^{}| \to \infty}\, \frac{43}{21}, 
\qquad
\Theta_{\kappa}^{\mathrm{MA}}
\,\xrightarrow{\, |\xi_d^{}| \to \infty}\, -2,   
\end{align}
and thus $[80/(7\pi^2)]\, C_{\kappa}^{\mathrm{MA}}
\,\xrightarrow{\, |\xi_d^{}| \to \infty}1/21$.

\subsection{$C_{L}^{\mathrm{MA}}$: order  $T^2$ term of  $L_{\mathrm{MA}}$}

The Lorenz number  
 $L_{\mathrm{MA}} \equiv 
\kappa_{\mathrm{MA}}^{}/(\sigma_{\mathrm{MA}}^{}\,T)$  
for magnetic alloys is defined 
as the ratio of 
the thermal conductivity  $\kappa_\mathrm{MA}^{}/T$ to 
 electrical conductivity $\sigma_\mathrm{MA}^{}$.  
It takes the universal Wiedemann-Franz value at zero temperature:  
$L_{\mathrm{MA}} \xrightarrow{T\to 0}\pi^{2}/(3e^{2})$. 
However, it deviates from this value as the temperature increases, 
showing the $T^2$ dependence  as described in Eq.\ \eqref{eq:L_MA_SUN_formula}. 
The precise expansion formula for the coefficient $C_{L}^{\mathrm{MA}}$ 
for  the order $T^2$ term is shown in Table \ref{tab:C_and_W_extended}. 
It is given by the difference,  
 $C_{L}^{\mathrm{MA}}=
 C_{\varrho}^{\mathrm{MA}}-C_{\kappa}^{\mathrm{MA}}$,   
between the order $T^2$ term of  
 the resistivities,  $1/\sigma_{\mathrm{MA}}^{}$ and 
$T/\kappa_{\mathrm{MA}}^{}$,  defined in   
Eqs.\ \eqref{eq:FL_rho_MA} and \eqref{eq:L_MA_SUN_formula}.  
The coefficient  $C_{L}^{\mathrm{MA}}$ for magnetic alloys 
and the quantum-dot counterpart $C_{L}^{\mathrm{QM}}$ 
are related to each other through Eq.\ \eqref{eq:CLMA_CLQD_relation}: 
 these two coefficients tend to have opposite signs.  
In Appendix \ref{sec:thermo_MA_U0},
we have also provide an analytic formula for $C_{L}^{\mathrm{MA}}$  
 in the noninteraction case, for comparison.

\begin{figure}[t]

\leavevmode
 \centering

\includegraphics[width=0.47\linewidth]{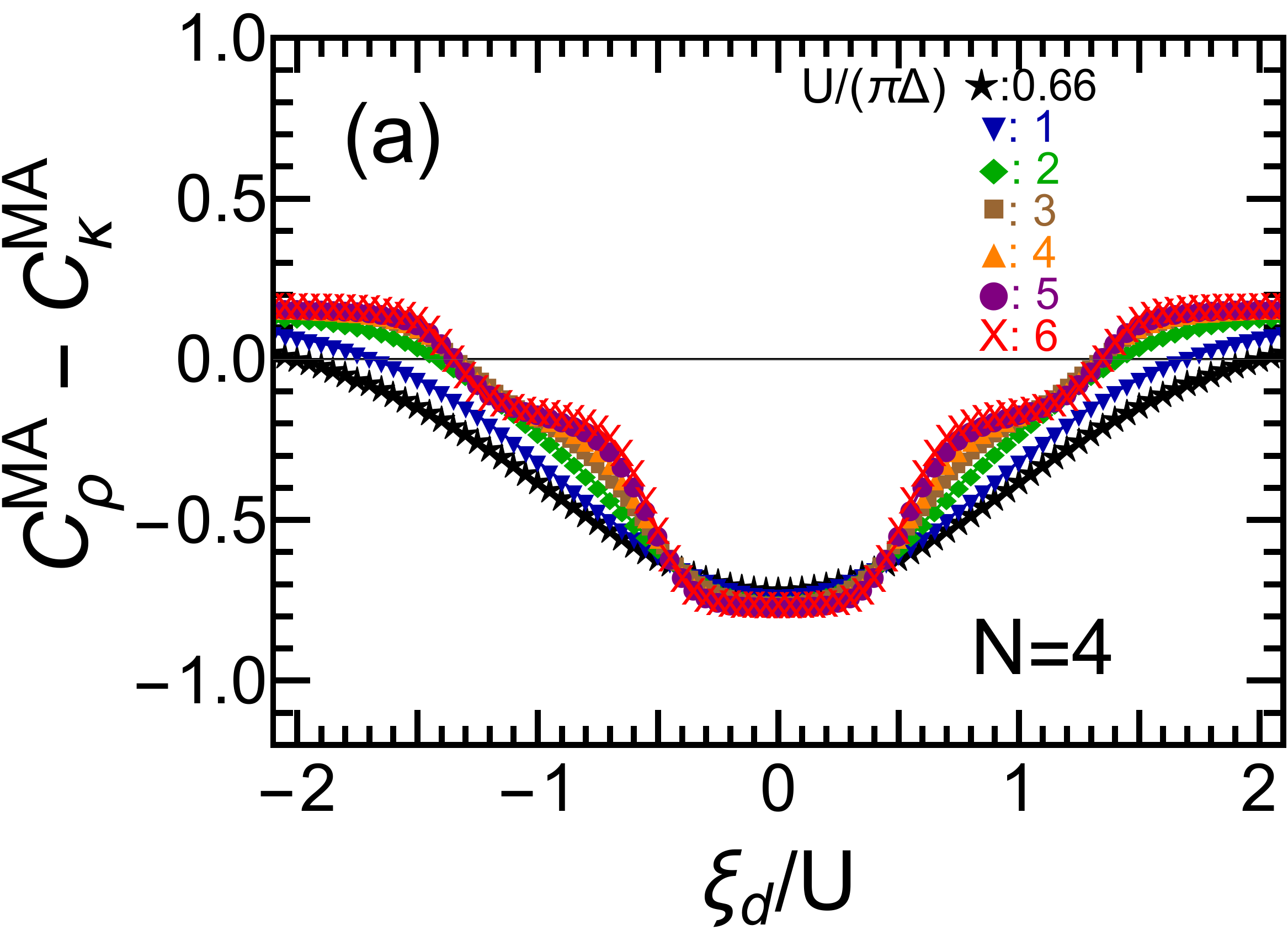}
 \hspace{0.01\linewidth} 
\includegraphics[width=0.47\linewidth]{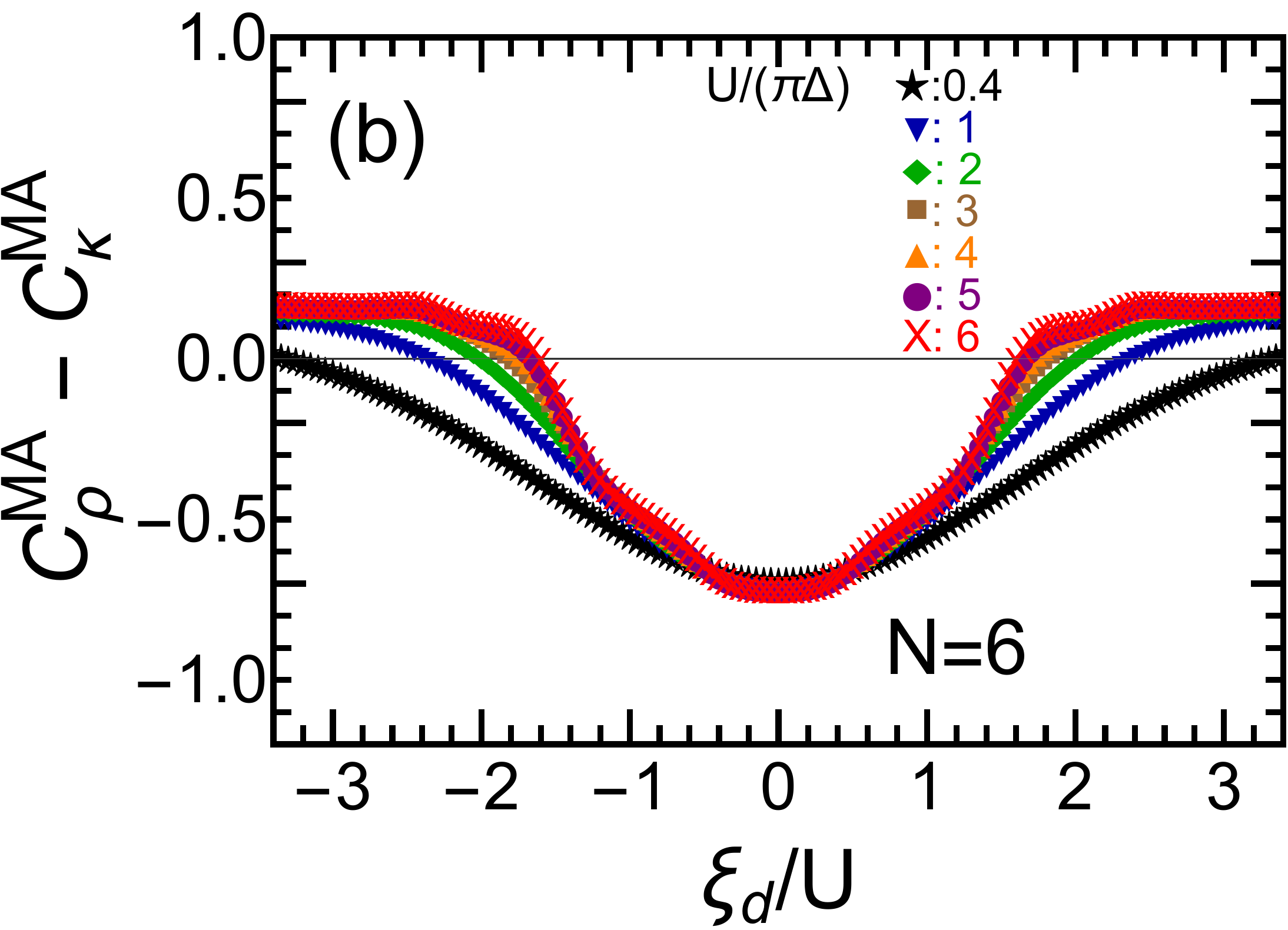}

\caption{Order $T^{2}$ term  
$C_{L}^{\mathrm{MA}}=
C_{\varrho}^{\mathrm{MA}}- C_{\kappa}^{\mathrm{MA}}$ 
of Lorenz number $L_{\mathrm{MA}}$ 
is plotted vs $\xi_{d}^{}$. 
Left panel: $N=4$,  for $U/(\pi\Delta) = 2/3(\star)$, 
$1(\blacktriangledown)$, $2(\blacklozenge)$, $3(\blacksquare)$, $4(\blacktriangle)$, 
$5(\bullet)$, $6(\times)$.
Right panel: $N=6$,  
for $U/(\pi\Delta) = 2/5(\star)$, $1(\blacktriangledown)$, $2(\blacklozenge)$, 
$3(\blacksquare)$,  $4(\blacktriangle)$, $5(\bullet)$, $6(\times)$.}
\label{fig:CL-MA-N4-6-U}
\end{figure}

The NRG results for $C_{L}^{\mathrm{MA}}$ 
are shown in
Figs.\ \ref{fig:CL-MA-N4-6-U}(a) and \ref{fig:CL-MA-N4-6-U}(b)  
for $N=4$ and $6$,  respectively.
Near half filling  $|\xi_d^{}| \lesssim U/2$, 
the coefficient  $C_{L}^{\mathrm{MA}}$ is determined 
by the two-body part $W_{L}^{\mathrm{MA}}$ as the three-body part, 
given by $-16\Theta_\mathrm{I}^{}$, vanishes at  $\xi_d^{} = 0$: 
\begin{align} 
C_{L}^{\mathrm{MA}}
\,\xrightarrow{\, \xi_d^{}=0}\, 
-\frac{\pi^2}{30}\left(  2+  \frac{\widetilde{K}^2}{N-1} \right)  \ < 0 \,. 
\end{align} 
This coefficient attains its greatest possible negative value 
in the limit of $U\to \infty$ where $\widetilde{K} \to 1$:   
 $C_{L}^{\mathrm{MA}}
\xrightarrow{\xi_d^{}=0\, \&\,U\to \infty}-0,767\cdots$ 
for $N=4$, and $-0.723\cdots$ for $N=6$.

In the strong-coupling region $|\xi_d^{}| \lesssim (N-1)U/2$, 
the $C_{L}^{\mathrm{MA}}$  also exhibits 
the plateau structures around the integer filling points 
$\xi_d^{}/U=0$, $\pm 1$, $\ldots$, $\pm (N-2)/2$ for large $U$, 
reflecting the structures that appear for both  $C_{\varrho}^{\mathrm{MA}}$ 
and $C_{\kappa}^{\mathrm{MA}}$. 
In particular,  the plateaus 
at  $N_d^{} \simeq \frac{N}{2} \pm 1$ fillings, 
seen in Figs.\ \ref{fig:CL-MA-N4-6-U}(a) and \ref{fig:CL-MA-N4-6-U}(b)  
for $U/(\pi \Delta) \gtrsim 3$, 
take a negative value for both $N=4$ and $N=6$, 
whereas  the other plateaus become positive for $N=6$   
as the electrical resistivity dominates, i.e.,   
$C_{\varrho}^{\mathrm{MA}} >  C_{\kappa}^{\mathrm{MA}}$.

As the impurity level goes far away from the Fermi level 
in the region $|\xi_d^{}| \gtrsim (N-1)U/2$,  
the occupation number approaches $N_d^{}\simeq 0$ 
or $N_d^{}\simeq N$.  
In the limit of $|\xi_{d}^{}| \to \infty$, 
 the two-body and three-body parts  take the noninteracting values,  
$W_{L}^{\mathrm{MA}}\xrightarrow{|\xi_d^{}| \to \infty} -28$ 
and  $-16 \Theta_\mathrm{I}^{} \xrightarrow{|\xi_d^{}| \to \infty} 32$,  
and the coefficient $C_{L}^{\mathrm{MA}}$ converges to the positive value,   
 \begin{align}
C_{L}^{\mathrm{MA}}
\,\xrightarrow{\, |\xi_d^{}| \to \infty}\, 
 \frac{\pi^2}{60}  \  = \ 
 0.164 \cdots \  > 0\,.
\end{align}

\section{Summary}
\label{sec:summary}

We have presented a comprehensive Fermi liquid description 
for nonlinear current and thermoelectric transport 
through quantum dots and magnetic alloys, 
which is asymptotically exact at low energies up to the next-leading order terms. 
Our formulation  is  based on the multilevel Anderson model 
and is applicable to  arbitrary impurity-level structures $\epsilon_{d\sigma}^{}$ 
 for $\sigma=1$, $2$, $\ldots$, $N$, including the spin degrees of freedom. 
The coefficients for the next-leading order terms have been shown to be expressed 
in terms of a set of the correlation functions defined with respect 
to the equilibrium ground state:  
the phase shift $\delta_{\sigma}^{}$, 
the static susceptibilities $\chi_{\sigma_1,\sigma_2}^{}$, and 
the three-body correlations $\chi_{\sigma_1,\sigma_2,\sigma_3}^{[3]}$   
which emerge when the system does not have both the electron-hole 
and time-reversal symmetries.   

Extending Yamada-Yosida's  field-theoretical approach,  we have obtained
 the formulas for the differential conductance $dI/dV$, 
current noise $S_{\mathrm{noise}}^{\mathrm{QD}}$ and 
thermal conductance $\kappa_{\mathrm{QD}}^{}$ of quantum dots, 
and also the electrical resistivity $\varrho_{\mathrm{MA}}^{}$ 
 and thermal conductivity $\kappa_{\mathrm{MA}}^{}$ of dilute magnetic alloys. 
 In the SU($N$) symmetric case,  these transport coefficients 
take simplified forms, as listed in  
Tables \ref{tab:low_energy_expansion} and \ref{tab:C_and_W_extended}, 
and the three-body correlations 
can be deduced from the derivatives of the susceptibilities: 
 $\partial \overline{\chi}_{C}^{}/\partial \epsilon_d^{}$, 
$\partial \overline{\chi}_{S}^{}/\partial \epsilon_d^{}$,  and $\chi_{B}^{[3]}$ 
through Eqs.\  \eqref{eq:chi3_SUN_1}--\eqref{eq:chi_B3_def_new}.

We have also calculated the correlation functions for the SU(4) and SU(6) cases,   
 using the NRG approach over the whole region of impurity-electron fillings $N_d^{}$,      
which includes the Kondo and the valence-fluctuation regimes. 
In the SU($N$) case, 
the three-body correlations have three linearly independent components    
that approach each other closely as the Coulomb interaction $U$ increases:       
 $\chi_{\sigma\sigma\sigma}^{[3]} 
\simeq - (N-1)\chi_{\sigma\sigma'\sigma'}^{[3]} 
\simeq \frac{(N-1)(N-2)}{2}\chi_{\sigma\sigma'\sigma''}^{[3]}$ 
for $\sigma \neq \sigma' \neq \sigma'' \neq \sigma$,     
in a wide filling range  $1 \lesssim N_{d}^{}\lesssim N-1$. 
This is caused by the suppression of 
the derivatives of the charge and spin susceptibilities,
 occurring at large  $U$: 
 $|\partial \overline{\chi}_{C}^{}/\partial \epsilon_d^{}| \ll (T^*)^{-2}$ and 
$|\partial \overline{\chi}_{S}^{}/\partial \epsilon_d^{}| \ll (T^*)^{-2}$,   
with  $T^* \equiv  1/(4\chi_{\sigma\sigma}^{})$ 
the characteristic energy scale of  the SU($N$) Fermi liquid.   
This property of three-body correlations  
is also related to a similar property of linear susceptibilities,  
  $\chi_{\sigma\sigma}^{} \simeq - (N-1)\chi_{\sigma\sigma'}^{}$, 
which reflects the suppression of charge fluctuations, 
 $\overline{\chi}_{C}^{}\simeq 0$, occurring at large $U$.

The coefficients  $C$'s for the next-leading order terms can be decomposed into 
the two-body part $W$'s and the three-body part $\Theta$'s,
as listed in Table \ref{tab:C_and_W_extended}.
The NRG results show that the three-body part $\Theta_{V}^{}$ 
of the coefficient $C_{V}^{}$ for  the order $(eV)^{2}$ term of $dI/dV$ 
exhibits  Kondo plateau structures  
 at integer filling points $N_d^{}= 1$, $2$, $\ldots$, $N-1$ for large $U$.
These plateaus of  $\Theta_{V}^{}$ complement 
 the two body part $W_V^{}$,    which decreases away from half filling,   
to form a wide ridge structure in $C_V^{}$   
 that spreads over the region of  $1 \lesssim N_d^{}\lesssim N-1$.

The linear-response term of  
current noise $S_{\mathrm{noise}}^{\mathrm{QD}}$ 
is maximized at quarter filling $N_d^{}/N=1/4$ and three-quarters $3/4$ filling,  
and the peaks exhibit flat structures for $N=4$, while they are round for $N=6$.  
This difference is caused by the fact that, at these fillings, 
the SU($N$) Kondo effects occur for $N \equiv 0$ (mod $4$),  
while  the intermediate valence fluctuations occur for $N \equiv 2$ (mod $4$).  
The coefficient $C_{S}^{}$ for  
the order $|eV|^{3}$ nonlinear term of $S_{\mathrm{noise}}^{\mathrm{QD}}$  
has a peak at half filling, which evolves into a plateau as $U$ increases. 
As the impurity level $\xi_d^{}$ deviates from the half filling point,
$C_{S}^{}$  decreases rapidly for $N=4$,   
whereas it varies more modestly for $N=6$.  
This is mainly due to the  higher-harmonic ``$\sin 4 \delta$" dependence  
of the three-body part $\Theta_{S}^{}$, 
which vanishes at the quarter and three-quarters filling points. 
As $|\xi_d^{}|$ increases further,  
 $\Theta_{S}^{}$ has a pronounced negative minimum  
 in the valence fluctuation regions at $\xi_d^{} \simeq \pm (N-1)U/2$ 
for both $N=4$ and $6$, 
which yield  the valley structures appearing in $C_{S}^{}$.
This is in marked contrast to the SU(2) case, 
where the valley structure emerging in $C_S^{}$ is caused by 
the two-body contributions $W_S^{}$, instead of $\Theta_S^{}$.

We have also studied the coefficient $C_{T}^{}$ for 
 the order $T^2$  term of the linear conductance 
 $g \equiv  \left. dJ/dV \right|_{eV=0}^{}$ 
and the coefficient  $C_{\kappa}^{\mathrm{QD}}$ for the order $T^3$ term of 
thermal conductance $\kappa_\mathrm{QD}^{}$, for SU($N$) quantum dots.  
The three-body part  $\Theta_{T}^{}$ of  $C_{T}^{}$ 
 is determined solely by  the derivatives of charge and spin susceptibilities, i.e.,  
$\partial \overline{\chi}_{C}^{}/\partial \epsilon_{d}^{}$ and 
$\partial \overline{\chi}_{S}^{}/\partial \epsilon_{d}^{}$, 
and is independent of $\chi_{B}^{[3]}$.  
Therefore, 
$\Theta_{T}^{}$ is significantly suppressed and almost vanishes 
 for strong interactions $U$ in a wide range of the electron filling 
$1 \lesssim N_{d}\lesssim N-1$. 
In contrast, the three-body part 
 $\Theta_{\kappa}^{\mathrm{QD}}$  for thermal conductance  
involves  $\chi_{B}^{[3]}$  and 
becomes comparable to the  two-body part $W_{\kappa}^{\mathrm{QD}}$, 
except for the plateau region near half filling. 
The Lorenz number 
  $L_{\mathrm{QD}} \equiv \kappa_{\mathrm{QD}}^{}/( g\, T)$ 
for quantum dots  takes the universal Wiedemann-Franz value $\pi^2/(3e^2)$ 
at $T=0$ and shows a $T^{2}$ dependence at low temperatures.  
The coefficient  for the order $T^2$ term of  $L_{\mathrm{QD}}$  
is given by  $C_{L}^{\mathrm{QD}}= C_{\kappa}^{\mathrm{QD}}-C_{T}^{}$, 
i.e., the difference between the next-leading order terms of 
$\kappa_{\mathrm{QD}}^{}$ and $ g$. 
This coefficient $C_{L}^{\mathrm{QD}}$ 
  also exhibits the Kondo plateau structures at integer filling 
points $N_d^{}= 1$, $2$, $\ldots$, $N-1$.

 The three-body correlations also play a significant role 
in the low-energy transport of magnetic alloys.
We have investigated the behaviors of 
the coefficient  $C_{\varrho}^{\mathrm{MA}}$ 
for the order $T^2$ resistivity  $\varrho_{\mathrm{MA}}^{}$ 
and  $C_{\kappa}^{\mathrm{MA}}$ for the order
 $T^{3}$ thermal conductivity  $\kappa_{\mathrm{MA}}^{}$.    
In the SU($N$) symmetric case,  
the three-body parts of  these coefficients become 
identical to the  related ones for quantum dots, i.e., 
$\Theta_{\varrho}^{\mathrm{MA}} \equiv \Theta_{T}^{}$ and 
 $\Theta_{\kappa}^{\mathrm{MA}} \equiv \Theta_{\kappa}^{\mathrm{QD}}$.  
Therefore, the difference between the coefficients  
 for MA and that for the QD counterparts arises from the two-body parts, 
more specifically,  from  the additional  $\cos^2 \delta$ terms appearing 
 in the right-hand side of Eqs.\ \eqref{eq:Crho_CT_relation} 
and \eqref{eq:CkappaMA_CkappaQD_relation}. 
 The additional  $\cos^2 \delta$ terms  make the coefficients 
  $C_{\varrho}^{\mathrm{MA}}$ and $C_{\kappa}^{\mathrm{MA}}$ 
positive definite and less sensitive to the impurity level position $\xi_d^{}$, 
compared to  the QD counterparts  $C_T^{}$  and $C_{\kappa}^{\mathrm{QD}}$. 
The coefficient $C_{\varrho}^{\mathrm{MA}}$ 
takes the maximum value  at the electron fillings of  $N_d^{}\simeq 1$ and $N-1$ 
for $N \geq 4$, 
while in the SU(2) case $C_{\varrho}^{\mathrm{MA}}$ 
has a single peak at half filling.
The coefficient 
$C_{L}^{\mathrm{MA}}
=C_{\varrho}^{\mathrm{MA}}-C_{\kappa}^{\mathrm{MA}}$ 
for the order $T^2$ term of the 
Lorenz number $L_{\mathrm{MA}} \equiv
\kappa_{\mathrm{MA}}^{}/(\sigma_{\mathrm{MA}}^{}\,T)$  
for magnetic alloys also  becomes less sensitive to the electron filling $N_d^{}$ 
 compared to $C_{L}^{\mathrm{QD}}$ for quantum dots.

The three-body correlation functions can be determined experimentally 
by measuring the coefficients $C$'s for the next leading order terms. 
These experimental values can then be used to infer 
the behaviors of other unmeasured transport coefficients.

\begin{acknowledgments}
 This work was supported by JSPS KAKENHI Grants No.\
 JP18K03495, No.\ JP18J10205, No. JP21K03415, and No.\ 23K03284 
and  JST CREST Grant No.\ JPMJCR1876.
K.\ M.\  was supported by JST Establishment of University 
Fellowships towards the Creation of Science Technology Innovation 
Grant No. JPMJFS2138.
Y.\ T.\  was supported by 
Sasakawa Scientific Research Grant from the Japan Science  Society No.\ 2021-2009, 
and  by  the Shigemasa and Shigeaki Nakazawa Fellowship of Graduate School 
of Science,  Osaka City University. 
\end{acknowledgments}

\appendix

\section{Fermi liquid parameters}
\label{sec:FL_parameters}

\subsection{Linear and nonlinear static susceptibilities}

The ground-state properties and 
the leading Fermi liquid corrections due to the low-lying excitations  
can be described in terms of   
the occupation number and the linear susceptibilities of the impurity level, 
derived from the free energy 
$\Omega \equiv  
- (1/\beta)\,\log \,\bigl[\,\mathrm{Tr}\,e^{-\beta \mathcal{H}}\,\bigr]$:  
\begin{align}
\bigl\langle n_{d\sigma}^{}\bigr\rangle\,
=& \   \frac{\partial \Omega}{\partial \epsilon_{d\sigma}^{}}
\,,  
\label{eq:ndFriedel}
\\
\chi_{\sigma\sigma'}^{}\, \equiv 
&\ -\frac{\partial^2 \Omega}{\partial \epsilon_{d\sigma}^{}
\partial\epsilon_{d\sigma'}^{}}
\,= \, 
\int_0^\beta \! d\tau\,
\bigl\langle \delta n_{d\sigma}^{}(\tau)\,\delta n_{d\sigma'}^{}\bigr\rangle\,.
\label{eq:chi_org}
\end{align}
Here, $\delta n_{d\sigma}^{} \equiv n_{d\sigma}^{}
-\langle n_{d\sigma}^{}\rangle$, 
and thermal-equilibrium averages are defined as 
$\langle  O \rangle
= \mathrm{Tr}\, \bigl[\,e^{-\beta \mathcal{H}}\,\mathcal{O}\,\bigr]
/\mathrm{Tr}\, e^{- \beta \mathcal{H}}$, 
with $\beta=1/T$ the inverse temperature.

In addition to the linear susceptibilities, 
the nonlinear susceptibilities $\chi_{\sigma_1\sigma_2\sigma_3}^{[3]}$ 
play an essential role in the next-leading order terms of the transport coefficients  
when the system does not have both the electron-hole and time-reversal symmetries: 
\begin{align}
&\chi_{\sigma_1\sigma_2\sigma_3}^{[3]} \,\equiv\, 
-\frac{\partial^3\Omega}{\partial\epsilon_{d\sigma_1}^{}
\partial\epsilon_{d\sigma_2}^{}\partial\epsilon_{d\sigma_3}^{}}
\,= \, 
\frac{\partial \chi_{\sigma_1\sigma_2}^{}}{\partial \epsilon_{d\sigma_3}^{}}
\nonumber
\\
=&-\int_0^\beta \! d\tau_1
\! \int_0^\beta \!d\tau_2\,
\bigl \langle T_\tau\delta n_{d\sigma_1}^{}(\tau_1)\,
\delta n_{d\sigma_2}^{}(\tau_2)\,
\delta n_{d\sigma_3}^{}\bigr\rangle\,.
\label{eq:canonical_correlation_3}
\end{align}
Here, $T_\tau$ is the imaginary-time ordering operator.
This  correlation function has the permutation symmetry: 
$\chi_{\sigma_1\sigma_2\sigma_3}^{[3]}
=\chi_{\sigma_2\sigma_1\sigma_3}^{[3]}
=\chi_{\sigma_3\sigma_2\sigma_1}^{[3]}
=\chi_{\sigma_1\sigma_3\sigma_2}^{[3]}=\cdots.$
Specifically, in our formulation  
we are using the ground state values 
for $\langle n_{d\sigma}^{}\rangle$,  
$\chi_{\sigma\sigma'}^{}$ and $\chi_{\sigma_1\sigma_2\sigma_3}^{[3]}$,  
determined at $T=0$.

The occupation number can be related to 
the phase shift $\delta_\sigma^{}$ through the Friedel sum rule:  
$\langle n_{d\sigma}^{}\rangle 
\xrightarrow{\,T\to 0\,}\delta_\sigma^{}/\pi$.    
The phase shift corresponds to the argument of the  Green's function, given by 
 $G_\sigma^r(0) = -\left| G_\sigma^r(0)\right| e ^{i \delta_{\sigma}^{}}$, 
at $\omega=T=eV=0$ \cite{ShibaKorringa},  
and determines the value of the spectral function $\rho_{d\sigma}^{}(\omega)$ 
at $\omega=0$: 
\begin{align}
\rho_{d\sigma}^{}(\omega) \,\equiv & \  
 A_{\sigma}^{}(\omega)\Big|_{T=eV=0}^{} \,, \\
\rho_{d\sigma}^{} \, \equiv &  \ \rho_{d\sigma}^{}(0)  
\,= \,  \frac{\sin^2\delta_\sigma^{}}{\pi\Delta}  
\,, \label{eq:rho_def} 
\end{align}
where $A_{\sigma}^{}(\omega)$ is the nonequilibrium 
spectral function defined in Eq.\ \eqref{eq:spectral_function}. 
The derivative of  $\rho_{d\sigma}^{}(\omega)$ also 
contributes to the next-leading order terms  
and can be expressed in terms of the diagonal susceptibility 
$\chi_{\sigma\sigma}$, using
Eq.\ \eqref{eq:YamadaYsidaRelationAppendix},    
\begin{align}
\rho_{d\sigma}'  \equiv  \left.
\frac{\partial \rho_{d\sigma}^{}(\omega)}{\partial \omega} 
\right|_{\omega=0}^{} 
=\,  -  
\frac{\partial \rho_{d\sigma}^{}}{\partial \epsilon_{d\sigma}^{}} 
\ = \ \frac{\chi_{\sigma\sigma}^{}}{\Delta}\,\sin2\delta_{\sigma}^{}.
\label{eq:rho_d_omega_2}
\end{align}

One of the most typical Fermi liquid corrections 
due to many-body scatterings arises    
 in the $T$-linear specific 
 heat $\mathcal{C}_\mathrm{imp}^{\mathrm{heat}}$  
of impurity electrons: 
\begin{align}
\mathcal{C}_\mathrm{imp}^{\mathrm{heat}} \,=\,   
\gamma_\mathrm{imp}^{}  \, T \,,   
\qquad 
\gamma_\mathrm{imp}^{} 
\,\equiv \, \frac{\pi^2}{3}\sum_\sigma \chi_{\sigma\sigma}^{} \,.    
\end{align}
The coefficient $\gamma_\mathrm{imp}^{}$ can be expressed in terms of  
the diagonal components of the linear susceptibility $\chi_{\sigma\sigma}^{}$ 
using the Ward identities \cite{YamadaYosida2,ShibaKorringa,Yoshimori}, 
 i.e., Eqs.\  
\eqref{eq:YamadaYsidaRelationAppendix} and 
\eqref{eq:YamadaYsidaRelationAppendixAdd},  
which follow from a relationship 
between the derivative of the self-energy with respect to 
$\omega$ and the derivative with respect to $\epsilon_{d\sigma}^{}$.

\subsection{Ward identities}

The local Fermi liquid state of quantum impurity systems can be microscopically 
described using  the retarded Green's function, 
defined in Eq.\ \eqref{eG}, which can also be written in the form,
\begin{align}
G_\sigma^r(\omega)\,=& 
\ \frac{1}{\omega-\epsilon_{d\sigma}^{}+i\Delta-\Sigma_\sigma^r(\omega)}\,.
\label{eG_selfEG}
\end{align}
The information about the low-lying energy states can be extracted 
from  the equilibrium self-energy 
$\Sigma_{\mathrm{eq},\sigma}^r(\omega)
\equiv \left. \Sigma_\sigma^r(\omega) \right|_{T=eV=0}^{}$, 
by expanding it, step by step,  around the Fermi energy $\omega=0$.
The expansion up to linear terms in $\omega$  
describes the renormalized resonance state of the form, 
\begin{align}
G_{\sigma}^{r}(\omega)\,\simeq & \ 
\frac{z_\sigma^{}}{\omega-\widetilde{\epsilon}_{d\sigma}^{}
+i\widetilde{\Delta}_{\sigma}^{}}
\,. 
\end{align}
Here, the renormalized parameters are defined by
 \begin{align}
\widetilde{\epsilon}_{d\sigma}^{}
\,\equiv & \ z_\sigma^{}
\left[\epsilon_{d\sigma}^{}+\Sigma_{\mathrm{eq},\sigma}^r(0)\right] 
\, = \,  
\widetilde{\Delta}_{\sigma}^{}  \cot \delta_\sigma^{}\,,
\nonumber 
\\
\widetilde{\Delta}_{\sigma}^{} \,\equiv & \  z_\sigma^{}\Delta, 
\qquad 
\frac{1}{z_\sigma^{}}\, \equiv \ 
1-\left.\frac{\partial \Sigma_{\mathrm{eq},\sigma}^r(\omega)}
{\partial \omega}\right|_{\omega=0}^{} .
\rule{0cm}{0.6cm}
\label{eq:zdefeddef}
\end{align}
The wavefunction renormalization factor  $z_\sigma^{}$  can be related  
to the derivative of $\Sigma_{\mathrm{eq},\sigma}^r(0)$ 
with respect to the impurity level $\epsilon_{d\sigma'}^{}$, 
using the Ward identity \cite{YamadaYosida2,ShibaKorringa,Yoshimori}: 
\begin{align}
\frac{1}{z_\sigma^{}}
\, = \, 
\widetilde{\chi}_{\sigma\sigma}^{}
\,, 
\qquad \quad 
\widetilde{\chi}_{\sigma\sigma'}^{} 
\,\equiv\,
\delta_{\sigma\sigma'}^{}+\frac{\partial \Sigma_{\mathrm{eq},\sigma}^r(0)}
{\partial \epsilon_{d\sigma'}} \,. 
\label{eq:YamadaYsidaRelationAppendix}
\end{align}
The coefficient $\widetilde{\chi}_{\sigma\sigma'}^{}$ determines 
the extent to which the susceptibility is enhanced at $T=0$:  
\begin{align}
\chi_{\sigma\sigma'}^{}\, =\,
- \frac{\partial \bigl\langle n_{d\sigma}^{}\bigr\rangle}
{\partial\epsilon_{d\sigma'}^{}}
\ \xrightarrow{\,T\to 0 \,}\  \rho_{d\sigma}^{} 
\widetilde{\chi}_{\sigma\sigma'}^{}\,.
\label{eq:YamadaYsidaRelationAppendixAdd}
\end{align}

Recently studies have been clarified that 
 the order $\omega^2$ real part  of  the self-energy 
can be expressed in terms of the 
the diagonal component of the three-body correlation function, 
$\chi_{\sigma\sigma\sigma}^{[3]}$, as 
\cite{FilipponeMocaWeichselbaumVonDelftMora,ao2017_1,ao2017_2_PRB,ao2017_3_PRB}  
\begin{align}
\!\!\! 
\left.
\frac{\partial^2}{\partial \omega^2}
\mathrm{Re}\,\Sigma_{\mathrm{eq},\sigma}^{r}(\omega)
\right|_{\omega \to 0}^{} 
\,=\, \frac{\partial^2 \Sigma_{\mathrm{eq},\sigma}^{r}(0)}
{\partial \epsilon_{d\sigma}^{2}}\,   
\ \ = \ 
\frac{\partial \widetilde{\chi}_{\sigma\sigma}^{}}{\partial \epsilon_{d\sigma}^{}} \,
.
\label{eq:self_w2}
\end{align}
Physically,  this coefficient determines  
the energy shifts of quasiparticles 
of order $\omega^2$, $T^2$ and $(eV)^2$, 
which affect  
the low-energy transport of the next-leading order.

\section{Low-energy asymptotic form of spectral function}
\label{sec:Low_energy_asymptotic_form_A}

The low-energy asymptotic form of the retarded self-energy $\Sigma_\sigma^r(\omega)$ 
for multilevel Anderson impurity model has been derived 
 up to terms of order $\omega^2$, $T^2$, and $(eV)^2$ 
in the previous works \cite{ao2021,TsutsumiSUNpaper}. 
For symmetric tunnel junctions with 
$\Gamma_L^{}=\Gamma_R^{}$ ($=\Delta/2$) and  
$\mu_L^{}=-\mu_R^{}$ ($=eV/2$), 
it takes the form, 
\begin{widetext}  
\begin{align}
\mathrm{Im}\, \Sigma_{\sigma}^r(\omega) 
\,  = & 
\ -\,   \frac{\pi}{2}\,   
\frac{1}{\rho_{d\sigma}^{}}
\sum_{\sigma' (\neq \sigma)} \chi_{\sigma\sigma'}^2
\,
\left[\,
\omega^2  +\frac{3}{4}\,(eV)^2  +(\pi T)^2  
\,\right] \ + \, \cdots \,, 
\label{eq:self_imaginary_N}
\\
\epsilon_{d\sigma}^{} + 
\mathrm{Re}\, \Sigma_{\sigma}^r(\omega) 
\,  = & \ \ 
\Delta\, \cot \delta_{\sigma}^{}
\,+ \bigl( 1-\widetilde{\chi}_{\sigma\sigma}^{} \bigr)\, \omega 
\, + \frac{1}{2}\,\frac{\partial \widetilde{\chi}_{\sigma\sigma}^{}}
{\partial \epsilon_{d\sigma}^{}}\, \omega^2 
\, +  \frac{1}{6}\,
\frac{1}{\rho_{d\sigma}^{}} 
\sum_{\sigma'(\neq \sigma)}
\chi_{\sigma\sigma'\sigma'}^{[3]}
\left[
\frac{3}{4} \,(eV)^2 
+
\left( \pi T\right)^2 
\right] 
\ + \,\cdots\,.
\label{eq:self_real_ev_mag_N}
\end{align}
Substituting these expansion results into Eq.\ \eqref{eG_selfEG}, 
we obtain the asymptotic form spectral function,
\begin{align}
\pi\Delta \, A_{\sigma}^{}(\omega) \, 
= & \ 
\sin^2 \delta_{\sigma}^{}
+ 
\frac{\pi^2}{3} 
\left(
\frac{3}{2}\cos 2 \delta_{\sigma}^{}\,
\sum_{\sigma'(\neq \sigma)}\chi_{\sigma\sigma'}^2
- 
\frac{\sin 2\delta_{\sigma}^{}}{2\pi}\,
\sum_{\sigma'(\neq \sigma)}
\chi_{\sigma\sigma'\sigma'}^{[3]}
\,\right)
\left[\,\frac{3}{4}
\left(eV\right)^2   + \left(\pi T\right)^2
\, \right]  
\nonumber \\
& \ 
+ \pi \sin 2\delta_{\sigma}^{}\, \chi_{\sigma\sigma}^{} 
\,\omega
+  
\pi^2
\left[\,
\cos 2\delta_{\sigma}^{}
\left(
\chi_{\sigma\sigma}^2
+ 
\frac{1}{2} 
\sum_{\sigma'(\neq \sigma)}\chi_{\sigma\sigma'}^2
\right)
- 
\frac{\sin 2\delta_{\sigma}^{}}{2\pi}
\,\chi_{\sigma\sigma\sigma}^{[3]}
\,\right] \, \omega^2 \ + \ \cdots .
\label{eq:A_including_T_eV_N_orbital}
\end{align}
Correspondingly, the inverse of the  spectral function  
that determines the  thermoelectric transport of magnetic alloys 
takes the following form, 
up to terms order $\omega^2$ and $T^2$ at $eV=0$, 
\begin{align}
\frac{1}{\pi \Delta A_{\sigma}(\omega)} \, 
\simeq   & \   
\frac{1}{\pi \Delta\rho_{d\sigma}^{}}
\Biggl[ 
1 
-
\frac{1}{6 \Delta \rho_{d\sigma}^{}} 
\left(\,
3\pi \cos 2 \delta_{\sigma}\sum_{\sigma'(\neq \sigma)}\chi_{\sigma\sigma'}^2
- \sin 2\delta_{\sigma}\,
\sum_{\sigma'(\neq \sigma)}
\chi_{\sigma\sigma'\sigma'}^{[3]}
\, \right)
\left(\pi T\right)^2 
- \frac{\sin 2\delta_{\sigma}^{}\, \chi_{\sigma\sigma}^{}}
{\Delta \rho_{d\sigma}^{}} 
\,\omega 
\nonumber \\
& \qquad \qquad 
+ 
\frac{\pi}{ \Delta\rho_{d\sigma}^{}}
\left\{\,
\left(
\cos 2\delta_{\sigma}^{} +2 \right)
\, \chi_{\sigma\sigma}^2
-
\frac{1}{2} \cos 2\delta_{\sigma}^{} 
\sum_{\sigma' (\neq \sigma)}\chi_{\sigma\sigma'}^2
+  
\frac{\sin 2\delta_{\sigma}^{}}{2\pi}
\,\chi_{\sigma\sigma\sigma}^{[3]}
\,\right\}
\,\omega^2  
\Biggr] 
+ \cdots \;.
\label{eq:inverse_A_result}
\end{align}
\end{widetext}

\section{Properties of $\chi_{\sigma_1\sigma_2\sigma_3}^{[3]}$ in SU($N$) case}
\label{sec:3body_SUN_properties}

We briefly describe here some relations between 
the three-body correlation functions and 
the derivative of the linear susceptibilities 
with respect to the center of mass coordinate of the impurity levels,
  $\epsilon_{d}^{}\equiv (1/N)\sum_{\sigma}\epsilon_{d\sigma}^{}$.

The derivative of the diagonal susceptibility $\chi_{\sigma\sigma}^{}$ 
can be written as   
\begin{align}
\frac{\partial \chi_{\sigma\sigma}^{}}{\partial  \epsilon_{d}^{}} 
&=  \  
\frac{\partial \chi_{\sigma\sigma}^{}}{\partial  \epsilon_{d\sigma}^{}} 
+ \sum_{\sigma'(\neq \sigma)}
\frac{\partial \chi_{\sigma\sigma}^{}}{\partial  \epsilon_{d\sigma'}^{}} 
\nonumber 
\\
& \xrightarrow{\,\mathrm{SU}(N) \,} \, 
\chi_{\sigma\sigma\sigma}^{[3]} + (N-1)\,\chi_{\sigma\sigma'\sigma'}^{[3]} 
\,,
\label{eq:SUN_chi_3_relation_1}
\end{align}
where  $\sigma \neq \sigma'$. Note that   
$\chi_{\sigma\sigma\sigma'}^{[3]} =  \chi_{\sigma\sigma'\sigma'}^{[3]}$ 
in the SU($N$) symmetric case.
Similarly, the derivative of the off-diagonal 
susceptibility $\chi_{\sigma\sigma'}^{}$ 
for  $\sigma \neq \sigma'$ takes the form 
\begin{align}
\frac{\partial \chi_{\sigma\sigma'}^{}}{\partial  \epsilon_{d}^{}} 
& = \  
\frac{\partial \chi_{\sigma\sigma'}^{}}{\partial  \epsilon_{d\sigma}^{}} 
+ \frac{\partial \chi_{\sigma\sigma'}^{}}{\partial  \epsilon_{d\sigma'}^{}} 
+ \sum_{\sigma''(\neq \sigma \atop 
\neq \sigma')}
\frac{\partial \chi_{\sigma\sigma'}^{}}{\partial  \epsilon_{d\sigma''}^{}} 
\nonumber 
\\
& \xrightarrow{\,\mathrm{SU}(N) \,} \,  
2 \,\chi_{\sigma\sigma'\sigma'}^{[3]} + (N-2)\,\chi_{\sigma\sigma'\sigma''}^{[3]} 
\;,
\label{eq:SUN_chi_3_relation_2}
\end{align}
for  $\sigma\neq \sigma' \neq \sigma'' \neq \sigma$.   
In the SU(2) symmetric case, 
Eqs.\ \eqref{eq:SUN_chi_3_relation_1} and \eqref{eq:SUN_chi_3_relation_2} 
provide enough information
to determine the two independent components  
$\chi_{\sigma\sigma\sigma}^{[3]}$ and 
$\chi_{\sigma\sigma'\sigma'}^{[3]}$ 
from the two differential coefficients 
$\partial \chi_{\sigma\sigma}^{}/\partial  \epsilon_{d}^{}$ and
$\partial \chi_{\sigma\sigma'}^{}/\partial  \epsilon_{d}^{}$.
However, for $N\geq 3$,
there are three independent three-body components, i.e., 
$\chi_{\sigma\sigma\sigma}^{[3]}$,  
$\chi_{\sigma\sigma'\sigma'}^{[3]}$, and  
$\chi_{\sigma\sigma'\sigma''}^{[3]}$,   
so that we need additional information to determine all these components.  

In order to obtain another independent relation,
we consider the derivative of the susceptibilities with respect to the magnetic field $b$, 
which induces the level splitting
 in impurity levels $\epsilon_{d\sigma}^{}$ in a such way that  
$\epsilon_{d,m,\uparrow}^{} = \epsilon_{d}^{} -  b$ and 
$\epsilon_{d,m,\downarrow}^{} = \epsilon_{d}^{} + b$, 
with $\sigma=(m,s)$ for $m=1,2,\ldots, N/2$ and $s=\uparrow, \downarrow$: 
\begin{align}
&
\!\!\!  
\frac{\partial \chi_{m_1 s_1,m_2 s_2}^{}}{\partial b}
\, =  \, 
\sum_{m_3=1}^{N/2} \sum_{s_3=\uparrow,\downarrow}
\frac{\partial \epsilon_{d,m_3,s_3}^{}}{\partial b}\, 
\frac{\partial \chi_{m_1 s_1,m_2 s_2}^{}}
{\partial \epsilon_{d,m_3,s_3}^{}}
\nonumber \\
& 
\!\!\!\!  
= \, 
-\sum_{m_3=1}^{N/2} \, \chi_{m_1 s_1,m_2 s_2, m_3 \uparrow}^{[3]}
\,+
\sum_{m_3=1}^{N/2} \chi_{m_1 s_1,m_2 s_2, m_3 \downarrow}^{[3]}. 
\end{align}
From this derivative, we obtain the following relation, 
taking $m_1 = m_2 \, (\equiv m)$,  
\begin{align}
& 
\!\!\!\!\!\!\!\!\! \!\!\!\!  
\!\!\!\!\!\!\!\!\! \!\!\!\! 
\frac{\partial}{\partial b}
\left(
\frac{\chi_{m\uparrow,m\uparrow}^{} - \chi_{m\downarrow,m\downarrow}^{}}{2} 
\right) 
\nonumber \\
=&  \ \ 
- \sum_{m'=1}^{N/2} 
\frac{1}{2}\,\biggl(  
\chi_{m\uparrow, m\uparrow, m'\uparrow}^{[3]} 
-\chi_{m\uparrow, m\uparrow, m'\downarrow}^{[3]} 
\nonumber \\
& \qquad \qquad \quad 
+ \chi_{m\downarrow, m\downarrow, m'\downarrow}^{[3]} 
-\chi_{m\downarrow, m\downarrow, m'\uparrow}^{[3]} 
\biggr)
\nonumber  
\\
\xrightarrow{\,b\to 0\,}& \, 
-
\sum_{m_3=1}^{N/2} 
\left[\,
\chi_{m\uparrow, m\uparrow, m_3\uparrow}^{[3]} 
\,-\,\chi_{m\uparrow, m\uparrow, m_3\downarrow}^{[3]} 
\,\right]
\nonumber \\
= &  \ 
-
\left(
\chi_{m\uparrow, m\uparrow, m\uparrow}^{[3]} 
\,-\,\chi_{m\uparrow, m\uparrow, m\downarrow}^{[3]} 
\right) .
\label{eq:SUN_chi_3_mag_1_new}
\end{align}
Here, we set the magnetic field to be zero, $b=0$, in the last two  lines. 
This relation can also be rewritten into the following form at $b=0$, 
using the original label $\sigma = (m,s)$,  
\begin{align}
\chi_{B}^{[3]}   \,\equiv \,    
\frac{\partial}{\partial b}
\left(
\frac{\chi_{m\uparrow,m\uparrow}^{} - \chi_{m\downarrow,m\downarrow}^{}}{2} 
\right)_{b=0}^{} 
=  \, 
- \chi_{\sigma\sigma\sigma}^{[3]}  + \chi_{\sigma\sigma'\sigma'}^{[3]}  ,
\label{eq:chi_B3_def_new_appendix}
\end{align}
for $\sigma' \neq \sigma$.


\begin{figure}[b]
\leavevmode
 \centering

\includegraphics[width=0.7\linewidth]{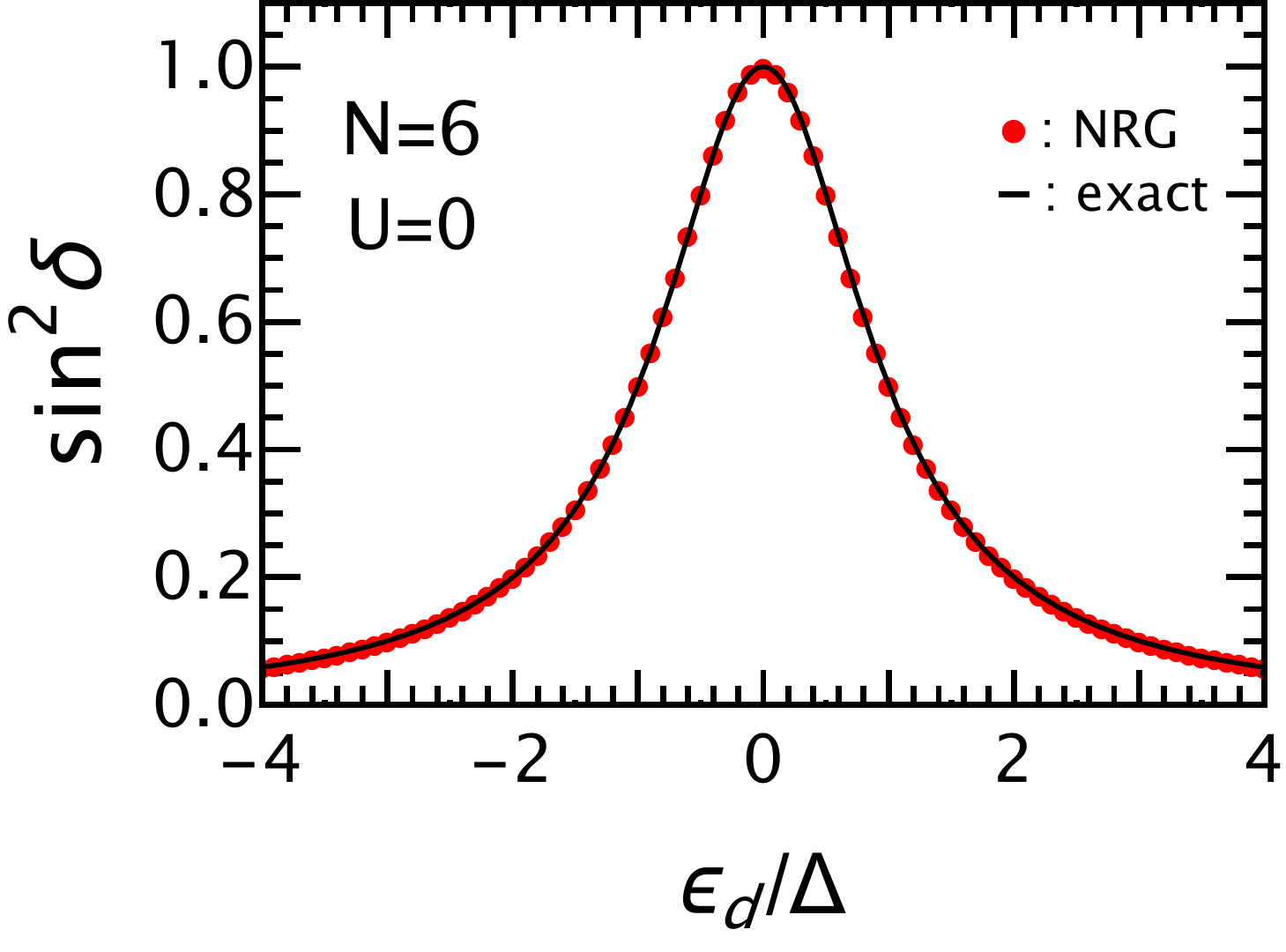}
\caption{Comparison of ($\bullet$)  NRG 
and (---) exact results of $\sin^2 \delta$ for $U=0$.  
The NRG calculation was performed for $N=6$, 
taking $\Lambda=20$ and keeping $N_\mathrm{trunc}=40 000$ low-energy states. 
}
\label{fig:U0_NRG_check}
\end{figure}

\section{NRG procedures 
} 
\label{sec:brief_exp_NRG}

We have carried out NRG calculations, 
dividing the $N$ conduction channels into $N/2$ pairs       
and using the SU(2) spin and U(1) charge symmetries for each of the pairs, 
i.e.,   $\prod_{k=1}^{\frac{N}{2}}
\left\{\mbox{SU(2)}\otimes \mbox{U(1)}\right\}_k$ symmetries. 
The discretization parameter $\Lambda$ 
and the number of retained low-lying excited states $N_\mathrm{trunc}$ are chosen 
to be $(\Lambda, N_\mathrm{trunc})= (6,10000)$ for $N=4$.  
Note that the SU(4) symmetry is preserved in our iteration scheme 
since the truncation of higher energy states has been 
carried out after adding all new states from the $N/2$ pairs. 

For $N = 6$, we have also exploited 
the method of Stadler {\it et al.} \cite{Stadler2016}. 
In this case, the truncation procedure is carried out  at each 
of the steps ($k=$1, $2$, $\ldots$, $N/2$) 
after adding the states constructed with one of the channel pairs,  
by using Oliveira's  $\mathcal{Z}$-trick \cite{Oliveira1994},     
choosing different $\mathcal{Z}$ values for each of the $N/2$ steps:  
$\mathcal{Z}_k^{} =  1/2 +k/N $ for the $k$-th pair. 
We have carried out calculations for $N=6$, 
taking rather large values for the NRG parameters,    
such that  $(\Lambda, N_\mathrm{trunc})=  (20,40000)$ 
for small interactions $U/(\pi \Delta) = 2/5$ and $1$, 
and $(20,30000)$ for large interactions $U/(\pi \Delta) = 2$, $3$, $4$, $5$, and  $6$. 
This method significantly reduces the computational cost for obtaining low-lying energy states and enables us to calculate the three-body correlation functions for $6$, 
although it does not faithfully preserve the SU($6$) symmetry.
We have checked whether this method reproduces the noninteracting results. 
Figure \ref{fig:U0_NRG_check} compares the NRG result for $\sin^2 \delta$ 
with the exact one for $U=0$; the results show reasonable agreement. 
It indicates that this truncation procedure works effectively   
for deducing the SU(6) FL parameters. 

In order to calculate $\chi_{B}^{[3]}$ defined in Eq.\ \eqref{eq:chi_B3_def_new}, 
we have also introduced a small external potential $\epsilon_{\mathrm{sp}, k}^{}$,   
which depends on the channel index $k=$1, $2$, $\ldots$, $N/2$ 
and shifts the impurity level from  $\epsilon_{d}^{}$.
Specifically, for $N=4$, it is applied 
in a way equivalent to the local Zeeman field:  
 $\epsilon_{\mathrm{sp}, 1}^{}=-b$ and $\epsilon_{\mathrm{sp}, 2}^{}=b$.  
For $N=6$, we have extended the potential  
such that   $\epsilon_{\mathrm{sp}, 1}^{}=-b$, 
 $\epsilon_{\mathrm{sp}, 2}^{}=0$, and  $\epsilon_{\mathrm{sp}, 3}^{}=b$, 
and have deduced  $\chi_{B}^{[3]}$ from the derivatives of the channel susceptibilities 
with respect to $b$.

\section{Two-body FL parameters for  \\ 
SU(4) \& SU(6) symmetric cases}

\label{sec:NRG_results_2body_functions_for_sun}

\begin{figure}[t]

 \leavevmode
 \centering

 \hspace{0.015\linewidth} 
\includegraphics[width=0.45\linewidth]{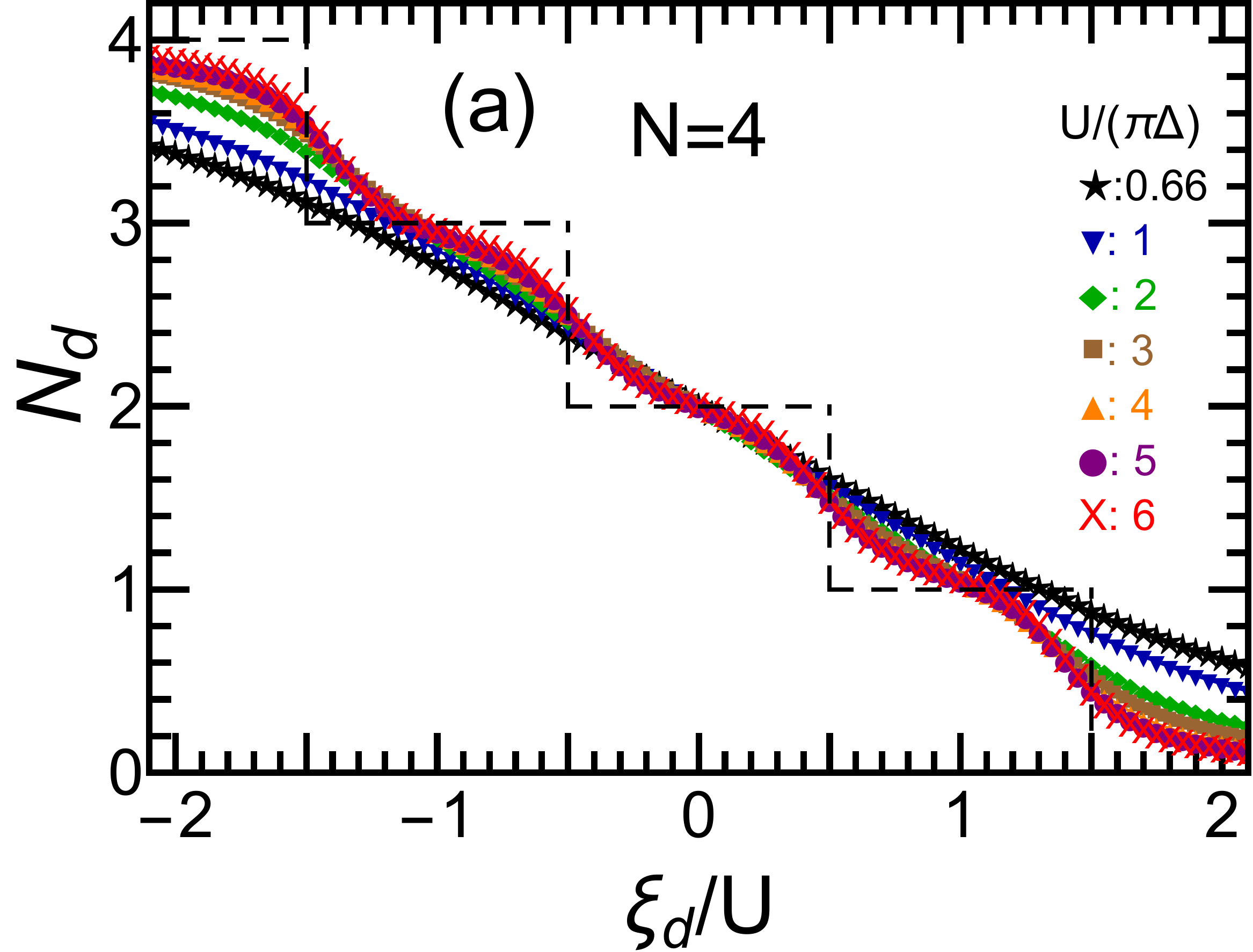}
 \hspace{0.01\linewidth} 
\includegraphics[width=0.47\linewidth]{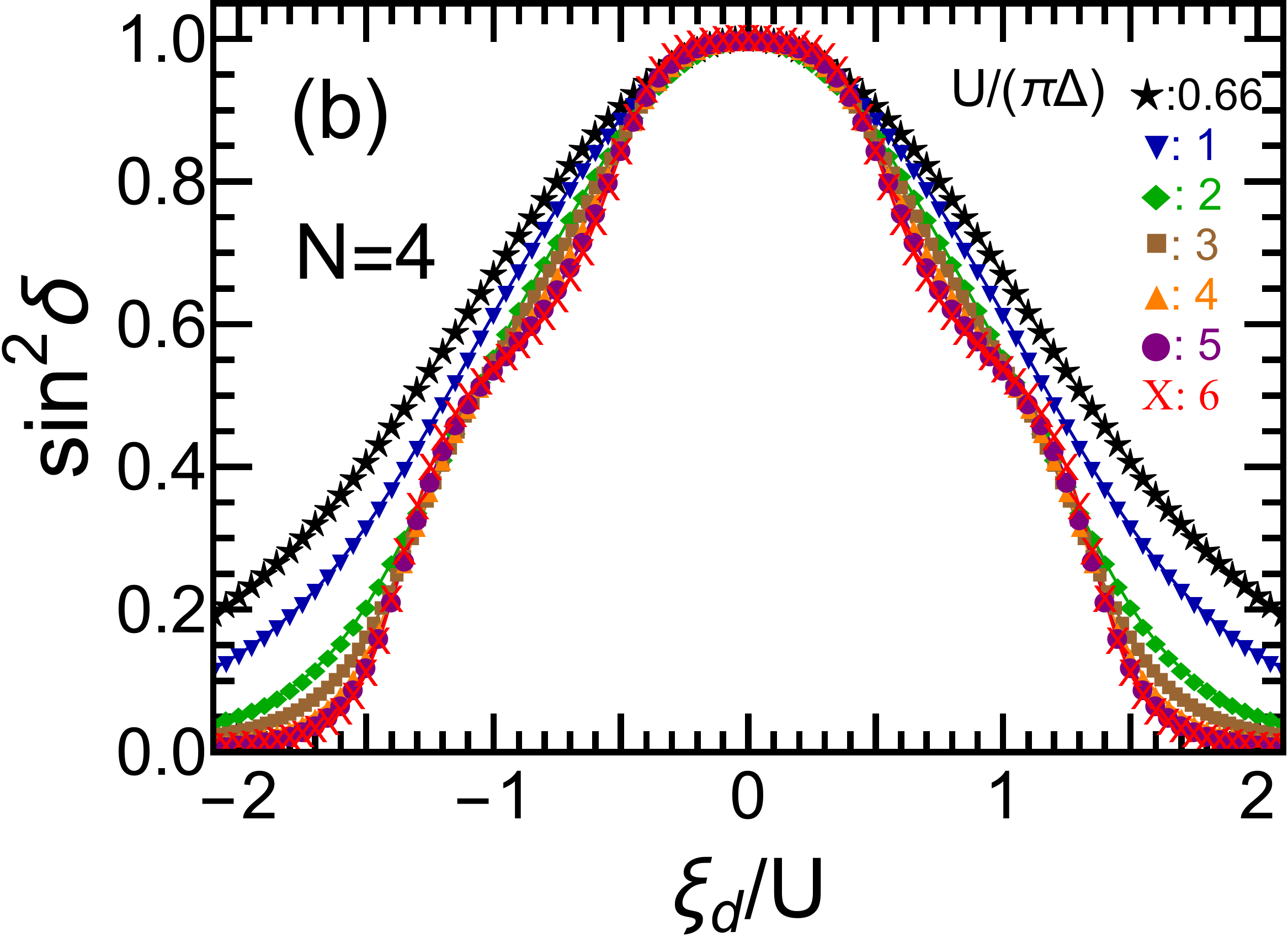}
\\
\includegraphics[width=0.49\linewidth]{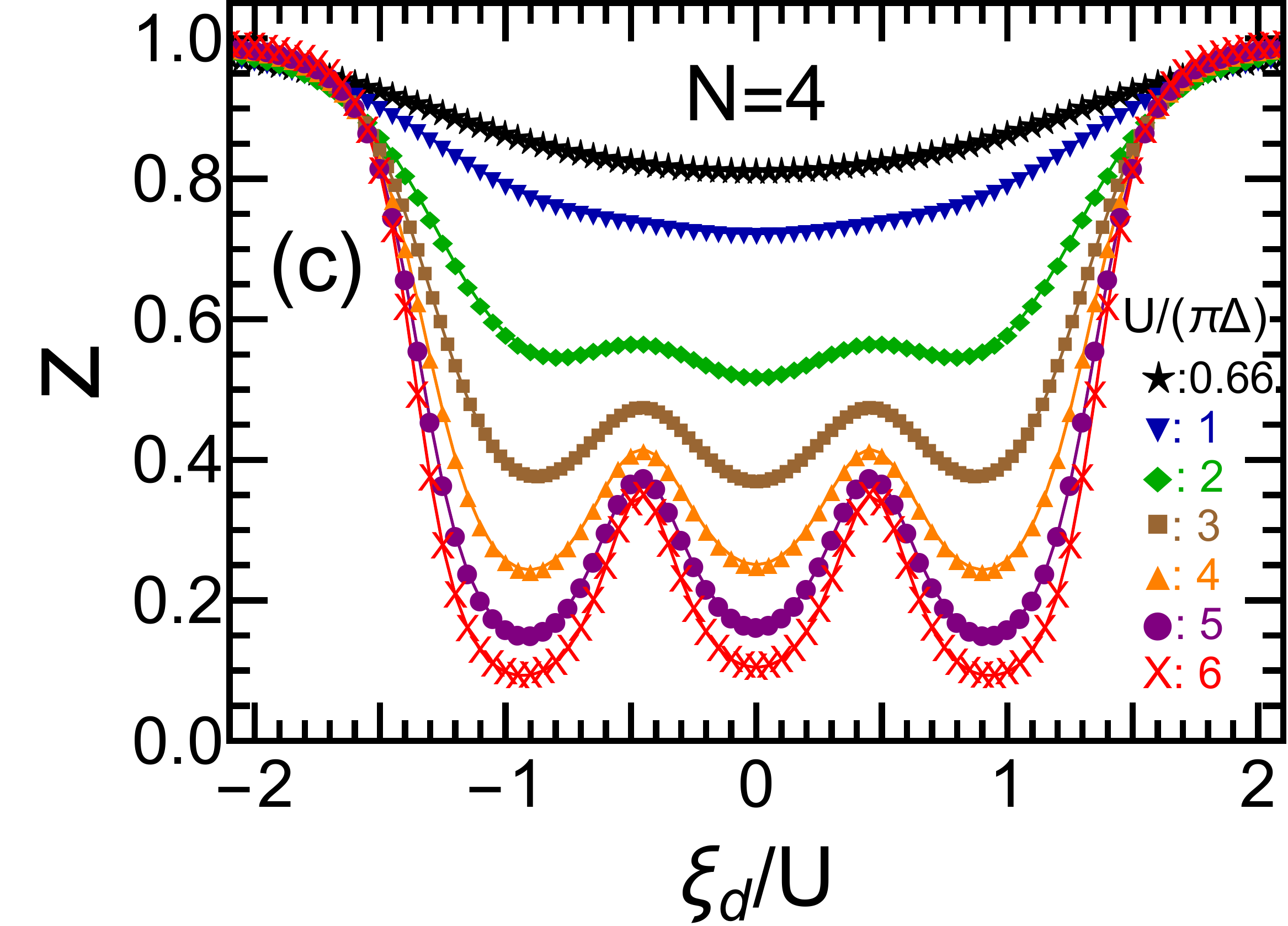}
 \hspace{0.01\linewidth} 
\includegraphics[width=0.47\linewidth]{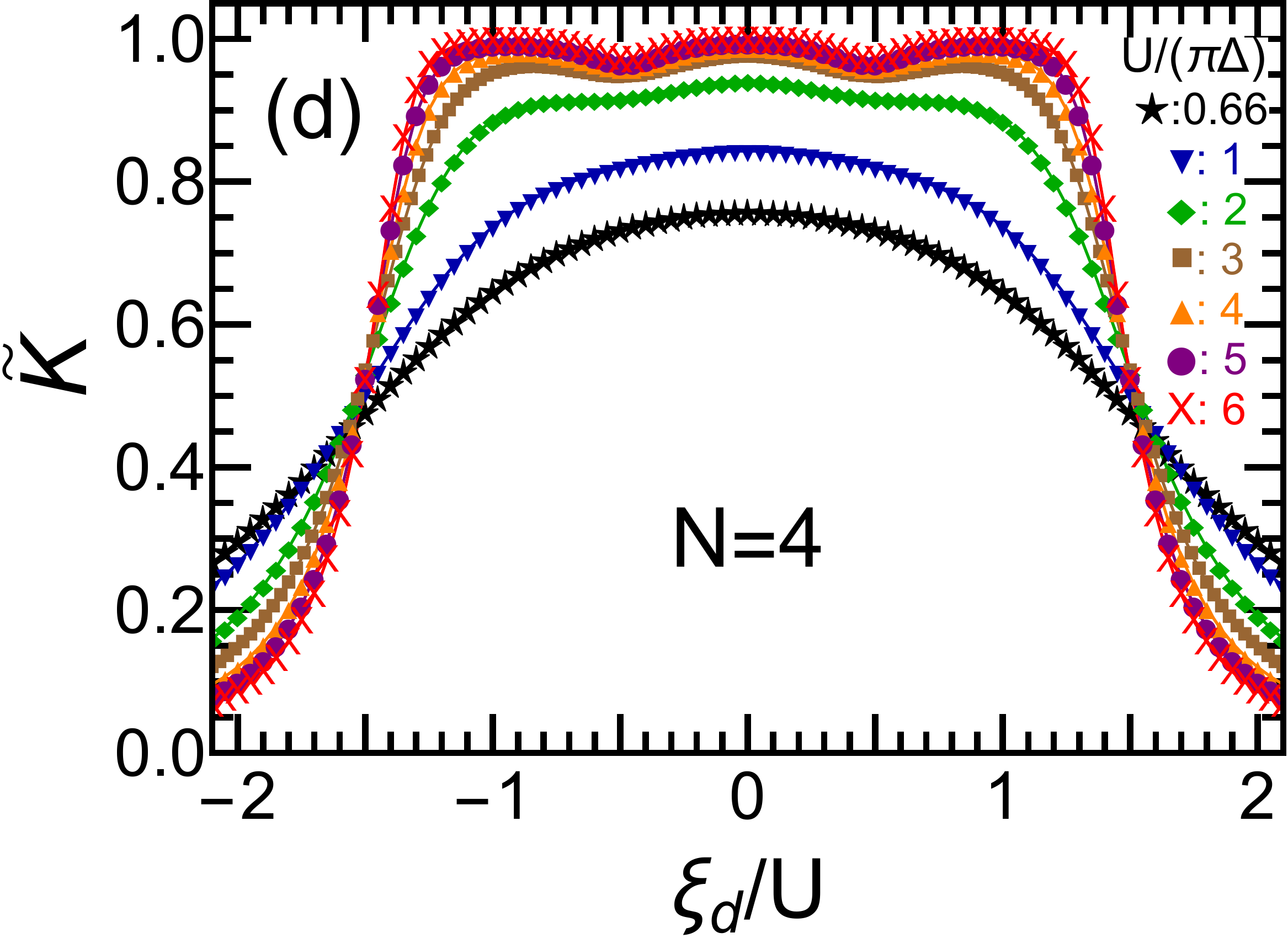}
\\
\includegraphics[width=0.49\linewidth]{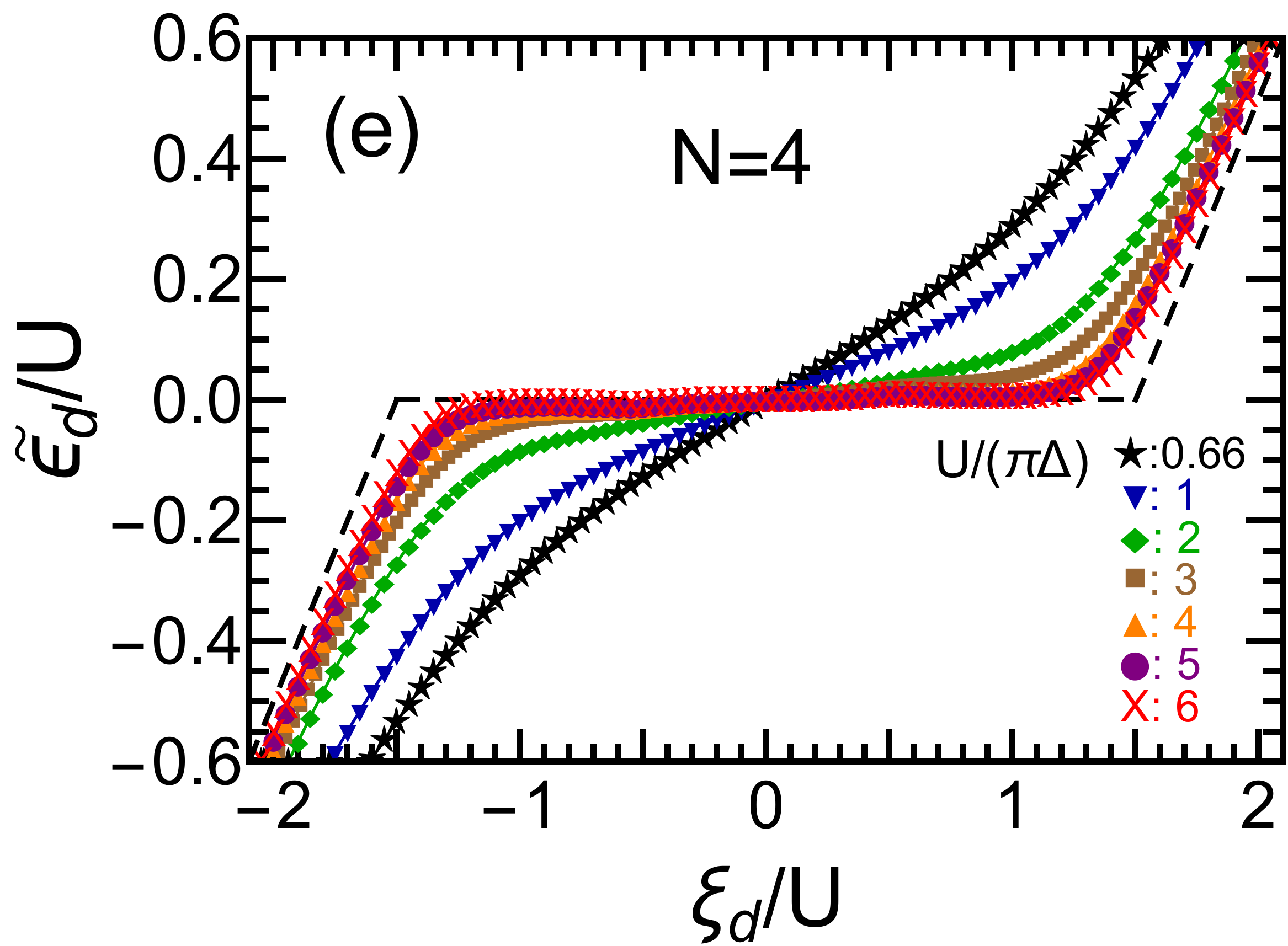}
 \hspace{0.01\linewidth} 
\includegraphics[width=0.47\linewidth]{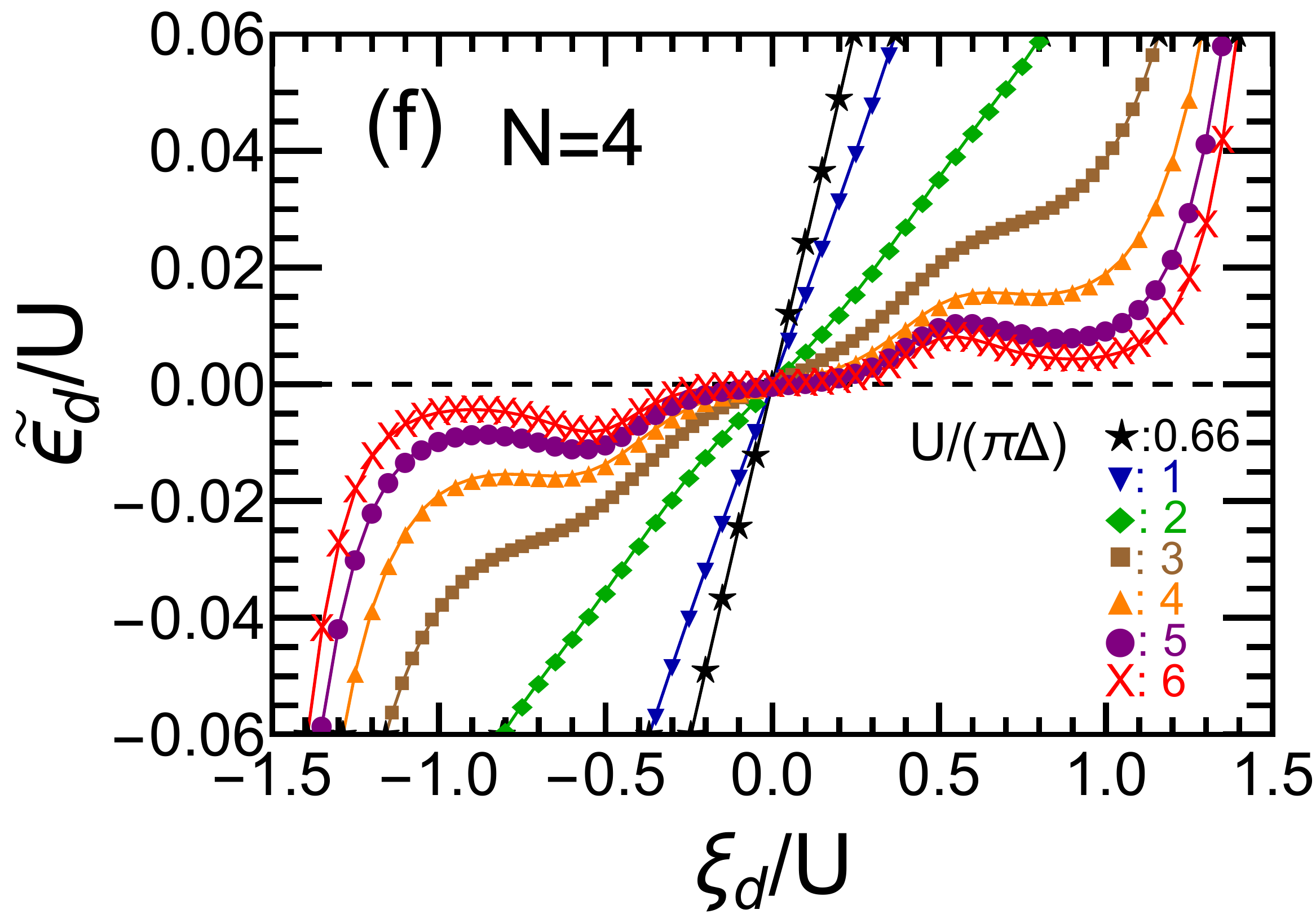}
\caption{Fermi liquid parameters 
 for the SU(4) symmetric Anderson model 
are plotted vs  $\xi_{d}^{}$: 
(a)  $N_{d}^{}$,
 (b)  $\sin^{2}\delta$,
(c) renormalization factor $z$, 
  (d) $\widetilde{K}=(N-1)(R-1)$.
  (e) renormalized level $\widetilde{\epsilon}_{d}^{}$, 
and 
  (f)  enlarged view of $\widetilde{\epsilon}_{d}^{}$, 
Interaction strengths are chosen to be 
 $U/(\pi\Delta) = 2/3(\star), 1(\blacktriangledown), 2(\blacklozenge), 3(\blacksquare),
 4(\blacktriangle), 5(\bullet), 6(\times)$. 
The dashed line  in (a) represents $N_{d}$ in the atomic limit $\Delta \to 0$.
} 
\label{fig:FL-parameters-N4-U-from-2-over-3-to-6}
\end{figure}

\begin{figure}[t]

 \leavevmode
 \centering

 \hspace{0.015\linewidth} 
\includegraphics[width=0.45\linewidth]{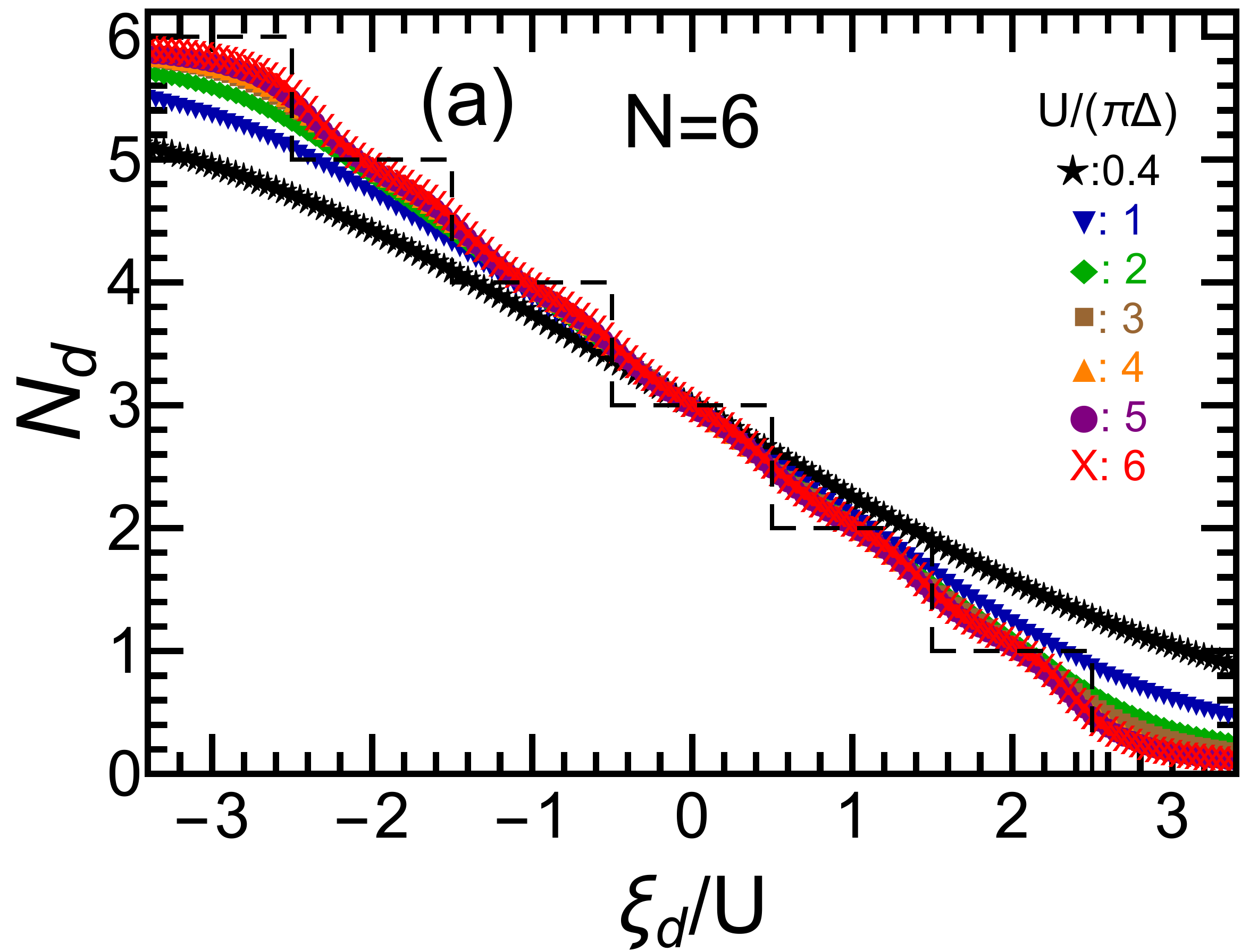}
 \hspace{0.01\linewidth} 
\includegraphics[width=0.47\linewidth]{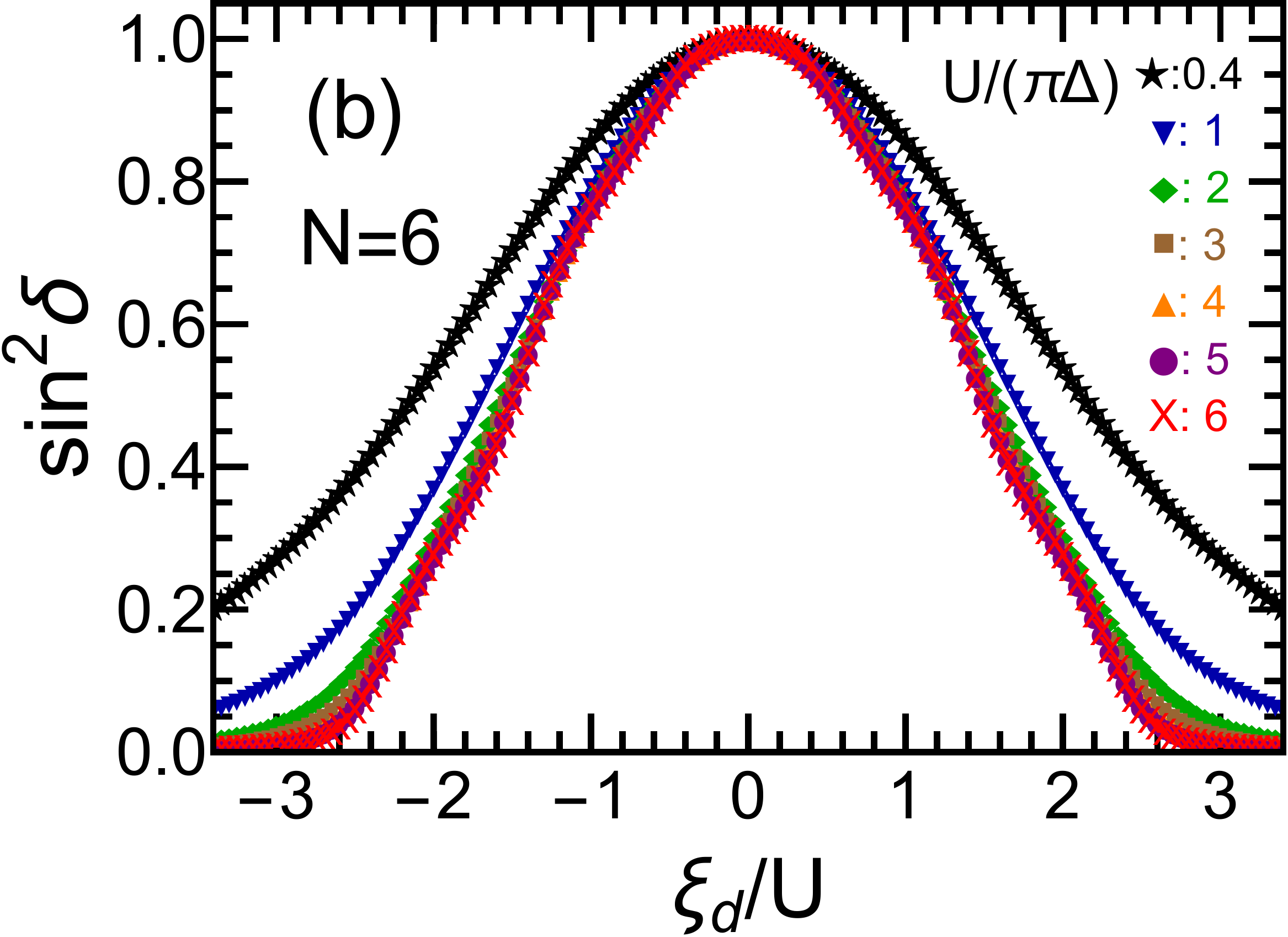}
\\
\includegraphics[width=0.49\linewidth]{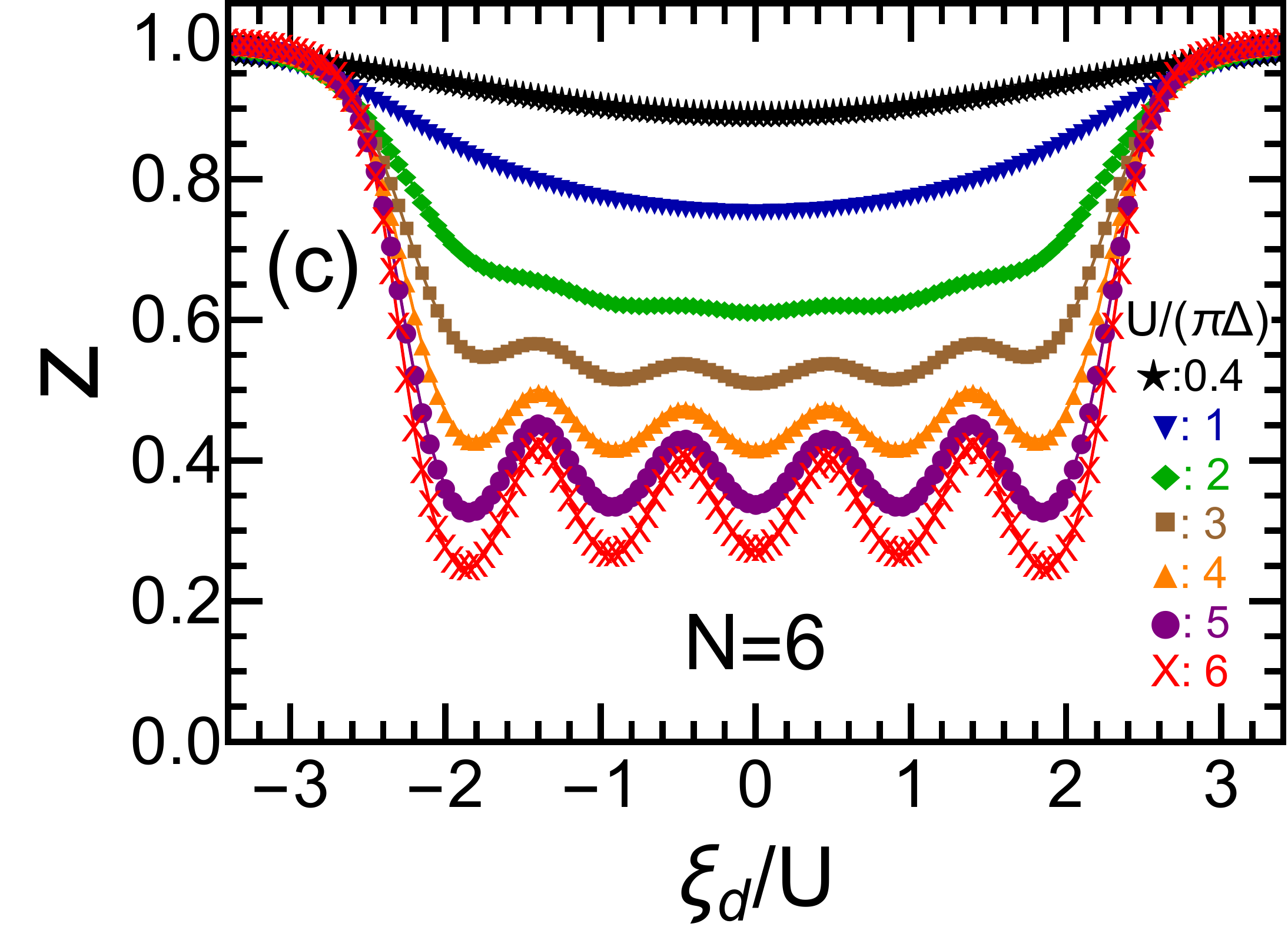}
 \hspace{0.01\linewidth} 
\includegraphics[width=0.47\linewidth]{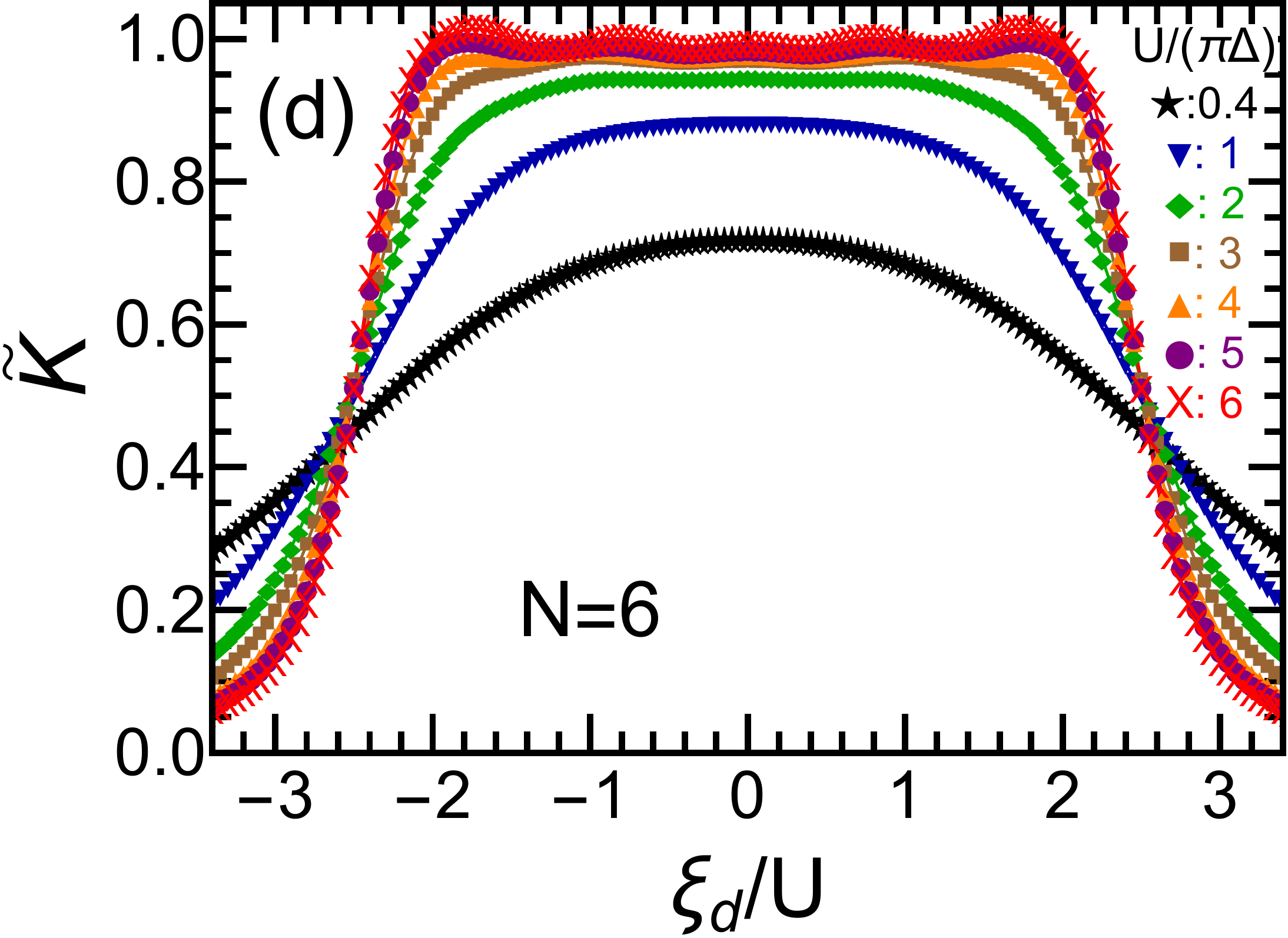}
\\
\includegraphics[width=0.49\linewidth]{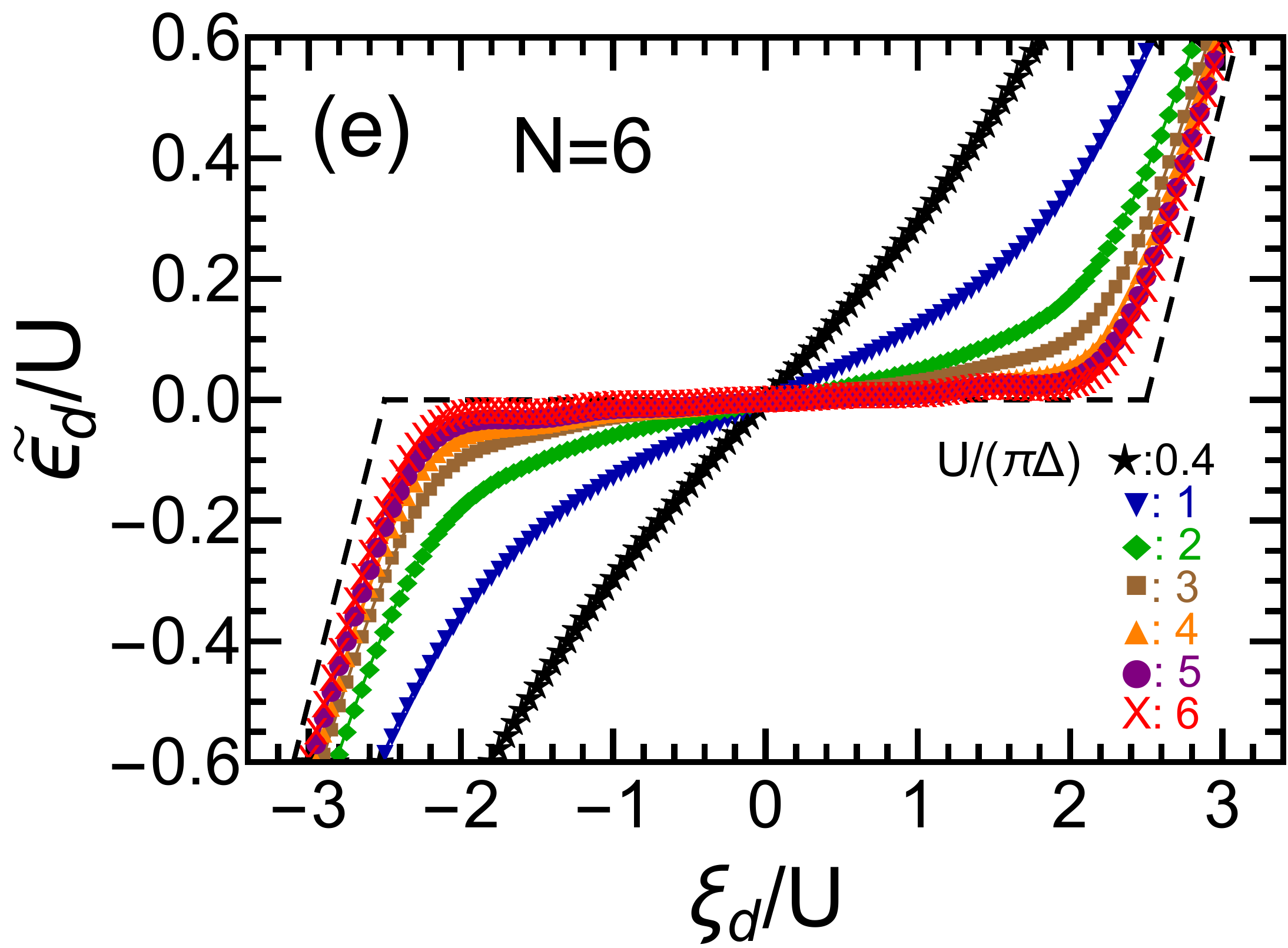}
 \hspace{0.01\linewidth} 
\includegraphics[width=0.47\linewidth]{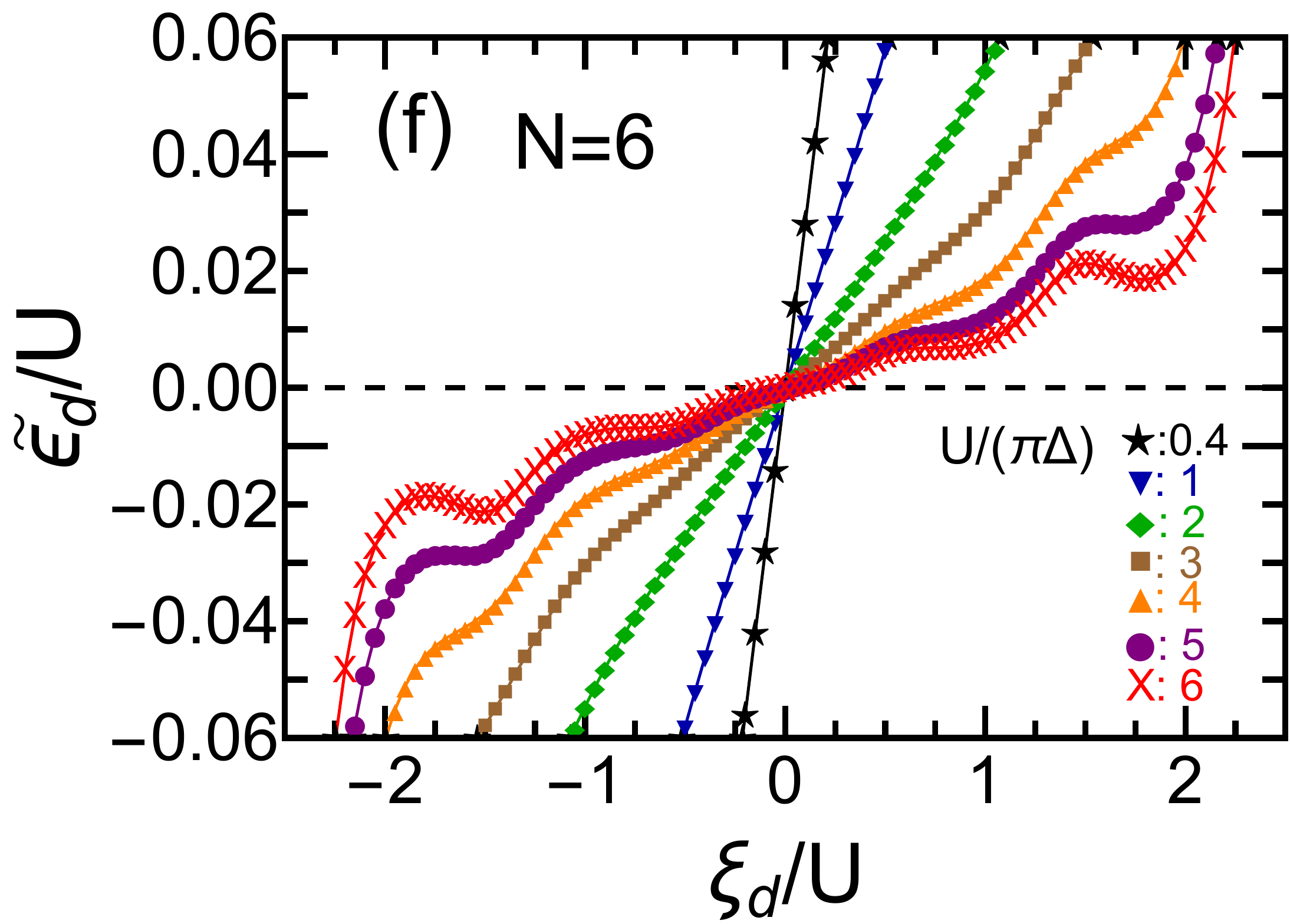}
\caption{Fermi liquid parameters 
 for the SU(6) symmetric Anderson model 
are plotted vs  $\xi_{d}^{}$: 
(a)  $N_{d}^{}$, 
 (b)  $\sin^{2}\delta$,
(c) renormalization factor $z$, 
  (d) $\widetilde{K}=(N-1)(R-1)$.
  (e) renormalized level $\widetilde{\epsilon}_{d}^{}$, 
and 
  (f)  enlarged view of $\widetilde{\epsilon}_{d}^{}$, 
Interaction strengths are chosen to be 
 $U/(\pi\Delta) = 2/5(\star), 1(\blacktriangledown), 2(\blacklozenge), 3(\blacksquare),
 4(\blacktriangle), 5(\bullet), 6(\times)$. 
The dashed line  in (a) represents $N_{d}$ in the atomic limit $\Delta \to 0$.
} 
\label{fig:FL-parameters-N6-U-from-2-over-5-to-6}
\end{figure}

 We provide  a quick overview of the behavior of  
the renormalized parameters in the SU(4) and SU(6) cases,  
which can be derived from the phase shift and the linear susceptibilities  
\cite{NishikawaCrowHewson1,NishikawaCrowHewson2,ao2011N,ao2012}.  
Our discussion here is based on the NRG results,   
plotted in Figs.\ \ref{fig:FL-parameters-N4-U-from-2-over-3-to-6}
and \ref{fig:FL-parameters-N6-U-from-2-over-5-to-6}
as functions of the impurity level position $\xi_{d}^{}$, 
for several different interaction strengths: 
from  weak to strong interactions up to $U/(\pi\Delta)=6.0$.

The SU($N$) Kondo effect occurs for strong interactions 
when the impurity levels are filled by an integer number of electrons, 
i.e., at the fillings of $N_d=1$, $2$, $\ldots$, $N-1$.  
It  takes place at  
 $\xi_d^{} \simeq 0$, $\pm U$, \ldots,  $\pm \frac{N-2}{2}U$, 
and gives an interesting variety in the low-energy properties. 
As  $N$ increases,  a greater interaction strength
 is required to clearly observe the Kondo behavior.
This is because the quantum fluctuations caused by the Coulomb interaction 
 are suppressed for large $N$.    
In particular, the mean-field theory becomes exact 
 in the limit $N\to \infty$ that is taken keeping  
 the scaled interaction $U^* \equiv (N-1)U$ constant \cite{ao2012}.


Figures \ref{fig:FL-parameters-N4-U-from-2-over-3-to-6}(a)
and  \ref{fig:FL-parameters-N6-U-from-2-over-5-to-6}(a) 
show the occupation number $N_d^{}$ 
for $N=4$ and $6$,  respectively. 
As $U$ increases, the Coulomb staircase structure emerges:   
 $N_d^{}$ varies steeply    
at  $\xi_d^{} \simeq  \pm U/2$, $\pm 3U/2$, $\ldots$,  $\pm (N-1)U/2$.  
Correspondingly, 
 as $U$ increases,   
the transmission probability $\sin^2\delta$, 
shown in Figs.\ \ref{fig:FL-parameters-N4-U-from-2-over-3-to-6}(b)
and \ref{fig:FL-parameters-N6-U-from-2-over-5-to-6}(b),   
exhibits a plateau structure that develops 
around  $\xi_d^{} \simeq 0$,  $\pm U$, $\ldots$,  $\pm (N-2)U/2$.  
In particular, the plateau at the half-filling point $\xi_d^{} \simeq 0$   
 reaches the unitary limit value $\sin^2\delta=1.0$.

The renormalization factor $z$,
shown in Figs.\ \ref{fig:FL-parameters-N4-U-from-2-over-3-to-6}(c)
and  \ref{fig:FL-parameters-N6-U-from-2-over-5-to-6}(c)  for $N=4$ and $6$, 
exhibits a broad valley structure at 
 $|\xi_{d}^{}| \lesssim (N-1)U/2$,  
where   $1 \lesssim N_{d} \lesssim N-1$. 
The valley becomes deeper as $U$ increases, 
 and the local minima emerge for $U/(\pi \Delta) \gtrsim 3.0$ 
at the integer-filling points,  
reflecting the occurrence  of the SU($N$) Kondo effects. 
The renormalization factor 
 also has local maxima at intermediate valence states 
in between two adjacent local  minima. 
Note that $z$ is significantly suppressed by the strong electron correlations 
even at these local maxima.
Figures \ref{fig:FL-parameters-N4-U-from-2-over-3-to-6}(d)
and  \ref{fig:FL-parameters-N6-U-from-2-over-5-to-6}(d) 
 show the rescaled  
Wilson ratio $\widetilde{K}\equiv (N-1)(R-1)$.  
For large interactions, 
$\widetilde{K}$ exhibits a wide flat structure 
at  $|\xi_d^{}|\lesssim (N-1)U/2$,  
 the height of which  
 approaches the saturation value $\widetilde{K}\simeq 1.0$,  
 especially for  $U/(\pi\Delta) \gtrsim 3.0$, 
reflecting the suppression of charge fluctuations in this region of $\xi_d^{}$.

Figures \ref{fig:FL-parameters-N4-U-from-2-over-3-to-6}(e) 
and \ref{fig:FL-parameters-N6-U-from-2-over-5-to-6}(e) 
show the renormalized resonance-level position 
$\widetilde{\epsilon}_{d}^{} \equiv z \Delta \cot \delta$ 
for $N=4$ and $6$, respectively,  as a function of $\xi_d^{}$.
For strong interactions, 
the renormalized level is almost locked at the Fermi level,  
$\widetilde{\epsilon}_{d}^{}\simeq 0.0$,   
in the strong-coupling region $|\xi_d^{}| \lesssim (N-1)U/2$,
which corresponds to the filling range of $1 \lesssim N_d \lesssim N-1$. 
Figures \ref{fig:FL-parameters-N4-U-from-2-over-3-to-6}(f)  
and \ref{fig:FL-parameters-N6-U-from-2-over-5-to-6}(f) show 
an enlarged view of $\widetilde{\epsilon}_{d}^{}$ in the vicinity of the Fermi level.
We see that $\widetilde{\epsilon}_{d}^{}$ 
exhibits a fine structure     
which reflects the staircase behavior of the phase shift $\delta$ 
and the oscillatory behavior of the renormalization factor $z$. 
Outside the strong-coupling region,  
$\widetilde{\epsilon}_{d}^{}$ approaches 
the bare value  
$\widetilde{\epsilon}_{d}^{} \simeq 
\epsilon_{d}^{}$ at  $\xi_{d}^{}\gg (N-1)U/2$, 
or the Hartree-Fock value  
$\widetilde{\epsilon}_{d}^{} 
 \simeq \epsilon_{d}^{}+(N-1)U$ at $\xi_{d}^{}\ll -(N-1)U/2$:
these asymptotic forms of  $\widetilde{\epsilon}_{d}^{}$ 
 are shown as the dashed lines 
in Figs.\ \ref{fig:FL-parameters-N4-U-from-2-over-3-to-6}(e) and 
\ref{fig:FL-parameters-N6-U-from-2-over-5-to-6}(e).

\begin{figure}[t]

 \leavevmode
 \centering

\includegraphics[width=0.47\linewidth]{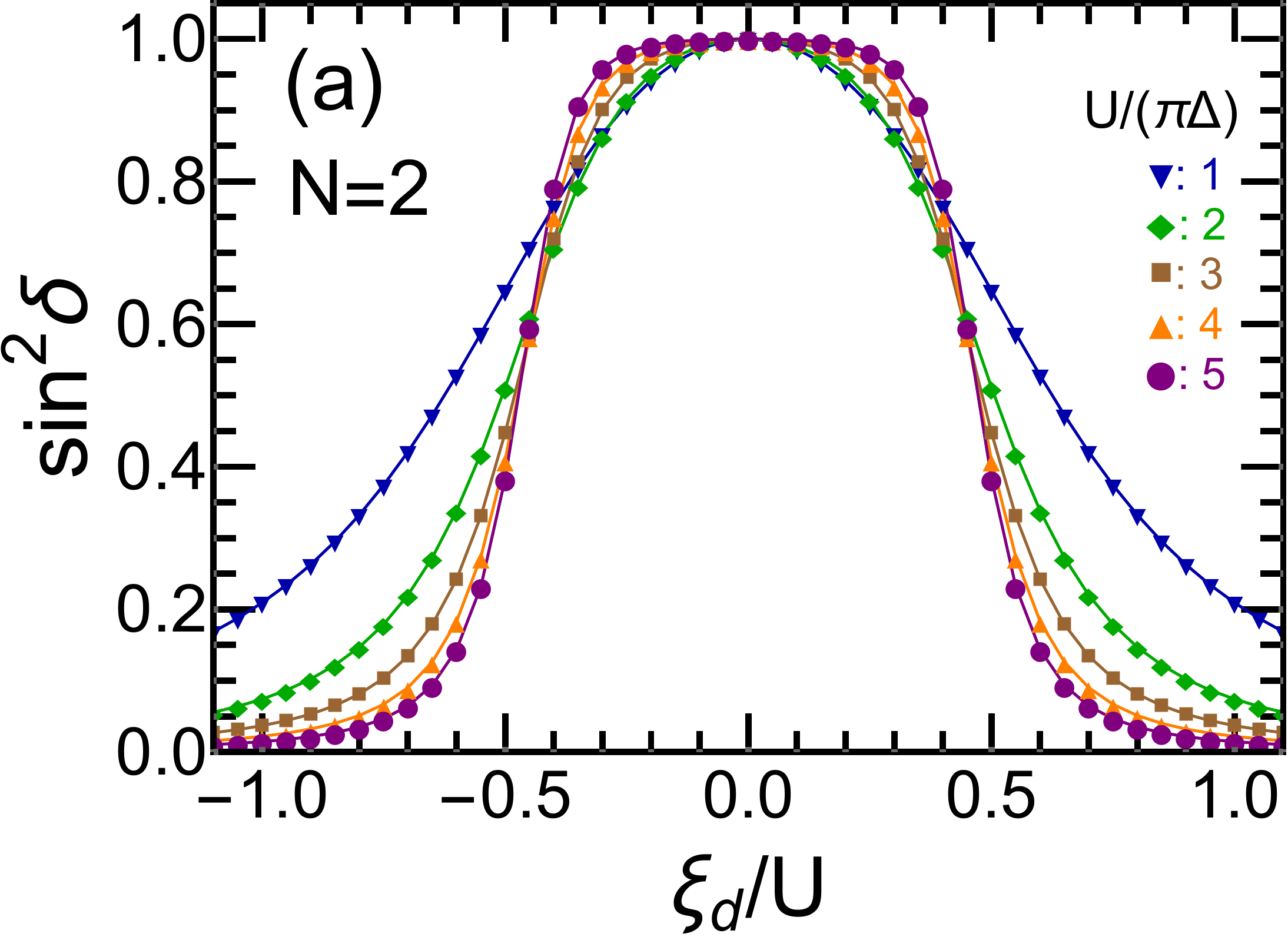}
 \hspace{0.01\linewidth} 
\includegraphics[width=0.47\linewidth]{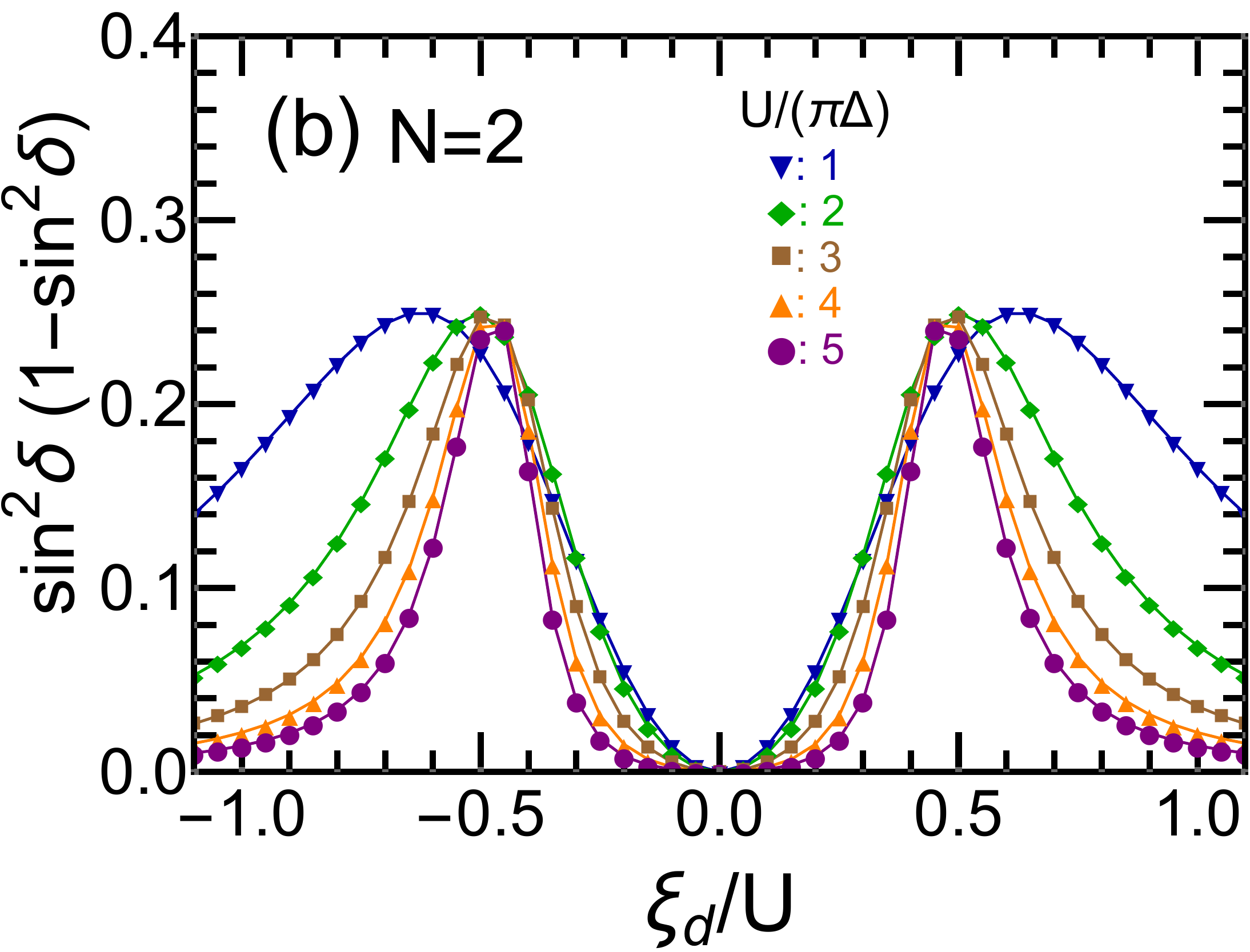}
\\
\includegraphics[width=0.47\linewidth]{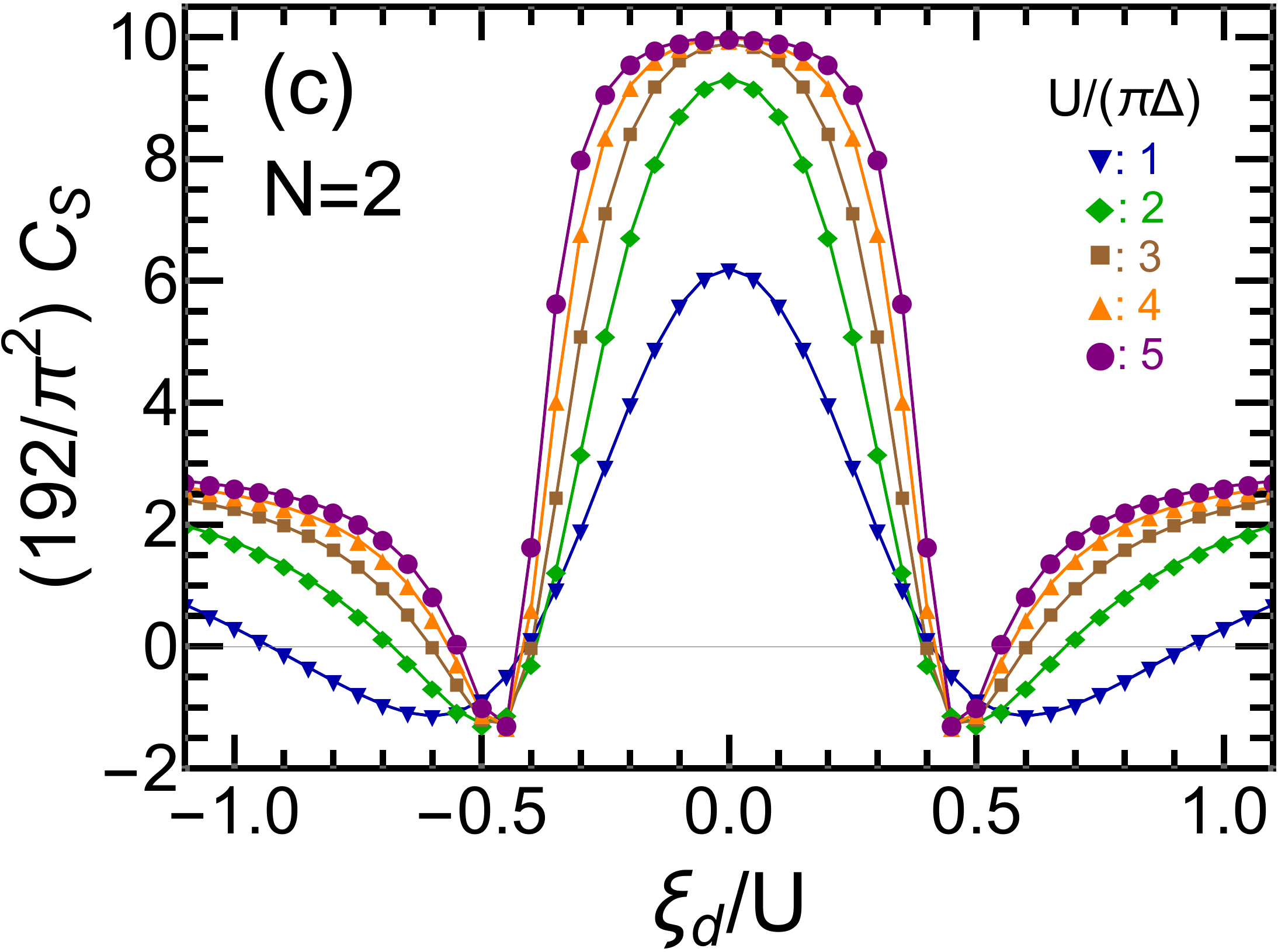}
 \hspace{0.01\linewidth} 
\includegraphics[width=0.47\linewidth]{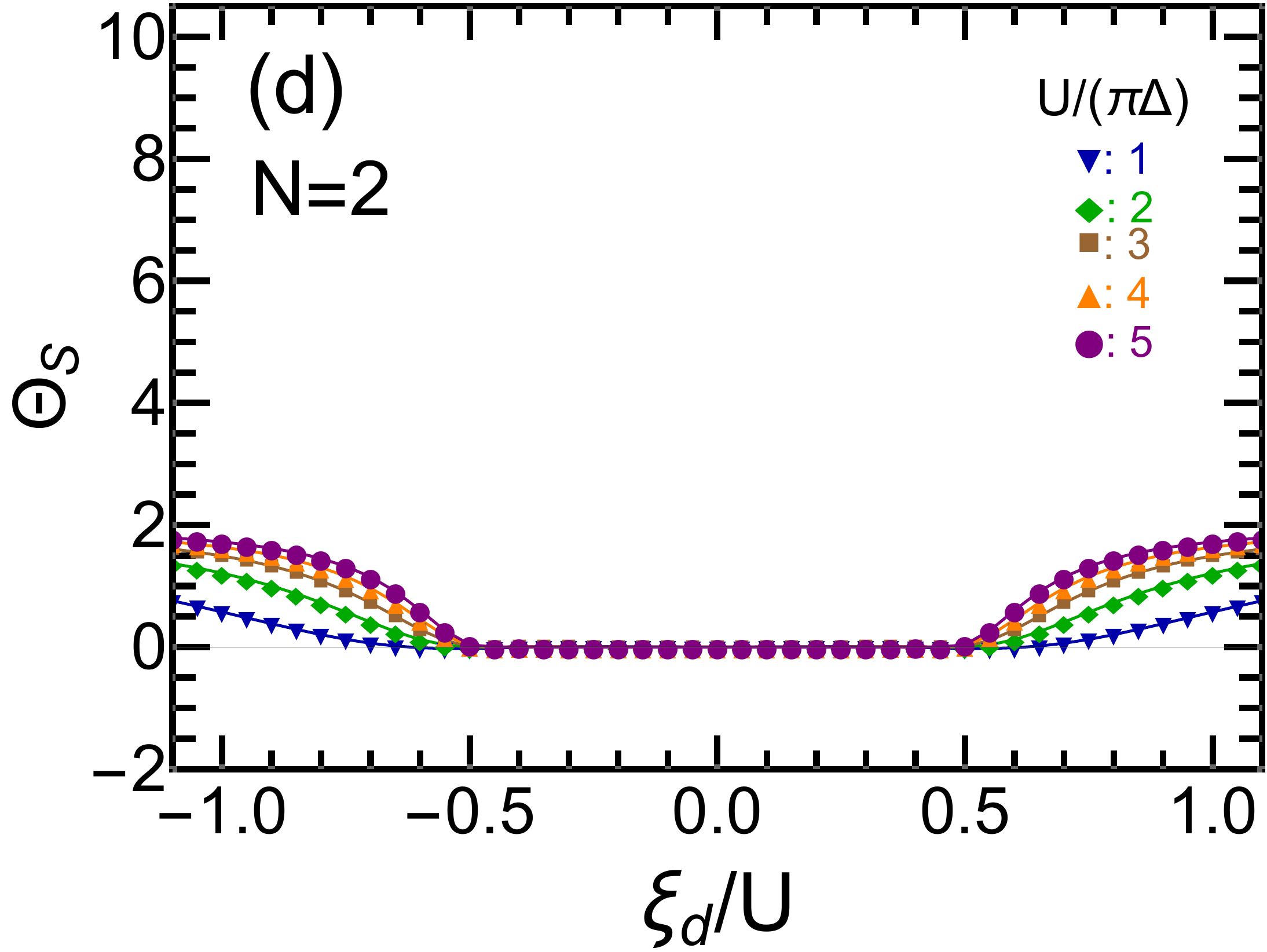}
\caption{Current noise for SU(2) symmetric quantum dots vs $\xi_{d}^{}$: 
 (a)  $\sin^{2}\delta$, 
(b) linear noise $\sin^{2}\delta\, (1-\sin^{2}\delta)$, 
  (c) $(192/\pi^2)\,C_S^{} = W_S^{}+\Theta_{S}^{}$,
(d)  $\Theta_{S}^{} 
 =  - \bigl(\Theta_\mathrm{I}^{} + 3 \,\widetilde{\Theta}_{\mathrm{II}} 
\bigr)  \cos 2\delta$  (see Table \ref{tab:C_and_W_extended}).
Interaction strengths are chosen to be 
 $U/(\pi\Delta) =  1(\blacktriangledown), 2(\blacklozenge), 3(\blacksquare),
 4(\blacktriangle), 5(\bullet)$.
} 
\label{fig:Noise_N2}
\end{figure}

\section{Current noise and thermoelectric transport for SU(2) symmetric case}
\label{sec:NoiseN2}

\begin{figure}[t]

 \leavevmode
 \centering

\includegraphics[width=0.47\linewidth]{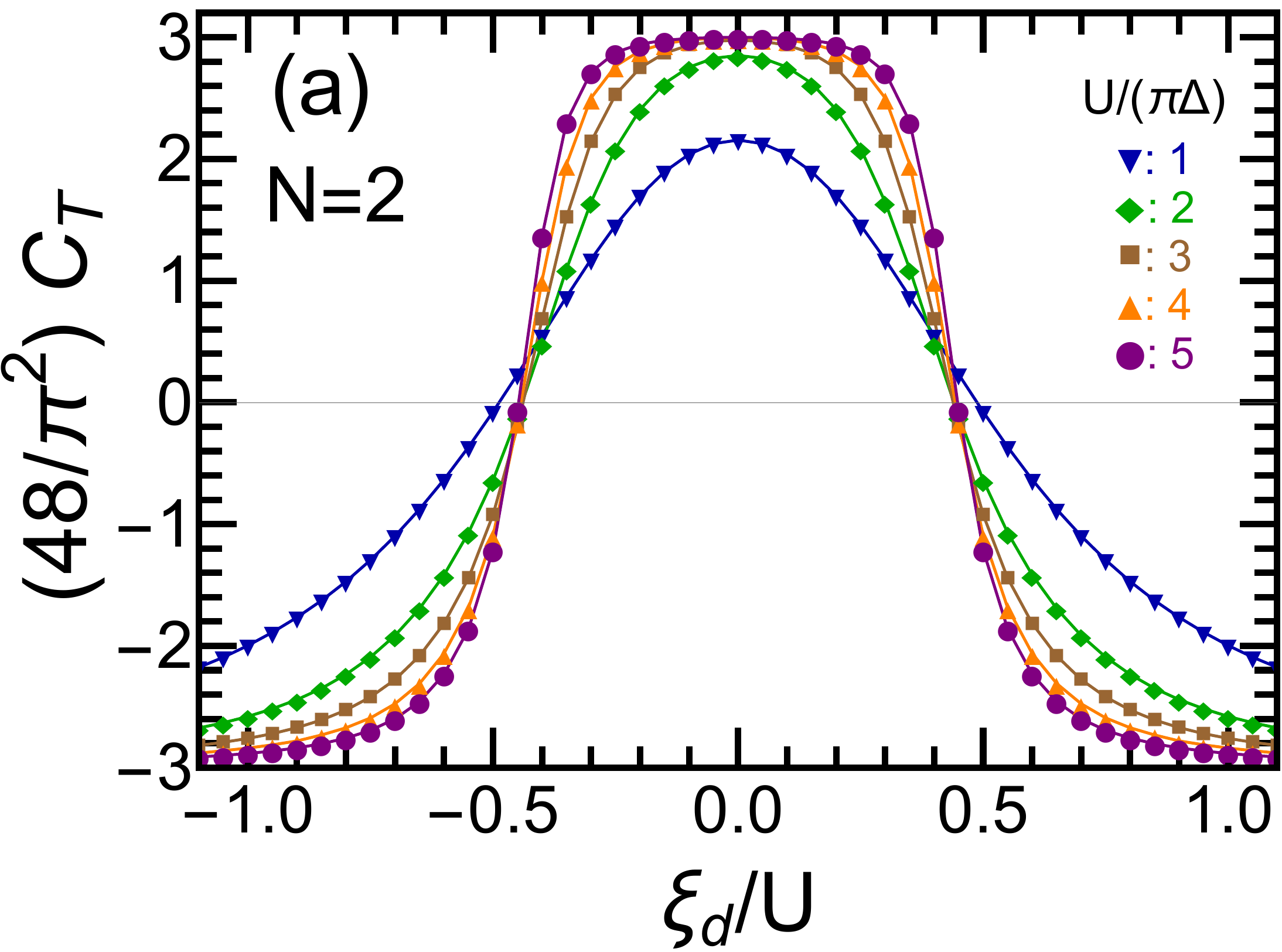}
 \hspace{0.01\linewidth} 
\includegraphics[width=0.47\linewidth]{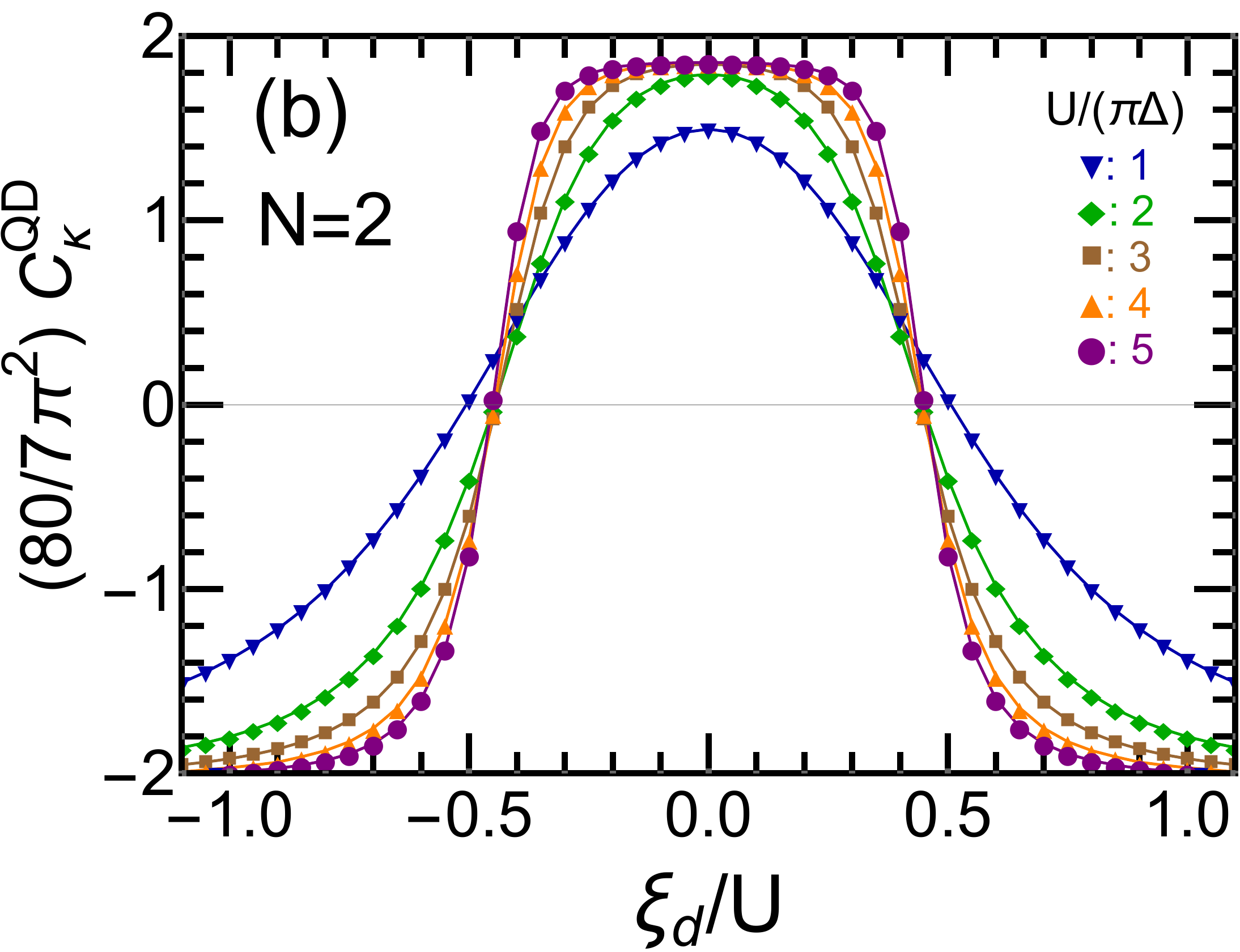}
\\
\includegraphics[width=0.47\linewidth]{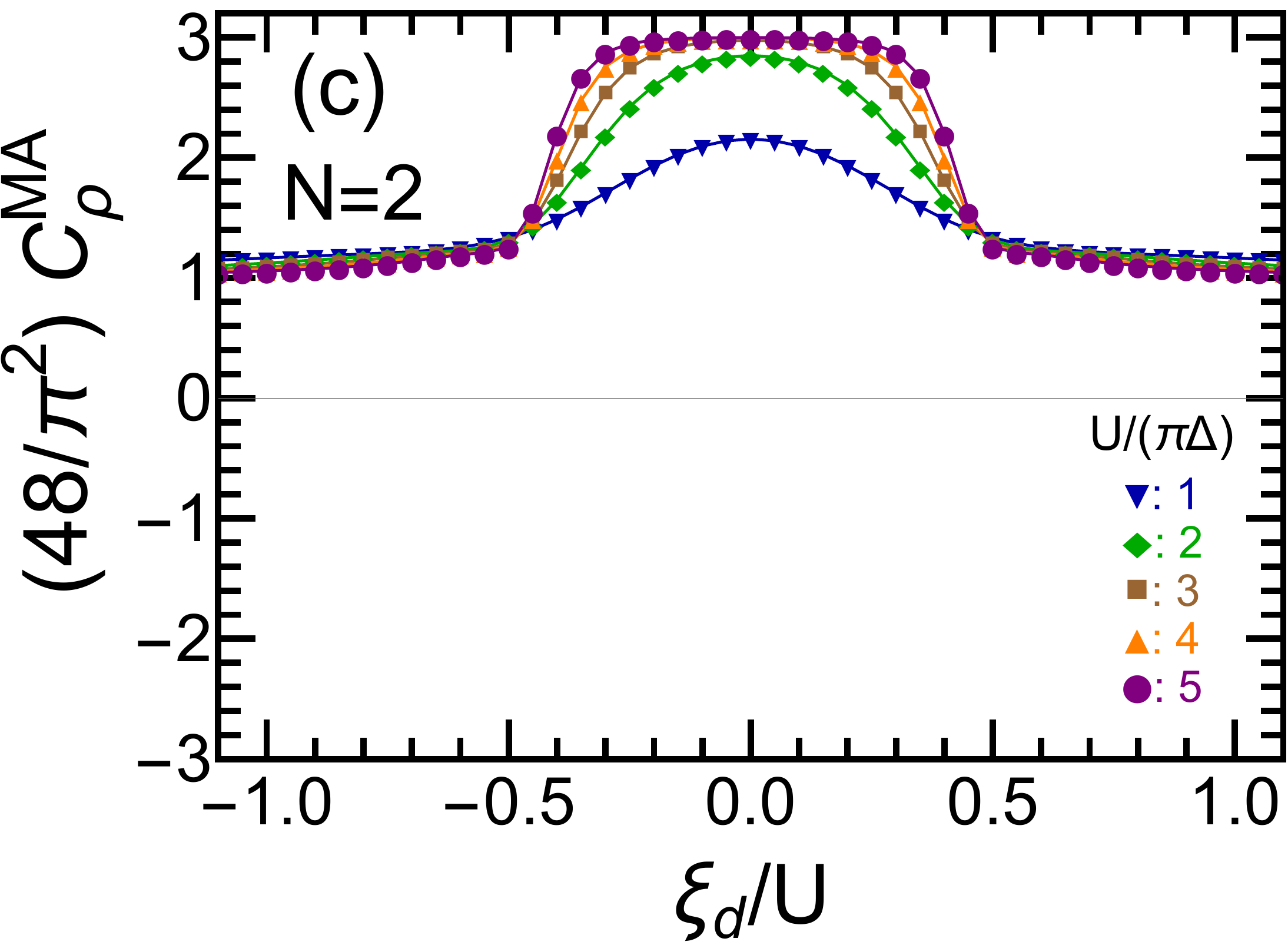}
 \hspace{0.01\linewidth} 
\includegraphics[width=0.47\linewidth]{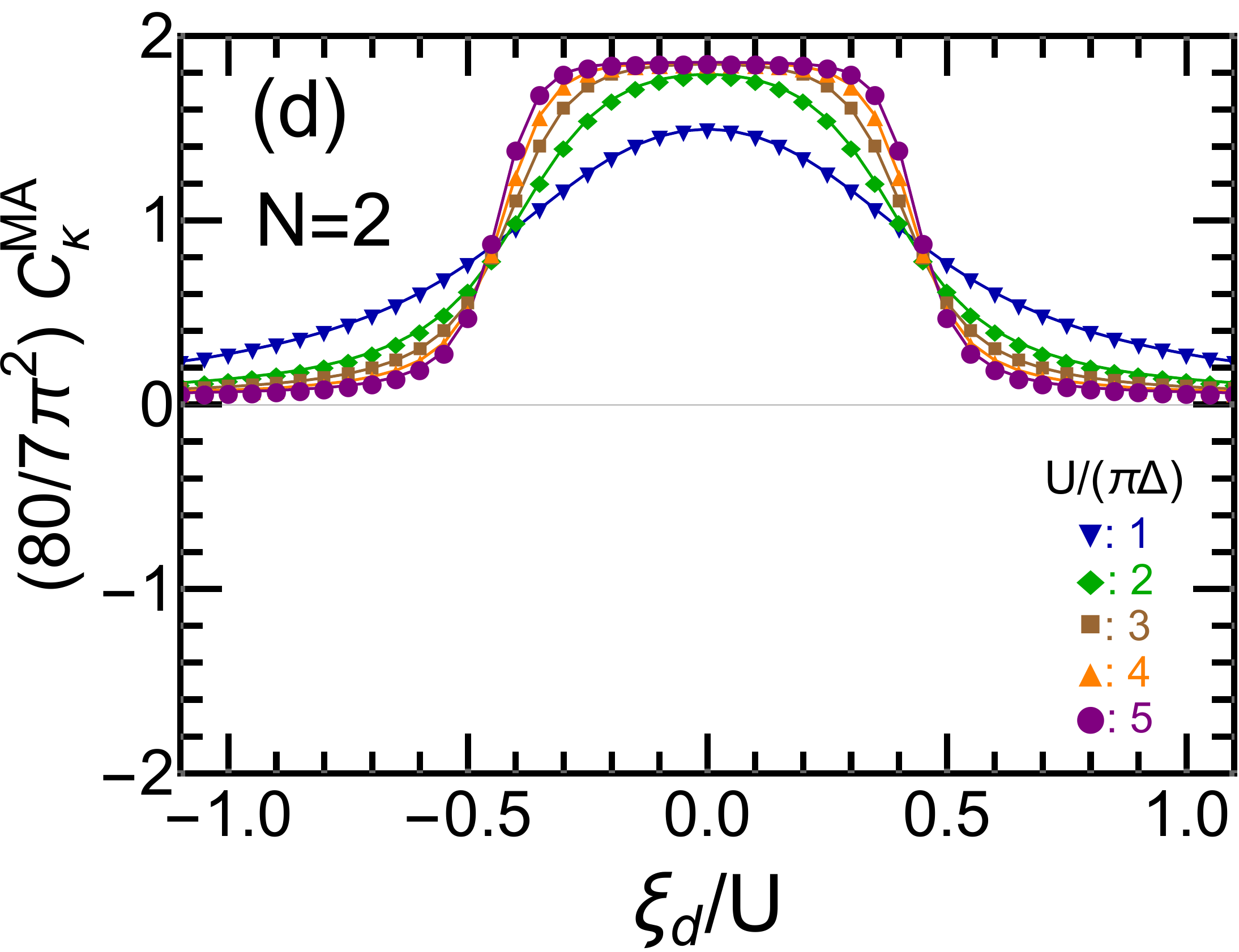}
\caption{Next-leading order terms 
of transport coefficients for an SU(2) symmetric Anderson impurity: 
(a) $(48/\pi^2) C_T^{}$, 
(b) $[80/(7\pi^2)] C_{\kappa}^{\mathrm{QD}}$, 
(c) $(48/\pi^2)C_{\varrho}^{\mathrm{MA}}$, and 
(b) $[80/(7\pi^2)] C_{\kappa}^{\mathrm{MA}}$, 
defined in Table \ref{tab:C_and_W_extended}. 
Interaction strengths are chosen to be 
 $U/(\pi\Delta) =  1(\blacktriangledown), 2(\blacklozenge), 3(\blacksquare),
 4(\blacktriangle), 5(\bullet)$.
} 
\label{fig:T2terms_N2}
\end{figure}

 We have discussed the behavior of the current noise 
and thermoelectric transport of the SU(4) and SU(6) Anderson model 
 in Secs.\ \ref{sec:NRG_CvCs},  \ref{sec:NRG_CtCkappa}, and \ref{sec:NRG_MA}.
 For comparison,  here we briefly describe the corresponding results in the SU(2) case, 
specifically, the current noise and the thermoelectric transport coefficients 
are plotted in Figs.\ \ref{fig:Noise_N2} and  \ref{fig:T2terms_N2}, respectively.

We can see in Fig.\ \ref{fig:Noise_N2}(b) that 
the linear noise is suppressed 
in the strong-coupling region $|\xi_d^{}| \lesssim U/2$ for large $U$,
where the transmission probability exhibits a wide 
Kondo plateau in Fig.\ \ref{fig:Noise_N2}(a). 
However, the linear noise increases and has the peaks, 
the width of which is of the order of $\Delta$ 
at valence fluctuation regions 
near the quarter and three-quarters filling points   
where the phase reaches $\delta = \pi/4$ or $3\pi/4$.   
At these filling points,  
 the coefficient $C_{S}^{}$ for the order $|eV|^3$ nonlinear current noise 
has the negative minima, as seen in Fig.\ \ref{fig:Noise_N2}(c).   
We see in Fig.\ \ref{fig:Noise_N2}(d) that  
the three-body part $\Theta_{S}^{}$ of  $C_{S}^{}$,  
 is also suppressed over a wide region $|\xi_d^{}| \lesssim U/2$.  
Therefore,  in this strong-coupling region,  the nonlinear noise 
$C_{S}^{}$ is determined solely by the two-body part $W_{S}^{}$ 
in the SU(2) case: 
\begin{align}
W_S^{}  \,=  \, 4 (R-1)^2 +  \Bigl[\, 1+5 (R-1)^2 \Bigr]\cos 4 \delta\,,
\end{align}
where $R$ is the Wilson ratio. 
In particular, the dip structures of  $C_{S}^{}$
at the quarter and three-quarters fillings reflect 
the mimina of $\cos 4 \delta$ in $W_{S}^{}$. 
In contrast,  at $|\xi_d^{}| \gtrsim U/2$, 
the three-body part $\Theta_{S}^{}$  becomes comparable to $W_{S}^{}$,
and contributes to $C_{S}^{}$.

Figure \ref{fig:T2terms_N2} 
shows the results for the next-leading order terms 
of the thermoelectric transport coefficients 
for the SU(2) quantum dots  (a) $C_T^{}$ and 
 (b) $C_{\kappa}^{\mathrm{QD}}$, 
and for SU(2) magnetic alloys 
 (c) $C_{\varrho}^{\mathrm{MA}}$  
and (d) $C_{\kappa}^{\mathrm{MA}}$.  
All these coefficients exhibit the plateau structures 
due to the Kondo effect 
in the strong-coupling region $|\xi_d^{}| \lesssim U/2$ for large $U$, 
where the phase shit is locked at  $\delta \simeq \pi/2$.   
In this region,  the three-body contributions, 
 $\Theta_\mathrm{I}^{}$ and  $\widetilde{\Theta}_\mathrm{II}^{}$, almost vanish 
(see Ref.\ \onlinecite{TsutsumiTerataniSakanoAO2021prb} for more details), 
and the height of these plateaus approach the saturation values for $U \to \infty$, i.e.,  
 $(48/\pi^2) C_T^{} \to 3$, 
 $(48/\pi^2) C_{\varrho}^{\mathrm{MA}}\to 3$, 
 $[80/(7\pi^2)] C_{\kappa}^{\mathrm{QD}}\to 13/7$, 
and 
 $[80/(7\pi^2)] C_{\kappa}^{\mathrm{MA}}\to 13/7$. 
The coefficients  $C_T^{}$ and $C_{\varrho}^{\mathrm{MA}}$ 
for charge transport show a similar behavior in the strong-coupling region. 
Furthermore, in this region, 
the coefficients for thermal conductivities,  
   $C_{\kappa}^{\mathrm{QD}}$ and  $C_{\kappa}^{\mathrm{MA}}$,  
also show a similar behavior. 
In contrast, at  $|\xi_d^{}| \gtrsim U/2$, 
the coefficients $C_{T}^{}$ and $C_{\kappa}^{\mathrm{QD}}$  for quantum dots 
change  sign and become negative 
as the occupation number approaches $N_d^{} \simeq 0.0$ or $2.0$,  
whereas the coefficients 
 $C_{\varrho}^{\mathrm{MA}}$ and  $C_{\kappa}^{\mathrm{MA}}$ 
 for magnetic alloys remain positive definite.

\section{Three-body Fermi liquid corrections to thermoelectric transport 
of magnetic alloys}

\label{sec:thermoelectric_transport_general_MA}

In this appendix, we describe   
the low-energy asymptotic form 
of  the electrical resistivity $\varrho_{\mathrm{MA}}^{}
=1/\sigma_{\mathrm{MA}}^{}$, 
 the thermopower $\mathcal{S}_\mathrm{MA}^{}$, 
and the 
thermal conductivity  $\kappa_{\mathrm{MA}}^{}$ of magnetic alloys: 
\begin{align}
\sigma_{\mathrm{MA}}^{}
\ = & \   
\sigma_\mathrm{MA}^{\mathrm{unit}}\,\frac{1}{N}\,
\sum_{\sigma} \mathcal{L}_{0,\sigma}^\mathrm{MA} \,, 
\label{eq:conductivity_MA_L_0_sigma} \\ 
%
\mathcal{S}_\mathrm{MA}^{}\ =&\ \frac{-1}{|e|T} 
\frac{\sum_{\sigma} \mathcal{L}_{1,\sigma}^\mathrm{MA}}
{\sum_{\sigma} \mathcal{L}_{0,\sigma}^\mathrm{MA}} \,,
\label{eq:thermo_power_MA_L_1_sigma}
\\ 
\kappa_\mathrm{MA}^{} \ = & \   
\frac{\sigma_\mathrm{MA}^{\mathrm{unit}}}{e^{2}T}\,\frac{1}{N}\,
\left[\, 
\sum_{\sigma}
\mathcal{L}_{2,\sigma}^\mathrm{MA} 
- 
\frac{ \left(
\sum_{\sigma}
\mathcal{L}_{1,\sigma}^\mathrm{MA}\right)^2}{\sum_{\sigma}
\mathcal{L}_{0,\sigma}^\mathrm{MA}} \,\right] .
\label{eq:kappa_MA_L0_L1_L2}
\end{align}
For magnetic alloys,  the response functions 
 $\mathcal{L}_{n,\sigma}^\mathrm{MA}$ for $n=0,1$ and $2$ are given by 
\begin{align}
\mathcal{L}_{n,\sigma}^\mathrm{MA} = 
\int_{-\infty}^{\infty}  
d\omega\, 
\frac{\omega^n}{\pi \Delta A_{\sigma}(\omega)}
\left( -\frac{\partial f(\omega)}{\partial \omega}\right)  .  
\label{eq:L_thermal_appendix}
\end{align}
 Equation \eqref{eq:conductivity_MA_L_0_sigma} 
defines the electrical conductivity relative to its unitary-limit value 
  $\sigma_\mathrm{MA}^{\mathrm{unit}}$. 
Correspondingly, 
 the prefactor for  $\kappa_\mathrm{MA}^{}$ is defined 
in such a way that the  $T$-linear thermal conductivity 
should take the form, 
\begin{align}
\lim_{T \to 0} \frac{\kappa_\mathrm{MA}^{}}{T}
 = \frac{\pi^{2}}{3\,e^{2}}\,\sigma_\mathrm{MA}^{\mathrm{unit}} \,.
\label{eq:kappa_0_unitary_limit}
\end{align}

The asymptotic form of 
 $\mathcal{L}_{n,\sigma}^\mathrm{MA}$ 
can be calculated,  using the low-energy expansion of  
the inverse spectral function $1/ A_{\sigma}(\omega)$ 
given in Eq.\ \eqref{eq:inverse_A_result}:
\begin{align}
\mathcal{L}_{0,\sigma}^\mathrm{MA}
=   & \  
\frac{1}{\sin^2\delta_{\sigma}}
\left[\,
1+
\frac{a_{0,\sigma}^{\mathrm{MA}}}{\sin^2\delta_{\sigma}} 
\left(\pi T\right)^2 \,\right]  \ + \ \cdots \,, 
\label{eq:L_0_sigma_MA} \\
\mathcal{L}_{1,\sigma}^\mathrm{MA} 
\,=& \  
- \frac{1}{3} 
\frac{\rho_{d\sigma}'}{\pi\Delta \rho_{d\sigma}^{2}} 
\,(\pi T)^2  \ +\  \cdots \,, 
\label{eq:L_1_sigma_MA} \\
\mathcal{L}_{2,\sigma}^\mathrm{MA} 
=  & \   
\frac{\left(\pi T\right)^2}{3\sin^2\delta_{\sigma}}
\left[\,
1 
+ 
\frac{a_{2,\sigma}^{\mathrm{MA}}}{\sin^2\delta_{\sigma}}
\left(\pi T\right)^2 
\,\right]  \ +\  \cdots \,.
\label{eq:L_2_sigma_MA}
\end{align}
Here, the coefficients $a_{0,\sigma}^{\mathrm{MA}}$ 
and $a_{2,\sigma}^{\mathrm{MA}}$ are  given by 
\begin{align}
\!\!\! 
a_{0,\sigma}^{\mathrm{MA}}
=& \,
\frac{\pi^2}{3}
\Biggl[
\left(\cos 2\delta_{\sigma} +2 \right)
\chi_{\sigma\sigma}^2
- 
2\cos 2 \delta_{\sigma}\,
\sum_{\sigma' (\neq \sigma)}
\chi_{\sigma\sigma'}^2
\nonumber \\
& \ \ 
+ 
\frac{\sin 2\delta_{\sigma}}{2\pi}\,
\left(
\chi_{\sigma\sigma\sigma}^{[3]}
+ 
\sum_{\sigma' (\neq \sigma)} 
\chi_{\sigma\sigma'\sigma'}^{[3]}
\right) \Biggr]
, \label{eq:c_0_sigma_MA_def} 
\\
\nonumber \\
\!\!\!
a_{2,\sigma}^{\mathrm{MA}} = & \, 
\frac{7\pi^2}{5}
\Biggl[
\left(\cos 2\delta_{\sigma} +2\right) 
\chi_{\sigma\sigma}^2
- \frac{6}{7}\cos 2\delta_{\sigma} 
\sum_{\sigma'(\neq \sigma)} \chi_{\sigma\sigma'}^2
\nonumber \\
& \ \    +  
\frac{\sin 2\delta_{\sigma}}{2\pi}
\left(
\chi_{\sigma\sigma\sigma}^{[3]}
+
\frac{5}{21}
\sum_{\sigma'(\neq \sigma)}
\chi_{\sigma\sigma'\sigma'}^{[3]}
\right) \Biggr]. 
\label{eq:c_2_sigma_MA_def}
\end{align}

We obtain the low-temperature expressions of 
$\varrho_\mathrm{MA}^{} = 1/\sigma_\mathrm{MA}^{}$, 
$\mathcal{S}_\mathrm{MA}$ and  $1/\kappa_\mathrm{MA}^{}$,  
substituting Eqs.\ \eqref{eq:L_0_sigma_MA}--\eqref{eq:L_2_sigma_MA} 
into Eqs.\ \eqref{eq:conductivity_MA_L_0_sigma}--\eqref{eq:kappa_MA_L0_L1_L2}:  
\begin{align}
\varrho_\mathrm{MA}^{} =&\,
\frac{1}{\sigma_{\mathrm{MA}}^{\mathrm{unit}}}
\, 
\left[\,
\bigl( \overline{\sin^2 \delta} \bigr)_\mathrm{HM}^{}
-c_{\varrho}^\mathrm{MA}\,\left(\pi T\right)^{2} + \cdots \right], 
\label{eq:R_MA_T_squared_general} \\
\mathcal{S}_\mathrm{MA}^{} =& \,  
\frac{\pi^2}{3} \, 
\frac{\sum_{\sigma} \rho_{d\sigma}'}{\sum_{\sigma}\rho_{d\sigma}^{}}
\,\frac{T}{|e|} \, + \, \cdots, \label{eq:S_MA_T_linear_general} \\
\frac{1}{\kappa_\mathrm{MA}}\,= & \,\frac{3\,e^{2}}{\pi^{2}\,
\sigma_\mathrm{MA}^{\mathrm{unit}}}\,\frac{1}{T}\,\Bigl[\,
\bigl( \overline{\sin^2 \delta} \bigr)_\mathrm{HM}^{} 
\,-\,c_{\kappa}^\mathrm{MA}\,\left(\pi T\right)^{2}
\, + \,  \cdots\,\Bigr].
\label{eq:thermal_resistivity_general}
\end{align}
Here, $\bigl( \overline{\sin^2 \delta} \bigr)_\mathrm{HM}^{}$
is  the harmonic mean (HM) of $\sin^2 \delta_\sigma^{}$, defined by
\begin{align}
\bigl( \overline{\sin^2 \delta} \bigr)_\mathrm{HM}^{}
\,\equiv \, 
\frac{1}{\frac{1}{N}\,\sum_{\sigma}\,\frac{1}{\sin^{2}\delta_{\sigma}}} \,. 
\label{eq:HM_sin2delta}
\end{align}
The coefficients  $c_{\varrho}^\mathrm{MA}$ and $c_{\kappa}^\mathrm{MA}$ 
of the next-leading order terms are given by 
\begin{align}
c_{\varrho}^\mathrm{MA}\,=&\,
\left\{\bigl( \overline{\sin^2 \delta} \bigr)_\mathrm{HM}^{}\right\}^2
\frac{1}{N}\,\sum_{\sigma}\,
\frac{a_{0,\sigma}^{\mathrm{MA}}}{\sin^{4}\delta_{\sigma}} \,
, \label{eq:C_R_MA_general} 
\\
c_{\kappa}^\mathrm{MA}\,=&\,
\left\{\bigl( \overline{\sin^2 \delta} \bigr)_\mathrm{HM}^{}\right\}^2
\left[\frac{1}{N}\,\sum_{\sigma}\,
\frac{a_{2,\sigma}^{\mathrm{MA}}}{\sin^{4}\delta_{\sigma}}\, \right. 
\nonumber \\
&\left. \,-\,\frac{\pi^{2}}{3}\,
\bigl( \overline{\sin^2 \delta} \bigr)_\mathrm{HM}^{}
\left\{\frac{1}{N}\,\sum_{\sigma}\,
\frac{\sin 2 \delta_\sigma}{\sin^4 \delta_\sigma}\,\chi_{\sigma\sigma}^{}
\right\}^{2}
\,\right]. 
\label{eq:C_kappa_MA_general}
\end{align}

Correspondingly, the electrical conductivity 
and the thermal conductivity take the following form, 
\begin{align}
\sigma_{\mathrm{MA}}\,=&\, 
\frac{\sigma_\mathrm{MA}^{\mathrm{unit}}}
{\bigl( \overline{\sin^2 \delta} \bigr)_\mathrm{HM}^{}}\,
\left[\,1\,+
\frac{c_{\varrho}^{\mathrm{MA}}}
{\bigl( \overline{\sin^2 \delta} \bigr)_\mathrm{HM}^{}} \,
\left(\pi T\right)^{2}\,+\,\cdots \, \right],
\label{eq:sigma_MA_T_squared_general} \\
\mathcal{\kappa}_{\mathrm{MA}}^{} =&\,
\frac{\pi^{2}\,\sigma_\mathrm{MA}^{\mathrm{unit}}}{3\,e^{2}}
\, \frac{T}{\bigl( \overline{\sin^2 \delta} \bigr)_\mathrm{HM}^{}} 
\nonumber \\
& \qquad  
\times 
\left[\,1\, 
+\,\frac{c_{\kappa}^{\mathrm{MA}}}
{\bigl(\overline{\sin^2 \delta}\bigr)_\mathrm{HM}^{}}\,
\left(\pi T\right)^{2}
\,+\,\cdots \, \right].
\label{eq:kappa_MA_T_squared_general} 
\end{align}
Furthermore, the Lorenz number 
$L_{\mathrm{MA}}\equiv \kappa_{\mathrm{MA}}/(\sigma_{\mathrm{MA}} T)$  
can also be deduced up to terms of order  $T^{2}$, as 
\begin{align}
L_{\mathrm{MA}}\,
\,=&\ \frac{\pi^{2}}{3\,e^{2}}\,\left[1\,-\,
\frac{c_{L}^{\mathrm{MA}}}
{\bigl( \overline{\sin^2 \delta} \bigr)_\mathrm{HM}^{}}
\,\left(\pi T\right)^{2} \ + \  \cdots \,\right] , 
\label{eq:L_MA_general_formula}
\\
c_{L}^{\mathrm{MA}}\,=& \ 
c_{\varrho}^{\mathrm{MA}}\,-\,c_{\kappa}^{\mathrm{MA}} \,.
\label{eq:C_L_MA_general_def}
\end{align}
In the limit of zero temperature, 
the Lorenz number takes a constant value 
$L_{\mathrm{MA}} \xrightarrow{\,T \to 0\,} \frac{\pi^{2}}{3\,e^{2}}$, 
and  the Wiedemann-Franz law holds.
However, it deviates  as temperature increases,  
showing the $T^2$ dependence.

\section{Thermoelectric transport coefficients for noninteracting magnetic alloys}
\label{sec:thermo_MA_U0}

We provide here the noninteracting results for 
the thermoelectric transport coefficients of magnetic alloys.
For $U=0$,  the response functions   
$\mathcal{L}_{n,\sigma}^{\mathrm{MA}}$ for $n=0,1,2$ 
can be calculated,  substituting the explicit form of the spectral function   
$\pi A_{\sigma}^{}(\omega ) \xrightarrow{\,U=0\,} 
\Delta /[(\omega_{}-\epsilon_{d\sigma})^2+\Delta^{2}]$  
into Eq.\ \eqref{eq:L_thermal_appendix}:  
\begin{align}
\mathcal{L}_{0,\sigma}^{\mathrm{MA}}
\,\xrightarrow{\,U=0\,}& \ 
\left[1+\left(\frac{\epsilon_{d\sigma}^{}}{\Delta}\right)^2 \right]
\, +\, \frac{1}{3} \left(\frac{\pi T}{\Delta} \right)^2 , 
\label{eq:L0_MA_U0} \\
\mathcal{L}_{1,\sigma}^{\mathrm{MA}}
\,\xrightarrow{\,U=0\,}& \ 
\frac{-2\, \epsilon_{d\sigma}^{}}{3} 
\left(\frac{\pi T}{\Delta} \right)^{2} , 
\label{eq:L1_MA_U0} \\ 
\mathcal{L}_{2,\sigma}^{\mathrm{MA}}
\,\xrightarrow{\,U=0\,}& \ 
\frac{\Delta^2}{3}  \left[1+\left(\frac{\epsilon_{d\sigma}^{}}{\Delta}\right)^2 
\right]
\left(\frac{\pi T}{\Delta} \right)^{2}
+\frac{7\Delta^2}{15}
\left(\frac{\pi T}{\Delta} \right)^{4} . 
\label{eq:L2_MA_U0} 
\end{align}

The transport coefficients, defined 
in Eqs.\ \eqref{eq:conductivity_MA_L_0_sigma}--\eqref{eq:kappa_MA_L0_L1_L2},  
can be deduced 
from these analytic expressions for $\mathcal{L}_{n,\sigma}^{\mathrm{MA}}$. 
In particular, in the SU($N$) symmetric case 
where $\epsilon_{d\sigma}^{} \equiv \epsilon_{d}^{}$,  
the electrical resistivity  $\varrho_{\mathrm{MA}}^{(0)}$ for $U=0$ is given by 
\begin{align}
\varrho_{\mathrm{MA}}^{(0)}
\,=& \  
 \frac{1}{\sigma_\mathrm{MA}^\mathrm{unit}}\,
\left[\,
\sin^2 \delta_0^{}  
\,-\,C_{\varrho}^{\mathrm{MA}(0)}\, \left(\frac{\pi T}{T^*_0} \right)^{2}
\,\right], 
\label{eq:RMA_U0}
\\ 
\sin^2 \delta_0^{} \,= &  \  
\frac{1}{1+(\epsilon_{d}^{}/\Delta)^2 } \,, 
\qquad 
C_{\varrho}^{\mathrm{MA}(0)}\,=\,   \frac{\pi^{2}}{48}\, .
\label{eq:C_rho_U0} 
\end{align}
Here,  $T_{0}^{*} \equiv \pi \Delta/(4 \sin^2 \delta_0^{})$ 
is the characteristic energy scale in the noninteracting case.
The coefficient $C_{\varrho}^{\mathrm{MA}(0)}$ for
 the $T^2$-resistivity is given by a constant 
$\pi^2/48$ ($= 0.205\cdots$) 
as the $\epsilon_{d}^{}$ dependence is absorbed into $T_{0}^{*}$.

The thermopower $\mathcal{S}_{\mathrm{MA}}^{0}$ 
and  thermal resistivity $1/\kappa_{\mathrm{MA}}^{(0)}$ for $U=0$  are given by 
\begin{align}
&\mathcal{S}_\mathrm{MA}^{(0)} 
\,= \,  
\frac{1}{|e|} \, 
\frac{2\pi^2}{3} \, 
\frac{\epsilon_{d}^{}}{\epsilon_{d}^{2}+\Delta^2} \,T  
\;,
\label{eq:S_MA_U0_Delta_scale}
\\
& 
\frac{1}{\kappa_{\mathrm{MA}}^{(0)}}\,=\, 
\frac{3\,e^{2}}{\pi^{2}\,\sigma_\mathrm{MA}^{\mathrm{unit}}}\,\frac{1}{T}\,
\left[\,
\sin^2 \delta_0^{}  
\,-\,C_{\kappa}^{\mathrm{MA}(0)}
\,\left(\frac{\pi T}{T_{0}^{*}}\right)^2\,\right],
\label{eq:kappa_MA_U0_T_star_scale}
\\
&C_{\kappa}^{\mathrm{MA}(0)}\,=\,
\frac{7\pi^2}{80}\,\frac{1+20\sin^2 \delta_{0}^{}}{21}  
\,.
\label{eq:C_kappa_MA_U0}
\end{align}
Furthermore, the dimensionless coefficient for the order $T^2$ term 
of the Lorenz number $L_{\mathrm{MA}}^{}$  is given by   
 $C_{L}^{\mathrm{MA}(0)}
 \equiv C_{\varrho}^{\mathrm{MA}(0)}-C_{\kappa}^{\mathrm{MA}(0)}$ 
and it takes the form,
\begin{align}
C_{L}^{\mathrm{MA}(0)}
\,=\, \frac{\pi^{2}}{60}\,\left(\, 1 -5 \sin^2 \delta_{0}^{}
\,\right)\,.
\label{eq:CL_MA_U0}
\end{align}

\begin{figure}[h]
\leavevmode
 \centering

\includegraphics[width=0.7\linewidth]{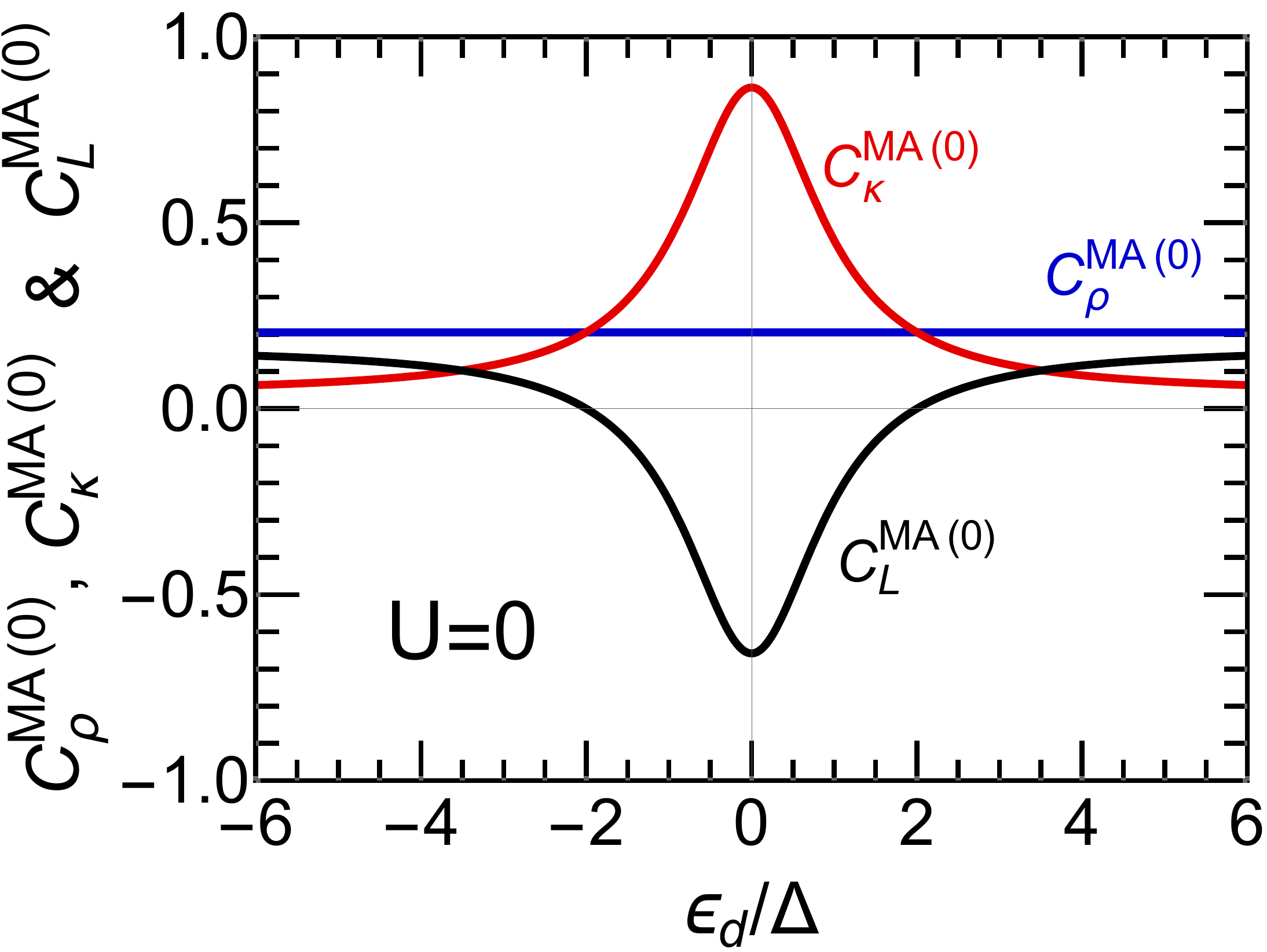}

\caption{Coefficients $C_{\varrho}^{\mathrm{MA}(0)}$,   
$C_{\kappa}^{\mathrm{MA}(0)}$, and 
$C_{L}^{\mathrm{MA}(0)}$ 
 for noninteracting $U=0$ magnetic alloys  plotted vs  $\epsilon_{d}^{}$.
}
\label{fig:Crho-Ckappa_MA_free}
\end{figure}

These results for the next-leading order terms in the noninterating case 
are plotted as functions of $\epsilon_{d}^{}$ in 
Fig.\  \ref{fig:Crho-Ckappa_MA_free}.
The coefficient  $C_{\kappa}^{\mathrm{MA}(0)}$
takes a Lorentzian form with an offset value  of   
$C_{\kappa}^{\mathrm{MA}(0)} 
 \xrightarrow{\,|\epsilon_d^{}| \to \infty\,}  \pi^2/240$ ($=0.041\cdots$). 
Correspondingly,  $C_{L}^{\mathrm{MA}(0)}$  
has a dip at  $\epsilon_{d}^{}=0$, 
and vanishes at the points $\epsilon_{d}^{} = \pm 2\Delta$ 
where  $C_{\kappa}^{\mathrm{MA}(0)}$ and 
 $C_{\varrho}^\mathrm{MA(0)}$ give equal contributions 
to the Lonrez number $L_{\mathrm{MA}}^{}$.
As the impurity level  moves further 
away from the Fermi level  $|\epsilon_{d}^{}| \gg \Delta$,
it approaches the value of $C_{L}^{\mathrm{MA}(0)} 
 \xrightarrow{\,|\epsilon_d^{}| \to \infty\,}  \pi^2/60$ ($=0.164\cdots$).

%

\end{document}